\documentclass[11pt,a4paper]{article}
\usepackage{epsfig}
\usepackage[T1]{fontenc}    
\usepackage{graphics}
\usepackage{graphicx}
\usepackage{pstricks,pst-coil,pst-fill,pst-plot}
\usepackage[fleqn]{amsmath}    
\usepackage{amssymb}    
\usepackage{amsfonts}   
\usepackage{verbatim}   
\usepackage{mathrsfs}   
\usepackage{dsfont}
\usepackage{euscript}
\usepackage{yfonts}
\usepackage{enumerate}     
\usepackage{txfonts}
\usepackage{marvosym}
\usepackage{vmargin}        

\setmarginsrb{1.8cm}{2cm}{1.8cm}{2cm}{1cm}{1cm}{1cm}{1.6cm}
 \makeatletter
 \@addtoreset{equation}{section}
 \makeatother

\providecommand{\bysame}{\leavevmode\hbox to3em{\hrulefill}\thinspace}
\providecommand{\MR}{\relax\ifhmode\unskip\space\fi MR }

\providecommand{\href}[2]{#2}

\newcommand{\bra}[1]{\left\langle\,#1\,\right|}
\newcommand{\ket}[1]{\left|\,#1\, \right\rangle}

\let\ua=\uparrow
\let\da=\downarrow
\let\tend=\rightarrow

\long\def\symbolfootnote[#1]#2{\begingroup%
\def\thefootnote{\fnsymbol{footnote}}\footnote[#1]{#2}\endgroup}

\newtheorem{theorem}{Theorem}[section]
\newtheorem{prop}{Proposition}[section]
\newtheorem{cor}{Corollary}[section]
\newtheorem{defin}{Definition}[section]

\newtheorem{conj}{Conjecture}[section]

\newtheorem{lemme}{Lemma}[section]

\def\Proof{\medskip\noindent {\it Proof --- \ }}

\def\qed{\hfill\rule{2mm}{2mm}}

\newcommand\beq{\begin{equation}}
\newcommand\enq{\end{equation}}
\newcommand\bem{\begin{multline}}
\newcommand\enm{\end{multline}}

\def\beqa{\begin{eqnarray}}
\def\eeqa{\end{eqnarray}}
\def\ba{\begin{array}}
\def\ea{\end{array}}

\def\det{\operatorname{det}}

\newcommand{\f}[2]{{\ensuremath{%
    \mathchoice%
    {\dfrac{#1}{#2}}
    {\dfrac{#1}{#2}}
    {\frac{#1}{#2}}
    {\frac{#1}{#2}}
}}}
\newcommand{\tf}[2]{\ensuremath{#1/#2}}
\newcommand{\pa}[1]{\ensuremath{\left(#1\right)}}
\newcommand{\paa}[1]{\ensuremath{\left\{#1\right\}}}
\newcommand{\pac}[1]{\ensuremath{\left[#1\right]}}
\newcommand{\paf}[2]{\ensuremath{\left(\f{#1}{#2}\right)}}
\newcommand{\pab}[2]{\ensuremath{\pa{\ba{c} #1 \\ #2 \ea }}}
\newcommand{\pabb}[3]{\ensuremath{\pa{ #1 \left| \ba{c} #2 \\ #3 \ea \right .}}  }

\def\a{\alpha}
\def\al{\aleph}
\def\be{\beta}
\def\ga{\gamma}
\def\Ga{\Gamma}
\def\de{\delta}

\def\De{\Delta}
\def\eps{\epsilon}
\def\veps{\varepsilon}
\def\la{\lambda}

\def\vsg{\varsigma}

\def\Ups{\Upsilon}
\def\th{\theta}

\def\Om{\Omega}
\def\om{\omega}
\def\vp{\varphi}

\newcommand{\mc}[1]{\ensuremath{\mathcal{#1}}}
\newcommand{\mf}[1]{\ensuremath{\mathfrak{#1}}}
\newcommand{\msc}[1]{\ensuremath{\mathscr{#1}}}

\newcommand{\bs}[1]{\ensuremath{\boldsymbol{#1}}}

\newcommand{\ov}[1]{\ensuremath{\overline{#1}}}
\newcommand{\wt}[1]{\ensuremath{\widetilde{#1}}}
\newcommand{\wh}[1]{\ensuremath{\widehat{#1}}}

\newcommand{\Int}[2]{\ensuremath{\int\limits_{#1}^{#2}}}
\newcommand{\Oint}[2]{\ensuremath{\oint\limits_{#1}^{#2}}}

\newcommand{\sul}[2]{\ensuremath{\sum\limits_{#1}^{#2}}}
\newcommand{\pl}[2]{\ensuremath{\prod\limits_{#1}^{#2}}}

\newcommand{\R}{\ensuremath{\mathbb{R}}}
\newcommand{\Cx}{\ensuremath{\mathbb{C}}}

\newcommand{\Dp}[1]{\ensuremath{\partial_{#1}}}

\newcommand{\limit}[2]{\ensuremath{\underset{#1 \tend #2}{\longrightarrow} }}
\newcommand{\J}[1]{\ensuremath{J_{\{#1 \,\}}}  }

\newcommand{\ex}[1]{\ensuremath{\e{e}^{#1}}}

\newcommand{\ddet}[2]{\ensuremath{\det_{#1}\pac{#2}}}

\newcommand{\abs}[1]{\ensuremath{\left| #1 \right|}}

\newcommand{\norm}[1]{\ensuremath{\left\|#1\right\|}}

\newcommand{\moy}[1]{\ensuremath{\langle #1 \rangle}}

\newcommand{\dd}{\mathrm{d}}
\newcommand{\e}[1]{\ensuremath{\mathrm{#1}}}

\newcommand{\intff}[2]{\ensuremath{\left [ \, #1 \,; #2 \, \right ] }}
\newcommand{\intfo}[2]{\ensuremath{\left [ \, #1 \,; #2 \, \right [ }}

\newcommand{\intoo}[2]{\ensuremath{\left ] \, #1 \,; #2 \, \right [ }}

\newcommand{\intn}[2]{\ensuremath{[\![ \, #1 \,;\, #2 \,]\!]}}

\begin{document}

\begin{flushright}
DESY 10-230
\end{flushright}
\par \vskip .1in \noindent

\vspace{14pt}

\begin{center}
\begin{LARGE}
{\bf Large-distance and long-time asymptotic behavior of the reduced density matrix in the non-linear Schr\"{o}dinger model.}
\end{LARGE}

\vspace{30pt}

\begin{large}

{\bf K.~K.~Kozlowski}\footnote[1]{DESY, Hamburg, Deutschland,
 karol.kajetan.kozlowski@desy.de}~~
\par

\end{large}

\vspace{40pt}

\centerline{\bf Abstract} \vspace{1cm}
\parbox{12cm}{\small Starting from the form factor expansion in finite volume, we derive
the multidimensional generalization of the so-called Natte series for the time and distance dependent reduced density 
matrix  at zero-temperature in the non-linear Schr\"{o}dinger model.
This representation allows one to \textit{read-off} straightforwardly the long-time/large-distance asymptotic behavior
of this correlator. This method of analysis reduces the complexity of the computation of the asymptotic behavior of correlation functions 
in the so-called interacting integrable models, to the one appearing in free fermion equivalent models. We compute explicitly the first 
few terms appearing in the asymptotic expansion. Part of these terms stems from excitations lying away from the Fermi boundary,
and hence go beyond what can be obtained by using the CFT/Luttinger liquid  based predictions.}

\end{center}

\vspace{40pt}

\section{Introduction\label{INT}}

One-dimensional quantum models with a gapless spectrum are believed to be critical at zero temperature.
In other words, in these models, the ground state expectation values of products of local operators should decay, for large distances of separation
between the operators, as some power-law in the distance.
It is also believed that, for a generic class of Hamiltonians,  the actual value of the exponents governing this power-law decay, the so-called critical
exponents,  does not depend on the microscopic details of the interactions in the model, but only on its overall symmetries
\cite{FisherIntroductionUniversalityHypothesis,GriffithsUniversalityAndExponentsParameterDepending}.
Therefore two models belonging to the same universality class should be characterized by the same critical exponents.

It has been argued that the equal-time correlation functions in quantum critical one-dimensional models exhibit conformal invariance in the
large-distance regime \cite{PolyakovArgumentAboutCFTInvarianceCriticalCorrelators}.
Hence, its appears plausible to infer their large-distance asymptotics from those of the associated conformal field theory (CFT).
The central charge of the CFT lying in the universality class of the model can be deduced from the finite-size corrections to the ground state energy
\cite{BloteCardyNightingalePredictionL-1correctionsEnergyAscentralcharge,CardyConformalExponents}.
The possibility to compute  such finite-size corrections for many integrable models
allowed the identification of the central charge and scaling dimensions leading to
the predictions for the critical exponents
\cite{BogoluibovIzerginKorepinCriticalExponentsforXXZ,DestriDeVegaAsymptoticAnalysisCountingFunctionAndFiniteSizeCorrectionsinTBAFirstpaper,
DestriDeVegaAsymptoticAnalysisCountingFunctionAndFiniteSizeCorrectionsinTBAFiniteMagField,
KlumperBatchelorPearceCentralChargesfor6And19VertexModelsNonLinIntEQns,KlumperWehnerZittartzConformalSpectrumofXXZCritExp8Vertex,
WojnaorwiczFiniteSizeCorrections} of the long-distance asymptotics. We remind that it is also sometimes possible
to give predictions for the critical exponents
\cite{HaldaneLuttingerLiquidCaracterofBASolvableModels,HaldaneCritExponentsAsymptBoseGas,LutherPeschelCriticalExponentsXXZZeroFieldLuttLiquid}
by putting the model in correspondence with a  Luttinger liquid  \cite{HaldaneStudyofLuttingerLiquid}.

Due to their wide applicability and relative simplicity, it is more than desirable to test these CFT/Luttinger
liquid based predictions versus some exact  calculations
carried out on such models; this starting from first principle and in such a way that no approximation
(apart from assuming a large distance) is made to the very end.
Such exact computations have been carried out in the 70's and 80's on various two-point functions appearing in
free fermion equivalent models such as the Ising
\cite{ChengWuMagnetization2DIsingAsymptoticsAtBelowAndAboveTc,McCoyWuMagnetization2DBoundaryIsingWithMagAtBound,McCoyWu2DIsingModel}, the XY model at the
critical magnetic field \cite{McCoyPArkShrockSpinTimeSpaceAutoCorrAsModSineKernel,McCoyPArkShrockSpinTimeAutoCorrAsModSineKernel} or the
impenetrable Bose gas \cite{TracyVaidyaRedDensityMatrixSpaceAsymptImpBosonsT=0}.
The latter approaches were then made much more systematic (and also simplified) with the occurrence of a Riemann--Hilbert based approach to free-fermion
models' asymptotics \cite{ItsIzerginKorepinSlavnovDifferentialeqnsforCorrelationfunctions} together with the development of the non-linear
steepest-descent method \cite{DeiftZhouSteepestDescentForOscillatoryRHPmKdVIntroMethod}. Indeed, the latter constitutes a
relatively simple and systematic tool for carrying out the asymptotic analysis
\cite{ItsIzerginKorepinTemperatureLongDistAsympBoseGas,ItsIzerginKorepinSlavnovTempCorrFctSpinsXY,ItsIzerginKorepinVarguzinTimeSpaceAsymptImpBoseGaz} of
Riemann--Hilbert problems associated with Fredholm determinants representing the correlators in free fermionic models.

However, obtaining long-distance asymptotic expansions of two-point functions for models not equivalent to
free fermions faced several additional problems of technical nature.
This fact takes its roots in that even obtaining explicit expressions for the correlation functions in the so-called interacting integrable models
demands to overcome new types of combinatorial intricacy that disappears when dealing with free fermion equivalent models.
The first approach to the problem of computing correlators out of the free fermion point can be attributed to Izergin and Korepin
\cite{IzerginKorepinQISMApproachToCorrFns2SiteModel,IzerginKorepinQISMApproachToCorrNextDiscussionDeterminantCurrentCurrentCorrFnction}.
These authors managed to construct certain series representations for the correlation functions of the non-linear Schr\"{o}dinger model
and the XXZ spin-$\tf{1}{2}$ chain.  However, the $n^{\e{th}}$ summand appearing in these series was only defined implicitly by induction.
Low-$n$ calculation allowed them for an effective perturbative characterization of a vicinity of the free fermion point.
First manageable expressions for correlators at zero temperature in an interacting integrable model were obtained by Jimbo, Miki, Miwa, Nakayashiki
through the vertex operator approach. They have provided multiple integral representations for the matrix elements of the so-called
elementary blocks\symbolfootnote[2]{these constitute a basis on which it is possible to decompose all correlation functions of the model} in the massive \cite{JimboMikiMiwaNakayashikiElementaryBlocksXXZperiodicDelta>1} regime of the infinite XXZ chain.
These results where later extended to the
massless regime of this chain \cite{JimboMiwaElementaryBlocksXXZperiodicMassless} or to a half-infinite chain subject to a longitudinal magnetic field
acting on one of its ends \cite{JimboKedemKonnoMiwaXXZChainWithaBoundaryElemBlcks}.
The multiple integral representations were then reproduced, in the framework of the algebraic Bethe Ansatz by Kitanine, Maillet and Terras
for the massive and massless regime of the periodic XXZ chain \cite{KMTElementaryBlocksPeriodicXXZ}.
These two series of works opened a way towards a systematic and effective computation of various types of multiple integral and/or combinatorial
representations for the correlation functions in numerous integrable models. In particular, it was possible to derive effective
representations in the case of finite temperature \cite{GohmannKlumperSeelFinieTemperatureCorrelationFunctionsXXZ}, non-equal times
\cite{KMNTDynamicalCorrelationFunctions}, models in finite volume
\cite{DamerauGohmannHasencleverKlumperDensityMatrixXXZFiniteLength}, higher spin chains \cite{DeguchiMatsuiElementBlocksHigherSpinXXZ},...
These results should be seen as of uttermost importance from the conceptual point of view: the multiple integral representations for the
correlators of interacting integrable models naturally provides an interpretation for these objects as a new class of special functions
(of the distance, time, coupling constants, ...).
However, the complexity of the integrands appearing in such multiple integral representations
makes the thorough description (computation at certain specific values of the
distance/coupling or extracting their large-distance/long-time behavior \textit{etc}...)
of these new special functions a quite challenging problem.
Many investigations that followed where oriented towards a better understanding of these special functions.
In particular, it was observed that the multiple integral representations for the  elementary blocks
of the XXZ chain can be reduced to one dimensional integrals by a case-by-case analysis
\cite{BoosKorepinEvaluationofEFPforXXX4sites,KMNTPeriodicEmptinessPiOver3,KMNTDelta1/2SpinSpinExact,SatoShiroishiTakahashiElementaryBlocks6sites}.
This observation led to the proof that it is possible to separate the multiple integrals
representing the elementary blocks of the XXZ chain on the algebraic
level \cite{BoosJimboMiwaSmironovTakeyamaAlgebraicRepresentationofCorrelationFunctions=SepofIntegrals}. In its turn,
this led to the discovery of a Grassmann structure in
the XXZ chain \cite{BoosJimboMiwaSmironovTakeyamaBasisforFermionicOpsXXZ,BoosJimboMiwaSmironovTakeyamaGrassMannStructureinXXZ,
BoosJimboMiwaSmironovTakeyamaFermionicStructureinXXZCreationOps}. Among many other developments
such as the possibility to compute the one-point functions of the sine-Gordon model
\cite{JimboMiwaSmirnovHiddenFermionsAndMappingToSineGordonOnePointFunctionsDescendents,JimboMiwaSmirnovOnePointfunctionssineGordon},
the existence of such a Grassmann structure constitutes a promising direction towards bringing the complexity of the analysis
of the correlation function in the XXZ chain to the one of a free fermion problem.
 A completely other method for reducing the complexity of the multiple integral representations for the correlation functions was the so-called
dual field approach \cite{KorepinFirstIntroDualFields}. It led to representations for the correlators in terms of expectation values in an
infinite dimensional Hilbert space of unbounded operator valued Fredholm determinants
\cite{KojimaKorepinSlavnovNLSEDeterminatFormFactorAndDualFieldTempeAndTime,KorepinSlavnovApplicationDualFieldsFredDets}.
However, apart from convergence issues posed by such an
infinite dimensional framework, the main problem of that method was posed by the non-commutativity of the dual field's
vacuum expectation values and the asymptotic expansion of a dual-field valued Fredholm determinant.
Its and Slavnov \cite{ItsSlavnovNLSTimeAndSpaceCorrDualFields}  carried out, on a formal level, such a dual field-based analysis for the
large-distance/long-time decay of the so-called one-particle reduced density matrix  at finite temperature in the non-linear Schr\"{o}dinger model (NLSM).
They have been able to provide operator valued expressions for the correlation length.
The dual field vacuum expectation values where computed in \cite{SlavnovComputationDualFieldVaccumExpLongTimeDistTempeRedDensNLSE}, which led
to a description of the correlation length in terms of a solution to a non-linear integral equation.
We would like to mention that until recently, although formal, the dual field approach was the only approach
 alternative to a
CFT/Luttinger liquid based correspondence that allowed one to write down certain predictions for the critical exponents.

There have also been developments aiming at obtaining alternative types of effective series of multiple integral representations for the
distance dependent two-point functions \cite{KMNTOriginalSeries,KMNTMasterEquation}.
The guideline being a construction of such a representation that would
allow one to carry out a long-distance  asymptotic analysis of the two-point functions. This project has met a success in
 \cite{KozKitMailSlaTerXXZsgZsgZAsymptotics}. This article developed a first fairly rigorous method
allowing one to compute, starting from "first principles",
the long-distance asymptotic behavior of the spin-spin correlation function in the massless regime of the XXZ spin-$\tf{1}{2}$ chain.
This method relied on a few conjectures relative to the permutation of symbols, summability of the remainders, convergence of the
obtained series representations, but was rigorous otherwise.
These last results not only confirmed the CFT/Luttinger liquid-based predictions for the critical exponents in this model but also
provided explicit expressions for the amplitudes in front of the power-law which, in their turn, cannot be predicted by universality arguments.
These explicit formulae for the amplitudes were then identified with
certain, properly normalized in respect to the size of the system, form factors of the spin operators
\cite{KozKitMailSlaTerThermoLimPartHoleFormFactorsForXXZ,KozKitMailSlaTerEffectiveFormFactorsForXXZ}.
This identification allowed one to point out the universality in the power of the system-size that one uses for normalizing the form
factor associated with the amplitudes. The aforementioned method of asymptotic analysis was applied recently to study the long-distance asymptotic
behavior of the correlation functions at finite temperature in the NLSM \cite{KozMailletSlaLongDistanceTemperatureNLSE}.

The large-distance/long-time asymptotic behavior of the correlation functions in massless one dimensional quantum models
goes beyond the predictions stemming from a correspondence with a CFT/Luttinger liquid. Hence, this constitutes a clear motivation
for obtaining such an asymptotic behavior from exact considerations on some integrable model as this could help to understand their 
structure and origin in the general case where exact computations are not feasible.
We would like to mention that there already exists several exact results relative to this regime of the asymptotics in the case
of free fermion equivalent models
\cite{ItsIzerginKorepinVarguzinTimeSpaceAsymptImpBoseGaz,McCoyPArkShrockSpinTimeSpaceAutoCorrAsModSineKernel,McCoyPArkShrockSpinTimeAutoCorrAsModSineKernel,
MullerShrockDynamicCorrFnctsTIandXXAsymptTimeAndFourier}. We also would like to remind that there has been 
proposed recently \cite{ImambekovGlazmanEdgeSingInDSFBoseGas} a non-linear Luttinger liquid theory allowing one 
to predict the leading power-law behavior near the edges of the purely hole or particle specta
for dynamic structure factors and spectral functions\symbolfootnote[2]{These quantities refer to space and time Fourier
transforms of particular two-point functions} at low energy and momentum. 
This approach has been combined with Bethe Ansatz considerations to propose values for the amplitudes in front of this behavior
\cite{CauxGlazmanImambekovShashiAsymptoticsStaticDynamicTwoPtFct1DBoseGas}.

\vspace{2mm}

This article develops a method allowing one to compute the zero-temperature asymptotic behavior of the correlation functions
in integrable models starting from the form factor expansion for two-point functions.
The fact that we build our method on the form factor expansion allows us to include the time-dependence and hence access to the
large-distance and long-time asymptotic behavior. The method has been introduced recently on the example 
of the current-current correlators \cite{KozKitMailTerNatteSeriesNLSECurrentCurrent}.
Here, we provide many elements of rigor to the method and treat the example of the one particle reduced density matrix in the non-linear 
Schr\"{o}dinger model. We would like to stress that this method of asymptotic analysis 
not only allows us to carry out the analysis in the large-distance/long-time regime but also
constitutes an important technical and computational simplification of the approach
proposed in \cite{KozKitMailSlaTerXXZsgZsgZAsymptotics}. It has also the advantage of being applicable to a much wider class of algebraic Bethe Ansatz
solvable models as it solely  relies on the universal structure of the form factors in these models. All the more than the number of models where these have been  determined is  constantly growing
\cite{DeguchiMatsuiFFsHigherSpinXXZ,IzerginKorepinQISMApproachToCorrNextDiscussionDeterminantCurrentCurrentCorrFnction,
KMTFormfactorsperiodicXXZ,KorepinSlavnovFormFactorsNLSEasDeterminants,OotaInverseProblemForFieldTheoriesIntegrability}.
The main result of this paper can be summarized as follows. We provide a method for constructing a new type of series
representation for the correlation  functions of integrable models, that we call multidimensional Natte series.
This representation is THE one that is fit for an asymptotic analysis, as the first few terms of the asymptotic expansion can be simply
\textit{read-off} without any effort by looking at the terms of the series. Moreover, the computation of the higher order asymptotics
effectively boils down to the case of a free fermionic model (\textit{ie} computation of subleading asymptotics of the Fredholm determinant
of an integrable integral operator)  and thus bears the same combinatorial complexity.
The main implication of our result for physics is that the asymptotics in the time-dependent case are not only driven by excitations
on the Fermi boundary (the latter coincides with the
region of the spectrum that can be taken into account by using CFT/Luttinger liquid-based predictions), but also by excitations around the saddle-point
$\la_0$ of the "plane-wave" combination $xp\pa{\la}-t\veps\pa{\la}$ of the dressed momentum $p$ and dressed energy $\veps$ of the excitations.
Also, we provide explicit expressions and identify the associated amplitudes with the infinite volume limit of the
properly normalized in the size of the system form factors of the field.

We stress that although we have been able to set our method in a more rigorous framework then it was done in
\cite{KozKitMailSlaTerXXZsgZsgZAsymptotics,KozMailletSlaLongDistanceTemperatureNLSE}, we still have to rely on a few conjectures.
More precisely, we have been able to split the asymptotic analysis part from the one of
proving the convergence of certain series of multiple integrals representing the correlators.
The part related to asymptotic analysis has been set into a rigorous framework. However, in order to raise the results of this
asymptotic analysis to the level of the two-point function of interest, we still need to assume the convergence of
the series of multiple integrals we obtain.

The main novelty of this method is that it provides a systematic way for carrying out the asymptotic analysis of multiple integrals or series thereof
whose integrands contain some large-parameter dependent driving term being dressed up by coupled functions of the integration variables.
We provide a setting that allows one to interpret the "coupled" case as some deformation of the "uncoupled" one.
This deformation is such that, provided one is able to carry out the analysis in the "uncoupled" case (but with a sufficiently rich range  of functions involved), one is able to
deform the "uncoupled" asymptotics back to the "coupled" case of interest.
It is in this respect that the analysis carried out in this article strongly relies on the results obtained in \cite{KozTimeDepGSKandNatteSeries}
(where the relevant "uncoupled" series of multiple integrals of interest has been analyzed)
as well as on the fact that correlation functions of generalized
free fermionic models (which correspond to the "uncoupled" case) are naturally representable in terms of Fredholm determinants
\cite{KorepinSlavnovTimeDepCorrImpBoseGas}.


This paper is organized as follows. In section \ref{Section NLSE}, we remind the definition and main properties of the model.
We also introduce all the necessary notations allowing us to present the asymptotic behavior of the reduced density matrix.
In section \ref{Section method and main results}, we present our result and discuss the strategy of our method.
Then, in section \ref{Section Facteurs de Formes series}, we outline the main properties of the form factors in the model and
write down the form factor series for the reduced density matrix. We explain how this series can be re-summed into the so-called multidimensional
deformation of the Natte series. Once that such a representation is built, thanks to the very properties of the Natte series, it is possible to
literally \textit{read-off} the first few terms of the asymptotic expansion.
We gather all the auxiliary and technical results in several appendices. We discuss the large size-behavior of the form factors of the fields
in appendix \ref{Appendix Thermo lim FF}. In appendix \ref{Appendix section Free fermionic correlators}, we derive finite-size Fredholm minor
representations for the form factor based expansions of certain two-point functions in generalized free fermion models. In appendix \ref{Appendix
multidimensional Fredholm series}, we prove the existence of the thermodynamic (\textit{ie} infinite volume) limit for certain quantities of interest.
We also provide various alternative expressions for this limit. In appendix \ref{Appendice section functional translation}, we develop the theory
of functional translation in spaces of holomorphic functions.
The results established in this appendix
constitute the main tools of our analysis. They allow for an effective separation of variables in the intermediate steps so that one is able to
carry out various re-summations of the formulae by building on the results stemming from the generalized free fermion model studied in
appendix \ref{Appendix section Free fermionic correlators}.

\section{The non-linear Schr\"{o}dinger model}
\label{Section NLSE}

\subsection{The eigenstates and Bethe equations}
\label{Subsection solutions of BAE}

The non-linear Schr\"{o}dinger model corresponds to the Hamiltonian
\beq
\bs{  H} _{NLS} =\Int{0}{L} \paa{ \Dp{y} \Phi^{\dagger}\!\pa{y} \Dp{y} \Phi\pa{y} + c \, \Phi^{\dagger}\!\pa{y}\Phi^{\dagger}\!\pa{y} \Phi\pa{y} \Phi\pa{y}
- h \, \Phi^{\dagger}\!\pa{y} \Phi\pa{y} } \dd y\;.
\label{definition Hamiltonien Bose}
\enq
The model is defined on a circle of length $L$, so that the canonical Bose fields $\Phi$, $\Phi^{\dagger}$ are subject to $L$-periodic boundary
conditions. In the following, we will focus on the repulsive regime $c>0$ in the presence of a positive chemical potential $h>0$.
The Hamiltonian $\bs{  H} _{NLS}$ commutes with the number of particles operator, and thus can be diagonalized independently in every sector with
a fixed number of particles $N$. In each of these sectors, the model is equivalent to a N-body gas of bosons subject to $\de$-like
repulsive interactions.
The corresponding model of interacting bosons was first proposed and studied by Girardeau \cite{GirardeauIntroductionImpBosons}
in the  $c=+\infty$ case and then introduced and solved, through the coordinate Bethe Ansatz, by Lieb and  Liniger \cite{LiebLinigerCBAForDeltaBoseGas}
in the case of arbitrary $c$.
It is also possible to build the eigenstates of the Hamiltonian by means of the algebraic Bethe Ansatz.
This was first done by Sklyanin \cite{SklyaninNLSEbyQISMDirectQuantizationOfFieldsByNormalOrder} directly in the infinite
volume. In the case of finite volume $L$, as observed by Izergin and Korepin \cite{IzerginKorepinLatticeVersionsofQFTModelsABANLSEandSineGordon},
it is possible to put the continuous model on a lattice in such a way that the standard construction \cite{FaddeevABALesHouchesLectures}
of the algebraic Bethe  Ansatz holds. At the end of the computations, it is then possible to send the lattice spacing to zero and recover the spectrum and
eigenstates of the continuous model. The fact that this manipulation is indeed fully rigorous has been shown by Dorlas
\cite{DorlasOrthogonalityAndCompletenessNLSE}.

In the algebraic Bethe Ansatz approach, the Hamiltonian \eqref{definition Hamiltonien Bose} appears as a member of a one-parameter commuting 
family of operators $\la\mapsto\mc{T}\pa{\la}$. It is sometimes useful to consider a $\be$-deformation of this family $\mc{T}_{\be}\pa{\la}$, 
such that $\mc{T}_{\be}\pa{\la}_{\mid\be=0} = \mc{T}\pa{\la}$. The common eigenstates $\mid \psi_{\be}\pa{\paa{\mu}} \rangle $ of 
$\mc{T}_{\be}\pa{\la}$
in the $N_{\kappa}$-particle sector are parameterized by a set of real numbers $\paa{\mu_{\ell_a}}_{a=1}^{N_{\kappa}}$ which are the unique 
solutions to the $\be$-deformed logarithmic Bethe equations \cite{BogoliubiovIzerginKorepinBookCorrFctAndABA,Yang-YangNLSEThermodynamics}
\beq
L p_0\pa{\mu_{\ell_a}} + \sul{b=1}{N_{\kappa}}\th\pa{\mu_{\ell_a}-\mu_{\ell_b}} = 2\pi \pa{\ell_a - \f{N_{\kappa}+1}{2}}  + 2i\pi \be \;
\qquad \e{with} \quad
p_0\pa{\la}=\la \quad \e{and} \quad \th\pa{\la} = i \ln \paf{ic + \la}{ic - \la} \;.
\label{definition eqn Bethe log}
\enq
$p_0$ is called the bare momentum and $\th$ the bare phase. The set of solutions corresponding to all choices of integers $\ell_a \in \mathbb{Z}$ such
that  $\ell_1 < \dots < \ell_{N_{\kappa}}$ yield the complete set of eigenstates in the $N_{\kappa}$-particle sector \cite{DorlasOrthogonalityAndCompletenessNLSE}.

In each sector with a fixed number of particles $N_{\kappa}$,
the so-called ground state's Bethe roots are given by the solution to \eqref{definition eqn Bethe log}
corresponding to the choice of $N_{\kappa}$ consecutive integers $\ell_a=a$, with $a=1,\dots,N_{\kappa}$ and $\be=0$.
The number $N_{\kappa}$ corresponding to the number of particles in the overall ground state of $\bs{  H} _{NLS}$
is imposed by the chemical potential $h$ and scales with $L$. It will be denoted by $N$ in the following.
One shows that in the thermodynamic limit ($N,L \tend +\infty$ so that $\tf{N}{L}\tend D$) the parameters $\{ \la_j \}_{1}^N$
associated to this ground state condensate on a symmetric interval $\intff{-q}{q}$ called the Fermi zone.

All other choices of sets of integers $\ell_a$ lead to ($\be$-deformed) excited states. In principle, these excited states
can also be found in sectors with a
different number $N_{\kappa}\not= N$ of particles. It is convenient to describe the excited states in the language of particle-hole excitations above the
$N_{\kappa}$-particle $\be$-deformed ground state\symbolfootnote[2]{ the $\be$-deformed ground state corresponds to the choice $\ell_a=a$, with $a=1,\dots,N_{\kappa}$}.
Namely, such an excited state corresponds to a choice of integers $\ell_j$ in \eqref{definition eqn Bethe log}
such that
\beq
\ell_j=j \quad \e{for} \; \; j \in \intn{1}{N_{\kappa}} \setminus{h_1,\dots,h_n}  \quad \e{and}  \quad
\ell_{h_a}=p_a \quad \e{for} \;\;  a=1,\dots,n \; .
\label{definition correpondance entiers ella et particules-trous}
\enq
The integers $p_a$ and $h_a$ are such that $p_a \not \in \intn{1}{N_{\kappa}}\equiv \{ 1,\dots, N_{\kappa} \}$ and $h_a\in \intn{1}{N_{\kappa}}$.
There is thus
a one-to-one correspondence between integers $\ell_j$ and the integers $h_a$ and $p_a$ describing particle-hole excitations.

In this picture, the integers $h_a$ correspond to holes in the increasing sequence of integers defining the $\be$-deformed ground state roots, whereas
$p_a$ correspond to extra integers appearing in the equation and can be seen as defining some new position of "particles".
Given a solution $\paa{\mu_{\ell_a}}_{1}^{N_{\kappa}}$ corresponding to a fixed choice of integers $\ell_1<\dots<\ell_{N_{\kappa}}$ it is convenient to
introduce their counting function:
\beq
\wh{\xi}_{\paa{\ell_a}} \pa{\om} \equiv \wh{\xi}_{\paa{\ell_a}} \pa{\om \mid \paa{\mu_{\ell_a}}_{1}^{N_{\kappa}} } = \f{p_0\pa{\om}}{2\pi} +\f{1}{2\pi L} \sul{a=1}{N_{\kappa}} \th\pa{\om-\mu_{\ell_a}} + \f{N_{\kappa}+1}{2L}  \; - \;  i \f{ \be }{ L } \;.
\label{definition counting function mu}
\enq
By construction, it is such that $\wh{\xi}_{\paa{\ell_a}}\pa{\mu_{\ell_a}}=\tf{\ell_a}{L}$, for $a=1,\dots, N_{\kappa}$.
Actually, $\wh{\xi}_{\paa{\ell_a}}\pa{\om }$ defines\symbolfootnote[2]{Note that different sets of roots $\paa{\mu_{\ell_a}}$ and
$\{ \mu_{\ell^{\prime}_a} \}$  lead to different sets of background parameters} a set of background parameters $\paa{\mu_a}$, $a\in \mathbb{Z}$,
as the unique\symbolfootnote[3]{The uniqueness of solutions follows from the fact that the solution to \eqref{definition eqn Bethe log}
are such that $\mu_{\ell_a}\pa{\be}=\mu_{\ell_a}\pa{0} +2i\pi\tf{\be}{L}$.
This allows one to show that $\wh{\xi}_{\paa{\ell_a}}\pa{\om}$ is strictly
increasing on $\R+2i\pi\tf{\be}{L}$ and maps it onto $\R$.
Moreover, one can check that $\Im\pa{\wh{\xi}_{\paa{\ell_a}}}\not=0$ on $\Cx\setminus \pa{\R+2i\pi\tf{\be}{L}}$.}
solutions to $\wh{\xi}_{\paa{\ell_a}}\pa{\mu_a}=\tf{a}{L}$. The latter allows one to define the rapidities $\mu_{p_a}$, resp. $\mu_{h_a}$, of the
particles, resp. holes, entering in the description of $\paa{\mu_{\ell_a}}_1^{N_{\kappa}}$.

\subsection{The thermodynamic limit}
\label{Subsection thermo limit}

When the thermodynamic limit of the model is considered, it is possible to provide a slightly more precise description of the solution to the Bethe
equations for the  ground state $\{ \la_a \}_{a=1}^{N}$
as well as for any particle-hole type $\be$-deformed excited states $\paa{\mu_{\ell_a}}_{a=1}^{N_{\kappa}}$ above it
with $N_{\kappa}-N$ being fixed and not depending on $L$ or $N$. Introducing the counting function for the ground state
\beq
\wh{\xi}\pa{\om} \equiv \wh{\xi} \pa{\om \mid \paa{\la_{a}}_{1}^{N} } = \f{p_0\pa{\om}}{2\pi} +\f{1}{2\pi L} \sul{a=1}{N_{\kappa}} \th\pa{\om-\la_a}
\; +  \; \f{N+1}{2L} \;, \quad  ie \;\; \wh{\xi}\pa{\la_a}=\f{a}{L} \;,
\label{definition counting function la}
\enq
it can be shown that, in the thermodynamic limit, it behaves as
\beq
\wh{\xi}\pa{\om } = \xi\pa{\om} + \e{O}\pa{L^{-1}} \quad \e{where} \quad
\xi\pa{\om}= \f{p\pa{\om}}{2\pi} + \f{D}{2} \qquad \e{and} \quad N/L\tend D \:.
\label{ecriture limite thermo fction comptage}
\enq
There, the $\e{O}\pa{L^{-1}}$ is uniform and holomorphic in $\om$ belonging to a strip of some fixed width around the real axis,
$p$ is the so-called dressed momentum, defined as the unique solution to the integral equation
\beq
p\pa{\la} - \Int{-q}{q} \th\pa{\la-\mu} p^{\prime}\pa{\mu} \f{\dd \mu}{2\pi} = p_0\pa{\la} \;.
\label{definition eqn integral p}
\enq
The parameter $q$ corresponds to the right end of the Fermi interval $\intff{-q}{q}$ on which the ground state's Bethe roots condensate.
It is fixed by the value of the chemical potential $h$ by demanding that the dressed energy $\veps\pa{\la}$, defined as the unique solution
to the below integral equation,  vanishes
at $\pm q$:
\beq
\veps\pa{\la}- \Int{-q}{q} K\pa{\la-\mu} \veps\pa{\mu} \f{\dd \mu}{2\pi}= \veps_{0}\pa{\la} \qquad \e{with} \; \; \veps_0\pa{\la}=\la^2-h
\qquad \e{and} \;\; \veps\pa{\pm q}=0 \;.
\label{definition eqn int eps}
\enq
We also remind the relation $ p_F= \pi D$ where $p_F=p\pa{q}$ is the Fermi momentum.

In the following, we will focus on the excited states in the $N_{\kappa}=N+1$-particle sector only.
In order to describe the thermodynamic properties of such $\be$-deformed excited states,
it is convenient to introduce the associated shift function
\beq
\wh{F}_{\paa{\ell_a}}\pa{\om}\equiv  \wh{F}\pa{\om\mid \paa{\mu_{\ell_a}}_1^{N+1}} = L \pac{\wh{\xi}\pa{\om}  - \wh{\xi}_{\paa{\ell_a}}\pa{\om} }=
\f{1}{2\pi} \sul{a=1}{N} \th\pa{\om-\la_a} \; - \; \f{1}{2\pi}\sul{a=1}{N+1} \th\pa{\om-\mu_{\ell_a}} \; - \f{1}{2} +i \be \;.
\label{definition shift function discrete}
\enq
It can be shown that this counting function admits a thermodynamic limit $F_{\be}$ that solves the linear integral equation
\beq
F_{\be}\pabb{\la}{ \paa{\mu_{p_a}} } {\paa{\mu_{h_a}} }  - \Int{-q}{q} K\pa{\la-\mu} F_{\be}\pabb{\mu}{ \paa{\mu_{p_a}} } {\paa{\mu_{h_a}} }
\f{\dd \mu}{2\pi} =
i\be - \f{1}{2}   \; - \; \f{1}{2\pi}\th\pa{\la-q}
\; - \; \f{1}{2\pi}\sul{a=1}{n} \pac{ \th( \la-\mu_{p_a} ) -\th\pa{\la-\mu_{h_a}}  } \;.
\nonumber
\enq
There $\mu_{p_a}$, resp. $\mu_{h_a}$, are to be understood as the unique solutions to $\xi\big( \mu_{p_a} \big)=\tf{p_a}{L}$, resp.
$\xi\pa{\mu_{h_a}}=\tf{h_a}{L}$, where $\xi$ is given by \eqref{ecriture limite thermo fction comptage}.
Note that we have explicitly insisted on the auxiliary dependence of the thermodynamic limit of the shift function on the positions of the
particles/holes. However, in the following, whenever the value of $\{ \mu_{p_a} \}$ and $\paa{\mu_{h_a}}$ will be dictated by the context,
we will omit it. We also remind that the above shift function measures the spacing between the ground state roots $\la_a$ and the 
background parameters $\mu_a$ defined by
$\wh{\xi}_{\paa{\ell_a}}$\,: \;  $\mu_a-\la_a = F_{\be}\pa{\la_a}\cdot \pac{L \xi^{\prime}\!\pa{\la_a}}^{-1} \pa{1+ \e{O}\pa{L^{-1}}}$.

The integral equation for the thermodynamic limit of the shift function $F_{\be}$ can be solved in terms of the dressed phase 
$\phi\pa{\la,\mu}$ and dressed charge $Z\pa{\la}$
\beq
\phi\pa{\la,\mu}- \Int{-q}{q} K\pa{\la-\tau} \phi\pa{\tau,\mu} \f{\dd \tau}{2\pi} =  \f{1}{2\pi}\th\pa{\la-\mu} \qquad \e{and} \qquad
Z\pa{\la}- \Int{-q}{q} K\pa{\la-\tau} Z\pa{\tau} \f{\dd \tau}{2\pi} =  1 \;.
\label{definition eqn int Z et phi}
\enq
Namely,
\beq
F_{\be}\pa{\la} \equiv  F_{\be}\pabb{\la}{ \{ \mu_{p_a} \} } {\paa{\mu_{h_a}} }
 = \pa{i\be -\tf{1}{2}}Z\pa{\la} \; -\; \phi\pa{\la,q} \; - \; \sul{a=1}{n} \pac{\phi ( \la,\mu_{p_a}) - \phi\pa{\la,\mu_{h_a}}}
\label{ecriture limite thermo fction shift}
\enq
Here, we also remind two very nice relationships that exist between the dressed phase and dressed charge
\beq
Z\pa{\la}=1+\phi\pa{\la,-q}-\phi\pa{\la,q} \quad \e{and} \quad
 Z^{-1}\pa{q}=1+\phi\pa{-q,q}-\phi\pa{q,q}  \;.
\label{ecriture relation Z moins 1 et phi}
\enq
The first one is easy to obtain and the second one has been obtained in
\cite{KorepinSlavnovNonlinearIdentityScattPhase,SlavnovNonlinearIdentityScattPhase}.

The shift function allows one to compute many thermodynamic limits involving the parameters $\paa{\mu_{\ell_a}}$. For instance, introducing the
combination of bare momentum and energy $u_0\!\pa{\la}=p_0\!\pa{\la}-\tf{t\veps_{0}\!\pa{\la}}{x}$, one readily sees that
for a $n$ particle/hole excited state $\paa{\mu_{\ell_a}}$ at $\be=0$
\beq
\lim_{N,L\tend +\infty} \paa{ \sul{a=1}{N+1} u_0\pa{\mu_{\ell_a}} \; - \;  \sul{a=1}{N} u_0\pa{\la_a} }_{\mid \be = 0} =
\sul{a=1}{n} u( \,\mu_{p_a} )- u(\, \mu_{h_a} ) \;.
\label{ecriture limite thermo exposant osc}
\enq
Above and in the following, $u$ stands for the combination of dressed momenta and energies $u\pa{\la}=p\pa{\la}-\tf{t\veps\pa{\la}}{x}$. 
It admits the integral representation
\beq
u\pa{\la} = u_0\pa{\la} - \Int{-q}{q} \!\!   u_0^{\prime}\pa{\mu} \phi\pa{\mu,\la}   \dd \mu   \;.
\label{ecritutre representation integrale u}
\enq
The function $u_0^{\prime}$ admits a unique zero of first order on $\R$. It is believed that this property is preserved for $u$.
Clearly, in virtue of Rouch\'{e}'s theorem, this holds true for $c$ large enough.
We will not purse the discussion of this property here as it goes out of the scope of this paper and will
use it as a working hypothesis.  In other words, we assume that given a fixed ratio $\tf{t}{x}$, there exists a unique $\la_0$ such that
$u^{\prime}\!\pa{\la_0}=0$ and $u^{\prime \prime}\!\pa{\la_0}<0$.

We do stress however that this working hypothesis should not be considered as a restriction
but a simplification of the exposition at most. Indeed, it follows from  $\abs{u^{\prime}\!\pa{\la}} \tend + \infty$ when
$\Re\pa{\la} \tend \pm\infty$ that, for any value of $c>0$, $u^{\prime}$ has a finite number of real zeroes.
The case when $u^{\prime}$ has multiple real zeroes of arbitrary order could be treated within out method but would make the analysis heavier.


As a concluding remark to this section, we would like to stress that all functions that have been introduced above
(the dressed momentum $p$, the dressed energy $\veps$, the dressed charge $Z$ and the dressed phase $\phi$) are holomorphic in the strip
\beq
U_{\de}=\paa{  z \in \Cx \; :\; \abs{\Im\pa{z}} < 2\de}
\label{definiton voisinage ouvert de R pour holomorphie Gn}
\enq
around the real axis. The parameter $\de$ satisfies $\tf{c}{8}>\de >0$ and is chosen sufficiently small so that $p$ is injective on 
$U_{\de}$ and that one has $\inf_{\la \in U_{\de}} \pac{\Re\pa{Z\!\pa{\la}} }>0$. We will tacitly assume such a choice in the following
each time the strip $U_{\de}$ will be used.


\section{The method and main results}
\label{Section method and main results}

The zero-temperature one-particle reduced density matrix in finite volume refers to the below
ground state expectation value:
\beq
\hspace{2cm} \rho_N \! \pa{x,t}\equiv \bra{\psi\pa{\paa{\la_a}_1^N}}\Phi\pa{x,t}\Phi^{\dagger}\!\pa{0,0}\ket{\psi\pa{\paa{\la_a}_1^N}}
\cdot  \norm{ \psi\pa{\paa{\la_a}_1^{N} } }^{-2}  \;.
\label{definition reduced density matrix}
\enq
The parameters $\paa{\la_a}_1^N$  correspond to the set of Bethe roots parameterizing the ground state of \eqref{definition Hamiltonien Bose}.
 We recall that the fields evolve in space and time according to
\beq
\Phi\!\pa{x,t} = \ex{ - ix P + i t \bs{H}_{NLS} } \Phi\!\pa{0,0} \ex{ix P - i t \bs{H}_{NLS}  } \; ,
\enq
where $\bs{H}_{NLS} $ is the Hamiltonian of the model given in \eqref{definition Hamiltonien Bose} and $P$ is the total momentum operator. The action of
$P$ on the eigenstates of $\bs{H}_{NLS} $ has been computed in \cite{BogoliubiovIzerginKorepinBookCorrFctAndABA}.

We denote by $\rho\!\pa{x,t} = \lim_{N,L \tend +\infty} \rho_{N}\!\pa{x,t}$ the, presumably existing, thermodynamic limit of $\rho_N\!\pa{x,t}$. We will
not develop further on the existence of this limit, and take this as a
quite reasonable working hypothesis.

\subsection{Description of the method}

In this article, we carry out several manipulations that lead us to propose a series representation for $\rho\!\pa{x,t}$
giving a straightforward access to its leading  large-distance/long-time
asymptotic  behavior.

The starting point of our analysis is the model in finite volume.
We will first provide certain re-summation formulae for $\rho_{N}\!\pa{x,t}$ starting from the form factor expansion of
\eqref{definition reduced density matrix}.
The latter involves a summation over all the excited states (\textit{ie} over all solutions to
\eqref{definition eqn Bethe log}-\eqref{definition correpondance entiers ella et particules-trous} at $\be=0$).
This sum has a very intricate structure which prevents us from analyzing its thermodynamic limit rigorously from the
very beginning. We therefore introduce a simplifying hypothesis. Namely, denoting the energy of an excited state by $E_{\e{ex}}$ and
the one of the ground state by $E_{\e{gs}}$ we argue that all contributions issued from excited states such that
$E_{\e{ex}}-E_{\e{gs}}$ scales with  $L$ do not contribute to the thermodynamic limit of the form factor expansion of
$\rho_N \! \pa{x,t}$. In the light of these arguments, we are led to analyze an effective form factor series $\rho_{N;\e{eff}}\!\pa{x,t}$
and a certain $\ga$-deformation $\rho_{N;\e{eff}}\!\pa{x,t\mid \ga}$ thereof.
Our conjecture is that $\rho_{N;\e{eff}}\!\pa{x,t\mid \ga=1}=\rho_{N;\e{eff}}\!\pa{x,t}$ has the \textit{same} thermodynamic limit as $\rho_N\!\pa{x,t}$.

We study $\ga \mapsto \rho_{N;\e{eff}}\pa{x,t\mid \ga}$ by means of its Taylor coefficients at $\ga=0$:
\vspace{-2mm}
\beq
\hspace{1cm}\rho_{N;\e{eff}}^{\pa{m}}\!\pa{x,t} \equiv   \f{ \Dp{}^m }{ \Dp{}\ga^m } \left. \rho_{N;\e{eff}}\pa{x,t\mid \ga} \right|_{\ga=0} \;.
\enq

\vspace{-2mm} All rigorous, conjecture-free, results of this paper are relative to these Taylor coefficients.
We show that these admit a well defined thermodynamic
limit $\rho_{\e{eff}}^{\pa{m}}\!\pa{x,t}$. In addition, we provide two different representations for this limit, each being a
 finite sum of multiple integrals.
\begin{itemize}
\item  The first representation is in the spirit of the ones obtained in
\cite{KozKitMailSlaTerXXZsgZsgZAsymptotics,KozMailletSlaLongDistanceTemperatureNLSE}.
It corresponds to some truncation of a multidimensional deformation of a Fredholm series for a Fredholm minor.
\item The second representation is structured in such a way that it allows one to  \textit{read-off} straightforwardly
the first few terms of the asymptotic expansion of $\rho_{\e{eff}}^{\pa{m}}\!\pa{x,t}$.
 The various terms appearing in this representation
are organized in such a way that the identification of those that are negligible (\textit{eg} exponentially small) in the $x\tend +\infty$ limit is
trivial.
\end{itemize}

The above two results are derived rigorously without any approximation or additional conjecture.
However, in order to push the analysis a little further and provide results that would have applications to physics,
we need to rely on several conjectures.
Namely,  we assume that
\begin{enumerate}
\item  the series of multiple integrals that arises upon summing up the thermodynamic limits of the Taylor coefficients
$\sum_{m=0}^{+\infty} \tf{ \rho_{\e{eff}}^{\pa{m}}\!\pa{x,t} }{m!}$  is convergent;
\item  this sum moreover coincides with the thermodynamic limit of $\rho_{N;\e{eff}}\!\pa{x,t \mid \ga=1}$  and hence, due to our first
conjecture, with $ \rho\pa{x,t}$.
\end{enumerate}
These conjectures allow us to claim that $\rho\pa{x,t}$ can be represented in terms of a series of multiple integrals.
The latter series corresponds to a multidimensional deformation
of the  Natte series expansion for Fredholm minors of integrable integral operators \cite{KozTimeDepGSKandNatteSeries}. This multidimensional Natte
series has all the virtues in respect to  the computation of the long-time/large-distance asymptotic behavior of $\rho\!\pa{x,t}$; it is structured in
such a way that one readily \textit{reads-off} from its very form, the sub-leading and the first few leading terms of the asymptoics.

So as to conclude the description of our method, we would like to stress that the aforementioned conjectures of convergence
are supported by the fact that they can be proven to hold in the limiting case of a generalized free fermion model \cite{KozTimeDepGSKandNatteSeries}.
Unfortunately, the highly coupled nature of the integrands involved in our representations does not allow
one for any simple check of the convergence properties in the general $+\infty>c>0$ case.

\subsection{Large-distance/long-time asymptotic behavior of the one-particle reduced density matrix}
\label{Theorem asymptotique rho}

We have now introduced enough notations so as to be able to present the physically interesting part of our analysis.

\vspace{2mm}

Let $x>0$ be large and the ratio $\tf{x}{t}$ is fixed.  Let $\la_0$ be the associated, presumably unique
(\textit{cf} \eqref{ecritutre representation integrale u}), saddle-point
of $u\pa{\la}=p\pa{\la}-\tf{t \veps\pa{\la}}{x}$. Assume in addition that $\la_0 \not = \pm q$ and $\la_0 > -q$.
Then, \textit{under the validity of the aforementioned conjectures}, the thermodynamic limit of the zero-temperature one-particle reduced density matrix
$\rho\pa{x,t}$ admits the asymptotic expansion
\bem
\rho\pa{x,t} = \sqrt{   \f{ -2i\pi }{ t \veps^{\prime\prime}\!\pa{\la_0} -x p^{\prime\prime}\!\pa{\la_0}  }   }
 \f{ p^{\prime}\!\pa{\la_0} \; \ex{ix \pac{ u\pa{\la_0} - u\pa{q} } } \;  \abs{ \mc{F}_q^{\la_0} }^2  }
 {  \pac{i\pa{x+v_F t}}^{ \big[ F^{\la_0}_q\pa{-q}  \big]^2  }   \pac{ -i\pa{x-v_F t}}^{ \big[ F^{\la_0}_q\pa{q}  \big]^2  } }
\pa{ \bs{1}_{\intoo{q}{+\infty}}\pa{\la_0}  +\e{o}\pa{1} } \\
\hspace{5cm} +\;   \f{ \ex{-2ixp_F} \abs{ \mc{F}_q^{-q} }^2   }
{  \pac{i\pa{x+v_F t}}^{ \big[ F^{-q}_q\pa{-q} -1 \big]^2  }   \pac{ -i\pa{x-v_F t}}^{ \big[ F^{-q}_q\pa{q} \big]^2  } } \pa{ 1+\e{o}\pa{1} }   \\
+ \; \f{ \abs{ \mc{F}_{\emptyset}^{\emptyset} }^2    }
{  \pac{i\pa{x+v_F t}}^{ \big[ F_{\emptyset}^{\emptyset}\pa{-q} \big]^2  }
\pac{ -i\pa{x-v_F t}}^{ \big[ F_{\emptyset}^{\emptyset}\pa{q} +1\big]^2  } } \pa{ 1+\e{o}\pa{1} } 
\;\;  + \hspace{-2mm} \sul{ \substack{ \ell^{+}, \ell^{-} \in \mathbb{Z} \\ \bs{\eta} (\ell^+ + \ell^-) \geq 0}  }{ }  \hspace{-5mm} ^{*} \; 
 C_{\ell^+;\ell^-} \f{ \ex{i x \vp_{\ell^+;\ell^-} } }{ x^{\De_{\ell^+,\ell^-} } } \pa{1+\e{o}\pa{1}}
\label{equation fondamentale asymptotiques rho x et t}
\end{multline}
The critical exponents governing the algebraic decay in the distance of separation are expressed in terms of the thermodynamic limit $F_{\mu_h}^{\mu_p}$
of the shift function (at $\be=0$) associated with an excited state of \eqref{definition Hamiltonien Bose}
having one particle at $\mu_p$ and one hole at $\mu_h$, namely,
\beq
F_{\emptyset}^{\emptyset}\pa{\la} = -\f{ Z\pa{\la} }{ 2}  - \phi\pa{\la, q} \qquad
F^{-q}_{q}\pa{\la} = -\f{ Z\pa{\la} }{ 2}  - \phi\pa{\la, -q} \qquad
F^{ \la_0 }_{ q }\pa{\la} = -\f{ Z\pa{\la} }{ 2}  - \phi\pa{\la, \la_0} \;.
\label{definition les fonction shift asymptotiques}
\enq

The type of algebraic decay in the explicit terms in \eqref{equation fondamentale asymptotiques rho x et t} 
can be organized in two classes.
There is a square root power-law decay $\pa{t\veps^{\prime\prime}\!\pa{\la_0}- x p^{\prime\prime}\!\pa{\la_0} }^{-\f{1}{2}}$ stemming 
from the saddle-point $\la_0$. All other sources of algebraic decay appear
in the so-called relativistic combinations $x\pm v_F t$ and exhibit non-trivial critical exponents driven by the shift function of the 
underlying type of excitation. We recall that $\pm v_F$ corresponds to the velocity of the
excitations on the right/left Fermi boundary: $v_F= \tf{ \veps^{\prime}\!\pa{q} }{ p^{\prime}\!\pa{q} }$.

Each of the three explicit terms in these asymptotics has its amplitude ($\big| \mc{F}_q^{\la_0} \big|^2$,
$\big|  \mc{F}_{\emptyset}^{\emptyset} \big|^2$ or $\big| \mc{F}_q^{-q} \big|^2$) given by the thermodynamic limit of
properly normalized in the length $L$  moduli squared of form factors of the conjugated field $\Phi^{\dagger}$.
More precisely,

\begin{itemize}

\item  $\abs{\mc{F}_q^{ \la_0 } }^2$ involves the form factor of $\Phi^{\dagger}$
taken between the $N$-particle ground state and an excited state above the $N+1$ particle ground state
with one particle at $\la_{0}$ and one hole at $q$.

\item $\abs{\mc{F}_q^{-q}}^2$ corresponds to the case when one considers an excited state above the $N+1$-particle ground state with one 
particle at $-q$ and one hole at $q$.

\item $\abs{\mc{F}_{\emptyset}^{\emptyset}}^2$ corresponds to the case where the form factor average of $\Phi^{\dagger}$ is taken between
the N and the N+1-particle ground state.

\end{itemize}

The explicit (but rather cumbersome) expressions for the amplitudes together with a more precise definition
are postponed to appendix \ref{Appendix Thermo lim FF Section Formules FF part/trou}.

\vspace{1mm}

Also, $ \bs{1}_{\intoo{q}{+\infty}}$ stands for the characteristic function of the interval $\intoo{q}{+\infty}$.
It is there so as to indicate that, to the leading order, the contribution stemming from the saddle-point only appears in the
space-like regime $\la_0> q$. We stress however that hole-type excitations in a vicinity of the
saddle-point also contribute in the time-like regime where $\la_0 \in \intoo{-q}{q}$. 
This fact follows from the structure of the terms present in the sum over $\ell^{+}, \ell^{-}$.

\vspace{1mm}

We would now like to discuss the sum over the integers $\ell^+, \ell^-$ in \eqref{equation fondamentale asymptotiques rho x et t}. 
The latter represents the contributions to the asymptotics associated to the so-called quicker harmonics. 
Ineed, every term in this sum oscillates with a phase 
\beq
\vp_{\ell^+  ; \ell^-} = \ell^+ u\pa{q} + \ell^- u\pa{-q} \; - \;  (\ell^+ + \ell^-) u(\la_0) \;. 
\enq
It is also caracterized by its own critical exponents 
\beq
\De_{\ell^+;\ell^-} = \pa{1+\ell^+ + \De_+}^2 + \pa{\De_- - \ell^-}^2  + \f{ |\ell^+ + \ell^- | }{2} \;,
\enq
where, 
\beq
\De_{\pm} = - \f{Z\pa{\pm q}}{2} \; - \; \ell^- \phi\pa{\pm q ,-q } \; - \; (\ell^+ + 1) \phi\pa{\pm q ,q }
 \; + \; (\ell^+ + \ell^-) \phi(\pm q , \la_0) \;. 
\enq
Our method of analysis only allows us to prove that the only harmonics present in the asymptotics are those 
oscillating with one of the frequencies $\vp_{\ell^+,\ell^-}$ and that they decay, to the leading order (\textit{ie}
up to $\e{o}\pa{1}$ terms), with the critical exponent $\De_{\ell^+,\ell^-}$. We are however unable to 
give an explicit prediction for the amplitudes $C_{\ell^+,\ell^-}$. Note that the 
sum runs over all integers $\ell^{\pm}$  subject to the 
constraint $\bs{\eta}(\ell^+ + \ell^-)\geq 0$.  The parameter $\bs{\eta}$
depends on the regime: $\bs{\eta}=1$ in the space-like regime ($\la_0 > q$) and $\bs{\eta}=-1$ in the time-like regime
$(\abs{\la_0}<q$). Finally, the $*$ in the sums indicates that one should not sum up over those integers $\ell^+, \ell^-$ giving rise to 
the frequencies that are present in the first three lines of \eqref{equation fondamentale asymptotiques rho x et t}.

Note that we have organized the large $x$ (with $\tf{x}{t}$ fixed)  asymptotic expansion in respect to the various oscillating 
phases. Each phase appears with its own exponent driving the power-law decay in $x$. Our computations allowed us to compute the 
leading (\textit{ie} up to $\e{o}\pa{1}$ corrections) behavior of each harmonic. Note that the $\e{o}\pa{1}$ terms stemming from one of 
the harmonics may be dominant even in respect to the leading terms coming from another harmonic.

\vspace{4mm}
\subsubsection*{Remarks}
The oscillating phases and amplitudes appearing in \eqref{equation fondamentale asymptotiques rho x et t} are reminiscent of the type of
excitations that give rise to their associated contribution. Each term in \eqref{equation fondamentale asymptotiques rho x et t} can be 
associated with
some macroscopic state of the model. For instance, the one occuring\symbolfootnote[2]{The $\e{o}\pa{1}$ corrections being excluded} in the 
first line of \eqref{equation fondamentale asymptotiques rho x et t} corresponds to a macroscopic state characterized by one particle at 
$\la_0$ and one hole at $q$. There are infinitely many microscopic realizations of such a macroscopic state. For instance, any excited
state realized as one particle at $\la_0$, one hole at $q$,
\begin{itemize}
\item  $n_+$ particles $\mu_{p_a}^{\pa{r}}$ and holes $\mu_{h_a}^{\pa{r}}$ located at $q$  in the thermodynamic limit:
$\mu_{p_a}^{\pa{r}}, \mu_{h_a}^{\pa{r}} \! \limit{N,L}{+\infty}\! q$ for $a=1,\dots, n_+$,
\item  $n_-$ particles $\mu_{p_a}^{\pa{l}}$ and holes $\mu_{h_a}^{\pa{l}}$ located at $-q$  in the thermodynamic limit:
$\mu_{p_a}^{\pa{l}}, \mu_{h_a}^{\pa{l}} \limit{N,L}{+\infty} \!\! -q$ for $a=1,\dots, n_-$.
\end{itemize}
would give rise to the same (from the point of view of energy $\mc{E}=\veps\!\pa{\la_0}$ , momentum $\mc{P}=p\!\pa{\la_0}-p\!\pa{q}$,...) macroscopic
state. In a joint collaboration with Kitanine, Maillet, Slavnov and Terras we have shown \cite{KozKitMailSlaTerRestrictedSums}
that indeed, in the zero-time case, the contribution of a given
macroscopic state to the asymptotics is obtained by summing up over all such zero-momentum
excitations on each of the Fermi boundaries. Clearly, this picture persists in the time-dependent case as well. The only difference
being that, in the time-dependent case, the number of relevant macroscopic states contributing to the asymptotics is bigger (one has to include
the contributions of excitations around the saddle-point in addition to the excitations on the Fermi boundary).
Moreover, we would like to draw the reader's attention
to the fact that it is precisely the sum over such zero momentum excitations on the Fermi boundary that gives rise, through
some intricate microscopic mechanism of summation, to the relativistic combinations $\pa{x+v_F t}^{\a_+}$ (in what concerns the left Fermi boundary) and
$\pa{x-v_F t}^{\a_-}$ (in what concerns the right Fermi boundary) arising in the asymptotics.
This mechanism can be considered as yet another manifestation of conformal field theory on the level of asymptotics.

Our analysis leads us to propose an alternative interpretation of the universality hypothesis. Namely, when dealing with asymptotics
(large-distance, \textit{etc})
of correlation functions, one is brought to the analysis of the contributions of "relevant" saddle-points. As one can expect from the saddle-point
type analysis of one-dimensional integrals, the leading asymptotics are only depending on the local behavior around the saddle-point of the
driving term. All other details of the integrand do not matter for fixing the exponent governing the algebraic decay. Therefore,
it is quite reasonable to expect that models sharing the same types of saddle-points exhibit the same type of critical behavior.
The universality hypothesis \cite{GriffithsUniversalityAndExponentsParameterDepending} stating that models sharing the same symmetry class have the same
value for their critical exponents can be now re-interpreted as the fact that the symmetries of a model uniquely determine the structure of the
driving terms in the saddle-points that are relevant for the asymptotics. As a consequence, the leading power-law decay stemming from the local
analysis around these saddle-points is always characterized by the same critical exponents regardless of the fine, model dependent, function
content of the integrals describing the  correlation functions.

\vspace{2mm}

We draw the reader's attention to the fact that  the terms appearing in the $2^{\e{nd}}$ and $3^{\e{rd}}$ lines of \eqref{equation fondamentale asymptotiques rho x et t} correspond solely to excitations on the Fermi boundaries and
confirm the CFT/Luttinger liquid-based predictions for the long-distance asymptotics\symbolfootnote[3]{Taking the $t\tend 0$ limit of
\eqref{equation fondamentale asymptotiques rho x et t} is slightly subtle. The first line produces a contribution proportional to
$t^{-\f{1}{2}} \ex{i \f{x^{2}}{4t}}$.  In the $t\tend 0$ limit, this function approaches, in the sense of distributions, a Dirac $\de\pa{x}$ function.
The presence of this $\de\pa{x}$ function is expected from the form of the commutation relations between the fields.
However, in the large-$x$ limit of interest to us, it does not contribute.} at $t=0$ due to the identifications
following from \eqref{ecriture relation Z moins 1 et phi}:
\beq
F^{\emptyset}_{\emptyset}\pa{q}+1= \f{Z^{-1}\!\pa{q}}{2}  \; , \; F^{\emptyset}_{\emptyset}\pa{-q}=-\f{Z^{-1}\!\pa{q}}{2} \; , \;
F^{-q}_{q}\pa{q}= \f{Z^{-1}\!\pa{q}}{2} - Z\pa{q}  \; , \; F^{-q}_{q}\pa{-q}-1=-\f{Z^{-1}\pa{q}}{2} -Z\pa{q} \;  .
\enq
However, we do stress that \eqref{equation fondamentale asymptotiques rho x et t} clearly shows the  need to go beyond  the CFT/Luttinger
liquid picture so as to provide the correct long-time/large-distance asymptotic behavior of the correlation functions in gapless
one-dimensional quantum Hamiltonians. In particular, our results contain additional terms in respect to the predictions obtained in
\cite{BerkovichMurthyWrongCFTBasedPredictionTimeMultiCorrNLSE}. 
Our result has a strong structural resemblance with the non-linear Luttinger liquid based predictions for the 
edge exponents \cite{ImambekovGlazmanEdgeSingInDSFBoseGas} and amplitudes \cite{CauxGlazmanImambekovShashiAsymptoticsStaticDynamicTwoPtFct1DBoseGas}
arizing in the low momentum $k$ and low energy $\om$ behavior of the spectral 
function\symbolfootnote[1]{The latter corresponds to the space and time Fourier transform of 
$\moy{\Phi\pa{x,t}\Phi^{\dagger}\pa{0,0}} \bs{1}_{\intoo{0}{+\infty}}\pa{t}  + 
\moy{\Phi^{\dagger}\pa{0,0}\Phi\pa{x,t}} \bs{1}_{\intoo{-\infty}{0}}\pa{t} $}.




\section{The form factor series}
\label{Section Facteurs de Formes series}

In this section, we will provide two new representations for the zero-temperature reduced density matrix
\eqref{definition reduced density matrix} starting from its form factor issued expansion:
\beq
\rho_N\pa{x,t} =
\sul{ \substack{\ell_1 < \dots < \ell_{N+1} \\ \ell_a \in \mathbb{Z} } }{}
\f{  \pl{a=1}{N+1} \ex{ixu_0\pa{\mu_{\ell_a}}}  }{  \pl{a=1}{N} \ex{ixu_0\pa{\la_a}}   }
\f{ \abs{ \bra{ \psi\pa{\paa{\mu_{\ell_a}}_1^{N+1}} } \Phi^{\dagger}\!\pa{0,0} \ket{ \psi\pa{\paa{\la_a}_1^N} } }^2 }
{ \norm{ \psi\pa{\paa{\mu_{\ell_a}}_1^{N+1} } }^2  \cdot \norm{ \psi\pa{\paa{\la_a}_1^{N} } }^2   } \;.
\label{equation developement correlateur serie FF}
\enq
The above series runs through all the possible  choices of integers $\ell_a$, $a=1, \dots, N+1$  such that $\ell_1<\dots<\ell_{N+1}$.

Below, we shall argue in favor of several reasonable approximations that allow us to
reduce  the form factor series  to another, effective one, whose structure is simple enough
so as to be able to continue the calculations directly on it.

\subsection{The effective form factors}
\label{Subsection effective FF}

It has been shown in \cite{KozFFConjFieldNLSELatticeSpacingGoes0} (slightly different determinant representations
for these form factors have already appeared in \cite{KorepinSlavnovFormFactorsNLSEasDeterminants,OotaInverseProblemForFieldTheoriesIntegrability})
that the  form factors of the operator $\Phi^{\dagger}$  taken between the $N$-particle ground state $\paa{\la_a}_1^N$ and
any particle-hole type  excited state $\paa{\mu_{\ell_a}}_1^{N+1}$
as described in \eqref{definition correpondance entiers ella et particules-trous} takes the form
\beq
\f{  \abs{ \bra{\psi\pa{\paa{\mu_{\ell_a}}_1^{N+1}}} \Phi^{\dagger}\!\pa{0,0} \ket{ \psi\pa{\paa{\la_a}_1^N}  } }^2 }
{ \norm{\psi\pa{\paa{\mu_{\ell_a}}_{1}^{N+1}}}^2  \norm{\psi\pa{\paa{\la_a}_{1}^{N}}}^2     } =
\wh{\mc{G}}_{N;1} \pab{ \! \!   \paa{p_a}_1^n \! \!  }{ \! \!  \paa{h_a}_1^n \! \!  } \pac{\wh{F}_{\paa{\ell_a}}, \wh{\xi}_{\paa{\ell_a}}, \wh{\xi} \, }
\cdot \wh{D}_{N}\pab{ \! \!   \paa{p_a}_1^n \! \!  }{ \! \!  \paa{h_a}_1^n \! \!  } \pac{\wh{F}_{\paa{\ell_a}}, \wh{\xi}_{\paa{\ell_a}}, \wh{\xi} \, }
 \;.
\label{ecriture FF exact}
\enq
This representation involves two functionals, the so-called smooth part of the form factor $\wh{\mc{G}}_{N;1}$ and the so-called discreet part
$\wh{D}_{N}$. These are functionals of the counting function $\wh{\xi}$ for the ground state, of the counting function $\wh{\xi}_{\paa{\ell_a}}$
for the excited state and of the associated shift function $\wh{F}_{\paa{\ell_a}}$.

It has been show in \cite{KozFFConjFieldNLSELatticeSpacingGoes0}, that, in the large $L$-limit and for
 any $n$ particle-hole type excited state, with $n$ bounded independently of $L$, these functionals
satisfy
\beq
\pa{ \wh{\mc{G}}_{N;1} \wh{D}_{N}}\pab{ \! \!   \paa{p_a}_1^n \! \!  }{ \! \!  \paa{h_a}_1^n \! \!  }
\pac{\wh{F}_{\paa{\ell_a}}, \wh{\xi}_{\paa{\ell_a}}, \wh{\xi} \, }
%
%
=
\wh{\mc{G}}_{N;1} \pab{ \! \!   \paa{p_a}_1^n \! \!  }{ \! \!  \paa{h_a}_1^n \! \!  } \pac{F_0, \xi, \xi_{F_0} }
\cdot \wh{D}_{N}\pab{ \! \!   \paa{p_a}_1^n \! \!  }{ \! \!  \paa{h_a}_1^n \! \!  } \pac{F_0, \xi, \xi_{F_0} }
\pa{1+\e{O}\paf{\ln L}{L}} \;.
\label{ecriture comportement FF part trou a grand L}
\enq
We stress that the functionals appearing on the \textit{rhs} of the above equation act on
\begin{itemize}
\item[i)]  \textit{the thermodynamic limit} $F_{0}\!\pa{\la}$ of the shift function at $\be=0$ associated to the excited state labeled by
the set of integers $\{ \ell_a \}_1^{N+1}$ \eqref{ecriture limite thermo fction shift},
\item[ii)]  \textit{the thermodynamic limit} $\xi\pa{\la}$ of the counting function \eqref{ecriture limite thermo fction comptage},
\item[iii)] the counting function associated  with $F_0$: $\xi_{F_0}\!\pa{\la} = \xi\!\pa{\la} + \tf{F_0\!\pa{\la}}{L}$.
\end{itemize}
We do stress that the shift function $F_{0}$ depends implicitly on the rapidities of the particles
$\{ \mu_{p_a} \}_1^n$ and holes $\paa{\mu_{h_a}}_1^n$ entering in the description of the excited state of interest, \textit{cf}
\eqref{ecriture limite thermo fction shift}.
We chose not to write this dependence explicitly in \eqref{ecriture comportement FF part trou a grand L}
as the auxiliary arguments of $F_{0}$ are undercurrent by those of the functionals
$\wh{D}_{N}$ and $\wh{\mc{G}}_{N,1}$.
Given any holomorphic function $\nu\pa{\la}$ in a neighborhood of $\R$, the explicit expressions for $\wh{D}_{N}\pac{\nu, \xi, \xi_{\nu}}$
(and $\wh{\mc{G}}_{N}\pac{\nu, \xi, \xi_{\nu}}$) involves two sets of parameters $\{ \la_a \}_1^N$ and $\{\mu_{\ell_a}\}_1^{N+1}$
which are \textit{defined} as follows
\begin{itemize}
\item $\mu_k$, $k\in \mathbb{Z}$ is the unique\symbolfootnote[1]{The uniqueness follows from the fact that the dressed momentum $p\pa{\la}$
is a biholomorphism on some sufficiently narrow strip $U_{\de}$ around the real axis and that $p\pa{\la} \in \R \Rightarrow \la \in \R$.}  solution to $\xi\pa{\mu_{k}}=\tf{k}{L}$, \textit{ie}
the \textit{second} argument of the functionals;
\item $\la_k$, $k \in \intn{1}{N}$ is the unique\symbolfootnote[2]{The uniqueness follows from Rouch\'{e}'s theorem when L is large enough.}
 solution to $\xi_{ \nu }\!\pa{\la_k}=\tf{k}{L}$, \textit{ie}
the \textit{third} argument of the functionals.
\end{itemize}
We insist that here and in the following, the parameters $\mu_k$ or $\la_p$ entering in the explicit expressions for these functionals are \textit{always}
to be understood in this way. Also, we remind that the integers $\ell_a$ are obtained from the integers $\{p_a\}_1^n$ and $\{h_a\}_1^n$ as explained in
\eqref{definition correpondance entiers ella et particules-trous}.

\subsubsection*{$\bullet$ The discreet part}

\vspace{2mm}

The functional $\wh{D}_{N}$ represents the universal part of the form-factor:
\beq
\wh{D}_{N}\pab{ \!\! \paa{p_a}_1^n \!\! }{ \!\! \paa{h_a}_1^n \!\! } \pac{F_{0}, \xi, \xi_{F_{0}} }  =
\f{ \prod_{k=1}^{N} \paa{4 \sin^{2}\!\pac{\pi F_{0}\!\pa{\la_k}}}  }
{ \pl{a=1}{N+1} 2\pi L \xi^{\prime}_{}\!\pa{\mu_{\ell_a}}  \pl{a=1}{N}  2\pi L \xi^{\prime}_{F_{0}}\! \pa{\la_a}  }
\pl{a=1}{N} \paf{\mu_{\ell_a}-\mu_{\ell_{N+1}} }{ \la_a-\mu_{\ell_{N+1}} }^2 \det_{N}^2\pac{ \f{1}{\mu_{\ell_a}-\la_b} }  \;.
\label{ecriture explicite fonctionnelle D+}
\enq
The large $N,L$  behavior of \eqref{ecriture explicite fonctionnelle D+} can be computed explicitly and
is given in \eqref{appendix thermo lim FF ecriture explicite thermo lim Dn hat}-\eqref{definition fonctionelle RNn}. However, it is the above
finite product representation of $\wh{D}_{N}$ that is suited for carrying out resummations.

\subsubsection*{  $\bullet$ The smooth part}

The functional $\wh{\mc{G}}_{N,1}$ represents the so-called smooth part of the form factor:
\bem
\hspace{-1cm}\wh{\mc{G}}_{N;\ga} \pab{ \! \!   \paa{p_a} \! \!  }{ \! \!  \paa{h_a} \! \!  } \pac{F_0, \xi, \xi_{F_0} } =
\f{  V_{N;1} \!\pa{\mu_{N+1}} V_{N;-1} \!\pa{\mu_{N+1}}   }{ \ddet{N+1}{ \Xi^{\pa{\mu}} \pac{\xi} } \ddet{N}{ \Xi^{\pa{\la}}\pac{\xi_{F_0}} } }
W_n\pab{ \{\mu_{p_a} \}_1^n }{ \{\mu_{h_a} \}_1^n }  \pl{a=1}{n} \pl{\eps=\pm}{} \paa{  \f{ V_{N;\eps} (\mu_{p_a}) }{ V_{N;\eps} (\mu_{h_a}) }
 \f{ \mu_{h_a}-\mu_{N+1} +i \eps c }{ \mu_{p_a}-\mu_{N+1} +i \eps c  } }
 \\
\times   W_N\pab{ \{\la_a \}_1^N }{ \{\mu_a \}_1^N }
 \ddet{N}{ \de_{jk}+ \ga \wh{V}_{\! jk}\!\pac{F_0}\!\pa{ \{\la_a \}_1^N ; \{\mu_{\ell_a} \}_1^{N+1} }  }
 \ddet{N}{ \de_{jk}+ \ga \wh{\ov{V}}_{\! jk}\! \pac{F_0}\!\pa{ \{\la_a \}_1^N ; \{\mu_{\ell_a} \}_1^{N+1} }  }
 \;.
%
\label{ecriture fonctionnelle hat GN gamma}
\end{multline}
Above, we have introduced several functions.  For any set of generic parameters
$\pa{ \{ z_a \}_1^n ; \{ y_a \}_1^{n} } \in U_{\de}^n\times U_{\de}^{n}$
\beq
W_n\pab{ \{ z_a \}_1^n}{ \{ y_a \}_1^{n} } = \pl{a,b=1}{n} \f{ \pa{z_a-y_b-ic}\pa{y_a-z_b-ic} }{ \pa{y_a-y_b-ic}\pa{z_a-z_b-ic} }
\qquad \e{and} \qquad
V_{N;\eps} \pa{\om} = \pl{a=1}{N} \f{ \om -\la_b +i\eps c }{ \om -\mu_b +i\eps c }
\label{definition fonction WN et VNepsilon}
\enq
 Also we have set
\beq
\Xi^{\pa{\mu}}_{jk} \pac{\xi} = \de_{jk} - \f{  K \pa{\mu_{\ell_a} -\mu_{\ell_b} } }{ 2\pi L \, \xi^{\prime}\!\pa{\mu_{\ell_b}} }
\qquad \e{and} \qquad
\Xi^{\pa{\la}}_{jk} \pac{\xi_{F_0}} = \de_{jk} - \f{  K \pa{\la_{a} -\la_{b} } }{ 2\pi L \, \xi^{\prime}_{F_0}\!\!\pa{\la_{b}} }
\enq

Finally, for any set of generic parameters $\pa{ \{ z_a \}_1^n ; \{ y_a \}_1^{n+1} } \in U_{\de}^n\times U_{\de}^{n+1}$
the entries of the two determinants in the numerator read
\beqa
\wh{V}_{\! k\ell}\!\pac{\nu}\pa{ \{ z_a \}_1^n ; \{ y_a \}_1^{n+1} }  &=& - i\,  \f{ \pl{a=1}{n+1} \pa{ z_k-y_a}}{ \pl{a\not= k }{ n} \pa{ z_k-z_a} }
\f{ \pl{a=1}{n} \pa{ z_k-z_a +ic} }{ \pl{a=1}{n+1} \pa{z_k-y_a+ic}}  \f{ K\pa{z_k-z_{\ell}} }{ \ex{-2i\pi \nu\pa{z_k}}-1 }  \nonumber \\
\wh{\ov{V}}_{\! k\ell}\!\pac{\nu}\pa{ \{ z_a \}_1^n ; \{ y_a \}_1^{n+1} }  &=&  i \, \f{ \pl{a=1}{n+1} \pa{z_k-y_a}}{ \pl{a\not= k }{ n} \pa{z_k-z_a} }
\f{ \pl{a=1}{n} \pa{z_k-z_a - ic} }{ \pl{a=1}{n+1} \pa{z_k-y_a - ic}}  \f{ K\pa{z_k-z_{\ell}} }{ \ex{2i\pi  \nu\pa{z_k}}-1 }
\label{appendix thermo lim FF definition entree V et Vbar chapeau}
\eeqa
Note that the singularities of the associated determinants at $z_k=z_j$, $j \not=k$ are only apparent, \textit{cf}
\cite{KozKitMailSlaTerXXZsgZsgZAsymptotics,KozFFConjFieldNLSELatticeSpacingGoes0}.

\subsection{Arguments for the effective form factors series}
\label{Subsection effective FF series}

It is believed\symbolfootnote[2]{ The computations presented in appendices \ref{Appendix Subsection generating function FF sums} and
\ref{Appendix Subsection thermo limit of the Fredholm Minor} can be seen as a proof of this statement in the case of a generalized free fermion model.}
that when computing the $T=0K$ form factor expansion of a two-point function $\bra{G.S.}O_1 O_2 \ket{G.S.}$
on the intermediate excited states (as in \eqref{equation developement correlateur serie FF}), the contribution of those excited states
whose energies differ macroscopically  from the ground state's one (\textit{ie} by a quantity scaling as some positive power of $L$) vanishes in the
$L\tend +\infty$ limit. This can, for instance, be attributed to an extremely quick oscillation of the phase factors
and the decay of form factors for states having large excitation momenta and energies.
Therefore, we shall assume in the following that the only part of the form factor expansion in \eqref{equation developement correlateur serie FF}
that has a non-vanishing contribution to the thermodynamic limit $\rho\pa{x,t}$ of $\rho_N\pa{x,t}$ corresponds to a summation over all those
excited states which are realized as some finite (in the sense that not scaling with $L$) number $n$, $n=0,1,\dots,$  of particle-hole excitations above
the $\pa{N+1}$-particle ground state. Indeed, these are the only excited states that can have a finite (\textit{ie} not scaling with $L$) energy gap
above the ground state in the $N$-particle sector.

\vspace{2mm}

Even when dealing with excited states realized as a finite number $n$ of particle-hole excitations above the $\pa{N+1}$-particle ground state, it is still
possible to generate a macroscopically different energy from the one of the $N$-particle ground state
if the rapidities of the particles become very large (\textit{ie} scale with $L$).
This case corresponds, among others, to integers $p_a$ becoming very large and scaling with $L$. We will drop the
contribution of such excited states  in the following.

\vspace{2mm}

Limiting the sum over all the excited states in the $\pa{N+1}$-particle sector to those having the same per-site energy
that the ground state means that one effectively neglects correcting terms in the lattice size $L$.
It thus seems very reasonable to assume that, on the same ground, only the leading large-$L$ asymptotic behavior
of the form factors will contribute to the thermodynamic limit of $\rho_{N}\pa{x,t}$. It is clearly so
when focusing on  states with a low number $n$ of particle/hole excitations.
However, in principle, problems could arise when the number $n$ becomes of the order of $L$.
Our assumption lead to the following consequences:

\begin{itemize}

\item we discard all summations over the excited states having a too large excitation energy. This means that we introduce
a "cut-off" in respect to the range of the integers entering in the description of the rapidities of the particles. Namely, we assume that
the integers $p_a$ are restricted to belong to the set\symbolfootnote[3]{Note that we could choose $w_L$ to scale as $L^{1+\eps}$, where $\eps>0$ is small enough
but arbitrary otherwise. We choose $\eps=\tf{1}{4}$ for definiteness. \textit{cf} appendix \ref{Appendix computation of singular sums} for a better
discussion of the origin of such a property.}
\beq
 \mc{B}^{\e{ext}}_L \equiv \Big\{ n \in \mathbb{Z} \; : \;  -w_L < n < w_L  \Big\} \setminus \intn{1}{N+1} \qquad \e{where} \qquad
w_L \sim  L^{1+\f{1}{4}}   \;.
\enq

\item The oscillating exponent $\sul{a=1}{N+1}u_0\pa{\mu_{\ell_a}}- \sul{a=1}{N}u_0\pa{\la_a} $
is replaced by its thermodynamic limit as given in \eqref{ecriture limite thermo exposant osc}.


\item We drop the contribution of the $\e{O}\pa{L^{-1} \cdot \ln L }$ terms in the large-size behavior of form factors
given in \eqref{ecriture comportement FF part trou a grand L}. 


\end{itemize}

Note that, within our approximations, the localization  of the Bethe roots $\paa{\mu_{\ell_a}}_1^{N+1}$ for an excited state whose particles' (resp. holes')
rapidities are labeled by the integers  $\paa{p_a}_{a=1}^{n}$ (resp. $\paa{h_a}_{a=1}^{n}$)
\textit{does not depend} on the specific choice of the excited state one considers.
Hence, we effectively recover a description of the excitations that is in the spirit of a free fermionic model.

Our simplifying hypothesis suggest to raise the below conjecture
\begin{conj}
\label{Conjecture Fondamentale}
The thermodynamic limit of the reduced density matrix $\rho_N\pa{x,t}$ coincides with the thermodynamic limit
of the effective reduced density matrix $\rho_{N;\e{eff}}\pa{x,t}$:
\beq
\lim_{N,L \tend +\infty} \rho_N\pa{x,t}=
\lim_{N,L \tend +\infty}   \rho_{N;\e{eff}}\pa{x,t}
\enq
where $\rho_{N;\e{eff}}\pa{x,t}$ is given by the series
\beq
\hspace{-5mm} \rho_{N;\e{eff}}\pa{x,t} =  \sul{n = 0}{ N+1 }
\sul{ \substack{p_1<\dots < p_n \\ p_a \in \mc{B}^{\e{ext}}_{L}  } }{}
\sul{ \substack{h_1<\dots < h_n \\ h_a  \in \mc{B}^{\e{int}}_{L}  } }{}
\pl{a=1}{n} \f{ \ex{ -ixu\pa{\mu_{h_a}} } }{ \ex{-ixu\pa{\mu_{p_a}} } }  \cdot
\pa{ \wh{D}_{N} \; \wh{\mc{G}}_{N; 1 } } \pab{ \!\! \paa{p_a}_1^n \!\!  }{  \!\! \paa{h_a}_1^n \!\!  }
\pac{  F_{0}\pabb{ * }{ \{ \mu_{p_a} \} } {\paa{\mu_{h_a}} } \; ; \xi_{} \; ; \xi_{ F_{0}}  } \;.
\label{ecriture series FF effective pour sommation}
\enq
There $\mc{B}_L=\paa{n\in \mathbb{Z} \; : \; -w_L < n < w_L}$, \; $\mc{B}_L^{\e{ext}}=\mc{B}_L\setminus \intn{1}{N+1}$ and
$\mc{B}_L^{\e{int}}=\intn{1}{N+1}$. Also, the $*$ refers to the running variable of $F_0$ on which the two functionals act.

%
%
%
%
%
%
%
%
%

\end{conj}

The effective form factor series  \eqref{ecriture series FF effective pour sommation}
possesses several
different features in respect to the form factor expansion-based series that would appear in a generalized free fermion model (\textit{cf}
\eqref{definition Fonction generatrice X_N}). Namely,

\begin{itemize}

\item the shift function $F_{0}$ depends parametrically on the
rapidities of the particles and holes entering in the description of \textit{each} excited state one considers, \textit{cf}
\eqref{ecriture limite thermo fction shift}.  It is thus summation \textit{dependent}.

\item  Each summand is weighted by the factor  $\wh{\mc{G}}_{N;1}$  that takes into account the more complex structure of the scattering and of the scalar
products in the interacting model.  This introduces a strong coupling between the summation variables $\{p_a\}_1^n$ and $\{h_a\}_1^n$.
Indeed, the explicit expression for  $\wh{\mc{G}}_{N;1}$ involves complicated functions of the rapidities $\{ \mu_{p_a} \}_1^n$ and
$\{ \mu_{h_a} \}_1^n$, which, in their turn, depend on the aforementioned integers.

\end{itemize}

\vspace{2mm}

A separation of variables that would allow one for a resummation of \eqref{ecriture series FF effective pour sommation} is not
possible for precisely these two reasons. To overcome this problem, we proceed in several steps.
First, we introduce a $\ga$-deformation of the effective form factor series such that
 $\rho_{N;\e{eff}}\pa{x,t\mid \ga}_{\mid \ga=1}=\rho_{N;\e{eff}}\pa{x,t}$:
\beq
\rho_{N;\e{eff}}\!\pa{x,t\mid \ga} =  \sul{n = 0}{ N +1}
\sul{ \substack{p_1<\dots < p_n \\ p_a \in \mc{B}^{\e{ext}}_{L}  } }{}
\sul{ \substack{h_1<\dots < h_n \\ h_a  \in \mc{B}^{\e{int}}_{L}  } }{}
\pl{a=1}{n} \f{ \ex{ -ix u\pa{\mu_{h_a}} } }{ \ex{-ix u\pa{\mu_{p_a}} } }
\pa{ \wh{D}_{N} \; \wh{\mc{G}}_{N;\ga} } \pab{ \!\! \paa{p_a}_1^n \!\!  }{  \!\! \paa{h_a}_1^n \!\!  }
\pac{ \ga F_{0}\pabb{ * }{ \{ \mu_{p_a} \}_1^n } {\paa{\mu_{h_a}}_1^n } \; ; \xi_{} \; ; \xi_{\ga F_{0}}  }
%
%
\;.
\label{ecriture series FF effective gamma deformee}
\enq
For any finite $N$ and $L$, it is readily checked by using the explicit representations \eqref{ecriture explicite fonctionnelle D+} for $\wh{D}_{N}$
and \eqref{ecriture fonctionnelle hat GN gamma} for $\wh{\mc{G}}_{N;\ga}$
that the $\ga$-deformation $\rho_{N;\e{eff}}\pa{x,t\mid \ga}$ is holomorphic in $\ga$ belonging to an open neighborhood of the
closed unit disc\symbolfootnote[2]{The apperent singularity of the determinants at $\ex{\pm 2i\pi F_0\,(\la_k)}-1=0$,
\textit{cf} \eqref{appendix thermo lim FF definition entree V et Vbar chapeau}, are candelled by the pre-factors 
$\sin^{2}\![ \pi F_0(\la_k) ]$ present in $\wh{D}_N$, \textit{cf} \eqref{ecriture explicite fonctionnelle D+}.}. 
Hence, its Taylor series around $\ga=0$ converges up to $\ga=1$.
We will then show in theorem \ref{Theorem representation series Fredholm multidim} that, given any fixed $m$, the $m^{\e{th}}$ Taylor coefficient of
$\rho_{N;\e{eff}}\!\pa{x,t\mid \ga}$ at $\ga=0$:
\beq
\rho_{N;\e{eff}}^{\pa{m}}\!\pa{x,t} = \left. \f{\Dp{}^m}{\Dp{}\ga^m} \rho_{N;\e{eff}}\!\pa{x,t\mid \ga}   \right|_{\ga=0} \;,
\label{definition rho N eff derivee m ieme}
\enq
can be re-summed into a representation where the existence of the  thermodynamic limit
$\rho_{\e{eff}}^{\pa{m}}\!\pa{x,t}$ is readily seen. This fact is absolutely not-clear on the level of
\eqref{definition rho N eff derivee m ieme} as, due to \eqref{AppendixThermoLimD+zero}-\eqref{definition fonctionelle RNn},
each individual summand vanishes as a complicated power-law in L that depends on the excited state considered.
We will then show that one can represent
the thermodynamic limit $\rho_{\e{eff}}^{\pa{m}}\!\pa{x,t}$ in another way. This representation is given in terms of a finite sum of multiple integrals
and corresponds to a truncation of the so-called multidimensional Natte series that we introduce below.
 The latter description of $\rho_{\e{eff}}^{\pa{m}}\!\pa{x,t}$ gives a straightforward access to its asymptotic expansion.

The proof of the existence of the thermodynamic limit and the construction of the truncated multidimensional Natte series for
$\rho^{\pa{m}}_{\e{eff}}\!\pa{x,t}$ constitute the rigorous and conjecture free part of our
analysis. This is summarized in theorem \ref{Theorem comportement asympt coeff Taylor limite Thermo}.

 Working on the Taylor coefficients $\rho^{\pa{m}}_{N;\e{eff}}\!\pa{x,t}$ instead of the full function $\rho_{N;\e{eff}}\!\pa{x,t\mid \ga}$
taken at $\ga=1$ has the advantage of separating all questions of convergence of the representations we
 obtain from the question of well-definiteness of the various re-summations and deformation procedures that we carry out
on $\rho_{N;\e{eff}}^{\pa{m}}\!\pa{x,t}$ (and subsequently on $\rho_{\e{eff}}^{\pa{m}}\!\pa{x,t}$ once that the thermodynamic limit is taken).
Indeed, by taking the $m^{\e{th}}$ $\ga$-derivative at $\ga=0$, we always end up dealing with a finite number of sums.
However, if we had carried out the forthcoming re-summation directly on the level of $\rho_{\e{eff}}\!\pa{x,t}$, we would have ended up with a series of
multiple integrals instead of a finite sum. The convergence of such a series constitutes a separate question that deserves, in its own right,
another study. Nonetheless, in the present paper, in order to provide physically interesting results, we will take this convergence as a
reasonable conjecture in a subsequent part of the paper.

\subsection{An operator ordering}
\label{Subsection Operator ordering for functional translation}

Prior to carrying out the re-summation of the form factor expansion for $\rho_{N;\e{eff}}^{\pa{m}}\!\pa{x,t}$,
we need to discuss  a way of representing functional translations and generalizations thereof. These objects will allow us
to separate the variables in the sums occuring in \eqref{definition rho N eff derivee m ieme}, and carry out the various re-summations.
A more precise analysis and discussion of these constructions is postponed to appendix \ref{Appendice section functional translation}.
In the following, we denote by $\msc{O}\pa{W}$, the ring of holomorphic functions in $\ell$ variables on $W \subset \Cx^{\ell}$.
Also, here and in the following $f \in \msc{O}\pa{W}$, with $W$ non-open means that $f$ is a holomorphic function
on some open neighborhood of $W$. Finally, for a set $S$ on which the function $f$ is defined
we denote $\norm{f}_S = \sup_{s \in S} \abs{f\!\pa{s}}$.

\vspace{2mm}
Throughout this paper we will deal with various examples ($\wh{D}_{N}$ , $\wh{\mc{G}}_{N}^{\pa{\be}}$, $\dots$)
of functionals $\mc{F}\pac{\nu}$ acting on holomorphic functions $\nu$.
The function $\nu$ will always be defined on some compact subset $M$ of $\Cx$ whereas the explicit expression for $\mc{F}\pac{\nu}$
will only involve the values taken by $\nu$  on a smaller compact\symbolfootnote[3]{Here and in the following, 
$\e{Int}\pa{M}$ stands for the interior of the set M.} $K \subset \e{Int}\pa{M}$.
In fact, all the functionals that we will consider share the regularity property below:

\begin{defin}
\label{Definition Fonctionelle reguliere}
Let $M$, $K$ be compacts in $\Cx$ such that $K \subset \e{Int}\pa{M}$. Let $ W_{z}$ be a compact in $ \Cx^{\ell_{z}}$, $\ell_z \in \mathbb{N}\equiv\paa{0,1,\dots}$. An $\ell_z$-parameter family of functionals $\mc{F}\pac{\cdot}\pa{\bs{z}}$
depending on a set of auxiliary variables $\bs{z} \in W_{z}$
is said to be regular (in respect to the pair $\pa{M,K}$) if

\begin{enumerate}[i)]

\item there exists constants  $C_{\mc{F}}>0$ and $C^{\prime}>0$ such that for any $f,g \in \msc{O}\pa{M}$
\beq
\norm{f}_K +\norm{g}_K <C_{\mc{F}} \quad \Rightarrow \quad
\norm{ \mc{F}\pac{f}\pa{\cdot}-\mc{F}\pac{g} \pa{\cdot} }_{ W_z } < C^{\prime} \norm{f-g}_K \;,
\label{Appendice transl fonct Conte fnelle}
\enq
where the $\cdot$ indicates that the norm is computed in respect to the set of auxiliary variables $ \bs{z} \in W_{ z }$.

\item Given any open neighborhood  $W_y$ of $0$ in $\Cx^{\ell_y}$, for some $\ell_y \in \mathbb{N}$,
if $\nu\pa{\la,\bs{y} } \in \msc{O}\pa{M \times W_y}$ is such that $\norm{\nu}_{K\times W_y} <C_{\mc{F}}$, then
the function $\pa{\bs{y},\bs{z}} \mapsto \mc{F}\pac{\nu\pa{ *, \bs{y}}}\pa{\bs{z}}$ is holomorphic on $W_y\times W_z$.
Here, the $*$ indicated the running variable $\la$ of $\nu\pa{\la,\bs{y}}$ on which the functional $\mc{F}\pac{\cdot}\pa{\bs{z}}$ acts.
\end{enumerate}

\noindent The constant $C_{\mc{F}}$ appearing above will be called constant of regularity of the functional.

\end{defin}

This regularity property is at the heart of the aforementioned representation for the functional translation and generalizations thereof
that we briefly discuss below. However, prior to this discussion we need to define the discretization of the boundary of a compact.

\begin{defin}
\label{Definition point discretisation}
Let $M$ be a compact with $n$ holes (\textit{ie} $\Cx\setminus M$ has $n$ bounded connected components) and such that
$\Dp{}M$ can be realized as a disjoint union of $n+1$ smooth Jordan curves $\ga_{a} : \intff{0}{1} \tend \Dp{}M$, \textit{ie}
$\Dp{}M=\bigsqcup_{a=1}^{n+1} \ga_a\pa{ \intff{0}{1}}$. A discretization (of order $s$) of $\Dp{}M$ will correspond to
a collection of $\pa{n+1}\pa{s+2}$ points $t_{j,a}=\ga_a(x_j)$ with $j=0,\dots,s+1$ and  $a=1,\dots,n+1$
 where $x_0=0\leq x_1<\dots<x_{s}\leq 1=x_{s+1}$ is a partition of $\intff{0}{1}$
of mesh $\tf{2}{s}$: $\abs{x_{j+1}-x_j} \leq \tf{2}{s}$.

\end{defin}

\subsubsection{Translations}

Suppose that one is given a compact $M$ in $\Cx$ without holes whose boundary is a smooth Jordan curve $\ga :  \intff{0}{1} \tend \Dp{}M$.
Let $K$ be a compact such that $K \subset \e{Int}\pa{M}$ and  $\mc{F}$ a regular functional (\textit{cf} definition
\ref{Definition Fonctionelle reguliere}) in respect to $\pa{M,K}$, for simplicity, not depending on auxiliary parameters $\bs{z}$.

It is shown in proposition \ref{Proposition translation pure avec dependence parametre auxiliaire} that,
then, for $\abs{\ga}$ small enough one has the identity
\beq
  \mc{F}\pac{ \ga W_n \pabb{ * }{  \{y_a \}_1^n  } {\paa{ z_a }_1^n } }   =
 \lim_{s\tend +\infty} \; \paa{  \pl{a=1}{n} \ex{ \wh{g}_s\pa{ y_a }- \wh{g}_s\pa{ z_a } } \; \cdot \;
 \mc{F}\pac{\ga f_s}    }_{ \left| \vsg_k=0  \right. }   \hspace{-3mm}.
\label{ecriture action operateur translation sur fnelle generique}
\enq
The function $W_n$ appearing above is defined in terms of an auxiliary function $\psi\pa{\la,\mu}$ that is holomorphic
on $M \times M$
\beq
W_n \pabb{ \la }{  \{y_a \}_1^n  } {\paa{ z_a }_1^n }   = \sul{a=1}{n} \psi ( \la , y_a  ) - \psi\pa{\la,z_a}
\quad \e{whereas} \quad
f_{s}\pa{\la\mid \paa{\vsg_a}_1^s}\equiv f_s\pa{\la}=  \sul{j=1}{s} \f{ \pa{t_{j+1}-t_j} }{t_j-\la}  \cdot \f{\vsg_j}{2i\pi} \; .
\label{definition fonction W et fs dans section bulk papier}
\enq
Finally, $\wh{g}_s\pa{\la} $  is a differential operator in respect to $\vsg_a$, with $a=1,\dots, s$:
\beq
\wh{g}_s\pa{\la} =  \sul{j=1}{s}  \psi\big( t_j,\la \big) \f{ \Dp{} }{ \Dp{}\vsg_j }  \;.
\enq
The definitions of $\wh{g}_s$ and $f_s$  involve a set of $s+1$ discretization points $t_j$ of $\Dp{}M$.

\vspace{2mm}
The limit in \eqref{ecriture action operateur translation sur fnelle generique} is uniform in the parameters $y_a$ and $z_a$ belonging to $M$
and in $\abs{\ga}$ small enough. Actually, the magnitude of $\ga$ depends on
the value of the constant of regularity $C_{\mc{F}}$. If the latter is large enough, one can even set $\ga=1$.
The limit in \eqref{ecriture action operateur translation sur fnelle generique} also holds uniformly in respect to any finite
order partial derivative of the auxiliary parameters. In particular,
\beq
\pl{a=1}{n} \bigg\{ \f{\Dp{}^{p_a} }{ \Dp{}y_a^{p_a}} \f{\Dp{}^{h_a} }{ \Dp{}z_a^{h_a}}  \bigg\} \cdot   \f{ \Dp{}^m }{ \Dp{}\ga^m }
\mc{F}\pac{ \ga W_n \pabb{ *  }{  \{\mu_{p_a} \}  } {\paa{\mu_{h_a}} } } _{ \mid \ga=0} =
 \lim_{s\tend +\infty} \paa{  \pl{a=1}{n} \bigg\{ \f{\Dp{}^{p_a} }{ \Dp{}y_a^{p_a}} \f{\Dp{}^{h_a} }{ \Dp{}z_a^{h_a}}  \bigg\}
 \pl{a=1}{n} \ex{ \wh{g}_s\pa{y_a}- \wh{g}_s\pa{z_a} } \; \cdot \;
\f{ \Dp{}^m }{ \Dp{}\ga^m } \mc{F}\pac{\ga f_s}    }_{ \left| \substack{ \vsg_k=0 \\ \ga=0} \right. }   \;.
\enq
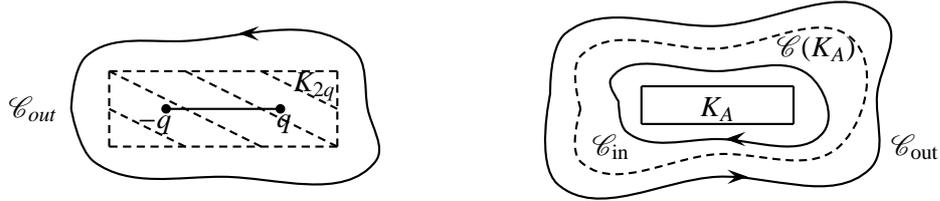
\begin{figure}[h]
\begin{center}

\begin{pspicture}(12.5,4)




\psline[linestyle=dashed, dash=3pt 2pt](2,2.5)(2,1.5)
\psline[linestyle=dashed, dash=3pt 2pt](2,1.5)(5,1.5)
\psline[linestyle=dashed, dash=3pt 2pt](5,1.5)(5,2.5)
\psline[linestyle=dashed, dash=3pt 2pt](5,2.5)(2,2.5)

\pscurve(1.5,2)(1.7,2.7)(3,2.9)(5.2,2.8)(5.4,2)(5.5,1.3)(3.5,1.1)(1.8,1.3)(1.5,2)

\psline[linestyle=dashed, dash=3pt 2pt](2,2)(3,1.5)
\psline[linestyle=dashed, dash=3pt 2pt](2,2.5)(4,1.5)
\psline[linestyle=dashed, dash=3pt 2pt](3,2.5)(5,1.5)
\psline[linestyle=dashed, dash=3pt 2pt](4,2.5)(5,2)

\psline(2.75,2)(4.25,2)
\psdots(2.75,2)(4.25,2)
\rput(2.6,1.8){$-q$}
\rput(4.3,1.8){$q$}

\rput(4.7,2.3){$K_{2q}$}
\rput(1,2){$\msc{C}_{out}$}

\psline[linewidth=2pt]{->}(3.8,3)(3.7,3)


\psline(9,2.3)(9,1.8)
\psline(9,1.8)(11,1.8)
\psline(11,1.8)(11,2.3)
\psline(11,2.3)(9,2.3)

\pscurve(8.7,2.1)(8.8,1.7)(10,1.6)(11.4,1.7)(11.2,2.45)(10,2.5)(8.6,2.4)(8.7,2.1)

\pscurve[linestyle=dashed, dash=3pt 2pt](8.2,2)(8.2,1.3)(10,1.4)(11.6,1.5)(11.8,2.9)(10,2.8)(8.2,2.7)(8.2,2)

\pscurve(7.8,2.2)(7.9,1)(10,1.1)(12,1.2)(12.1,2.2)(12.2,3.2)(10,3.1)(8,3)(7.8,2.2)

\psline[linewidth=2pt]{->}(10.3,1.1)(10.4,1.1)

\psline[linewidth=2pt]{->}(10.2,1.6)(10.1,1.6)

\rput(10,2){$K_A$}
\rput(11.3,2.8){$\msc{C}\!\pa{K_A}$}

\rput(8.6,1.5){$\msc{C}_{\e{in}}$}
\rput(12.6,1.5){$\msc{C}_{\e{out}}$}

\end{pspicture}

\caption{Example of discretized contours. In the $lhs$ the compact $M$ is located inside of its boundary $\msc{C}_{out}$
whereas the compact $K$ corresponds to $K_{2q}$ as defined in \eqref{definition compact KA}. In this case $M$ has no holes.
In the $rhs$ the compact $M$ is  delimited by the two Jordan curves $\msc{C}_{in}$ and $\msc{C}_{out}$ depicted in solid lines.
The associated compact $K$ (of definition \ref{Definition Fonctionelle reguliere}) corresponds to the loop $\msc{C}\pa{K_A}$ depicted by dotted lines.
The compact $M$ depicted in the \textit{rhs} has one hole. This hole contains a compact $K_A$ inside.
\label{contour exemple de courbes encerclantes} }
\end{center}
\end{figure}

We refer to appendix \ref{Appendice section functional translation} for a proof of the above statement.
Here, we would like to describe in words how formula \eqref{ecriture action operateur translation sur fnelle generique} works.
By properly tuning the value of $\ga$ and invoking the regularity property of the functional $\mc{F}\pac{\ga f_s}$
one gets that, for any $s$, $\paa{\vsg_a}_1^s \mapsto \mc{F}\pac{\ga f_s}$ is holomorphic in a sufficiently large neighborhood of $0 \in \Cx^s$.
This allows one to act with the translation operators $\prod_{b=1}^n\ex{\wh{g}_s(y_b) -\wh{g}_s(z_b) }$.
Their action replaces each variable $\vsg_a$  occurring in $f _s$ by the combination
$\sum_{b=1}^{n} \pac{ \psi\pa{t_a,y_b}-\psi (t_a,z_b) } $.
Taking the limit $s\tend +\infty$ changes the sum over $t_a$ occurring in $f_s$ into a contour integral over $\msc{C}_{out}$, \textit{cf} \textit{lhs} of
Fig.~\ref{contour exemple de courbes encerclantes}. Due to the presence
of a pole at $t=\la$, this contour integral exactly reproduces the function $W_n$
that appears in the \textit{rhs} of \eqref{ecriture action operateur translation sur fnelle generique}.

\vspace{2mm}

Note that such a realization of the functional translation can also be build in the case of compacts $M$ having several holes as depicted in
the \textit{rhs} of Fig.~\ref{contour exemple de courbes encerclantes}. Also, there is no problem to consider regular functionals
$\mc{F}\pac{\cdot} \pa{\bs{z}}$ that depend on auxiliary sets of parameters $\bs{z}$.

\subsubsection{Generalization of translations}
\label{Subsubsection generalization of translations}

In the course of our analysis, in addition to dealing with functional translations as defined above,
we will also have to manipulate more involved expressions involving series of partial derivatives.
Namely, assume that one is given a regular functional $\mc{F}\pac{f,g}$ of two arguments $f$ and $g$.
Then, the expression $ \bs{:} \Dp{\ga}^m \mc{F}\pac{\ga f_s, \wh{g}_s }_{\mid \ga=0} \bs{:}$ is to be understood as the left substitution of the various $\Dp{\vsg_a}$ derivatives symbols stemming from $\wh{g}_s$.

More precisely, let $\wt{g}_s$ be the below holomorphic function of $a_1,\dots, a_s$
\beq
\wt{g}_s\pa{\la}=  \sul{j=1}{s} \psi\big( t_j,\la \big) \, a_j \;.
\label{definiton fonction gs scalaire}
\enq
The regularity of the functional $\mc{F}$ ensures that the function $\{ a_p\} \mapsto \Dp{\ga}^m \mc{F}\pac{\ga f_s, \wt{g}_s}$ is holomorphic
in $a_1,\dots, a_s$ small enough. As a consequence, the  below multi-dimensional series is convergent for $a_j$ small enough:
\beq
\f{\Dp{}^m}{\Dp{}\ga^m} \cdot  \mc{F}\pac{\ga f_s, \wt{g}_s} _{\mid \ga=0}=
\sul{n_j \geq 0}{}   \pl{j=1}{s} \paa{ \f{ a_j^{n_j} }{n_j!} \f{ \Dp{}^{n_j} }{ \Dp{}a_j^{n_j}  }  }
\f{\Dp{}^m}{\Dp{}\ga^m} \cdot  \mc{F}\pac{\ga f_s, \wt{g}_s}_{\left| \substack{\ga=0 \\ a_j=0} \right. } \;.
\label{ecriture developpement F nu et g en puissance des aj}
\enq
%
%
%
We stress that as $f_s$ \eqref{definition fonction W et fs dans section bulk papier} is a holomorphic function of $\vsg_1,\dots,\vsg_s$, the functional of $f_s$ coefficients of the above series give rise to a family
of holomorphic functions  in the variables $\vsg_1,\dots, \vsg_s$. This analyticity follows, again, from the regularity of
the functional $\mc{F}\pac{f, g}$ and the smallness of $\abs{\ga}$.

The $\bs{:} \cdot \bs{:}$ ordering constitutes in substituting $a_j  \hookrightarrow \Dp{\vsg_j}$, $j=1,\dots,s$ in such a way that
all differential operators appear to the left. That is to say,
\beq
\bs{:} \f{\Dp{}^m}{\Dp{}\ga^m} \cdot  \mc{F}\pac{\ga f_s, \wh{g}_s} _{ \mid \ga=0} \bs{:} \; \equiv  \;
\sul{n_j \geq 0}{} \pl{j=1}{s}  \paa{  \f{\Dp{}^{n_j} }{\Dp{}\vsg_j^{n_j} } } \cdot \pl{j=1}{s} \paa{ \f{1}{n_j!} \f{ \Dp{}^{n_j} }{ \Dp{}a_j^{n_j}  }  } \cdot
\f{\Dp{}^m}{\Dp{}\ga^m} \cdot \mc{F}\pac{\ga f_s, \wt{g}_s} _{ \left| \substack{\ga=0 \\ a_j=0; \vsg_j=0} \right. } \;.
\label{definition substitution operatorielle}
\enq
Although there where no convergence issues on the level of expansion \eqref{ecriture developpement F nu et g en puissance des aj},
these can \textit{a priori} arise  on the level of the $rhs$ in \eqref{definition substitution operatorielle}.
Clearly, convergence depends on the precise form of the functional $\mc{F}$, and  should thus be
studied on a case-by-case basis. However, in the case of interest to us, this will not be a problem due to
the quite specific class of functionals that we will deal with.

At this point, two observations are in order.
\begin{itemize}
 \item \eqref{definition substitution operatorielle}
bears a strong resemblance with an $s$-dimensional Lagrange series.
\item The functional (of $f_s$) coefficients appearing in the $rhs$ of
\eqref{definition substitution operatorielle} are completely determined by the functional $ \mc{F}\pac{\ga f_s, \wt{g}_s} $
whose expression only involves standard (\textit{ie} non-operator valued) functions. Should this functional have two (or more) equivalent representations, then
any one of them can be used as a starting point for computing the coefficients in \eqref{ecriture developpement F nu et g en puissance des aj}
and then carrying out the substitution \eqref{definition substitution operatorielle}.
\end{itemize}

Actually, for the class of functionals that we focus on, no convergence issues arise. Indeed, in all of the cases,
the $m^{\e{th}}$ $\ga$-derivative at $\ga=0$ of the $\bs{:}\cdot \bs{:}$ ordered functionals of interest appears as a finite linear combinations
(or integrals thereof) of expressions of the type
\beq
\wh{\msc{E}}_m=
\bs{:} \f{ \Dp{}^m }{  \Dp{}\ga^m} \paa{ \pl{a=1}{r} \ex{ \eps_a  \wh{g}_s (\la_{\a_a}) }  \cdot \pl{b=1}{\wt{r} } \ex{ \upsilon_b  \wh{g}_s\pa{ y_b } }
\cdot \mc{F}\pac{\ga f_s}}_{\mid \ga=0} \hspace{-2mm}\bs{:}  \qquad \e{where} \;\; \a_a \in \intn{1}{N} \quad \e{and}
\quad  \eps_a, \upsilon_b \in \paa{\pm 1}\;.
\label{equation explication prise gamma derivee}
\enq
Above $y_a$ are some auxiliary and generic parameters whereas $\la_{\a_a}$ are implicit functions of $\ga$ and $\vsg_1,\dots,\vsg_s$.
For $L$-large enough, $\la_{\a_a}$ is the unique solution  to the equation $\xi_{\ga f_s}\!\pa{\la_{\a_a}}=\tf{\a_a}{L}$.

The prescription that we have agreed upon implies that one should first
substitute $\wh{g}_s \hookrightarrow \wt{g}_s$ as defined in \eqref{definiton fonction gs scalaire}.  Then, one
computes the  $m^{\e{th}}$ $\ga$-derivative at $\ga=0$ of \eqref{equation explication prise gamma derivee}, this
in the presence of non-operator valued functions $\wt{g}_s$.
In the process, one has to differentiate in respect to $\ga$ the functional $\mc{F}\pac{\ga f_s}$ and the arguments of $\wt{g}_s\pa{\la_{\a_a}}$.
Using that ${\la_{\a_a}}_{\mid \ga=0} = \mu_{\a_a}$, one arrives to
\beq
\wt{\msc{E}}_m  \equiv
\f{ \Dp{}^m }{  \Dp{}\ga^m} \paa{ \pl{b=1}{r} \ex{ \eps_b  \wt{g}_s (\la_{\a_b}) }  \pl{b=1}{\wt{r} } \ex{ \upsilon_b  \wt{g}_s\pa{ y_b } }
\cdot \mc{F}\pac{\ga f_s}}_{\mid \ga=0} =   \pl{b=1}{r} \ex{\eps_b \wt{g}_s(\mu_{\a_b})} \pl{b=1}{\wt{r} } \ex{ \upsilon_b  \wt{g}_s\pa{ y_b } }
\sul{n_1,\dots, n_s =0}{m} \pl{j=1}{s}  a_j^{n_j}    \; \cdot  \;  c_{\paa{n_j}}\pac{ f_s} \;.
\label{ecriture mise en place ordre operatoriel}
\enq
The sum is truncated at most at $n_j=m$, $j=1,\dots,m$ due to taking the $m^{\e{th}}$ $\ga$-derivative at $\ga=0$.
It is readily verified that the $\{n_j\}-$dependent coefficients $c_{\paa{n_j}}\pac{ f_s}$ are regular functionals of $f_s$
with sufficiently large constants of regularity.
It remains to impose the operator substitution on the level of \eqref{ecriture mise en place ordre operatoriel}
$a_j \hookrightarrow \Dp{\vsg_j}$ with all differential operators $\Dp{\vsg_k}$, $k=1,\dots,s$ appearing to the left. It is clearly not a problem
to impose such an operator order on the level of the polynomial part of the above expression.
Indeed, the regularity of the functionals $c_{\paa{n_j}}\pac{f_s}$ implies that these are holomorphic in $\vsg_1,\dots,\vsg_s$
belonging to an open neighborhood $\mc{N}_{0}$ of $0\in\Cx^s$. Hence, $\prod_{k=1}^{s} \Dp{\vsg_k}^{m_k} \cdot  c_{\paa{n_j}}\pac{f_s}_{\mid \vsg_k=0} $
is well-defined for any set of integers $\{ m_k \}$. In fact, in all the cases of interest for us, the neighborhood $\mc{N}_0$ is always large enough
so as to make the Taylor series issued from the products of translation operators
 $\prod_{a=1}^{r} \ex{\eps_a \wh{g}_s\pa{\mu_{\a_a}}} \prod_{b=1}^{\wt{r} } \ex{ \upsilon_b  \wh{g}_s\pa{ y_b } } $ convergent. Their  action can then be incorporated by a re-definition of $f_s$ leading to
\beq
\msc{E}_m= \sul{n_1,\dots, n_s =0}{m} \pl{j=1}{s} \bigg\{  \f{ \Dp{}^{n_j} }{ \Dp{}\vsg_j^{n_j}  }  \bigg\} \cdot
c_{\paa{n_j}}\big[  \wt{f}_s \big]_{\mid \vsg_k = 0}
\enq
 \e{with}
\beq
\wt{f}_s\pa{\la} = f_s\pa{\la} \;  + \;  \sul{b=1}{s}
 \f{ \pa{t_{b+1}-t_b} }{ 2i\pi \pa{t_b-\la} }
\paa{ \sul{k=1}{r} \eps_k \psi\pa{t_b,\mu_{\a_k} }  +  \sul{k=1}{\wt{r}} \upsilon_k \psi\pa{t_b,y_{k} } }  \;.
\nonumber
\enq
 In this way, one obtains a (truncated to a finite number of terms) s-dimensional Lagrange series. The procedure for dealing with such series
and taking their $s\tend +\infty$ limits is described in proposition \ref{Proposition
construction derivee fnelle complexe}. In the following, all operator valued expressions ordered by $\bs{:} \cdot \bs{:}$
should be understood in this way.

\subsection{Resummation of the finite-volume Taylor coefficients}
\label{Subsection resum effective FF series}

In order to carry out the re-summation of the effective form factor expansion with the help of functional translations and generalizations thereof,
we need to regularize the expression for the functional $\wh{\mc{G}}_{N;\ga}$ with the help of an additional parameter $\be$.
This regularization will allow us to represent it as a regular functional that, moreover, has
a form suitable for carrying out the intermediate calculations.

\subsubsection*{The parameter $\be$}

It is easy to see that
\beq
\pa{ \wh{D}_{N} \; \wh{\mc{G}}_{N;\ga} } \pab{ \!\! \paa{p_a}_1^n \!\!  }{  \!\! \paa{h_a}_1^n \!\!  }
\pac{ \ga F_{0} \; ; \xi_{} \; ; \xi_{\ga F_0}  }
= \lim_{\be \tend 0} \Bigg\{  \wh{D}_{N}  \pab{ \!\! \paa{p_a}_1^n \!\!  }{  \!\! \paa{h_a}_1^n \!\!  }
\pac{ \ga F_{\be} \; ; \xi_{} \; ; \xi_{\ga F_{\be}}  }
 \wh{\mc{G}}_{N;\ga} \pab{ \!\! \paa{p_a}_1^n \!\!  }{  \!\! \paa{h_a}_1^n \!\!  }
\pac{ \ga F_{\be} \; ; \xi_{} \; ; \xi_{\ga F_{\be}}  } \Bigg\}
\enq
We now introduce  a prescription for taking the $\be \tend 0$ limit.
When considered as a separate object from $\wh{D}_{N}$, the functional $\wh{\mc{G}}_{N;\ga}$ may exhibit singularities
should it happen that $ \ga^{-1}\{ \ex{2i\pi \ga F_{\be}(\la_j)}-1 \} =0$, \textit{cf}
\eqref{ecriture fonctionnelle hat GN gamma}-\eqref{appendix thermo lim FF definition entree V et Vbar chapeau}.
For $\abs{\ga}$ small enough, as it will always be the case for us, such potential zeroes correspond to the existence of
solutions to $F_{\be}(\la_j)=0$. For $\be\in  \bs{\wt{U}}_{\be_0}$ with
\beq
 \bs{\wt{U}}_{\be_0} = \paa{ z \in \Cx \; :\; 10\, \Re\pa{\be_0} \geq \Re\pa{z} \geq  \Re\pa{\be_0}  \;\; \e{and} \;\; \abs{\Im\pa{z}}\leq \Im\pa{\be_0}}
\enq
$\Re\pa{\be_0}>0$ large enough and $\Im\pa{\be_0}>0$ small enough, there are no solutions of
\beq
F_\be\!\pabb{\om}{ \{ y_a \}_1^n }{ \{ z_a \}_1^n }=0 \quad \e{for} \quad \om\in U_{\de} \; ,
\;\; \e{this} \; \e{uniformly} \; \e{in} \quad
 0\leq n \leq m \; \e{and}\; \pa{\be,  \{ y_a \}_1^n, \{  z_a \}_1^n } \in \bs{\wt{U}}_{\be_0} \times U_{\de}^n \times U_{\de}^n \;.
\nonumber
\enq
It is clear that the optimal value of $\be_0$ preventing the existence of such solutions depends on the width $\de$ of
the strip $U_{\de}$ and on the integer $m$.

Hence, our strategy is as follows. We will always start our computations on a representation that is holomorphic in
the half-plane $\Re\be \geq 0$, as for instance \eqref{ecriture series FF effective gamma deformee}-\eqref{definition rho N eff derivee m ieme}.
In the intermediate calculations whose purpose is to allow one to relate
the initial representation to another one, we will assume that $ \be \in \bs{\wt{U}}_{\be_0}$.
This will allow us to avoid the problem of the aforementioned poles and represent $\wh{\mc{G}}_{N;\ga}$ in terms of a regular functional
that is moreover fit for carrying out the intermediate calculations.
Then, once that we obtain the final expression,
we will check that this new representation is in fact holomorphic in the half-plane $\Re\pa{\be} \geq 0$
and has thus a unique extension from $ \bs{\wt{U}}_{\be_0}$ up to
$\be=0$. As the same property holds for the initial representation, both will be equal at $\be=0$.

\vspace{3mm}

Having agreed on such a prescription for dealing with the $\be$-regularization and treating the $\be\tend 0$ limit, the effective form factor expansion-based representation
for $\rho_{N; \e{eff}}^{\pa{m}}\! \pa{x,t}$ \eqref{ecriture series FF effective gamma deformee} can be simplified with the use of the two properties below.

\subsubsection*{The functional $\wh{\mc{G}}_{N;\ga}$}

Given $A \in \R^+$, we define the compact $K_A$ contained in $U_{\de}$:
\beq
K_A = \paa{ z \in \Cx \; : \; \abs{\Im z} \leq \de \;, \; \abs{\Re z} \leq A} \;,
\label{definition compact KA}
\enq
and denote the open disk of radius $r$ by $\mc{D}_{0,r}= \paa{ z  \in \Cx \; : \; \abs{z} < r }$.

As follows from lemma \ref{Lemme fonctionnelle hat GA pour partie lisse hat FF}, given $A>0$ and large and  $m\in \mathbb{N}^*$ fixed, there exists
\begin{itemize}
\item a complex number
$\be_0$ with a sufficiently large real part and an imaginary part small enough
\item a positive number $\wt{\ga}_0>0$ small enough
\item a regular functional $\wh{\msc{G}}_{\ga;A}^{\,\pa{\be}}$
\end{itemize}
such that, uniformly in  $0\leq n \leq m$, $\pa{\ga,\be, \{ \mu_{p_a}\}_1^n, \{ \mu_{h_a}\}_1^n } \in \mc{D}_{0,\wt{\ga}_0} \times \bs{\wt{U}}_{\be_0} \times K_A^n \times K_A^n$ one has
\beq
 \wh{\mc{G}}_{N;\ga}  \pab{ \!\! \paa{p_a}_1^n \!\!  }{  \!\! \paa{h_a}_1^n \!\!  }
\pac{ \ga F_{\be} \; ; \xi_{} \; ; \xi_{\ga F_{\be}}  }  =
\wh{\msc{G}}_{\ga; A}^{\,\pa{\be}}\pac{ H\pabb{ * }{ \{ \mu_{p_a} \}_1^n  } {  \{ \mu_{h_a} \}_1^n  } }
\quad \e{with} \quad H\pabb{ \la }{ \{ \mu_{p_a} \}_1^n  } {  \{ \mu_{h_a} \} _1^n } = \sul{a=1}{n} \f{1}{\la-\mu_{p_a}} - \f{1}{\la - \mu_{h_a}} \;.
\label{definition facteur forme lisse simple fonctionnelle}
\enq
The $*$ in the argument of $\wh{\msc{G}}_{\ga;A}^{\,\pa{\be}}$ appearing above indicates the running variable of $H$ on which this functional acts.
The explicit expression for the functional $\wh{\msc{G}}_{\ga;A}^{\,\pa{\be}}$ is given in lemma \ref{Lemme fonctionnelle GA pour partie lisse FF}.
The main advantage of such a representation is that all the dependence on the auxiliary parameters is now solely contained in the function $H$ given in
\eqref{definition facteur forme lisse simple fonctionnelle}.
The constant $\wt{\ga}_0$ is such that
\beq
 \abs{ \ga F_{\be}\pabb{\om}{ \{ y_a \}_1^n }{  \{ z_a  \}_1^n  }  } < \f{1}{2} \quad \e{uniformly} \; \e{in} \quad
\pa{\ga,\be, \{ y_a \}_1^n, \{ z_a \}_1^n } \in \mc{D}_{0,\wt{\ga}_0} \times \bs{\wt{U}}_{\be_0} \times K_A^n \times K_A^n
\;\; \e{and} \;\; 0\leq n \leq m \;.
\label{ecriture condition Fbeta et gamma petit}
\enq

The functional $\wh{\msc{G}}_{\ga;A}^{\, \pa{\be}}$ is regular in respect to the to the pair $\pa{M_{\msc{G}_A}, \msc{C}\pa{K_A} }$
where $\msc{C}\pa{K_A}$ in a loop in $U_{\de}$ around $K_A$ as depicted in the $rhs$  of Fig.~\ref{contour exemple de courbes encerclantes}
and $M_{\msc{G}_A}$ corresponds to the compact with one hole that is delimited by $\msc{C}_{in}$ and $\msc{C}_{out}$.
This hole contains $K_{A}$. Finally, the parameters $\be_0 \in \Cx$ and $\wt{\ga}_0>0$ are such that the
 constant of regularity  $C_{\msc{G}_A}$ of $\wh{\msc{G}}^{\pa{\be}}_{\ga;A}$ satisfies to the estimates
\beq
 C_{\msc{G}_A}   \f{\pi d \!\pa{ \Dp{}M_{\msc{G}_A}, \msc{C}\!\pa{K_A} } }
 {  \abs{\Dp{}M_{\msc{G}_A}}+2\pi d \!\pa{ \Dp{}M_{\msc{G}_A}, \msc{C}\!\pa{K_A} }} >A \;,
\label{ecriture condition grandeur constante de regularite GAkappa}
\enq
where $\abs{\Dp{}M_{\msc{G}_A}}$ stands for the length of the boundary $\Dp{}M_{\msc{G}_A}$ and 
$d \!\pa{ \Dp{}M_{\msc{G}_A}, \msc{C}\!\pa{K_A} }>0$ stands for the distance of $\msc{C}\pa{K_A}$ to $\Dp{}M_{\msc{G}_A}$.

\vspace{2mm}
Similarly to the discussion carried out in section \ref{Subsubsection generalization of translations} and
according to proposition \ref{Proposition translation pure avec dependence parametre auxiliaire}, one has that, uniformly in $n,p\in \paa{0,\dots,m}$,
and $z_j$, $y_j$, $j=1,\dots, m$ belonging to $K_A$:
\beq
 \f{\Dp{}^p}{\Dp{}\ga^p }\cdot   \wh{\msc{G}}_{\ga;A}^{\, \pa{\be}} \pac{ H \pabb{ * }{  \{ z_j \}_1^{n} \vspace{1mm} }{ \{y_j\}_1^{n}  } }
_{\left| \ga=0 \right.}
=  \lim_{r \tend + \infty } \Bigg\{
\pl{j=1}{n} \ex{\wh{g}_{2,r}\pa{z_j}-\wh{g}_{2,r}\pa{y_j}}  \cdot
 \f{\Dp{}^p}{\Dp{}\ga^p } \wh{ \msc{G} }_{\ga;A}^{\, \pa{\be}}\pac{ \varpi_r}  \Bigg\}_{ \left| \substack{ \eta_{a,p}=0 \\ \ga=0 }\right. }
\hspace{-5mm} .
\label{ecriture decomposition Archeologue G+m+1}
\enq

The compact $M_{\msc{G}_A}$ has one hole. Hence, as discussed in section \ref{subsection Pure translations}
one has to consider two sets of discretization points  $t_{1,p}$, $p=1,\dots, r+1$
for $\msc{C}_{in}$ and $t_{2,p}$, $p=1,\dots,r+1$ for $\msc{C}_{out}$.
The function $\varpi_r$ appearing in \eqref{ecriture decomposition Archeologue G+m+1} is a linear polynomial
in the variables $\eta_{a,p}$ with $a=1,2$ and $p=1,\dots,r$:
\beq
\varpi_r \big( \la \mid \{  \eta_{a,p} \}  \big) =  \sul{p=1}{r}  \f{ t_{1,p+1}-t_{1,p} }{ 2i\pi \pa{t_{1,p}-\la} } \eta_{1,p}
\; + \; \sul{p=1}{r}  \f{ t_{2,p+1}-t_{2,p}  }{ 2i\pi \pa{t_{2,p}-\la} } \eta_{2,p} \;.
\label{definition fonction varpi r}
\enq
Finally, $\wh{g}_{2,r} \pa{\la}$ is a differential operator in respect to $\eta_{a,p}$ with $a=1,2$ and $p=1,\dots,r$:
\beq
\wh{g}_{2,r} \pa{\la} = \sul{p=1}{r} \f{ 1 }{ t_{1,p} -\la } \f{ \Dp{} }{ \Dp{}\eta_{1,p} }
\; + \; \sul{p=1}{ r } \f{ 1 }{ t_{2,p} -\la } \f{ \Dp{} }{ \Dp{}\eta_{2,p} }  \; .
\enq

\subsubsection*{The functional $\wh{D}_{N}$}

One can draw a small loop  $\msc{C}_{out}$ around $K_{2q}$ in $U_{\de}$ as depicted in the $lhs$ of Fig.~\ref{contour exemple de courbes encerclantes}.
Let $M_{\wh{D}}$ be the compact without holes whose boundary is delimited by $\msc{C}_{out}$.
Then, given $L$ large enough, the functional $\wh{D}_{N}$, as defined by \eqref{ecriture explicite fonctionnelle D+}, is a regular
functional (in respect to the pair $(M_{\wh{D}},K_{2q})$) of $\ga F_{\be}$ with $\be \in \bs{\wt{U}}_{\be_0}$ and $\abs{\ga} \leq \wt{\ga}_0$.
The parameters $\be_0$ and $\wt{\ga}_0$ are as defined previously.
This regularity is readily seen by writing down the integral representation:
\beq
\la_{j} = \Oint{ \Dp{}K_{2q}}{}   \f{ \xi_{\ga F_{\be} }^{\prime} \!\pa{\om} }{  \xi_{\ga F_{\be} }\!\pa{\om} - \tf{j}{L} } 
\cdot \f{\dd \om}{2i\pi}  \;,
\qquad j=1,\dots,N
\enq
which holds provided that $L$  is large enough (indeed then all $\la_j$'s are located in a very small vicinity of the interval 
$\intff{-q}{q}$).
Therefore, according to the results developed in appendix \ref{Appendice section functional translation}
and outlined in section \ref{Subsection Operator ordering for functional translation}, one has that, uniformly in $\be \in \bs{\wt{U}}_{\be_0}$
and $0\leq p,n \leq m$
\bem
 \f{\Dp{}^p}{\Dp{}\ga^p}  \Bigg\{ \wh{D}_{N} \pab{\!\! \paa{p_a}_1^n \!\! }{ \!\! \paa{h_a}_1^n \!\!  }
\pac{ \ga F_{\be}\pabb{\cdot}{  \{\mu_{p_a} \}_1^n  } {\paa{\mu_{h_a}}_1^n } \; ; \xi_{} \; ; \xi_{\ga F_{\be}}  }   \Bigg\}_{\mid\ga=0}  \\
=
 \lim_{s\tend +\infty} \Bigg[  \pl{a=1}{n} \ex{ \wh{g}_{1,s}\pa{\mu_{p_a}}- \wh{g}_{1,s}\pa{\mu_{h_a}} } \; \cdot \;
 \f{ \Dp{}^p }{ \Dp{}\ga^p } \paa{  \wh{D}_{N} \pab{ \!\! \paa{p_a}_1^n \!\! }{ \!\! \paa{h_a}_1^n \!\! }
 \pac{ \ga \nu_s \; ; \xi_{} \; ; \xi_{\ga \nu_s}  }    } _{ \left|  \substack{ \ga=0 \\ \vsg_k=0} \right.   }    \Bigg] \;.
\label{ecriture fonctionnelle Dn+ Fonct Trans}
\end{multline}

The function $\la \mapsto \nu_s\pa{\la}$ appearing above is holomorphic in some open neighborhood of $K_{2q}$ in $M_{\wh{D}}$ and given by
\beq
\nu_{s}\pa{\la\mid \paa{\vsg_a}_1^s}\equiv \nu_s\pa{\la}= \pa{i\be-\tf{1}{2}}Z\pa{\la} -\phi\pa{\la,q}+ \sul{j=1}{s}
 \f{ \pa{t_{j+1}-t_j} }{t_j-\la}  \cdot \f{\vsg_j}{2i\pi} \; .
\label{definition fonction nus}
\enq
The parameters $t_j$, $j=1,\dots,s$ correspond to a discretisation (\textit{cf} definition \ref{Definition point discretisation})
of the loop $\msc{C}_{out}$ around $K_{2q}$ in $U_{\de}$
that has been depicted in the $lhs$ of Fig.~\ref{contour exemple de courbes encerclantes}. $\vsg_j$ are some
sufficiently small complex numbers and $\wh{g}_{1,s}\pa{\la} $  is a differential operator in respect to $\vsg_a$:
\beq
\wh{g}_{1,s}\pa{\la} = - \sul{j=1}{s}  \phi\big(t_j,\la \big) \f{ \Dp{} }{ \Dp{}\vsg_j }  \;.
\label{definition widehat gs}
\enq

We remind that the parameters $\la_a$ appearing
in the second line of \eqref{ecriture fonctionnelle Dn+ Fonct Trans} through the expression  \eqref{ecriture explicite fonctionnelle D+} for
$\wh{D}_{N}$, are the unique\symbolfootnote[2]{Here, as previously, the uniqueness follows from Rouch\'{e}'s theorem. By writing down an integral
representation
for $\xi^{-1}_{\ga \nu_s}$, one readily convinces oneself that,  for $\ga$ small enough and given any fixed $s$,
$\la_a$ is holomorphic in $\paa{\vsg_a}_1^s$. It is also holomorphic in $\ga$ belonging to some open neighborhood of $\ga=0$.}
 solutions to $\xi_{\ga\nu_s}\pa{\la_a}=\tf{a}{L}$.
As such, the $\la_a$'s become holomorphic functions of $\{ \vsg_a\}_1^s$ when these belong to a sufficiently small neighborhood of the origin
in $\Cx^{s}$.


\subsubsection*{Representation for the Taylor coefficients}

To implement the simplifications induced by the functional translations on the level of $\rho_{N;\e{eff}}^{\pa{m}}\!\pa{x,t}$, we first
observe that all of the rapidities $\mu_{p_a}$ and $\mu_{h_a}$ occurring in the course of summation in \eqref{ecriture series FF effective gamma deformee}
belong to the interval $\intff{-A_L}{B_L}$ with $L\xi\pa{-A_L}=-w_L-\tf{1}{2}$ and $L\xi\pa{B_L}=w_L+\tf{1}{2}$ ($A_L>B_L$).
Hence,  a fortiori, they belong to the compact $K_{2A_L}$. We can thus represent
the smooth part functional as $\wh{\msc{G}}_{\ga;2A_L}^{ \, \pa{\be}}$.
We are interested solely in the $m^{\e{th}}$  $\ga$-derivative of \eqref{ecriture series FF effective gamma deformee}
at $\ga=0$. As $\wh{D}_{N} \big( \{p_a\}_1^n, \{h_a\}_1^n \big) \propto \ga^{2 \pa{n-1} }$ and $\wh{\msc{G}}_{\ga;2A_L}^{ \, \pa{\be}} \pac{\varpi_r}$
has no singularities around $\ga=0$, all terms issuing from  $n$ particle/hole excitations with $n\geq m$
will not contribute to the value of the derivative. Hence, we can truncate the sum over $n$ in
\eqref{ecriture series FF effective gamma deformee} at $n=m$.
Once that the sum is truncated, we
represent the functional $\Dp{\ga}^m \cdot \Big\{ \wh{D}_{N} \cdot \wh{\msc{G}}_{\ga;2A_L}^{\,\pa{\be}} \Big\}_{\mid \ga=0}$
with the help of identities \eqref{ecriture fonctionnelle Dn+ Fonct Trans}
and \eqref{ecriture decomposition Archeologue G+m+1}. This leads to
\bem
 \rho_{N;\e{eff}}^{\pa{m}}\!\pa{x,t} =   \lim_{\be \tend 0} \lim_{s \tend +\infty} \lim_{r\tend +\infty} \Bigg[  \; \sul{n = 0}{m}
\sul{ \substack{p_1<\dots < p_n \\  p_a \in \mc{B}^{\e{ext}}_{L}  } }{}
\sul{ \substack{h_1<\dots < h_n \\ h_a  \in \mc{B}^{\e{int}}_{L}  } }{}
 \pl{a=1}{n} \f{\wh{E}_-^{\, 2}\pa{\mu_{h_a}}  }{ \wh{E}_-^{\, 2} (\mu_{p_a}) } \\
\cdot  \f{ \Dp{}^m }{ \Dp{}\ga^m } \Bigg\{ \wh{D}_{N} \pab{ \!\! \paa{p_a}_1^n \!\! }{ \!\! \paa{h_a}_1^n \!\!  }
\pac{ \ga \nu_{s}  \; ; \xi \; ; \xi_{\ga \nu_s}  }
\wh{\msc{G}}_{\ga;2A_L}^{\, \pa{\be}}\pac{\varpi_r} \;   \Bigg\}_{ \left| \substack{ \ga=0 \\ \vsg_p=0= \eta_{a,p} } \right.  } \Bigg]  \;.
\label{ecriture serie rho eff completement factorisee}
\end{multline}
We have set
\beq
\wh{E}_ {-}^{\, 2}\pa{\la} = \ex{-ix u\pa{\la} -\wh{g}\pa{\la}} \qquad \e{with} \qquad
\wh{g}\pa{\la} \equiv \wh{g}_{1,s}\pa{\la} +  \wh{g}_{2,r}\pa{\la} \;.
\label{definition fonction E hat chapeau}
\enq
Above, in order to lighten the notation we have not written explicitly the dependence of $\nu_s$, $\varpi_r$ on the auxiliary parameters
$\vsg_p$, $\eta_{a,p}$ nor the one of $\wh{E}_-\pa{\la}$ on the
discretization indices $r$ and $s$.
However, we have kept the hat so as to insist on the operator nature of $\wh{E}_-$.
We do insist that \eqref{ecriture serie rho eff completement factorisee} has to be understood as it was discussed in
section \ref{Subsection Operator ordering for functional translation}.

Starting from representation \eqref{ecriture serie rho eff completement factorisee},
$\rho_{N;\e{eff}}^{\pa{m}}\!\pa{x,t}$ can be related with the $m^{\e{th}}$ $\ga$-derivative of the
form factor like representation of the functional\symbolfootnote[2]{The latter is a functional of $\nu_s$ and $\wh{g}$
as discussed in subsection \ref{Subsubsection generalization of translations}} $X_N\pac{ \ga \nu_{s}, \wh{E}_-^{\,2}  }$
given in \eqref{definition Fonction generatrice X_N}. Namely, for such an identification to hold, one has
to extend the upper bound in the summation over $n$ from $m$ up to $N+1$. This does not alter the result as it corresponds to
adding up a finite amount of terms that are zero due to the presence of $\ga$-derivatives.
Then, one should use the identity
\bem
 \f{ \Dp{}^m }{ \Dp{}\ga^m } \Bigg\{ \wh{D}_{N} \pab{ \!\! \paa{p_a}_1^n \!\! }{ \!\! \paa{h_a}_1^n \!\!  }
\pac{ \ga \nu_{s}  \; ; \xi \; ; \xi_{\ga \nu_s}  }
\wh{\msc{G}}_{\ga;2A_L}^{\, \pa{\be}}\pac{\varpi_r} \;   \Bigg\}_{ \left| \substack{ \ga=0 \\ \vsg_p=0= \eta_{a,p} } \right.  }   \\
= \bs{:}  \f{ \Dp{}^m }{ \Dp{}\ga^m } \Bigg\{
\f{ \prod_{a=1}^{N+1}  \wh{E}_-^{\, 2} \pa{\mu_{a}}    }{  \prod_{a=1}^{N}  \wh{E}_-^{\, 2} \pa{\la_{a}}   } \cdot
 \f{  \prod_{a=1}^{N}  \wh{E}_-^{\, 2} \pa{\la_{a}}   }{  \prod_{a=1}^{N+1}  \wh{E}_-^{\, 2} \pa{\mu_{a}} }
\wh{D}_{N} \pab{ \!\! \paa{p_a}_1^n \!\! }{ \!\! \paa{h_a}_1^n \!\!  }
\pac{ \ga \nu_{s}  \; ; \xi \; ; \xi_{\ga \nu_s}  }
\wh{\msc{G}}_{\ga;2A_L}^{\, \pa{\be}}\pac{\varpi_r} \;   \Bigg\}_{ \left| \substack{ \ga=0 \\ \vsg_p=0= \eta_{a,p} } \right.  }
\, \hspace{-5mm} \bs{:}
\label{ecriture identie operatorielle produit egal a 1}
\end{multline}
Just as it is the case for the parameters $\la_j$ appearing in the expression for $\wh{D}_N$, the ones appearing in
the pre-factors of the \textit{rhs} in \eqref{ecriture identie operatorielle produit egal a 1} are the unique solutions to
$\xi_{\ga \nu_s}\!\pa{\la_s}=\tf{s}{L}$. \eqref{ecriture identie operatorielle produit egal a 1}
is an expression of the type \eqref{equation explication prise gamma derivee},
and to deal correctly with it one should implement a  $\bs{:}\cdot \bs{:}$ prescription for the way the
differential operators $\Dp{\vsg_a}$ or $\Dp{\eta_{a,p}}$ should  be substituted in the \textit{rhs} of
\eqref{ecriture identie operatorielle produit egal a 1}.

With the help of identity \eqref{ecriture identie operatorielle produit egal a 1}, one is able to force
the appearance of the product of function $\wh{E}_-$ whose presence is necessary for identifying  the sum
over the particle-hole type labeling of integers in \eqref{ecriture serie rho eff completement factorisee} with the functional
$\Dp{\ga}^m X_N\pac{\ga \nu_s \; \wh{E}_-^{\,2}}_{ \mid_{\ga=0} }$ given in \eqref{definition Fonction generatrice X_N}.
This leads to the below representation:
\beq
\rho^{\pa{m}}_{N;\e{eff}} \!\pa{x,t} =   \; \lim_{\be \tend 0}\; \lim_{s \tend +\infty}  \; \lim_{r \tend +\infty}  \bs{: }
 \f{\Dp{}^m}{\Dp{} \ga^m } \; \Bigg\{  \f{ \prod_{a=1}^{N+1}  \wh{E}_{-}^{\,2} \pa{\mu_a} }{\prod_{a=1}^{N}   \wh{E}_{-}^{\,2} \!\pa{\la_a} }
X_N\pac{\ga \nu_{s}, \wh{E}_-^{\,2} } \, \wh{\msc{G}}_{\ga;2A_L}^{\, \pa{\be} }\!\pac{\varpi_r}\;
\Bigg\}_{\left| \substack{\ga=0 \\ \vsg_{p}=0=\eta_{a,p} } \right.  }  \hspace{-2mm}  \bs{ : } \;.
\label{equation exprimant Q cal en terme moyenne XN}
\enq

\subsection{Taking the thermodynamic limit}

It is shown in appendix  \ref{Appendix multidimensional Fredholm series}, theorem \ref{Theorem representation series Fredholm multidim}
that $\rho^{\pa{m}}_{N;\e{eff}}\!\pa{x,t}$ admits a well defined thermodynamic limit that we denote $\rho^{\pa{m}}_{\e{eff}}\!\pa{x,t}$.
This limit is given in terms of a multidimensional analogue of a (truncated) Fredholm series. This series is close in spirit to the
type of series that have appeared in \cite{KozKitMailSlaTerXXZsgZsgZAsymptotics,KozMailletSlaLongDistanceTemperatureNLSE}. It is also shown in that
appendix (proposition \ref{Proposition permutation limit thermo rho eff}) that it is allowed to exchange
\begin{itemize}
\item the thermodynamic limit $N,L \tend +\infty$, $\tf{N}{L} \tend D$
\end{itemize}
with
\begin{itemize}
\item the $\Dp{\ga}^m$ differentiation along with its associated operator substitution,
\item the computation of the  translation generated by $\wh{g}_{2,r}$,
\item the computation of the $s$-dimensional Lagrange series associated with $\wh{g}_{1,s}$,
\item the computation of the $r \tend +\infty$ and $s\tend +\infty$ limits,
\item the analytic continuation in $\be$ from $\wt{\bs{U}}_{\be_0}$ up to $\be=0$.
\end{itemize}
The result of such an exchange of symbols is that $\rho_{\e{eff}}^{\pa{m}}\!\pa{x,t}$ admits the representation
\beq
\hspace{-4mm}\rho^{\pa{m}}_{\e{eff}}\!\pa{x,t} = \lim_{w \tend +\infty}  \; \lim_{\be \tend 0} \;  \lim_{s \tend + \infty}
\; \lim_{r \tend + \infty}
\bs{:}  \f{ \Dp{}^m }{ \Dp{}\ga^m } \; \Bigg\{
\wh{E}_-^{\,2}\!\pa{q} \cdot \ex{- \Int{-q}{q} \pac{ix u^{\prime}\!\pa{\la} + \wh{g}^{\, \prime}\!\pa{\la} } \ga \nu_s\pa{\la} \dd \la }
 X_{\msc{C}_E^{\pa{w}}}\!\pac{\ga \nu_s,\wh{E}_-^{\,2}} \, \msc{G}_{\ga;2w}^{\pa{\be}}\!\pac{\varpi_r}
\Bigg\}_{\mid \ga=0}  \hspace{-5mm} \bs{: }   \;.
\label{equation exprimant rho m en terms moyenne X limit thermo}
\enq
This formula deserves a few comments. In the case of complex valued functions $\wt{E}_-$,
the functional $X_{\msc{C}_E^{\pa{w}}}\!\big[\ga \nu_s, \wt{E}_-^{2} \big]$ appearing
in \eqref{equation exprimant rho m en terms moyenne X limit thermo} corresponds to a Fredholm minor
\eqref{appendix ecriture limite thermo XN} of an integrable integral operator $I+V$ acting on $L^{2}\pa{\intff{-q}{q}}$.
The kernel $V$ of this operator is given by \eqref{definition noyau integrable volume infini}.

\begin{figure}[h]
\begin{center}

\begin{pspicture}(6,4)

\psline[linestyle=dotted, dash=3pt 2pt]{->}(-1,2)(7.8,2)
\psline[linestyle=dotted, dash=3pt 2pt]{->}(-1,0.5)(7.8,0.5)
\psline[linestyle=dotted, dash=3pt 2pt]{->}(-1,3.5)(7.8,3.5)

\rput(7.2,2.2){$\R$}
\rput(7.2,3.8){$\R + i\de$}
\rput(-0.7,0.8){$\R - i\de$}


\psdots(2,2)(4,2)(4.6,2)(0,2)(6.5,2)

\rput(2.1,1.8){$-q$}
\rput(3.8,1.8){$q$}
\rput(4.7,1.7){$\la_0$}
\rput(0,1.8){$-w$}
\rput(6.5,1.8){$w$}

\pscurve(0,3.3)(0.7,3)(0.9,2.9)(1,2.8)(1.2,2)(1.5,2)(2,2.6)(4,2.8)(4.5,2.6)(5,2)(5.5,2)(6,1.4)(6.5,1)


\pscurve[linestyle=dashed, dash=3pt 2pt](-1,3.3)(-0.5,3.3)(0,3.3)

\pscurve[linestyle=dashed, dash=3pt 2pt](6.5,1)(7,0.8)(7.5,0.9)

\rput(0.6,2.5){$\msc{C}_E^{\pa{w}}$}

\rput(7.3,1.2){$\msc{C}_E^{\pa{\infty}}$}

\end{pspicture}

\caption{The contour $\msc{C}_{E}^{\pa{w}}$ consists of the solid line. The contour $\msc{C}_{E}^{\pa{\infty}}$
corresponds to the union of the solid and dotted lines. The localization of the saddle-point $\la_0$ corresponds to the space-like regime.
Both contours lie in $U_{\tf{\de}{2}}$.
\label{contour CE et sa restriction CEw}}
\end{center}
\end{figure}
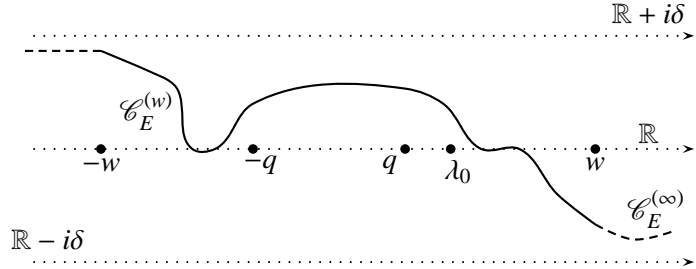

The subscript $\msc{C}_E^{\pa{w}}$ in $X_{\msc{C}^{\pa{w}}_E} \big[\ga \nu_s , \wt{E}^{\,2}_- \big] $
refers to an auxiliary compact contour entering in the definition of the kernel $V$. The parameter $w$
delimiting the size of this contour plays the role of a regularization.
The limit of an unbounded contour $\msc{C}_E^{\pa{\infty}}$ can only be taken after $r$ and $s$ are sent to infinity
and the analytic continuation up to $\be=0$ is carried out.
Finally, in \eqref{equation exprimant rho m en terms moyenne X limit thermo} also appears the
functional $\msc{G}_{\ga;2w}^{\pa{\be}}$. It can be thought of as the thermodynamic limit of the functional
$\wh{\msc{G}}_{\ga;2w}^{\,\pa{\be}}$. Its precise expression and properties are discussed in lemma \ref{Lemme fonctionnelle GA pour partie lisse FF}.

We also would like to stress that the parameter $\be_0$ defining the region $\bs{\wt{U}}_{\be_0}$
from which one should carry out the analytic continuation up to $\be=0$ depends on $2w$ as stated in
lemma \ref{Lemme fonctionnelle GA pour partie lisse FF}. This dependence is chosen in such a way that
the constant of regularity $C_{\msc{G}_A}$ for the functional $\msc{G}_{\ga;2w}^{\pa{\be}}$
is large enough so as to make licit all the necessary manipulations with the translation operators and generalizations
thereof.


\vspace{2mm}

We stress that formula \eqref{equation exprimant rho m en terms moyenne X limit thermo} constitutes the most important result of appendix
\ref{Appendix multidimensional Fredholm series}. Indeed, it provides one with a convenient representation for the thermodynamic limit
$\rho^{\pa{m}}_{\e{eff}}\!\pa{x,t}$. The latter constitutes  the first step towards extracting the large-distance $x$ and long-time $t$ asymptotic
behavior of $\rho^{\pa{m}}_{\e{eff}}\!\pa{x,t}$. The proof of such a representation for the thermodynamic limit
is however quite technical and lengthy. It can definitely be skipped on a first reading.
Moreover, should one be solely interested in a "short path" to extracting the asymptotics,  we stress that  formula
\eqref{equation exprimant rho m en terms moyenne X limit thermo} can be readily obtained without the use of any complicated and
long computations. It is enough to take the thermodynamic limit formally on the level of formula  \eqref{equation exprimant Q cal en terme moyenne XN}.
Such a formal manipulation leads straightforwardly to the representation \eqref{equation exprimant rho m en terms moyenne X limit thermo}.


\subsection{The multidimensional Natte series and asymptotics}
\label{subsection multidim Natte}

\begin{theorem}
\label{Theorem comportement asympt coeff Taylor limite Thermo}
The thermodynamic limit of
the Taylor coefficients $\rho_{\e{eff}}^{\pa{m}}\!\pa{x,t}$ admits the below truncated multidimensional Natte series representation
\bem
\rho_{\e{eff}}^{\pa{m}}\!\pa{x,t}   = \f{ \Dp{}^m }{ \Dp{}\ga^m }
\left\{
\f{ \bs{1}_{\intoo{q}{+\infty}}\pa{\la_0}    }{ \sqrt{-2\pi x u^{\prime \prime}\!\!\pa{\la_0} } } \times
\f{  \ex{ix\pac{u\pa{\la_0}-u\pa{q}} }   \mc{B}\pac{\ga F^{\la_0}_{q} ; p }\mc{A}_{0} \pac{\ga F^{\la_0}_{q}  }  }
{  \pa{x-t v_F + i0^+ }^{ \big[ \ga F^{\la_0}_{q} \!\pa{q} \big]^2 }   \pa{x + t v_F }^{  \big[ \ga F^{\la_0}_{q}\!\pa{-q} \big]^2 } }
\mc{G}_{1;\ga}^{\pa{0}}\pab{ \la_0 }{ q } \right. \\
+ \f{ \ex{ix\pac{u\pa{-q}-u\pa{q}} }   \pa{\mc{B}\mc{A}_-}\pac{\ga F^{-q}_{q} ; p}
 \mc{G}_{1;\ga}^{\pa{0}}\pab{ -q }{ q } }
{  \pa{x-t v_F + i0^+ }^{ \big[ \ga F_{q}^{-q} \!\pa{q} \big]^2}    \pa{x + t v_F }^{ \big[ \ga F^{-q}_{q}\!\pa{-q}-1\big]^2 } }
\; + \;
\f{ \pa{\mc{B}\mc{A}_+}\pac{\ga F^{\emptyset}_{\emptyset} ;p }
\mc{G}_{0;\ga}^{\pa{0}}\pab{ \emptyset }{ \emptyset } }
{  \pa{x-t v_F + i0^+ }^{  \pac{ \ga F^{\emptyset}_{\emptyset}\!\pa{q} +1 }^2}  \pa{x + t v_F }^{\pac{\ga F^{\emptyset}_{\emptyset}\!\pa{-q}}^2 } }
\\
\vspace{5mm}
\hspace{-1cm} + \ex{-ix u\pa{q} }  \sul{n=1}{m} \sul{ \mc{K}_n }{} \sul{ \mc{E}_n (\vec{k}\, ) }{}
\int_{ \msc{C}_{\eps_{\bs{t}} }^{\pa{w}} }{}  \hspace{-2mm}
\left.
\f{  H_{n;x}^{\pa{\paa{\eps_{\bs{t}}}}}\pa{  \paa{u\pa{z_{\bs{t}}} } ; \paa{z_{\bs{t}}}  } \big[ \ga  F^{ z_{\bs{+}} }_{ z_{\bs{-}} } \big]
\mc{B}\big[\ga F^{ z_{\bs{+}} }_{ z_{\bs{-}} }  ; p \big]      }
{  \pa{x-t v_F + i0^+ }^{ \pac{ \ga F^{ z_{\bs{+}} }_{ z_{\bs{-}} } \!\pa{q}}^2}
						\pa{x + t v_F }^{ \pac{ \ga F^{ z_{\bs{+}} }_{ z_{\bs{-}} }\!\pa{-q}}^2 } }
\mc{G}_{ \abs{\paa{z_{\bs{+}} } } ; \ga }^{\pa{0}} \pab{ \paa{z_{\bs{+}}} }{ \paa{z_{\bs{-}}} \cup\paa{q} }
 \f{ \dd^n z_{\bs{t}} }{ \pa{2i\pi}^n }     \right\} _{\mid \ga=0}  \hspace{-3mm}.
\label{ecriture serie Natte multDim coeff taylor rho eff}
\end{multline}
There, we have introduced the notations
\beq
\big\{  z_{\bs{+}} \big\}  = \big\{ z_{\bs{t}} \; , \;  \bs{t} \in \J{ \vec{k} } \; : \; \eps_{\bs{t}} =1 \big\} \;,  \qquad
\big\{  z_{\bs{-}}  \big\}  = \big\{  z_{\bs{t}} \; , \;  \bs{t} \in \J{\vec{k}} \; : \; \eps_{\bs{t}} = - 1 \big\}  \; ,
\quad \abs{ \big\{  z_{\bs{+}} \big\} } \equiv \# \big\{ z_{\bs{t}} \; , \;  \bs{t} \in \J{ \vec{k} } \; : \; \eps_{\bs{t}} =1 \big\}   \;.
\label{definition des ensembles z plus et moins}
\enq
$F^{\emptyset}_{\emptyset}$, $F^{\la_0}_{q}$, $F^{-q}_{q}$ have been defined in \eqref{definition les fonction shift asymptotiques} and, in general,
we agree upon
\beq
F^{ z_{\bs{+}} }_{ z_{\bs{-}} }\pa{\la} \equiv F\pabb{ \la  }{ \paa{z_{\bs{+}}}  }{ \paa{z_{\bs{-}}} \cup\paa{q}  }  = -\f{Z\pa{\la}}{2}
\; - \; \sul{ \substack{ \bs{t} \in \J{k}  \\   \eps_{\bs{t}}=1 } }{} \phi\pa{\la,\,z_{\bs{t}}}
 \; + \;  \sul{ \substack{ \bs{t} \in \J{k}  \\   \eps_{\bs{t}}=-1 } }{} \phi\pa{\la,\,z_{\bs{t}}} \;.
\label{definition fction shift multi-parametres z}
\enq
The function $\mc{G}_{n;\ga}^{\pa{0}}$ is related to the thermodynamic limit of the smooth part of the form factor. Its expression can be
found in \eqref{formule explicite G+ thermo}. The functionals $\mc{B}$, $\mc{A}_{\pm}$ and $\mc{A}_0$ are given by
\beq
\mc{B}\pac{\nu,p}=  \f{ \pac{\varkappa\pac{\nu}\pa{-q}}^{\nu\pa{-q}} }{ \pac{\varkappa\pac{\nu}\pa{q}}^{\nu\pa{q}} }
\f{ G^2\!\pa{ 1+\nu\pa{q} } G^2\!\pa{ 1-\nu\pa{-q} }  \ex{i\f{\pi}{2} \pa{\nu^2\!\pa{q}-\nu^2\!\pa{-q}}}  }
{ \pac{2q p^{\prime}\!\pa{q} }^{\nu^2\!\pa{q}} \pac{2q p^{\prime}\!\pa{-q} }^{ \nu^2\!\pa{-q} }   \pa{2\pi}^{ \nu\pa{q} - \nu\pa{-q}}  }
\; \ex{ \f{1}{2} \Int{-q}{q} \f{ \nu^{\prime}\!\pa{\la}\nu\pa{\mu}-\nu^{\prime}\!\pa{\mu}\nu\pa{\la} }{\la-\mu}  \dd \la \dd \mu } \; ,
\label{definition fonctionnelle B}
\enq
where $G$ is the Barnes double Gamma function,
\beq
\mc{A}_+\!\pac{\nu,p} = \f{ - 2q \varkappa^{-2}\!\pac{\nu}\pa{q} }{ \pac{ 2q p^{\prime}\!\pa{q} }^{2\nu\pa{q} + 1} }
\Ga\pab{ 1+\nu\pa{q} }{ - \nu\pa{q} }  \f{1}{\ex{-2i\pi \nu\pa{q}}-1}
\quad , \qquad \varkappa\pac{\nu}\pa{\la}=\exp\paa{-\int_{-q}^{\,q} \f{ \nu\pa{\la}-\nu\pa{\mu} }{ \la-\mu } \dd \mu}   \; ,
\label{definition fonctionnelle A+ et kappa}
\enq
and
\beq
\hspace{-5mm}\mc{A}_-\pac{\nu,p} = \f{ - 2q}{ \varkappa^{2}\!\pac{\nu}\!\pa{-q} }  \Ga\pab{\!\! 1-\nu\pa{-q} \!\! }{\nu\pa{-q}}
\f{\pac{2q p^{\prime}\!\pa{-q} }^{2\nu\pa{-q} - 1} }{\ex{-2i\pi \nu\pa{-q}}-1}
\quad \e{and} \quad
\mc{A}_0\pac{\nu} = \ex{-i\f{\pi}{4}} \varkappa^{-2}\!\pac{\nu}\!\pa{\la_0}
\paf{ \la_{0}-q }{ \la_{0}+q }^{2\nu\pa{\la_0} } \hspace{-3mm}.
\label{definition fonctionnelle A- et A0}
\enq
\vspace{2mm}
The second sum appearing in the last line of \eqref{ecriture serie Natte multDim coeff taylor rho eff} runs through
all the elements $\vec{k}$ belonging to
\beq
\mc{K}_n  =  \Bigg\{\vec{k} = \pa{k_1,\dots, k_{n+1}} \; : \; k_{n+1} \in \mathbb{N}^*  \; \; \e{and} \;\; k_a \in \mathbb{N} \; , a=1,\dots, n
\quad \e{such} \; \e{that} \quad  \sul{a=1}{n} a k_a  + k_{n+1}  =n \Bigg\} \;.
\enq
Once that an element of $\mc{K}_n$ has been fixed, one defines the associated set of triplets $\J{ \vec{k} }$:
\beq
\J{ \vec{k} }=\bigg\{ \pa{\bs{t}_1,\bs{t}_2,\bs{t}_3 } \, , \, \bs{t}_1\in \intn{1}{n+1} \, , \, \bs{t}_2 \in \intn{1}{k_{\bs{t}_1}} \, ,
\, \bs{t}_3 \in \intn{1}{\bs{t}_1 - n\de_{\bs{t}_1,n+1}} \bigg\} \; .
\enq
The third sum runs through all the elements $\{ \eps_{\bs{t}} \}_{\bs{t} \in \J{ \vec{k} } }$  belonging to the set
\beq
\hspace{-3mm} \mc{E}_n ( \vec{k} \, ) = \Bigg\{  \paa{\eps_{\bs{t}}}_{\bs{t} \in \J{ \vec{k} } } \: : \;
\eps_{\bs{t}} \in \paa{\pm 1, 0} \; \forall \bs{t} \in \J{ \vec{k} }
 \quad \e{with} \quad \sul{\bs{t}_3=1}{\bs{t}_1}\eps_{\bs{t}} =0 \quad \e{for}\;  \bs{t}_1=1,\dots, n \quad \e{and} \quad
\sul{p=1}{k_{n+1}}\eps_{n+1,p,1} =1 \Bigg\} \;.
\nonumber
\enq
In other words, $\mc{E}_n (\vec{k}\, )$ consists of $n$-uples of parameters $\eps_{\bs{t}}$ labeled by triplets
$\bs{t}=\pa{\bs{t}_1,\bs{t}_2,\bs{t}_3}$ belonging to $\J{ \vec{k} }$. Each element of such an $n$-uple takes its values in
$\paa{\pm 1,0}$. In addition, the components of this $n$-uple are subject to summation constraints.
These hold for any value of $\bs{t}_1$ or $\bs{t}_{2}$ and are different whether one deals with $\bs{t}_1=1,\dots,n$  or with $\bs{t}_{1}=n+1$.

The integral appearing in the $n^{\e{th}}$ summand occurring in the third line of 
\eqref{ecriture serie Natte multDim coeff taylor rho eff} is $n$-fold.
The contours of integration $\msc{C}_{\eps_{\bs{t}}}^{\pa{w}}$ depend on the choices of elements in $\mc{E}_n( \vec{k}\, )$
and are realized as $n$-fold Cartesian products of one-dimensional compact curves that correspond to various deformations of
the base curve $\msc{C}^{\pa{w}}_E$ depicted in Fig.~\ref{contour CE et sa restriction CEw}. In the $w\tend +\infty$ limit, these curves go to analogous
deformations of the base curve $\msc{C}_E^{\pa{\infty}}$. All these contours lie  in $U_{\tf{\de}{2}}$

The integrand  $H_{n;x}^{\pa{\paa{\eps_{\bs{t}}}}}\pa{  \paa{u\pa{z_{\bs{t}}} } ; \paa{z_{\bs{t}}}  } \pac{ \nu }$ is a regular functional of $\nu$, that
is  simultaneously a function of $u\pa{z_{\bs{t}}}$ and $z_{\bs{t}}$  with $\bs{t}$ running through the set $\J{ \vec{k} }$.
This functional depends on the  choice of an element $\paa{\eps_{\bs{t}}}_{\bs{t}\in \J{ \vec{k} } }$ from
$\mc{E}_n( \vec{k} \, )$ and on $x$. It appears originally as a building block of the Natte series
(\textit{cf} appendix \ref{subsection Natte series Fred minor} for more details).
\end{theorem}

We stress that all summands involving the functional $\mc{G}_{n;\ga}^{\pa{\be}}$ are well defined at $\be=0$. The potential singularities
present in $\mc{G}_{n;\ga}^{\pa{\be}}$ are canceled by the zeroes of the pre-factors.

\Proof

As a starting point for the proof, we need to introduce the below set of functions depending on
the auxiliary parameters $a_p$, $b_{1,p}$ and $b_{2,p}$. As it has been discussed in
section \ref{Subsubsection generalization of translations}, these functions will allow us to compute the (functional)
coefficients  necessary for carrying out the operator substitution. We set
\beq
\wt{E}_-^{\, 2} \pa{\la} = \ex{-ix u\pa{\la}-\wt{g}\pa{\la}} \quad  \e{with} \quad \wt{g}\pa{\la} = \wt{g}_{1,s}\pa{\la} + \wt{g}_{2,r}\pa{\la}\;,
\label{definition fction E-tilde}
\enq
where
\beq
\wt{g}_{1,s}\pa{\la} = - \sul{p=1}{s} \phi\big(t_p,\la\big) a_p \qquad  \e{and} \qquad
\wt{g}_{2,r}\pa{\la} =  \sul{p=1}{r}  \f{ b_{1,p} }{t_{1,p}-\la}  \; + \;  \sul{ p=1 }{ r } \f{ b_{2,p} }{t_{2,p}-\la} \;.
\label{definition fonctions tilde g1 et g2}
\enq
It is readily checked with the help of lemma \ref{Lemme fonctionnelle GA pour partie lisse FF} 
and proposition \ref{Proposition limit thermo Fred minor pur} that for $\ga$ small enough $\mc{F}$ given below is a regular functional of $\nu_s$, $\varpi_r$ and $\wt{g}$:
\beq
\mc{F}\pac{\ga \nu_s, \wt{g},\varpi_r}\!\pa{\ga} = \wt{E}_-^{\,2}\!\pa{q} \cdot
\ex{- \Int{-q}{q} \pac{ix u^{\prime}\!\pa{\la} + \wt{g}^{\, \prime}\!\pa{\la} } \ga \nu_s\pa{\la} \dd \la }
 X_{\msc{C}_E^{\pa{w}}}\!\big[ \ga \nu_s,\wt{E}_-^{\,2} \big] \, \msc{G}_{\ga;2w}^{\pa{\be}}\!\pac{\varpi_r} \;.
\label{definition fonctionnelle F qui mime substitution operatorielle dans serie Natte}
\enq
In particular $\mc{F}\pac{\ga \nu_s, \wt{g},\varpi_r}\!\pa{\ga}$  is holomorphic in $\ga$, at least for $\ga$ small enough.
In order to implement the operator substitution, we have to compute the Taylor coefficients of the series expansion of
$\mc{F}\pac{\ga \nu_s, \wt{g},\varpi_r}\!\pa{\ga}$ into powers of $b_{1,p}$, $b_{2,p}$ with $p=1,\dots, r$ and $a_p$ with $p=1,\dots,s$.
These Taylor coefficients are \textit{solely} determined by the functional $\mc{F}\pac{\ga \nu_s, \wt{g}, \varpi_r}\!\pa{\ga}$
depending on the \textit{classical} function $\wt{g}$ \eqref{definiton fonction gs scalaire}.
Therefore, one can use \textit{any} equivalent  representation for $\mc{F}\pac{\ga \nu_s, \wt{g},\varpi_r}\!\pa{\ga}$ as a starting point for
computing the various partial derivatives in respect to $b_{j,p}$ or $a_p$.
In other words, one can use \textit{any} equivalent series representation\symbolfootnote[2]{One natural representation that can be used as a starting point for taking the derivatives is the Fredholm series-like representation for
$X_{\msc{C}_E^{\pa{w}}}\big[\ga \nu_s, \wt{E}_-^{\,2} \big]$. In fact, it is this series representation that has been used for the
computations carried out in theroem \ref{Theorem representation series Fredholm multidim}.}
 for the Fredholm minor $X_{\msc{C}_E^{\pa{w}}}\big[\ga \nu_s, \wt{E}_-^{\,2} \big]$.
Clearly, different series representations for the Fredholm minor will lead to different type of expressions for the
Taylor coefficients. However, in virtue of the uniqueness of the Taylor coefficients, their \textit{values} coincide.
As shown in \cite{KozTimeDepGSKandNatteSeries}, the Fredholm minor we're interested in admits the  so-called Natte series representation.
The latter series of multiple integrals is built in such a way that it gives a quasi-direct access to the asymptotic behavior of
$X_{\msc{C}_E^{\pa{w}}}\big[\ga \nu, \wt{E}_-^{2} \big]$.  It is thus clear that this is THE series representation that is fit for providing the
large-distance/long-time asymptotic expansion of the two-point function.
We will thus take this series representation as a starting point for our calculations.

\vspace{2mm}

The first remarkable consequence of the use of the Natte series is that the exponential pre-factor in front of
$X_{\msc{C}_E^{\pa{w}}} \big[\ga \nu_s, \wt{E}^{\,2}_- \big]$ in
\eqref{definition fonctionnelle F qui mime substitution operatorielle dans serie Natte}  \textit{exactly compensates} the one appearing in
the Natte series \eqref{equation developpement det serie de Natte}.
Once that these pre-factors are simplified, one should take the $m^{\e{th}}$ $\ga$-derivative of the remaining part of
the Natte series representation \eqref{equation developpement det serie de Natte} for
$X_{\msc{C}_E^{\pa{w}}}\big[ \ga \nu_s, \wt{E}_-^{\,2}\big] \msc{G}_{\ga;2w}^{\pa{\be}}\pac{\varpi_r}$.
One of the consequences of taking the $m^{\e{th}}$-$\ga$ derivative is that the Natte series given in \eqref{equation developpement det serie de Natte}
becomes truncated  at $n=m$ due to the property $ii)$ of the functions $H_{n;x}^{ \pa{ \paa{ \eps_{\bs{t}} } } }$ (\textit{cf} appendix
\ref{subsection Natte series Fred minor}):
\bem
\hspace{-1cm} \f{ \Dp{}^m }{ \Dp{}\ga^m } \mc{F}\pac{\ga \nu_s, \wt{g},\varpi_r}\!\pa{\ga}_{\mid \ga=0} =
 \ex{ ix\pac{u\pa{\la_0}-u\pa{q}} } \ex{\wt{g}\pa{\la_0}-\wt{g}\pa{q}}
 \f{ \Dp{}^m }{ \Dp{}\ga^m } \paa{ \f{ \mc{B}\pac{\ga \nu_s; u+i0^+}\mc{A}_{0} \pac{\ga \nu_s} }
{ \sqrt{-2\pi u^{\prime \prime}\!\pa{\la_0} x  } \cdot  x^{\ga^2\!\nu_s^{2}\!\pa{q} + \ga^2\!\nu_s^{2}\!\pa{-q} } }
\msc{G}_{\ga;2w}^{\pa{\be}}\!\pac{\varpi_r} }_{\mid \ga=0}
\hspace{-4mm} \times  {\bf 1}_{\intoo{ q }{+\infty}}\pa{\la_0}   \\
 \hspace{-1cm} + \ex{ ix\pac{u\pa{-q}-u\pa{q}} } \ex{\wt{g}\pa{-q}-\wt{g}\pa{q}}
 \f{ \Dp{}^m }{ \Dp{}\ga^m } \paa{ \f{ \pa{\mc{B}\mc{A}_-}\pac{\ga \nu_s; u+i0^+}  } {  x^{ \pa{1-\ga\nu_s\!\pa{-q}}^2 + \ga^2\!\nu_s^{2}\!\pa{q} } }
 \msc{G}_{\ga;2w}^{\pa{\be}}\!\pac{\varpi_r} }_{\mid \ga=0}
\hspace{-2mm}+ \; \;  \f{ \Dp{}^m }{ \Dp{}\ga^m } \paa{
\f{ \pa{\mc{B}\mc{A}_+}\pac{\ga \nu_s; u+i0^+}    } {  x^{  \ga^2\!\nu_s^{2}\!\pa{-q} + \pa{\ga\nu_s\!\pa{q}+1}^2 } }
\msc{G}_{\ga;2w}^{\pa{\be}}\!\pac{\varpi_r} }_{\mid \ga=0}\\
\hspace{-1cm} + \ex{-\wt{g}\pa{q}-ixu\pa{q} } \;  \sul{n=1}{m} \sul{ \mc{K}_n }{} \sul{ \mc{E}_n(\vec{k}) }{}
 \Oint{ \msc{C}_{\eps_{\bs{t}} }^{\pa{w}} }{}  \pl{ \bs{t} \in \J{ k } }{}\paa{ \ex{\eps_{\mf{t}} \wt{g}\pa{z_{\bs{t}}} } }
\cdot \f{ \Dp{}^m }{ \Dp{}\ga^m }  \paa{  H_{n;x}^{\pa{\paa{\eps_{\bs{t}}}}}\pa{ \paa{u\pa{z_{\bs{t}}} } ; \paa{z_{\bs{t}}}  } \pac{ \ga \nu_s}
\f{ \mc{B}\pac{\ga \nu_s; u+i0^+}    } {  x^{  \ga^2\!\nu_s^{2}\!\pa{q} + \ga^2\!\nu_s^2\!\pa{-q} } } \msc{G}_{\ga;2w}^{\pa{\be}}\!\pac{\varpi_r}
}_{ \mid \ga=0}
     \f{ \dd^n z_{\bs{t}} }{ \pa{2i\pi}^n }   \;.
\label{ecriture serie Natte type tilda avant toute substitution}
\end{multline}
It follows from lemma \ref{Lemme fonctionnelle GA pour partie lisse FF}, representation \eqref{ecriture dependence serie Natte en nu}
and the explicit formulae for the functionals $\mc{B}$, $\mc{A}_0$ and $\mc{A}_{\pm}$
\eqref{definition fonctionnelle B}-\eqref{definition fonctionnelle A- et A0} that the functionals occurring in
\eqref{ecriture serie Natte type tilda avant toute substitution}  are all regular (\textit{cf} definition \ref{Definition Fonctionelle reguliere}).
Moreover, as follows from the previous discussion relative to the procedure of taking the $\be\tend 0$ limit, at this stage of the
calculations, $\Re\pa{\be}>0$  is large enough so that the constant of regularity $C_{\msc{G}_{2w}}$ of the functionals $\msc{G}_{\ga;2w}^{\pa{\be}}$
is sufficiently large to be able to apply proposition \ref{Proposition translation pure avec
dependence parametre auxiliaire} and corollary \ref{Corollaire translation dans le cas integral et derivee}
(due to the estimates \eqref{ecriture condition grandeur constante de regularite GAkappa} for $C_{\msc{G}_{2w}}$, the constant  $\ga_0$ occurring in
\eqref{appendix funct der definition ga0 pour translation pures} is greater then $1$ for $w$ large enough, which is the limit of interest)
to this functional. Proposition \ref{Proposition translation pure avec
dependence parametre auxiliaire} and corollary \ref{Corollaire translation dans le cas integral et derivee} are also directly applicable
to all functionals of $\ga \nu_s$ in as much as, at the end of the day, one sets $\ga=0$.

Clearly, there is no problem to implement the substitution $a_p \mapsto \Dp{\vsg_p}$ and $b_{i,p} \mapsto \Dp{\eta_{i,p}}$
on the level of \eqref{ecriture serie Natte type tilda avant toute substitution} in such a way that
all the partial derivative operators appear to the left of all $\eta_{i,p}$ and $\vsg_{p}$ dependent functions.
The first two lines in \eqref{ecriture serie Natte type tilda avant toute substitution} will give rise
to translation operators. In the case of the ultimate line in \eqref{ecriture serie Natte type tilda avant toute substitution},
this operator substitution will produce expressions of the type
\bem
\sul{ n_p \geq 0 }{ +\infty } \sul{  n_{a,p} \geq 0 }{ +\infty }
\pl{p=1}{s} \paa{\f{1}{n_p!} \f{ \Dp{}^{n_p} }{ \Dp{}\vsg_p^{n_p} } } \pl{p=1}{r}  \pl{a=1}{2}
\paa{ \f{1}{ ( n_{a,p} )!} \f{ \Dp{}^{n_{a,p}} }{ \Dp{}\eta_{a,p}^{n_{a,p}} } }
\Oint{ \msc{C}_{\eps_{\bs{t}} }^{\pa{w}} }{}  \pl{p=1}{s} \pac{\Om_{ p }\pa{\paa{z_{\bs{t}}} }}^{n_p}
\pl{p=1}{r} \pl{a=1}{2} \pac{\Om^{\prime}_{ a, p }\pa{\paa{z_{\bs{t}}} }}^{n_{a,p}}  \\
\times
\f{ \Dp{}^m }{ \Dp{}\ga^m }  \paa{  H_{n;x}^{\pa{\paa{\eps_{\bs{t}}}}}\pa{ \paa{u\pa{z_{\bs{t}}} } ; \paa{z_{\bs{t}}}  } \pac{ \ga \nu_s }
\f{ \mc{B}\pac{\ga \nu_s; u+i0^+}    } {  x^{  \ga^2\!\nu_s^{2}\!\pa{q} + \ga^2\!\nu_s^2\!\pa{-q} } } \msc{G}_{\ga;2w}^{\pa{\be}}\!\pac{\varpi_r}
}_{ \left| \substack{ \ga=0 \\  \vsg_p=0=\eta_{a,p} } \right. }
\f{ \dd^n z_{\bs{t}} }{ \pa{2i\pi}^n } \;.
\label{ecriture series des derivees substitues en dehors integral}
\end{multline}
Where $\Om_p$ and $\Om^{\prime}_{a,p}$ take the form
\beq
\Om_{p}\pa{\paa{z_{\bs{t}}} } = \phi\pa{t_p, q} - \sul{ \bs{t} \in \J{k} }{} \eps_{\bs{t}}\phi\pa{t_p,z_{\bs{t}}}
\quad \e{and} \quad
\Om^{\prime}_{ a,p}\pa{\paa{z_{\bs{t}}} } = \f{ - 1}{t_{a,p}- q} \; + \; \sul{ \bs{t} \in \J{k} }{} \eps_{\bs{t}} \f{1}{t_{a,p}-z_{\bs{t}}}  \;.
\label{definition fonctions Omega et Omega Prime}
\enq

One can compute the $r,s \tend +\infty$ limit of such series of integrals by applying corollary\symbolfootnote[2]{This corollary can be applied to
$\msc{G}_{\ga;2w}^{\pa{\be}}$ precisely because its constant of regularity is large enough.}
\ref{Corollaire translation dans le cas integral et derivee} and observing that $\msc{C}_{\eps_{\bs{t}}}^{\pa{w}}$
is a Cartesian product of a finite number of compact one dimensional curves that are contained in $U_{\tf{\de}{2}}$.
In fact, the result of this corollary allows one to carry out the operator substitution in
\eqref{ecriture serie Natte type tilda avant toute substitution} directly under the integration sign.
In other words, one is allowed to replace $\wt{g}_{1,s} \hookrightarrow \wh{g}_{1,s}$ and
$\wt{g}_{2,r} \hookrightarrow \wh{g}_{2,r}$
directly on the level of \eqref{ecriture serie Natte type tilda avant toute substitution}, this without pulling out the partial $\vsg_p$
or $\eta_{a,p}$ derivatives out of the integrals.
Hence, one is brought to computing the action of translation operators. The latter can be estimated by applying proposition
\ref{Proposition translation pure avec dependence parametre auxiliaire}.
Again, there is no problem to apply this proposition either because we compute the $m^{\e{th}}$ $\ga$-derivative at $\ga=0$ (so that
$\gamma$ can be as small as desired in the case of functionals of $\ga \nu_s$)
or because the constant of regularity is large enough for $\msc{G}_{\ga;2w}^{\pa{\be}}$.
As follows from this proposition, one can permute the partial $\ga$-derivative symbols at $\ga=0$ with
the action of the finite $s$ and $r$ translation operators.
It then remains to take the $r\tend +\infty$ and the $s\tend +\infty$ limits.
As in each case the convergence is uniform, the limit can be taken directly under the finite sum, compact integrals and
partial $\ga$-derivatives symbols.

Then, in order to compute the effect of the $s\tend +\infty$ limit we apply the identity \eqref{ecriture action operateur translation sur fnelle generique}
(also \textit{cf} appendix \ref{subsection Pure translations}):
\beq
\lim_{s\tend +\infty} \pl{a=1}{n} \ex{ \wh{g}_{1,s}\pa{z_a}-\wh{g}_{1,s}\pa{y_a} } \cdot \msc{F}\pac{\ga \nu_s}  
\; = \; \msc{F}\big[ \ga F_{\be} \big]
\quad \e{with} \quad   F_{\be}\pa{\la} \equiv F_{\be}\pabb{\la}{ \paa{z_a}_1^n }{ \paa{y_a}_1^n } \; ,
\enq
valid for any regular functional $\msc{F}$, $\abs{\ga}$ small enough and $z_a$, $y_a$ all lying in $U_{\de}$.
Here, we would like to remind that $F_{\be}$ appearing above corresponds to the
thermodynamic limit of the $\be$-deformed shift function,  \textit{cf} \eqref{ecriture limite thermo fction shift}.
Similarly,
\beq
\lim_{r\tend +\infty} \pl{a=1}{n} \ex{ \wh{g}_{2,r}\pa{z_a}-\wh{g}_{2,r}\pa{y_a} } \cdot \msc{G}_{\ga; 2w }^{\pa{\be}}\pac{\varpi_r}  \; = \;
\msc{G}_{\ga; 2w }^{\pa{\be}}\pac{ H \pabb{*}{ \paa{z_a}_1^n }{ \paa{y_a}_1^n } }
\quad \e{with} \quad  H \pabb{\la}{ \paa{z_a}_1^n }{ \paa{y_a}_1^n } =  \sul{a=1}{n} \f{1}{\la-z_a} - \f{1}{\la-y_a} \;.
\label{ecriture reconstruction G cal A kappa}
\enq
All this for $\pa{ \paa{z_a}_1^n ;\paa{y_a}_1^n} \in K_{2w}^{2n}$.
Then, by applying lemma \ref{Lemme fonctionnelle GA pour partie lisse FF} backwards, we get
\beq
 \msc{G}_{\ga; 2w }^{\pa{\be}}\pac{ H \pabb{*}{ \paa{z_a}_1^n }{ \paa{y_a}_1^n } }
=\mc{G}^{\pa{\be}}_{n;\ga}\pab{ \paa{z_a}_1^n }{ \paa{y_a}_1^n }\;.
\enq
The function $\mc{G}^{\pa{\be}}_{n;\ga}$
has been defined in \eqref{formule explicite G+ thermo}.

Therefore, we obtain
\bem
\rho^{\pa{m}}_{\e{eff}}\!\pa{x,t} = \lim_{w \tend +\infty} \lim_{\be \tend 0}  \f{ \Dp{}^m }{ \Dp{}\ga^m }
\Bigg\{  \f{ \mc{B}\big[\ga \wt{F}^{\la_0}_{q} ; u+i0^+ \big]\mc{A}_{0} \big[\ga \wt{F}^{\la_0}_{q}  \big] } { \sqrt{-2\pi u^{\prime \prime}\!\pa{\la_0} x } }
\f{  \ex{ix[u\pa{\la_0}-u\pa{q}] }  }
{  x^{ \big[  \ga \wt{F}^{\la_0}_{q} \!\pa{q} \big]^2 + \big[ \ga \wt{F}^{\la_0}_{q}\!\pa{-q} \big]^2 } }
\mc{G}_{1;\ga}^{\pa{\be}}\pab{ \la_0 }{ q } \\
 + \f{ \ex{ix[u\pa{-q}-u\pa{q}] }  }
{  x^{ \big[\ga \wt{F}_{q}^{-q} \!\pa{q}\big]^2 + \big[1- \ga \wt{F}^{-q}_{q}\!\pa{-q} \big]^2 } } \cdot
\pa{\mc{B}\mc{A}_-}\big[\ga \wt{F}^{-q}_{q} ; u+i0^+\big]  \mc{G}_{1;\ga}^{\pa{\be}}\pab{ -q }{ q }
\; \; \\
 + \; \; \f{ \pa{\mc{B}\mc{A}_+}\big[\ga \wt{F}^{\emptyset}_{\emptyset} ; u+i0^+\big]    }
{  x^{  \big[ 1+ \ga \wt{F}^{\emptyset}_{\emptyset}\!\pa{q} \big]^2  + \big[\ga \wt{F}^{\emptyset}_{\emptyset}\!\pa{-q}\big]^2 } }
\mc{G}_{0;\ga}^{\pa{\be}}\pab{ \emptyset }{ \emptyset }
\; + \;  \ex{-ixu\pa{q} }  \sul{n=1}{m} \sul{ \mc{K}_n }{} \sul{ \mc{E}_n(\vec{k}) }{}
 \Oint{ \msc{C}_{\eps_{\bs{t}} }^{\pa{w}} }{}
\f{  \mc{B}\big[\ga \wt{F}^{ z_{\bs{+}} }_{ z_{\bs{-}} }  ; u+i0^+\big]    }
{   x^{ \big[ \ga \wt{F}^{ z_{\bs{+}} }_{ z_{\bs{-}} } \!\pa{q} \big]^2 + \big[ \ga \wt{F}^{ z_{\bs{+}} }_{ z_{\bs{-}} }\!\pa{-q} \big]^2 }  }
\\
\times H_{n;x}^{\pa{\paa{\eps_{\bs{t}}}}} \!\!
\pa{ \big\{ u\pa{z_{\bs{t}}} \big\} ; \big\{ z_{\bs{t}} \big\} \,  } \! \big[ \ga  \wt{F}^{ z_{\bs{+}} }_{ z_{\bs{-}} }  \big]
\; \mc{G}_{ \abs{\paa{z_{\bs{+}} } };\ga }^{\pa{\be}} \pab{ \paa{z_{\bs{+}}} }{ \paa{z_{\bs{-}}} \cup \paa{q} }
 \f{ \dd^n z_{\bs{t}} }{ \pa{2i\pi}^n }     \Bigg\}_{\mid \ga=0}  \hspace{-3mm}.
\label{ecriture serie Natte operatorielle apres limite r et s vont infini}
\end{multline}
Here $\wt{F}^{z_{\bs{+}}}_{z_{\bs{-}}}\pa{\la}=F^{z_{\bs{+}}}_{z_{\bs{-}}}\pa{\la} + i \be Z\pa{\la}$ and $F^{z_{\bs{+}}}_{z_{\bs{-}}}$
has been defined in \eqref{definition fction shift multi-parametres z}.

Once that the functional translations have been computed, one should carry out the analytic continuation
of the expression in brackets from $\be \in \bs{\wt{U}}_{\be_0}$
up to $\be=0$ and then send $w$ to $+\infty$.
For this, we recall that the functions $H_{n;x}^{\pa{\paa{\eps_{\bs{t}}}}}$ admit the below decomposition (\textit{cf} \eqref{ecriture dependence serie Natte en nu}):
\beq
\hspace{-5mm} H_{n;x}^{\pa{\paa{\eps_{\bs{t}}}}} \!\!
\pa{ \big\{ u\pa{z_{\bs{t}}} \big\} ; \big\{ z_{\bs{t}} \big\} \,  } \! \big[ \ga  \wt{F}^{ z_{\bs{+}} }_{ z_{\bs{-}} }  \big]
=
\wt{H}_{n;x}^{\pa{\paa{\eps_{\bs{t}}}}}\pa{ \{ \ga \wt{F}^{ z_{\bs{+}} }_{ z_{\bs{-}} } \pa{z_{\bs{t}}}\}, \{ u\pa{z_{t}} \} , \{ z_{\bs{t}}\} }
\pl{ \bs{t} \in \J{k} }{}  \pa{ \varkappa [ \wt{F}^{ z_{\bs{+}} }_{ z_{\bs{-}} } ] \pa{z_{\bs{t}}} }^{-2\eps_{\bs{t}}}
\pl{z_{\bs{t}}\in \paa{z_{\bs{+}} } }{} \pa{\ex{-2i\pi \ga \wt{F}_{z_{\bs{-}}}^{z_{\bs{+}}}\pa{z_{\bs{t}}}} -1 }^2 .
\label{ecriture explicite decomposition Hnx}
\enq
It follows from the way $H_{n;x}^{\pa{\paa{\eps_{\bs{t}}}}}$ depends on the set of its $\nu$-type arguments \eqref{ecriture explicite decomposition Hnx}
and from the expression for the functional $\mc{B}\pac{\nu,u}$ \eqref{definition fonctionnelle B} and $\mc{G}^{\pa{\be}}_{n;\ga}$
\eqref{formule explicite G+ thermo} that all of the expressions one deals with contain the combination
\beq
G^{2}\pa{1-\ga \wt{F}_{z_{\bs{-}}}^{z_{\bs{+}}}\pa{-q} } G^{2}\pa{1+\ga \wt{F}_{z_{\bs{-}}}^{z_{\bs{+}}}\pa{q} }
\pl{z_{\bs{t}}\in \paa{z_{\bs{+}} } }{} \pa{\ex{-2i\pi \ga \wt{F}_{z_{\bs{-}}}^{z_{\bs{+}}}\pa{z_{\bs{t}}}} -1 }^2
\ddet{\msc{C}_{q+\eps} }{ I+\ga \mc{V}\big[ \ga \wt{F}_{z_{\bs{-}}}^{z_{\bs{+}}} \big]   }
\ddet{\msc{C}_{q+\eps} }{ I+\ga \ov{\mc{V}}\big[ \ga  \wt{F}_{z_{\bs{-}}}^{z_{\bs{+}}} \big]   }
\nonumber
\enq
In virtue of proposition \ref{Proposition Holomorphie en beta et rapidite part-trou det fred}, the function appearing above is holomorphic in
$\pa{\be, \ga , \{ z_{\bs{+}} \} , \{z_{\bs{-}} \} } \in
\{ \Re\pa{\be} \geq 0 \} \times \ov{\mc{D}}_{0,1} \times U_{\de}^{\abs{ \{ z_{\bs{+}} \} }} \times K_{q+\eps}^{\abs{ \{ z_{\bs{+}} \} }}$.

The function $\wt{H}_{n;x}^{\pa{\paa{\eps_{\bs{t}}}}}\pa{ \{ \ga \wt{F}^{ z_{\bs{+}} }_{ z_{\bs{-}} } \pa{z_{\bs{t}}}\}, \{ u\pa{z_{t}} \} , \{ z_{\bs{t}}\} }$ is analytic in $\pa{\ga,\be} \in \mc{D}_{0,\wt{\ga}_0} \times \bs{\wt{U}}_{\be_0}$
(here $\wt{\ga}_0$ is chosen so that $\abs{ \ga \wt{F}^{ z_{\bs{+}} }_{ z_{\bs{-}} } \pa{z_{\bs{t}}} }  < \tf{1}{2}$
uniformly in  the variables $z_{\bs{t}}$, $\bs{t} \in \J{\vec{k}}$  belonging to $U_{\de}$), and integrable in respect to
the $\{ z_{\bs{t}}\}_{\bs{t} \in \J{k_a}}$  variables. The remaining part of $\mc{G}_{n;\ga}^{\pa{\be}}$ has also the same properties.
As the integrals are compactly supported it follows that the whole expression appearing inside of the "big" brackets
in \eqref{ecriture serie Natte operatorielle apres limite r et s vont infini} is holomorphic in $\pa{\ga,\be} \in \mc{D}_{0,\wt{\ga}_0} \times
\bs{\wt{U}}_{\be_0}$. As a consequence, the $m^{\e{th}}$ $\ga$-derivative at $\ga=0$  can be continued up to $\be=0$.
To get the value of the analytic continuation at this point it is in fact enough to set $\be=0$ in \eqref{ecriture serie Natte operatorielle apres limite r et s vont infini}.

The last step consists in taking the limit $w \tend +\infty$.
This operation will result in an extension of the integration contours from  bounded ones $\msc{C}^{\pa{w}}_{\eps_{\bs{t}}}$
to ones going to infinity $\msc{C}^{\pa{\infty}}_{\eps_{\bs{t}}}$. Hence, one needs to check that the resulting integrals will be convergent.
Note that the function $F^{ z_{\bs{+}} }_{ z_{\bs{-}} }\pa{z_{\bs{t}}}$  are bounded whenever $z_{\bs{t}}$ or any of the variables belonging to the set
$\{z_{\bs{+}} \}$ or $\{z_{\bs{-}} \}$ goes to infinity.  Also, the function $\mc{G}_{n;1}^{\pa{\be}}$ is bounded at infinity by a polynomial
in $z_{\bs{t}}$ of degree $n$, this uniformly in respect to $\ga$-derivatives of order $0,\dots,m$.
Therefore, as the functions
$\wt{H}_{n;x}^{\pa{\paa{\eps_{\bs{t}}}}}\pa{ \{\wt{F}^{ z_{\bs{+}} }_{ z_{\bs{-}} } \pa{z_{\bs{t}}}\}, \{ u\pa{z_{t}} \} , \{ z_{\bs{t}}\} }$
go to zero exponentially fast in all directions where $\msc{C}^{\pa{\infty}}_{\eps_{\bs{t}}}$
goes to $\infty$, the integrals over $\msc{C}^{\pa{\infty}}_{\eps_{\bs{t}}}$ are indeed convergent.

\qed

\subsection{Some more conjectures leading to the dominant asymptotics of $\rho\!\pa{x,t}$}

Under the assumption that
\begin{enumerate}
\item  the Taylor series $\sum_{m=0}^{+\infty} \ga^m \tf{\rho_{\e{eff}}^{\pa{m}}\!\pa{x,t}}{m!}$ is convergent up to $\ga=1$,

\item its sum gives $\rho\!\pa{x,t}$,

\item  the multidimensional Natte series given below is convergent.

\end{enumerate}
We get that  $\rho\pa{x,t}$ is obtained from
\eqref{ecriture serie Natte multDim coeff taylor rho eff} by removing the $m^{\e{th}}$ $\ga$-derivative symbol and setting $\ga=1$.
 It then remains to identify the coefficients in the first two lines with the
properly normalized thermodynamic limit of form factors of the field as given in
\eqref{appendix thermo lim FF facteur forme lambda 0 q}, \eqref{appendix thermo lim FF facteur forme -q q}  and
\eqref{appendix thermo lim FF facteur forme 0 0}. One then obtains the below series of multiple integral representation for
the thermodynamic limit of the one-particle reduced density matrix:
\bem
\rho\pa{x,t} = \sqrt{ \f{-2 i \pi }{ t \veps^{\prime \prime}\pa{\la_0} - x p^{\prime \prime}\pa{\la_0} } } \times
\f{ p^{\prime}\!\pa{\la_0} \ex{ix [u\pa{\la_0}-u\pa{q}] }   \abs{\mc{F}_q^{\la_0}}^2  }
{  \pac{-i\pa{x-t v_F } }^{ \big[  F^{\la_0}_{q} \!\pa{q} \big]^2}   \pac{i\pa{x + t v_F} }^{  \big[  F^{\la_0}_{q}\!\pa{-q} \big]^2 } }
\bs{1}_{\intoo{q}{+\infty}}\pa{\la_0}
 \\
+ \f{ \ex{-2ixp_F}  \abs{\mc{F}_q^{-q}}^2 }
{  \pac{-i\pa{x-t v_F } }^{ \big[ F_{q}^{-q} \!\pa{q} \big]^2}    \pac{i\pa{x + t v_F} }^{ \big[  F^{-q}_{q}\!\pa{-q}-1\big]^2 } }
\; + \;
\f{ \abs{\mc{F}_{\emptyset}^{\emptyset}}^2  }
{  \pac{-i\pa{x-t v_F } }^{  \big[  F^{\emptyset}_{\emptyset}\pa{q}+1 ]^2}  \pac{i\pa{x + t v_F} }^{ \big[ F^{\emptyset}_{\emptyset}\pa{-q}\big]^2 } }  \vspace{5mm}\\
\hspace{-1cm}\; + \;  \ex{-ix u\pa{q} } \sul{n=1}{+\infty} \sul{ \mc{K}_n }{} \sul{ \mc{E}_n ( \vec{k} ) }{}
\oint_{ \msc{C}_{\eps_{\bs{t}} }^{\pa{w}} }^{}
 \mc{G}_{ \abs{ \paa{z_{\bs{+}}} };1  }^{\pa{0}} \pab{ \paa{z_{\bs{+}}} }{ \paa{z_{\bs{-}}} }
\f{ H_{n;x}^{\pa{\paa{\eps_{\bs{t}}}}}\!\pa{ \paa{u\pa{z_{\bs{t}}} } ; \paa{z_{\bs{t}}}  } \big[ F^{z_{\bs{+}}}_{z_{\bs{-}}} \big]
\mc{B}\big[ F^{z_{\bs{+}}}_{z_{\bs{-}}}  ; p\big]    }
{  \pa{x-t v_F + i0^+ }^{ \pac{ F^{z_{\bs{+}}}_{z_{\bs{-}}} \pa{q}}^2}  \pa{x + t v_F }^{ \pac{ F^{z_{\bs{+}}}_{z_{\bs{-}}}\pa{-q}}^2 } }
    \cdot     \f{ \dd^n z_{\bs{t}} }{ \pa{2i\pi}^n }  \; .
\label{formule serie de Natte multidimensionnelle rho}
\end{multline}

It follows from the  above representation  and from conjecture \ref{Conjecture structure fine forme detaillee Hn}
that

\begin{cor}

The reduced density matrix admits the asymptotic expansion as given in subsection \ref{Theorem asymptotique rho}.

\end{cor}

\Proof

The proof is immediate as far as the multidimensional Natte series defining $\rho\pa{x,t}$ is convergent. Indeed, then, the
fine structure of the functions $H_{n;x}^{ \pa{\paa{\eps_{\bs{t}}}} }$ given in \eqref{ecriture forme detaille fonction Hn}
implies that all the contributions stemming from integrations are subdominant in respect to the first two lines in 
\eqref{formule serie de Natte multidimensionnelle rho}, this provided that $ \abs{ F^{z_{\bs{+}}}_{z_{\bs{-}}}\pa{\pm q} } <\tf{1}{2}$
for all configurations of variables in $\{z_{\bs{\pm}}\}$ that belong to $\{\la_0, \pm q\}$. 
This condidtion is not satisfied, especially if $\abs{\{ z_{\bs{+}}\}}$ becomes large. 
One should then invoke conjecture \ref{Conjecture structure fine forme detaillee Hn} stating that, in fact, higher order
oscillating terms in the representation \eqref{ecriture forme detaille fonction Hn} for $H_{n;x}^{ \pa{\paa{\eps_{\bs{t}}}} }$ 
are more dampen than it is apparent from the sum in \eqref{ecriture forme detaille fonction Hn}. 
This is enough to show that, indeed, the contributions to the oscillating tems at $-2p_F$, $u(\la_0)-u(q)$ and 
$0$ frequencies that are stemming from 
$H_{n;x}^{ \pa{\paa{\eps_{\bs{t}}}} }$  are subdominant in respect to the terms appearing in the first two lines of \eqref{formule serie de Natte multidimensionnelle rho}. There will of course arize terms oscillating at higher multiples of these frequencies 
exactly as it happens in  \eqref{ecriture forme detaille fonction Hn}. These higher oscillating terms give rise
to other critical exponents. For instance, one can convince oneself that, from this structure, one recovers the 
whole expected tower of critical exponents for the terms corresponding purely to 
oscillations at integer multiples of $u\pa{q}-u\pa{-q}$ as predicted in \cite{BerkovichMurthyWrongCFTBasedPredictionTimeMultiCorrNLSE}
on the basis of CFT-based technique. 
 \qed



\section*{Conclusion}

In this article, we have continued developing  a new method allowing one to build two types of series of multiple integral representation
for the correlation functions of integrable models starting from their form factor expansion. One of these series
which we called the multidimensional Natte series yields a straightforward
access to the large-distance/long-time asymptotic behavior of the two-point functions. In this way, we were able to extract
the long-time/large-distance asymptotic behavior of the reduced density matrix for the non-linear Schr\"{o}dinger model.

In order to provide applications to physically pertinent cases, the method we have developed has to recourse to a few conjectures.
The first one is  relative to the convergence of the series of multiple
integrals representing the correlators. This conjecture is supported by the free fermion case, where the convergence is rather quick, especially in the
large-distance/long-time regime. The second conjecture concerns the possibility of using an effective series instead of the
one appearing in the form factors expansion of two-point functions. Both series have been assumed to have the same thermodynamic limit
$N,L \tend +\infty$. This conjecture is supported, on the physical ground, by the argument that sums over states whose energies scale
as some power of the system-size ought to give a vanishing contribution to the sum over form factors once that the thermodynamic limit is taken.
It would be very interesting and important from the conceptual point of view to prove these two conjectures in the case of models that are away from their
free fermion points.

However, we do insist that we have organized the analysis in such a way that all of the aforementioned convergence issues are
separated from the asymptotic analysis part. Therefore, all the part of this work related purely to the asymptotic analysis is rigorous.
Moreover, we do expect that the scheme of asymptotic analysis we have developed can be applied in full rigor
to many cases which are free of convergence issues. We do also stress that, for the moment, the proofs of convergence
of series of multiple integral representations for correlation functions of models away from their free-fermion point
are, in general, an open problem. Apart from very specific representations related to the spin-$\tf{1}{2}$ XXZ chain,
the proof of convergence of a series representation for two-point functions could have been carried out only in the case of the Lee-Yang model
by F. Smirnov.

We have chosen to develop our method on the example of the one-particle reduced density matrix in the non-linear Schr\"{o}dinger model. The case
of the current-to-current correlation functions in this model will appear in \cite{KozKitMailTerNatteSeriesNLSECurrentCurrent}.
It seems however that the method is quite general and applicable to a vast class of integrable models where the form factors of local operators are known.
In particular, it should be applicable not only to lattice models where the form factors admit determinant-like representations
\cite{DeguchiMatsuiFFsHigherSpinXXZ,KMTFormfactorsperiodicXXZ,KorepinSlavnovFormFactorsNLSEasDeterminants,OotaInverseProblemForFieldTheoriesIntegrability}
but also to integrable field theories where the form factors of local operators can be computed through the resolution of the so-called
bootstrap program. 
For instance, the method seems applicable to the analysis of certain two-point functions (and their short-distance asymptotics) in the sine-Gordon model whose form factors have been obtained in
 \cite{SmirnovFormFactors,SmirnovIntegralRepSolitonFFSineGordonBootstrap}. In the latter case, we expect to
deal with some multidimensional deformation of the $3^{\e{rd}}$ Painlev\'{e} transcendent, a new type of special function whose description and
asymptotic behavior is interesting in its own right.

\section*{Acknowledgment}

I acknowledge the support of the EU Marie-Curie Excellence Grant MEXT-CT-2006-042695.
I would like to thank N.Kitanine, J.-M. Maillet, N. Slavnov and V. Terras for numerous stimulating discussions.


\appendix




\section{Thermodynamic limit of the Form Factors of conjugated fields}
\label{Appendix Thermo lim FF}

\subsection{Thermodynamic limit of form factors}
\label{Subsection Thermo Lim FF}

It has been shown in \cite{KozFFConjFieldNLSELatticeSpacingGoes0} with the help of techniques introduced in \cite{KozKitMailSlaTerEffectiveFormFactorsForXXZ,SlavnovFormFactorsNLSE} that the normalized modulus squared of the form factor of the
conjugated field taken between the ground state $\{ \la_a \}_1^N$ and any finite $n$ particle/hole type excited state
$\{ \mu_{\ell_a}\}_1^{N+1}$ admits the below behavior in the thermodynamic limit $N,L \tend +\infty$, $\tf{N}{L} \tend D$
\beq
\hspace{-5mm}
\f{  \abs{ \bra{\psi\pa{\paa{\mu_{\ell_a}}_1^{N+1}}} \Phi^{\dagger}\!\pa{0,0} \ket{ \psi\pa{\paa{\la}_1^N}  } }^2 }
{ \norm{\psi\pa{\paa{\mu_{\ell_a}}_{1}^{N+1}}}^2  \norm{\psi\pa{\paa{\la}_{1}^{N}}}^2     }
=
D_{0;L} \big[ F_{0} \big] \; \mc{R}_{N,n}\pab{ \{ \mu_{p_a} \}_1^n; \{ p_a \}_1^n }{ \paa{\mu_{h_a}}_1^n; \paa{h_a}_1^n  }\big[ F_{0} \big]
 \; \mc{G}^{\pa{0}}_{n;1} \pab{ \!\! \{ \mu_{p_a} \}_1^n  \!\!  }{  \!\! \paa{\mu_{h_a}}_1^n \!\!  }  \pa{1+ \e{O}\paf{ \ln L}{ L } } \;.
\label{appendix thermo lim FF formule pple pour limite thermo}
\enq
Above, $F_0$ corresponds to the thermodynamic limit of the shift function associated to the excited state of interest (at $\be=0$).
The auxiliary parameters of $F_0$ are undercurrent by the various functionals appearing above.
We recall that the parameters $\mu_k$ appearing in the \textit{rhs} of \eqref{appendix thermo lim FF formule pple pour limite thermo} are defined as the unique solutions to $\xi\pa{\mu_a}=\tf{a}{L}$.

\subsubsection*{The discreet part}

The first two functionals appearing in \eqref{appendix thermo lim FF formule pple pour limite thermo} correspond to the leading in $L$
behavior of the so-called singular part $\wh{D}_{N}\big[ \wh{F}_{\paa{\ell_a}}, \wh{\xi}_{\paa{\ell_a}},\wh{\xi} \big]$ of the form factor, namely
\beq
\wh{D}_{N}\pab{  \{ p_a\}_1^n }{ \{ h_a\}_1^n }   \big[ \wh{F}_{\paa{\ell_a}}, \wh{\xi}_{\paa{\ell_a}},\wh{\xi} \big] =
D_{0;L} \big[ F_{0} \big] \; \mc{R}_{N,n}\pab{ \{ \mu_{p_a} \}_1^n; \{ p_a \}_1^n }{ \paa{\mu_{h_a}}_1^n; \paa{h_a}_1^n  }\big[ F_{0} \big]
\pa{1+ \e{O}\paf{ \ln L}{ L } } \;.
\label{appendix thermo lim FF ecriture explicite thermo lim Dn hat}
\enq
Given any function $\nu\!\pa{\la}$ holomorphic in some neighborhood of $\intff{-q}{q}$, one has
\beq
D_{0;L}\!\pac{\nu} =
\f{\pac{\varkappa\!\pac{\nu}\pa{-q} }^{\nu\pa{-q}} }{ \pac{\varkappa\pac{\nu}\pa{q} }^{\nu\pa{q} + 2} }
  \pl{a=1}{n} \paf{\la_{N+1}-\mu_{p_a}}{\la_{N+1}-\mu_{h_a}}^2
 \f{ q \, \tf{ G^2\!\pa{1-\nu\pa{-q}}G^2\!\pa{2+\nu\pa{q}} }{\pi} }{  \pa{2\pi}^{\nu\pa{q} - \nu\pa{-q}} \cdot
 \pac{ 2q L \xi^{\prime}\!\pa{q} }^{\pa{\nu\pa{q} + 1}^2 + \nu^2 \pa{-q}}  }
\cdot \ex{\f{1}{2} \Int{-q}{q} \f{\nu^{\prime}\!\pa{\la}\nu \pa{\mu}-\nu^{\prime}\!\pa{\mu}\nu\pa{\la} }{\la-\mu} \dd \la \dd \mu }
 \; ,
\label{AppendixThermoLimD+zero}
\enq
The parameter $\la_{N+1}$ appearing above is defined as the unique solution to $L \xi_{\nu}\pa{\la_{N+1}}= N+1$,
$\varkappa\pac{\nu}\pa{\la}$ is given by \eqref{definition fonctionnelle A+ et kappa} and $G$ stands for the Barnes double Gamma
function. Finally, we agree upon
\bem
\hspace{-4mm} \mc{R}_{N,n}\pab{ \{ \mu_{p_a} \}_1^n; \paa{p_a}_1^n }{ \paa{\mu_{h_a}}_1^n; \paa{h_a}_1^n  }\pac{\nu}  =
\pl{a=1}{n} \paa{ \f{ \vp\pa{\mu_{h_a},\mu_{h_a}}  \vp(\mu_{p_a},\mu_{p_a}) \ex{\al\pac{\nu}\pa{\mu_{p_a}}} }
 { \vp (\mu_{p_a},\mu_{h_a})  \vp(\mu_{h_a},\mu_{p_a}) \ex{\al\pac{\nu}\pa{\mu_{h_a}}}  }  }
\f{ \pl{a<b}{n} \vp^2(\mu_{p_a},\mu_{p_b}) \vp^2\pa{\mu_{h_a},\mu_{h_b}} }
                                        { \pl{a\not= b }{n} \vp^2(\mu_{p_a},\mu_{h_b}) }
\det^{2}_{n}\pac{ \f{ 1}{ h_a-p_b}}
\\
\hspace{1cm}\times \pl{a=1}{n} \paf{\sin \pac{ \pi \nu\pa{\mu_{h_a}} } }{\pi   }^2  \cdot
\pl{a=1}{n} \Ga^{2}\pab{  p_a-N-1+\nu(\mu_{p_a})  , \; p_a,  \;N+2-h_a -\nu\pa{\mu_{h_a}} , \; h_a +\nu\pa{\mu_{h_a}} }
{ p_a-N-1 , \;  p_a+\nu(\mu_{p_a}) , \; N+2-h_a ,  \; h_a } \;.
\label{definition fonctionelle RNn}
\end{multline}
There
\beq
\al\pac{\nu}\pa{\om} = 2 \nu\pa{\om} \ln \paf{\vp\pa{\om,q} }{ \vp\pa{\om,-q} }
+2 \Int{-q}{q}  \f{\nu\pa{\la}-\nu\pa{\om}}{\la-\om} \dd \la  \; \qquad  \e{and} \quad
\vp\pa{\la,\mu}= 2\pi \f{ \la-\mu }{ p\pa{\la} - p\pa{\mu} }  \; .
\label{definition fonction aleph et varphi}
\enq
Above, we have used the standard hypergeometric-type representation for products of $\Ga$-functions:
\beq
\Ga\pab{ a_1,\dots , a_n }{ b_1,\dots, b_{n} }= \pl{k=1}{n} \f{ \Ga\pa{a_k} }{ \Ga\pa{b_k} } \; .
\enq

\subsubsection*{Description of $\mc{G}_{n;1}^{\pa{\be}}$}

In order to give an explicit representation for $\mc{G}_{n;\ga}^{\pa{\be}}$ we need to introduce a few notations.
First, let $F_{\be}$ correspond to the thermodynamic limit of the $\be$-deformed shift function associated to the choice of the rapidities
$\{ \mu_{p_a} \}$ for the particles and $\{ \mu_{h_a} \}$ for the holes.  The auxiliary arguments of the shift function will be kept undercurrent.
Also, let $m\in \mathbb{N} $ and $U_{\de}$ be the open strip \eqref{definiton voisinage ouvert de R pour holomorphie Gn} around $\R$.

\noindent Then there exists $\wt{\ga}_0>0$ small enough and  $\be_0\in \Cx$ with $\Re\pa{\be_0}>0$ large enough and
$\Im\pa{\be_0}>0$ small enough such that
\beq
\ex{2i\pi \ga F_{\be}\pa{\om}}-1 \not=0 \quad \e{uniformly} \; \e{in} \;  n=0,\dots,m  \quad \e{and} \quad
\pa{\ga, \om , \be, \{\mu_{p_a} \}_1^{n} , \{ \mu_{h_a}\}_1^{n} } \in \mc{D}_{0,\wt{\ga}_0}  \times U_{\de} \times \bs{\wt{U}}_{\be_0} \times U^{n}_{\de} \times U_{\de}^{n} \;.
\label{appendix thermo lim FF condition validite rep G n beta}
\enq
Let all parameters $\mu_{h_a}$, $a=1, \dots, n$ belong to a compact $K_{q+\eps} \supset \intff{-q}{q}$ for some $\eps>0$ and let $\msc{C}_{q+\eps}$
be a small counterclockwise loop around this compact $K_{q+\eps}$, then $\mc{G}_{n,\ga}^{\pa{\be}}$ admits the below representation
\bem
\mc{G}_{n;\ga}^{\pa{\be}}\pab{ \!\! \{ \mu_{p_a} \}_1^n \!\! }{ \!\! \paa{\mu_{h_a}}_1^n \!\! } =
 \ex{- 2i\pi  \sul{\eps=\pm}{} C\pac{ \ga F_{\be}}\pa{q + \eps ic}    }
\pl{a=1}{n} \pl{\eps=\pm}{} \paa{ \f{   \mu_{h_a}-q+\eps ic }{ \mu_{p_a}-q+\eps ic   }
\f{ \ex{2i\pi   C\pac{\ga F_{\be}}\pa{\mu_{h_a} + \eps ic}    }    } {  \ex{2i\pi   C\pac{\ga F_{\be}}\pa{\mu_{p_a} + \eps ic} }  } }
 \ex{C_0\pac{\ga F_{\be}}}
 \\
\times \pl{a,b=1}{n} \f{\big(\mu_{p_a} - \mu_{h_b}-ic \big) \big(\mu_{h_a}-\mu_{p_b}-ic \big) }
		{  \big( \mu_{p_a}-\mu_{p_b}-ic\big) \pa{\mu_{h_a}-\mu_{h_b}-ic} } \;
\cdot  \f{  \ddet{\msc{C}_{q+\eps}}{I+ \ga \mc{V}\big[ \ga F_{\be} \big] } 
\ddet{ \msc{C}_{q+\eps}  }{I+ \ga \ov{\mc{V}} \big[ \ga F_{\be} \big]  }  }{   \det^2\pac{I-\tf{K}{2\pi}}  }  \; .
\label{formule explicite G+ thermo}
\end{multline}
There $C\big[ F_{\be} \big]$ stands for the Cauchy transform on $\intff{-q}{q}$ and $C_0\big[ F_{\be} \big]$ is given by a double integral
\beq
C\big[ F_{\be} \big]\pa{\la} = \Int{-q}{q} \f{\dd \mu}{2i\pi} \f{ F_{\be}\pa{\mu} }{\mu-\la} \qquad \e{and} \qquad
C_0\big[ F_{\be} \big] = -\Int{-q}{q} \f{ F_{\be}\pa{\la} F_{\be}\pa{\mu} }{  \pa{\la-\mu -ic}^2 }   \dd \la \dd \mu  \;.
\label{appendix themo FF definition transfo Cauchy et C0}
\enq
The integral kernels $\mc{V}$ and $\ov{\mc{V}}$ read
\beq
\mc{V} \pac{\nu}\!\pa{\om,\om^{\prime}} = \f{-1}{2\pi}\f{\om-q}{\om-q+ic}
\pl{a=1}{n} \paa{ \f{ \pa{\om- \mu_{p_a}} \pa{\om- \mu_{h_a}+ic} }
{\pa{\om- \mu_{h_a}} \pa{\om- \mu_{p_a}+ic} }   }  \cdot \ex{C\pac{2i\pi \nu}\pa{\om} - C\pac{2i\pi \nu}\pa{\om + ic}  }
 \f{ K\pa{\om-\om^{\prime}} }{ \ex{-2i\pi \nu\pa{\om}}-1 }
\label{formule noyau integral U}
\enq
and
\beq
\ov{\mc{V}}\pac{\nu}\!\pa{\om,\om^{\prime}}  =\f{1}{2\pi} \f{\om-q}{\om-q - ic}
\pl{a=1}{n} \paa{ \f{ \pa{\om- \mu_{p_a}} \pa{\om- \mu_{h_a}-ic} }
{\pa{\om- \mu_{h_a}} \pa{\om- \mu_{p_a}-ic} }   }  \cdot \ex{C\pac{2i\pi \nu}\pa{\om} - C\pac{2i\pi \nu}\pa{\om - ic}  }
 \f{ K\pa{\om-\om^{\prime}} }{ \ex{2i\pi \nu\pa{\om}}-1 } \;.
\label{formule noyau integral U bar}
\enq

The representation \eqref{formule explicite G+ thermo} is valid for $n=0,\dots,m$ and
$\pa{\ga , \be, \{\mu_{p_a} \}_1^{n} , \{ \mu_{h_a}\}_1^{n} } \in
\ov{\mc{D}}_{0,\ga_0} \times \bs{\wt{U}}_{\be_0} \times U_{\de}^{n} \times  K^{n}_{q+\eps} $ and defines a holomorphic function of these parameters belonging to this set.

It is also valid at $\ga=1$, provided that $\Re\pa{\be_0}>0$ is taken large enough for condition
\eqref{appendix thermo lim FF condition validite rep G n beta} to be fulfilled at $\ga=1$.

Finally as follows from proposition \ref{Proposition Holomorphie en beta et rapidite part-trou det fred} given below,
the product $\mc{D}_{0;L} [F_{\be}] \mc{R}_{N,n}[F_{\be}]\mc{G}^{\pa{\be}}_{n;1}$ is
holomorphic in $\Re{\be}\geq 0$,
and can thus be analytically continued from $\bs{\wt{U}}_{\be_0}$ up to $\be=0$. It is in this sense that the
formula \eqref{appendix thermo lim FF formule pple pour limite thermo} for the leading asymptotics
in the size $L$ of the form factors of $\Phi^{\dagger}$ is to be understood.

\begin{prop}
\label{Proposition Holomorphie en beta et rapidite part-trou det fred}
\cite{KozFFConjFieldNLSELatticeSpacingGoes0}

Let $m\in \mathbb{N}$,  $\de>0$ small enough define the width of the strip $U_{\de}$
around $\R$ and $\pa{\{ \mu_{p_a} \}_1^n ; \{ \mu_{h_a} \}_1^n } \in U_{\de}^n \times K^n_{q+\eps}$,
where $\eps>0$ and the compact $K_{q+\eps}$ is as defined by \eqref{definition compact KA}.

Let $\nu$, $h$ and $\tau$ be holomorphic function in the strip $U_{\de}$ around $\R$
and such that $h\pa{ U_{\de} }\subset \paa{ z \; : \; \Re\pa{z} >0 }$ and $\Im \pa{h\pa{z}} $ is bounded on $U_{\de}$.
Set $\nu_{\be}\pa{\la}=\nu\pa{\la}+ i \be h\pa{\la}$.

\noindent Then, there exists
\begin{itemize}

\item $\be_0 \in \Cx$ with $\Re\pa{\be_0}>0$ large enough  and $\Im\pa{\be_0}>0$ small enough
\item $\wt{\ga_0}>0$ and small enough
\item  a small loop $\msc{C}_{q+\eps} \subset  U_{\de}$ around the compact $K_{q+\eps}$
\end{itemize}
such that uniformly in $\be \in \bs{\wt{U}}_{\be_0}$ the function $\la \mapsto \ex{-2i\pi \ga \pa{\nu+ i\be h}\pa{\la}}-1$ has no roots inside of
$\msc{C}_{q+\eps}$. In addition, the function
\beq
\big( \{ \mu_{p_a} \}_1^n,\paa{\mu_{h_a}}_1^n, \ga, \be \big) \mapsto
G\pa{1- \ga \tau\pa{-q} } G\pa{2+  \ga \tau\pa{q}}  \pl{a=1}{n} \pa{ \ex{-2i\pi \ga \nu_{\be}\pa{\mu_{h_a}}} - 1 } \; \cdot \;
\ddet{ \msc{C}_{q+\eps} }{ I+ \ga \mc{V}\pac{   \ga \nu_{\be}}\big(  \{ \mu_{p_a} \}_1^n,\paa{\mu_{h_a}}_1^n \big)  } \;
\label{appendix thermo lim FF fonction avec limite themor reg}
\enq
is a holomorphic function in $U_{\de}^{n} \times K_{q+\eps}^n \times \mc{D}_{0,\wt{\ga}_0} \times \bs{\wt{U}}_{\be_0}$, this uniformly in $0\leq n\leq m$.

It admits a (unique) analytic continuation to $U_{\de}^{n} \times K_{q+\eps}^n \times \ov{\mc{D}}_{0,1} \times
\paa{ z  \in \Cx \; : \; \Re\pa{z} \geq 0}$.
In particular, it has a well defined $\be \tend 0$ limit.
The $\be\tend 0$ limit of this analytic continuation is still holomorphic in
$\big( \{ \mu_{p_a} \}_1^n,\paa{\mu_{h_a}}_1^n \big) \in U_{\de}^{n} \times K_{q+\eps}^n$.
\end{prop}

In \eqref{appendix thermo lim FF fonction avec limite themor reg} we have insisted explicitly on the dependence of the integral kernel $\mc{V}$
on the auxiliary parameters  $\big(  \{ \mu_{p_a} \}_1^n,\paa{\mu_{h_a}}_1^n \big)$, \textit{cf} \eqref{formule noyau integral U}.
The same proposition holds when the kernel $\mc{V}$ is replaced by $\ov{\mc{V}}$ as it has been defined in \eqref{formule noyau integral U bar}.

\subsubsection*{Alternative representation for $\mc{G}_{n;\ga}^{\pa{\be}}$}

\vspace{2mm}

It so happens that the smooth part of the form factor's asymptotics admits a representation as a functional acting on a unique
function $H$. More precisely,

\begin{lemme}
\label{Lemme fonctionnelle GA pour partie lisse FF}
Let $m\in \mathbb{N}$ and the strip $U_{\de}$ be fixed. Let $A>0$ be some constant defining the size of the compact $K_{A}$ \eqref{definition compact KA}.
Then, there exists A,$\de$, $m$ dependent parameters
\begin{itemize}
\item $\be_0\in \Cx$ with $\Re\pa{\be_0}>0$ large enough and $\Im\pa{\be_0}>0$ small enough
\item $ \wt{\ga}_0>0$ small enough
\end{itemize}
such that uniformly in $\pa{\{ y_a \}_1^{n} , \{ z_a \}_1^{n} } \in K_A^{n}\times K_A^n$, $\abs{\ga} \leq \wt{\ga}_0$, $\be \in \bs{\wt{U}}_{\be_0}$
and  $0\leq n \leq m$
\beq
\mc{G}_{n;\ga}^{\pa{\be}}\pab{  \!\! \{ y_a \}_1^n \!\! }{ \!\! \paa{ z_a }_1^n \!\! }  =
\msc{G}_{\ga;A}^{\pa{\be}} \pac{H \pabb{\cdot }{ \{ y_a \}_1^n }{ \paa{ z_a }_1^n } }
\quad \e{with} \quad H \pabb{ \la }{ \{ y_a \}_1^n }{ \paa{ z_a }_1^n }
=\sul{ a=1 }{ n }  \f{1}{\la - y_a } - \f{1}{ \la - z_a } \;.
\label{definition nouvelle fonctionnelle pure G smooth}
\enq
The functional $\msc{G}_{\ga;A}^{\pa{\be}}$ acts on a bounded loop $\msc{C}\pa{K_A}\subset U_{\de}$ around the compact $K_A$.  The functional
$\msc{G}_{\ga;A}^{\pa{\be}}\pac{\varpi}$ is a regular functional (\textit{cf} definition \ref{Definition Fonctionelle reguliere})
of $\varpi$ in respect to the pair $\pa{M_{\msc{G}_A},\msc{C}\pa{K_A} }$ where the compact $M_{\msc{G}_A}$ has its boundaries
given by $\msc{C}_{out}$ and $\msc{C}_{in}$ as depicted in the \textit{rhs} of Fig.~\ref{contour exemple de courbes encerclantes}.
For all $\varpi \in \msc{O}\pa{M_{\msc{G}_A}}$ such that  $\norm{\varpi}_{\msc{C}\pa{K_A}} \leq C_{\msc{G}_A}$, where $C_{\msc{G}_A}$
is a constant of regularity of the functional $\msc{G}_{\ga;A}^{\pa{\be}}$, one has
\bem
\msc{G}_{\ga;A}^{\pa{\be}}\pac{\varpi} = \ex{C_0\pac{\ga G_{\be}} }
\pl{\eps=\pm}{} \exp\Bigg\{ - \Oint{\msc{C}\pa{K_A}}{} \hspace{-3mm}\f{\dd z }{2i\pi} \varpi\pa{z}  \paa{2i\pi C\pac{\ga G_{\be}}\pa{z+i\eps c}
			+\ln\pa{z-q+i\eps c}}  \Bigg\} \\
\times  \exp\Bigg\{ -\Oint{\msc{C}\pa{K_A}}{} \hspace{-3mm}\f{ \dd y \dd z}{ \pa{2i\pi}^2 } \varpi\pa{y}\varpi\pa{z} \ln\pa{z-y-ic} \Bigg\}
\f{  \ddet{\msc{C}_{A}}{I+ \ga \msc{V}\pac{\ga G_{\be},\varpi}}
\ddet{ \msc{C}_{A}  }{I+ \ga \ov{\msc{V}} \pac{\ga G_{\be},\varpi}} }
{   \det^2\pac{I-\tf{K}{2\pi}} \; \exp \Big\{ 2i\pi \sul{\eps=\pm}{} C\big[ \ga G_{\be} \big] \pa{q + i \eps c} \Big\}   }  \;.
\label{appendix thremo lim FF definition fonctionnelle G super caligraphique}
\end{multline}
In the above formula, one should understand $G_{\be}$ as a one parameter family of functionals of $\varpi$ given by
\beq
G_{\be}\pa{\la} \equiv G_{\be}\pac{\varpi}\pa{\la} = \pa{i\be -\tf{1}{2}} Z\pa{\la} - \phi\pa{\la,q} -\Oint{\msc{C}\pa{K_A}}{} \!\! \f{\dd z}{2i\pi}
\varpi\pa{z} \phi\pa{\la,z} \; .
\label{appendix thermo lim FF fonctionelle G beta}
\enq
 In the second line of
\eqref{appendix thremo lim FF definition fonctionnelle G super caligraphique} there appear
Fredholm determinants of integral operators acting on a contour $\msc{C}_A$. The contour $\msc{C}_A$ corresponds to a loop around $\msc{C}\pa{K_A}$
such that $ \msc{C}_A  \subset U_{\de}$. The kernels read
\beq
\hspace{-7mm}\msc{V}\! \pac{\nu,\varpi}\pa{\om,\om^{\prime}} = \f{-1}{2\pi}\f{\om-q}{\om-q+ic}
\exp\Bigg\{\Oint{\msc{C}\pa{K_A} }{} \hspace{-1mm}\f{\dd z}{2i\pi} \varpi\pa{z} \ln\paf{\om-z}{\om-z+ic} \Bigg\}
\cdot \ex{C\pac{2i\pi \nu}\pa{\om} - C\pac{2i\pi \nu}\pa{\om + ic}  }
 \f{ K\pa{\om-\om^{\prime}} }{ \ex{-2i\pi \nu\pa{\om}}-1 }
\enq
and
\beq
\hspace{-7mm} \ov{\msc{V}}\! \pac{\nu,\varpi}\pa{\om,\om^{\prime}} = \f{1}{2\pi}\f{\om-q}{\om-q-ic}
\exp\Bigg\{\Oint{\msc{C}\pa{K_A} }{} \hspace{-1mm}\f{\dd z}{2i\pi} \varpi\pa{z} \ln\paf{\om-z}{\om-z-ic} \Bigg\}
\cdot \ex{C\pac{2i\pi \nu}\pa{\om} - C\pac{2i\pi \nu}\pa{\om - ic}  }
 \f{ K\pa{\om-\om^{\prime}} }{  \ex{2i\pi \nu\pa{\om}}-1 }
\label{appendix thermo lim FF formule noyau v cal bar}
\enq

\vspace{2mm}
The $A$, $m$ and $\de$-dependent parameters $\be_0$ and  $\wt{\ga_0}$ and the compacts $\msc{C}\pa{K_A}$, $M_{\msc{G}_A}$ 
are such that the constant of regularity $C_{\msc{G}_{A}}$ satisfies to the estimates given in
\eqref{ecriture condition grandeur constante de regularite GAkappa} and is such that one has

\beq
\forall \; \norm{\varpi}_{\msc{C}\pa{K_A}} < C_{\msc{G}_A} \quad \norm{ \wt{\ga}_0 G_{\be}\pac{\varpi} }_{ U_{\de} } < \tf{1}{2}
\quad \e{and} \quad \norm{H}_{\msc{C}\pa{K_A}} <  C_{\msc{G}_A} \;\; \e{uniformly} \; \e{in} \; \pa{\{y_a\}_1^n, \{ z_a\}_1^n} \in K_A^n\times K_A^n \;.
\nonumber
\enq
\end{lemme}

\Proof

\noindent We first check that $\msc{G}_{\ga;A}^{\pa{\be}}$ is a regular functional.
\begin{itemize}
\item $G_{\be}\pac{\varpi}$ is a regular functional as it is linear in $\varpi$ and $\msc{C}\pa{K_A}$ is compact.
\item the estimates $\abs{\ex{x}-\ex{y}} \leq \ex{\abs{x}+\abs{y}}\abs{x-y}$, majorations of integrals in terms of sup norm
and derivation under the integral sign theorems ensure that all of the exponential pre-factors in
\eqref{appendix thremo lim FF definition fonctionnelle G super caligraphique} are also regular functionals of $\varpi$.
\end{itemize}
The associated constants of regularity can be taken as large as desired. It thus remains to focus on the Fredholm determinants.
For this let us first assume that we are able to pick the contours $\msc{C}_{out/in}$ delimiting the boundary of the compact $M_{\msc{G}_A}$
in such a way that there exists
\beq
 \; \be_0 \in \Cx \;\; \e{and} \;\; \wt{\ga}_0 >0 \quad \e{such}\;\e{that}
\quad  \ex{2i\pi \ga G_{\be} \pac{\varpi(*,\bs{y})}\pa{\la}}-1 \not=0  \quad  \forall
\pa{\la,\bs{y}, \be ,\ga } \in U_{\de} \times W_y \times \bs{\wt{U}}_{\be_0}
\times \mc{D}_{0,\wt{\ga}_0}
\label{appendix thermo lim FF ecriture condition non-annulation facteur exponentiel}
\enq
this for any holomorphic function $\varpi\!\pa{\la,\bs{y}}$ on $ M_{\msc{G}_A} \times W_y$, $W_y \subset \Cx^{\ell_y}$,
that satisfies $ \norm{\varpi}_{\msc{C}\!\pa{K_A}\times W_y}<C_{\msc{G}_A}$.

If this condition is satisfied, then the integral kernels $\ga \msc{V}\pac{\ga G_{\be}, \ga G_{\be}, \varpi}\!\pa{\om,\om^{\prime}}$
and $\ga \ov{\msc{V}}\pac{\ga G_{\be}, \ga G_{\be}, \varpi} \!\pa{\om,\om^{\prime}}$ are holomorphic in $\om, \om^{\prime}$ belonging
to a small neighborhood of $\msc{C}_A$ and $\bs{y}\in W_y$.
The contour $\msc{C}_A$ being compact, the two integral operators
$\ga \mc{V}\pac{\ga G_{\be}, \ga G_{\be}, \varpi}$  and $\ga \ov{\mc{V}}\pac{\ga G_{\be}, \ga G_{\be}, \varpi}$ are trace class
operators that have an analytic dependence on $\bs{y}\in W_y$.

Recall that if $A$, $B$ are trace class operators ($\norm{\cdot}_1$ stands for the trace class norm) then
\beq
\abs{  \ddet{}{I+A} - \ddet{}{I+B} } \leq  \norm{A-B}_1 \ex{ \norm{A}_1+\norm{B}_1+1} \, .
\enq
Also \cite{SimonsInfiniteDimensionalDeterminants}, if $A\pa{\bs{y}}$,  $\bs{y}\in W_y \subset \Cx^{\ell_y}$, is an analytic
trace class operator then $\ddet{}{I+A\pa{\bs{y}}}$ is holomorphic on $W_y$
These two properties show that, indeed, in \eqref{appendix thremo lim FF definition fonctionnelle G super caligraphique},  the two Fredholm determinants
of integral operators acting on the contour $\msc{C}_A$ are regular functionals of $\varpi$.

Hence, it remains to prove the existence of $\wt{\ga}_0$ and $\be_0$ such that condition
\eqref{appendix thermo lim FF ecriture condition non-annulation facteur exponentiel} holds.
Given $\varpi\!\pa{\la,\bs{y}} \in \msc{O}\big(  M_{\msc{G}_A} \times W_y \big)$, the  function
$\om \mapsto  \ex{2i\pi \ga G_{\be} \pac{\varpi}\pa{\om}}-1$ has no zeroes provided that
\beq
\abs{ \wt{\ga}_0 G_{\be}\pac{\varpi(*,\bs{y})}\pa{\la} } < \tf{1}{2} \qquad \e{and} \qquad
\Im\pa{ G_{\be}\pac{\varpi(*,\bs{y})}\pa{\la} } >0 \quad \e{uniformly}\;\e{in} \;  \pa{\la,\bs{y},\be} \in U_{\de} \times W_y \times \bs{\wt{U}}_{\be_0} \;.
\label{appendix thermo lim FF bornes Gbeta de varpi}
\enq
One has that, for $\be \in \bs{\wt{U}}_{\be_0}$
\beq
\Im\pa{ G_{\be}\pac{\varpi}\pa{\la} }  > \Re\pa{\be_0} \inf_{\la \in U_{\de}} \pac{ \Re\pa{Z\pa{\la}} }
- \pa{\Im\pa{\be_0} + \tf{1}{2} }  \norm{ \Im\pa{Z} }_{U_{\de}} - \norm{\phi}_{U^{2}_{\de}}
- \norm{\varpi}_{\msc{C}\pa{K_A}\times W_y} \sup_{\la\in U_{\de}} \Oint{ \msc{C}\pa{K_A} }{} \!\! \f{ \abs{ \dd z} }{2\pi} \abs{\phi\pa{\la,z}} \;.
\nonumber
\enq
Hence, $\Im\pa{ G_{\be}\pac{\varpi}\pa{\la} } >0$ as soon as $\norm{\varpi}_{\msc{C}\pa{K_A}\times W_y} \leq C_{\msc{G}_{A}}$ with
\beq
C_{\msc{G}_{A}} = \Bigg\{ 2 \sup_{\la\in U_{\de}} \Oint{ \msc{C}\pa{K_A} }{} \!\! \f{ \abs{ \dd z} }{2\pi} \abs{\phi\pa{\la,z}}  \Bigg\}^{-1}
\cdot \paa{ \Re\pa{\be_0} \inf_{\la \in U_{\de}} \pac{ \Re\pa{Z\pa{\la}} }
- \pa{\Im\pa{\be_0} + \tf{1}{2} }  \norm{ \Im\pa{Z} }_{U_{\de}} - \norm{\phi}_{U^{2}_{\de}}
 } \;.
\enq
Here $\Re\pa{\be_0}>0$ is taken large enough for $C_{\msc{G}_A}$ as defined above to be positive.
Then, if $\norm{\varpi}_{\msc{C}\pa{K_A}\times W_y} \leq C_{\msc{G}_{A}}$ with $C_{\msc{G}_A}$ as given above, one has
\beqa
\sup_{ \substack{ \om \in U_{\de} \\ \bs{y} \in W_y} } \abs{ G_{\be}\pac{\varpi(*,\bs{y})}\pa{\om} } &\leq& \pa{10 \Re\pa{\be_0} + \Im\pa{\be_0} +\tf{1}{2}} \norm{Z}_{U_{\de}}
+ \norm{ \phi}_{U_{\de}^2} + \norm{\varpi}_{\msc{C}\pa{K_A}\times W_y}
\sup_{\la\in U_{\de}} \Oint{ \msc{C}\pa{K_A} }{} \!\! \f{ \abs{ \dd z} }{2\pi} \abs{\phi\pa{\la,z}}  \nonumber \\
& < & \pa{11 \Re\pa{\be_0} + 2 \Im\pa{\be_0} + 1 } \norm{Z}_{U_{\de}} \;.
\eeqa
Hence, if we take $\wt{\ga}_0^{-1} = 2 \pa{11 \Re\pa{\be_0} + 2 \Im\pa{\be_0} + 1 } \norm{Z}_{U_{\de}} $,
the condition $\abs{ \ga G_{\be}\pac{\varpi}} < \tf{1}{2}$ will be satisfied for all $\abs{\ga} \leq \wt{\ga}_0$
and $\be \in \bs{\wt{U}}_{\be_0}$.
It remains to tune $\Re\pa{\be_0}$ so that conditions
\beq
C_{\msc{G}_A} \cdot \f{ \pi  \e{d}\!\pa{\Dp{}M_{\msc{G}_A} , \msc{C}\pa{K_A} } }
{  \abs{\Dp{}M_{\msc{G}_A}} + 2\pi  \e{d}\!\pa{\Dp{}M_{\msc{G}_A} , \msc{C}\pa{K_A} } } >A
\qquad \e{and} \qquad \f{ 2m  }{ \e{d}\!\pa{ K_A , \msc{C}\pa{K_A} } } < C_{\msc{G}_{A}} \;.
\label{appendix thermo lim FF ecriture condition grandeur cste reg G cal A}
\enq
are satisfied.

One can always choose the contours $\msc{C}_{out/in}$ defining $\Dp{}M_{\msc{G}_A}$ in such a way that
$\e{d}\!\pa{\Dp{}M_{\msc{G}_A} , \msc{C}\pa{K_A} }>c$ this uniformly in A>0. These contours can also be chosen such that
there exists an $A$-independent constant $c_1$ with $\abs{\Dp{}M_{\msc{G}_A}}< c_1 A$. It is also clear that the contour $\msc{C}\pa{K_A}$
surrounding the compact $K_A$ can be chosen such that $\abs{ \msc{C}\pa{K_A} }< c_2 A$ for some $A$-independent constant $c_2$
and also $\e{d}\!\pa{ K_A , \msc{C}\pa{K_A} }>c^{\prime}$.
It is then enough to take $\Re\pa{\be_0}> c_{\be_0} A^3$ with $c_{\be_0}$ being properly tuned in terms of $c, c^{\prime}, c_1, c_2$ so that
conditions \eqref{appendix thermo lim FF ecriture condition grandeur cste reg G cal A} hold for any $A$ sufficiently large. 

Note that the second condition in \eqref{appendix thermo lim FF ecriture condition grandeur cste reg G cal A} guarantees that the function
$H$ as given in
\eqref{definition nouvelle fonctionnelle pure G smooth} satisfies $\norm{H}_{\msc{C}\pa{K_A}}< C_{\msc{G}_A}$ uniformly in the parameters
$\pa{\{\mu_{p_a}\}_1^n,\{\mu_{h_a}\}_1^n } \in K_A^n\times K_A^n$.

Having proved that $\msc{G}_{\ga;A}^{\pa{\be}}$ is a regular functional with a regularity constant $C_{\msc{G}_{A}}>0$ sufficiently large,
we can evaluate it on $H$. Then, it is readily seen that $G_{\be}\pac{\varpi}\pa{\la}$ coincides
 with the shift function $F_{\be}$ once upon taking $\varpi =H$ as given in
\eqref{definition nouvelle fonctionnelle pure G smooth}. All other integrals involving $\varpi=H$ are computed by the residues
at $\mu_{p_a}$ and $\mu_{h_a}$. All calculations done, one recovers the representation \eqref{formule explicite G+ thermo}
for the function $\mc{G}_{n;\ga}^{\pa{\be}}$.
We stress that the parameters $\wt{\ga}_0$ and $\be_0$ ensuring the regularity of the functional $\msc{G}_{\ga;A}^{\pa{\be}}$
are also such that  $\mc{G}_{n;\ga}^{\pa{\be}}$ is well defined due to conditions \eqref{appendix thermo lim FF bornes Gbeta de varpi}.
\qed

\subsubsection*{Regular functional for $\wh{\mc{G}}_{N;\ga}$}

A very similar representation to the one given in the previous lemma exists for the functional $\wh{\mc{G}}_{N;\ga}$.
\begin{lemme}
\label{Lemme fonctionnelle hat GA pour partie lisse hat FF}
Let $m\in \mathbb{N}$ and the strip $U_{\de}$ be fixed. Let $A>0$ be some constant defining the size of the compact $K_{A}$ \eqref{definition compact KA}.
Then, there exists $A$, $m$ and $\de$-dependent constants
\begin{itemize}
\item $\be_0\in \Cx$ with $\Re\pa{\be_0}>0$ large enough and $\Im\pa{\be_0}>0$ small enough,
\item $\wt{\ga}_0>0$ small enough,
\end{itemize}
such that for $L$ large enough and uniformly in $\pa{\{\mu_{p_a} \}_1^{n} , \{ \mu_{h_a}\}_1^{n} } \in K_A^{n}\times K_A^n$, $\abs{\ga} \leq \wt{\ga}_0$
 and  $0\leq n \leq m$
\beq
\wh{\mc{G}}_{N;\ga}\pab{  \!\! \{ p_a \}_1^n \!\! }{ \!\! \{ h_a \}_1^n \!\! }\big[ \ga F_{\be}, \xi, \xi_{\ga F_{\be}} \big] =
\wh{\msc{G}}_{\ga;A}^{ \, \pa{\be}} \pac{H \pabb{ *  }{ \{ \mu_{p_a} \}_1^n }{ \paa{\mu_{h_a}}_1^n } }
\quad \e{with} \quad H \pabb{ \la }{ \{ \mu_{p_a} \}_1^n }{ \paa{\mu_{h_a}}_1^n }
=\sul{ a=1 }{ n }  \f{1}{\la - \mu_{p_a} } - \f{1}{ \la - \mu_{h_a}} \;.
\label{appendix thermo lim FF definition fonctionnelle smooth hat cal G }
\enq
The functional $\wh{\msc{G}}_{\ga;A}^{\,\pa{\be}}$ acts on a bounded loop $\msc{C}\pa{K_A}\subset U_{\de}$ around the compact $K_A$.  The functional
$\wh{\msc{G}}_{\ga;A}^{\,\pa{\be}}\pac{\varpi}$ is a regular functional (\textit{cf} definition \ref{Definition Fonctionelle reguliere})
of $\varpi$ in respect to the pair $\pa{M_{\msc{G}_A},\msc{C}\pa{K_A} }$ where the compact $M_{\msc{G}_A}$ has its boundaries
given by $\msc{C}_{out}$ and $\msc{C}_{in}$ as depicted in the \textit{rhs} of Fig.~\ref{contour exemple de courbes encerclantes}.
For all $\varpi \in \msc{O}\pa{M_{\msc{G}_A}}$ such that  $\norm{\varpi}_{\msc{C}\pa{K_A}} \leq C_{\msc{G}_A}$, where $C_{\msc{G}_A}$
is a constant of regularity of the functional $\wh{\msc{G}}_{\ga;A}^{\,\pa{\be}}$,
\bem
\wh{\msc{G}}_{\ga;A}^{\, \pa{\be}}\pac{\varpi} = W_N\big[ \ga G_{\be}  \big] \pab{ \paa{\la_a}_1^N  }{  \paa{\mu_a}_1^N }
\pl{\eps=\pm}{} \exp\Bigg\{ - \Oint{\msc{C}\pa{K_A}}{} \hspace{-3mm}\f{\dd z }{2i\pi} \varpi\pa{z}
\paa{ - \ln \pa{ V_{N;\eps}\big[ \ga G_{\be}  \big] }\pa{z}		+\ln\pa{z- \mu_{N+1}+i\eps c} }  \Bigg\} \\
\times  \exp\Bigg\{ -\Oint{\msc{C}\pa{K_A}}{} \hspace{-3mm}\f{ \dd y \dd z}{ \pa{2i\pi}^2 } \varpi\pa{y}\varpi\pa{z} \ln\pa{z-y-ic} \Bigg\}
\f{  \ddet{\msc{C}_{A}}{I+ \ga \wh{\msc{V}}_N\pac{\ga G_{\be},\ga G_{\be},\varpi}}
\ddet{ \msc{C}_{A}  }{I+ \ga \wh{\ov{\msc{V}}}_N \pac{\ga G_{\be},\ga G_{\be},\varpi}} }
{   \det_{N+1}\pac{\Xi^{\pa{\mu}}\pac{ \xi } } \det_{N}\big[ \Xi^{\pa{\la}}\big[ \xi_{\ga G_{\be}} \big]  \big]
\; \pl{\eps=\pm }{} V_{N;\eps}^{-1}\pac{\ga G_{\be}[\varpi] }\pa{\mu_{N+1}} } \;.
\label{appendix thremo lim FF definition fonctionnelle G super caligraphique}
\end{multline}
In the above formula, one should understand $G_{\be}$ as the one-parameters family of regular functionals of $\varpi$
as defined by \eqref{appendix thermo lim FF fonctionelle G beta}. We did not make the functional dependence of $G_{\be}$
on $\varpi$ explicit in \eqref{appendix thremo lim FF definition fonctionnelle G super caligraphique}.
The functionals $W_N$ and $V_{N;\eps}$ have been defined in \eqref{definition fonction WN et VNepsilon}. We have added the
$\big[\ga G_{\be} \big]$ symbol so as to make it clear that the parameters $\{ \la_a \}_1^N$ entering in their definition are
 functionals of $\ga G_{\be}$ through the relation $\la_a = \xi_{\ga G_{\be}}^{-1}\pa{\tf{a}{L}}$.

 In the second line of
\eqref{appendix thremo lim FF definition fonctionnelle G super caligraphique} there appear
Fredholm determinants of integral operators acting on a contour $\msc{C}_A$. The contour $\msc{C}_A$ corresponds to a loop around $\msc{C}\pa{K_A}$
such that $ \msc{C}_A  \subset U_{\de}$. The kernels read
\beq
\hspace{-7mm}\wh{\msc{V}}_N\! \pac{\nu,\varpi}\pa{\om,\om^{\prime}} = \f{-1}{2\pi}\f{\om-\mu_{N+1}}{\om-\mu_{N+1}+ic}
\exp\Bigg\{\Oint{\msc{C}\pa{K_A} }{} \hspace{-1mm}\f{\dd z}{2i\pi} \varpi\pa{z} \ln\paf{\om-z}{\om-z+ic} \Bigg\}
\cdot \f{ V_{N;1}\!\pac{\nu}\pa{\om} }{ V_{N;0}\!\pac{\nu}\pa{\om}} \cdot
 \f{ K\pa{\om-\om^{\prime}} }{ \ex{-2i\pi \nu\pa{\om}}-1 }
\enq
and
\beq
\hspace{-7mm} \wh{\ov{\msc{V}}}_N\! \pac{\nu,\varpi}\pa{\om,\om^{\prime}} = \f{1}{2\pi}\f{\om-\mu_{N+1}}{\om-\mu_{N+1}-ic}
\exp\Bigg\{\Oint{\msc{C}\pa{K_A} }{} \hspace{-1mm}\f{\dd z}{2i\pi} \varpi\pa{z} \ln\paf{\om-z}{\om-z-ic} \Bigg\}
\cdot \f{ V_{N;-1}\!\pac{\nu}\pa{\om} }{ V_{N;0}\!\pac{\nu}\pa{\om}} \cdot
 \f{ K\pa{\om-\om^{\prime}} }{  \ex{2i\pi \nu\pa{\om}}-1 }
\label{appendix thermo lim FF formule noyau v cal bar}
\enq

\vspace{2mm}
\noindent The constant of regularity $C_{\msc{G}_{A}}$ satisfies to the estimates already given in
\eqref{ecriture condition grandeur constante de regularite GAkappa} and is such that
\beq
 \forall \; \norm{\varpi}_{\msc{C}\pa{K_A}} < C_{\msc{G}_A} \quad \norm{ \wt{\ga}_0 G_{\be}\pac{\varpi} }_{ U_{\de} } < \tf{1}{2}
\quad \e{and} \quad \norm{H}_{\msc{C}\pa{K_A}} <  C_{\msc{G}_A} \;\; \e{uniformly} \; \e{in} \; \pa{\{\mu_{p_a}\}_1^n, \{ \mu_{h_a}\}_1^n} \in K_A^n\times K_A^n \;.
\nonumber
\enq
\end{lemme}

\Proof

The proof is very similar to the one of lemma \ref{Lemme fonctionnelle GA pour partie lisse FF}. Hence, we only specify that
for $L$-large enough, and as soon as condition $\abs{\ga G_{\be}\pac{\varpi} \pa{\la}}<\tf{1}{2}$ for all $\la \in U_{\de}$ is satisfied,
the parameters $\la_j$ are seen to be regular functionals of $\varpi$ thanks to their integral representation
\beq
\la_j\pac{\varpi} = \Oint{\msc{C}_q}{} \f{\dd z}{ 2i\pi}
 \f{  \xi_{\ga G_{\be}[\varpi]}^{\prime}\!\pa{z} }{ \xi_{\ga G_{\be}[\varpi]}\!\pa{z} - \tf{j}{L}} \;.
\enq
All other details are left to the reader. \qed


\subsection{Specific values of the functionals $\msc{G}_{\ga;A}^{\pa{\be}}$ and $\wh{\msc{G}}_{\ga;A}^{ \, \pa{\be}}$}

In this subsection, we estimate the value of the functional $\msc{G}_{\ga;A}^{\pa{\be}}\pac{\varpi}$ for a specific type
of function $\varpi$. This result will play a role later on.

\begin{lemme}
\label{Proposition evaluation fonctionnelle G super caligraphique}
Let the function $\nu\pa{\la}\equiv\nu\big( \la\mid \paa{z_{k}}_1^n,\paa{y_k}_1^{n+1} \big)$
be the unique solution to the linear integral equation driven by the resolvent $R$ of the Lieb kernel
(\textit{ie} $\pa{I-\tf{K}{2\pi}}\pa{I+\tf{R}{2\pi}}=I$):
\beq
\nu\pa{\la} \; + \;  \ga \Int{-q}{q} \! \f{\dd \mu}{2\pi} R\pa{\la,\mu} \nu\pa{\mu}
=  \pa{i\be -\tf{1}{2}} Z\pa{\la}  \; + \; \sul{k=1}{n} \phi\pa{\la,z_k}  \; - \;  \sul{k=1}{n+1} \phi\pa{\la,y_k} \;.
\enq
Let $A>0$ be large enough and such that $ \pa{ \paa{z_{k}}_1^n,\paa{y_k}_1^{n+1} } \in K^{n}_A\times K_A^{n+1}$. Let $\be_0\in \Cx$ and $\wt{\ga}_0$
be the two numbers associated to the constant $A$ as stated in lemma \ref{Lemme fonctionnelle GA pour partie lisse FF}.
Then defining
\beq
\varpi\pa{\la} = \sul{a=1}{n+1} \f{ 1 }{ \la-y_a }  \; -\; \f{1}{\la-q} \; - \;  \sul{a=1}{n} \f{1}{\la-z_{a}}
-\Int{-q}{q} \f{\ga \nu\pa{\tau}}{ \pa{\la-\tau}^2} \dd \tau \; ,
\label{appendix thermo lim FF definition varpi fonction de nu}
\enq
the below identity holds
\bem
\msc{G}_{\ga;A}^{\pa{\be}}\pac{\varpi}
%
%
= -ic \f{ \pl{a=1}{n} \pl{b=1}{n+1} \pa{y_b-z_a-ic}\pa{z_a-y_b-ic} }{ \pl{a,b=1}{n+1} \pa{y_a-y_b-ic} \pl{a,b=1}{n}\pa{z_a-z_b-ic} }
\f{ \ddet{n}{ \de_{k\ell} + \ga \wh{V}_{k\ell}\pac{\ga \nu} }   \ddet{n}{ \de_{k\ell} + \ga \wh{\ov{V}}_{k\ell}\pac{\ga \nu} }  }{  \det^{2}\pac{I-\tf{K}{2\pi}}}
\; .
\label{appendix thermo lim FF fction G super caligraphe det rang fini}
\end{multline}
The non-trivial entries of the two determinants are given by \eqref{appendix thermo lim FF definition entree V et Vbar chapeau}.
%
%
%
%
%
%
%
%
%
%
%
%
The auxiliary variables $\big( \{ z_k \}_{1}^{n} , \{ y_k \}_1^{n+1} \big)$ on which these entries depend
are undercurrent by the set of auxiliary variables on which depends $\nu$.
\end{lemme}

\Proof

The function $\nu$ is bounded on the strip $U_{\de}$. As a consequence, the associated function  $\varpi$
\eqref{appendix thermo lim FF definition varpi fonction de nu} is also bounded by an $A$ independent constant.
The estimates \eqref{ecriture condition grandeur constante de regularite GAkappa} for
 the constant of regularity $C_{\msc{G}_A}$  for the functional $\msc{G}_{\ga;A}^{\pa{\be}}$
ensure that there exists $A$ large enough such that $\norm{\varpi}_{\msc{C}\!\pa{K_A}}<C_{\msc{G}_A}$
uniformly in $(\{z_k\}_1^n, \{y_k\}_1^{n+1}) \in K^{n}_A\times K_{A}^{n+1}$. Thus, $\msc{G}^{\pa{\be}}_{\ga;A}\pac{\varpi}$
is then well definied.

A direct calculation leads to
\beq
\exp\paa{ \Oint{\msc{C}_A}{}  \f{\dd z}{2i\pi} \varpi\pa{z} \ln\paf{\om - z}{\om-z \pm ic }} =
\f{ \om - q \pm i c }{ \om - q } \pl{a=1}{n+1} \f{ \om - y_a }{ \om - y_a \pm i c } \pl{a=1}{n} \f{ \om - z_a \pm i c }{ \om - z_a }
\ex{ C\pac{2i\pi \ga \nu}\pa{\om \pm i c} - C\pac{2i\pi \ga \nu}\pa{\om }  } \;.
\enq
By using the linear integral equation satisfied by $\nu$ and the representation \eqref{appendix thermo lim FF fonctionelle G beta} we get that
\beq
G_{\be}\pac{\varpi}\pa{\la} = \nu\pa{\la}.
\enq
As a consequence, the kernel $\ov{\msc{V}}$ and $\msc{V}$ simplify
\beq
\msc{V}\pac{ \ga G_{\be}\pac{\varpi}, \varpi }\pa{\om,\om^{\prime}} = -
\pl{a=1}{n+1} \f{ \om - y_a }{ \om - y_a + i c } \pl{a=1}{n} \f{ \om - z_a + i c }{ \om - z_a }
 \f{ K\pa{\om-\om^{\prime}} }{ 2\pi \pa{\ex{-2i\pi \ga \nu\pa{\om}}-1}}
\enq
and
\beq
\ov{\msc{V}}\pac{ \ga G_{\be}\pac{\varpi}, \varpi }\pa{\om,\om^{\prime}} = 
\pl{a=1}{n+1} \f{ \om - y_a }{ \om - y_a - i c } \pl{a=1}{n} \f{ \om - z_a - i c }{ \om - z_a }
 \f{ K\pa{\om-\om^{\prime}} }{ 2\pi \pa{\ex{2i\pi \ga \nu\pa{\om}}-1}}
\enq

The associated Fredholm determinants can now be reduced to finite-size determinants by computing the poles at $\om=z_a$ with $a=1,\dots, n$
(by definition of $\wt{\ga_0}$ and $\be_0$, since $\abs{\ga} \leq \wt{\ga_0}$ and $\be \in \bs{\wt{U}}_{\be_0}$,
there are no poles of $\ex{2i\pi \ga \nu\pa{\om}}-1$ inside of $\msc{C}_A$).

This leads to
\beqa
\ddet{\msc{C}_A}{I+ \ga \msc{V}\big[ \ga G_{\be}\pac{\varpi},\varpi \big] } &=&
\ddet{n}{\de_{k\ell} + \ga \wh{V}_{k\ell}\pac{\ga \nu} \pa{\{ z_a \}_1^n , \{ y_a \}_1^{n+1} } } \\
\ddet{\msc{C}_A}{I+ \ga \ov{\msc{V}}\big[ \ga G_{\be}\pac{\varpi},\varpi  \big] } &= &
\ddet{n}{\de_{k\ell} + \ga \wh{\ov{V}}_{k\ell}\pac{\ga \nu} \pa{\{ z_a \}_1^n , \{ y_a \}_1^{n+1}} } \;.
\eeqa

The claim then follows once upon applying the identity
\bem
-ic \f{ \pl{a=1}{n} \pl{b=1}{n+1} \pa{y_b-z_a-ic}\pa{z_a-y_b-ic} }{ \pl{a,b=1}{n+1} \pa{y_a-y_b-ic} \pl{a,b=1}{n}\pa{z_a-z_b-ic} }
=
\exp\Bigg\{ -\Oint{\msc{C}\pa{K_A}}{} \hspace{-3mm}\f{ \dd y \dd z}{ \pa{2i\pi}^2 } \varpi\pa{y}\varpi\pa{z} \ln\pa{z-y-ic} \Bigg\}  \\
\times \ex{C_0\pac{\ga G_{\be}[\varpi] } }  \ex{- 2i\pi \sul{\eps=\pm}{} C\pac{\ga G_{\be}[\varpi] }\pa{q + i \eps c} }
\pl{\eps=\pm}{} \exp\Bigg\{ - \Oint{\msc{C}\pa{K_A}}{} \hspace{-3mm}\f{\dd z }{2i\pi} \varpi\pa{z}
 \paa{2i\pi C\pac{\ga G_{\be}[\varpi]}\pa{z+i\eps c}
			+\ln\pa{z-q+i\eps c}}  \Bigg\}
\end{multline}
\qed

\begin{lemme}
\label{Proposition evaluation fnelle G hat Reduction vars determinant petit}

Let $\ga$ be small enough and $L$ large enough
such that $\nu^{\pa{L}}\pa{\mu}$ is the unique solution to the non-linear integral equation
\beq
\nu^{\pa{L}}\pa{\la} = \pa{i\be -\tf{1}{2}}Z\pa{\la} \; -\;  \phi\pa{\la, q} \; + \; \sul{a=1}{N+1} \phi\pa{\la,\mu_a} \;
- \; \sul{a=1}{n+1} \phi\pa{\la,y_a} \; - \; \sul{ \substack{a=1 \\ \not= i_1,\dots, i_n}}{N} \phi(\la,\wt{\la}_a) \qquad
\label{appendix thermo lim FF NLIE nuL et evaluation G lisse hat}
\enq
The parameters $\wt{\la}_a$ appearing above are functional of $\nu^{\pa{L}}$ through the relation  $\xi_{ \ga \nu^{\pa{L}} }( \wt{\la}_a) =  \tf{a}{L}$,
$\mu_a$ are such that $\xi\pa{\mu_a}=\tf{a}{L}$ and the parameters $y_a \in U_{\tf{\de}{2}}$ are arbitrary.
Finally, $L$ is assumed large enough so that all parameters $\pa{ \{\mu_a \}_1^{N+1}, \{ \wt{\la}_a \}_1^{N},  \{ y_a \}_1^{n+1}} $
of the $2N+2+n$-uple belong to $K_{2A_L}$.
Then, given  $\be_0$ and $\wt{\ga}_0$ as in lemma \ref{Lemme fonctionnelle hat GA pour partie lisse hat FF}, one has the identity
\bem
\hspace{-1cm}\wh{\msc{G}}^{\,\pa{\be}}_{\ga;2A_L} \pac{ H\pabb{* }{ \{y_a\}_1^{n+1} \cup \{ \wt{\la}_a \}_1^N }
{ \{\wt{\la}_{i_a}\}_1^n \cup \{ \mu_a \}_1^{N+1}  }   }=
-ic \f{ \pl{a=1}{n} \pl{b=1}{n+1} \pa{y_b-\wt{\la}_{i_a}-ic}\pa{ \wt{\la}_{i_a}-y_b-ic} }
{ \pl{a,b=1}{n+1} \pa{y_a-y_b-ic} \pl{a,b=1}{n}\pa{\wt{\la}_{i_a}-\wt{\la}_{i_b}-ic} }
\det_{N+1}^{-1}\Big[ \Xi^{\pa{\mu}}\pac{\xi}  \Big] \det_{N}^{-1} \Big[ \Xi^{\pa{\la}}\big[ \xi_{\ga\nu^{\pa{L}}} \big]  \Big] \\
\times  \det_{n} \Big[  \de_{k\ell} + \ga \wh{V}_{k\ell}\big[ \ga \nu^{\pa{L}} \big]
\pa{ \{\wt{\la}_{i_a}\}_1^n,\{ y_{a}\}_1^{n+1} }  \Big]
\det_{n} \Big[  \de_{k\ell} + \ga \wh{\ov{V}}_{k\ell}\big[ \ga \nu^{\pa{L}} \big]
\pa{ \{\wt{\la}_{i_a}\}_1^n,\{ y_{a}\}_1^{n+1} }  \Big]
\end{multline}

\end{lemme}

\Proof

It has been shown in  proposition \ref{Proposition serie Lagrange avec fonctionnelle Gamma L}
that for $\abs{\ga}$ small enough and $L$ large enough the solution $\nu^{\pa{L}}$ to the non-linear integral equation occurring in  the \textit{rhs} of
\eqref{appendix thermo lim FF NLIE nuL et evaluation G lisse hat} is unique and exists. Moreover this solution is bounded on $U_{\de}$
by an $L$-independent constant.

As discussed in the proof of lemma \ref{Lemme fonctionnelle GA pour partie lisse FF}, the contour $\msc{C}\!\pa{ K_{2A_L} }$ 
can always be taken such that, uniformly in $L$, $ \e{d} (\msc{C}\!\pa{K_{2A_L} } , K_{2A_L})> c^{\prime}>0 $ for some constant $c^{\prime}$.
Hence, the principal argument $\la$ of $H$ is uniformly away from the compact $K_{2A_L}$ where the auxiliary arguments of 
$H$ are located. As a consequence,
it follows from the expression for $H$ and the estimates for the spacing between the parameters $\mu_a$ and $\wt{\la}_a$
\beq
 \mu_a-\wt{\la}_a = 2 \pi \ga \tf{ \nu^{\pa{L}}\pa{\mu_a} }{ \pa{L p^{\prime}\!\pa{\mu_a}} } + \e{O}\pa{L^{-2}}\; , \quad \e{uniformly}\;
 \e{in} \;  a=1,\dots, N
\enq
that $\norm{H}_{ \msc{C}\pa{K_{2A_L}}}$ is bounded by an $L$-independent constant, this uniformly in $L$ large enough.
In particular, for $L$ large enough, due to the estimates \eqref{ecriture condition grandeur constante de regularite GAkappa}
for the constant $C_{\msc{G}_{2A_L}}$ of regularity for $\wh{ \msc{G} }_{\ga;2A_L}^{\, \pa{\be}}$, we get that
$\norm{H}_{\msc{C}\pa{K_A}}<C_{\msc{G}_{2A_L}}$.
One can thus acts with the functional $\wh{\msc{G}}^{\,\pa{\be}}_{\ga;2A_L}$ on $H$.
A straightforward residue calculation shows that
\beq
G_{\be} \pac{ H\pabb{* }{ \{y_a\}_1^{n+1} \cup \{ \wt{\la}_a \}_1^N }{ \{\wt{\la}_{i_a}\}_1^n \cup \{ \mu_a \}_1^{N+1}  }   }     = \nu^{\pa{L}}\!\pa{\la} \;.
\enq
This means that all the $\la_a$ appearing in the expression \eqref{appendix thremo lim FF definition fonctionnelle G super caligraphique}
for the functional $\wh{\msc{G}}^{\,\pa{\be}}_{\ga;2A_L}\pac{H}$  coincide with the parameters $\wt{\la}_a$ defined above.
The claim of the lemma then follows from straightforward residue computations and multiple cancelations.
The Fredholm determinants reduce to finite rank determinants that can be computed by the residues at
$\om= \wt{\la}_{i_a}$, $a=1,\dots, n$.  \qed


\subsection{Leading asymptotic behavior of one particle/one hole form factors}
\label{Appendix Thermo lim FF Section Formules FF part/trou}

We now build on the formulae for the leading asymptotic behavior of form factors so as to
provide, properly normalized in the size of the model, expressions for the large-$L$ limit of the form factors of the fields between
the $N$-particle ground state and $N+1$-particle excited states corresponding to one hole at one of the ends of the Fermi zone
and one particle either at the other end of the Fermi zone or at the saddle-point $\la_{0}$ of the function $u\pa{\la}$
given in \eqref{ecritutre representation integrale u}.
Such thermodynamic limits of properly normalized form factors appear as amplitudes
in the large-distance/long-time asymptotic expansion of the reduced density matrix.
The explicit expressions that we write down will allow for such an identification.
We do stress that all shift functions appearing below are taken at $\be=0$. The fact that
\eqref{appendix thermo lim FF facteur forme lambda 0 q}-\eqref{appendix thermo lim FF facteur forme 0 0}
are well-defined in this limit follows from
proposition \ref{Proposition Holomorphie en beta et rapidite part-trou det fred}.

In the following, let $\paa{\la}\equiv \paa{\la_a}_1^N$ stand for the Bethe roots corresponding to the ground state in the $N$-particle sector.
Let $\{ \mu^{\emptyset}_{\emptyset} \} \equiv \{ \mu^{\emptyset}_{\emptyset} \}_{1}^{N+1}$ stand for the Bethe roots corresponding to the ground state in the $\pa{N+1}$-particle sector.
Taking into account that $F_{\emptyset}^{\emptyset}$ stands for the thermodynamic limit of the corresponding shift
function \textit{cf} \eqref{definition les fonction shift asymptotiques}, we define
\beq
\abs{ \mc{F}_{\emptyset}^{\emptyset} }^2=  \lim_{N,L\tend +\infty}
\paf{L}{2\pi}^{ \big[ F_{\emptyset}^{\emptyset}\pa{q} + 1 \big]^2 + \big[ F_{\emptyset}^{\emptyset} \pa{-q} \big]^2 }
\Bigg|  \f{ \bra{ \psi\Big( \big\{\mu^{\emptyset}_{\emptyset} \big\} \Big) }\Phi^{\dagger}\!\pa{0,0}\ket{ \psi\Big(\big\{\la \big\} \Big)  }  }
{ \norm{  \psi\Big(\big\{\la \big\} \Big)   } \cdot \norm{ \psi\Big( \big\{ \mu^{\emptyset}_{\emptyset} \big\} \Big) } } \Bigg|^2  \;.
\enq
Similarly, given the set  $\{ \mu_q^{-q} \} \equiv \{ \mu_q^{-q} \}_{1}^{N+1}\! $ corresponding to a particle-hole excitation such that $p_1=0$ and
$h_1=N+1$, we denote by $F_{q}^{-q}$  the thermodynamic limit of the corresponding shift function \textit{cf} \eqref{definition les fonction shift asymptotiques}, and define
\beq
\abs{ \mc{F}_{q}^{-q} }^2=  \lim_{N,L\tend +\infty}
\paf{L}{2\pi}^{ \big[ F_{q}^{-q} \pa{q} \big]^2 + \big[ F_{q}^{-q}\pa{-q} -1 \big]^2 }
\Bigg|  \f{ \bra{ \psi\Big( \big\{\mu^{-q}_{q} \big\} \Big)  }\Phi^{\dagger}\!\pa{0,0}\ket{ \psi\Big(\big\{\la \big\} \Big) } }
{ \norm{ \psi\Big(\big\{\la \big\} \Big) }   \cdot \norm{ \psi\Big( \big\{\mu^{-q}_{q} \big\} \Big)  }  } \Bigg|^2 \;.
\enq

Finally, given the set   $\{ \mu_q^{\la_0} \} \equiv  \{ \mu_q^{\la_0} \}_{1}^{N+1}\!\!$ corresponding to a
particle-hole excitation such that $h_1=N+1$ and $\mu_{p_a}=\la_{0}$
we denote by $F_{q}^{\la_0}$ the thermodynamic limit of the corresponding
shift function \textit{cf} \eqref{definition les fonction shift asymptotiques}, and define
\beq
\abs{ \mc{F}_{q}^{\la_0} }^2=  \lim_{N,L\tend +\infty} \paf{L}{2\pi}^{ \big[ F_{q}^{\la_0}\pa{q} \big]^2 + \big[ F_{q}^{\la_0}\pa{-q} \big]^2 +1 }
\Bigg|  \f{ \bra{ \psi\Big( \big\{\mu^{\la_0}_{q} \big\} \Big)  }\Phi^{\dagger}\!\pa{0,0}\ket{  \psi\Big(\big\{\la \big\} \Big)  } }
{ \norm{  \psi\Big(\big\{\la \big\} \Big)  } \cdot \norm{ \psi\Big( \big\{\mu^{\la_0}_{q} \big\} \Big)   }    } \Bigg|^2  \;.
\enq
By using \eqref{appendix thermo lim FF formule pple pour limite thermo} and expressions
\eqref{AppendixThermoLimD+zero}-\eqref{definition fonction aleph et varphi} we are lead to
\beq
\abs{ \mc{F}_{q}^{ \la_0 } }^2= \f{\ex{i\f{\pi}{4}}}{ 2\pi p^{\prime}\pa{\la_0} }
\mc{A}_0\pac{ F_{q}^{\la_0}  }\mc{B}\pac{ F_{q}^{\la_0},p } \mc{G}_{1;1}^{\pa{0}}\pab{\la_{0}}{q}
\; \exp\paa{i\f{\pi}{2}   \pa{  \big[ F_{q}^{\la_0}\!\pa{-q} \big]^2 - \big[ F_{q}^{\la_0}\!\pa{q} \big]^2  }   }  \; ,
\label{appendix thermo lim FF facteur forme lambda 0 q}
\enq
\beq
\abs{ \mc{F}_{q}^{-q} }^2=  \mc{A}_-\pac{ F_{q}^{-q},p }\mc{B}\pac{F_{q}^{-q},p} \mc{G}_{1;1}^{\pa{0}}\pab{-q}{q}
\;\; \exp\paa{  i\f{\pi}{2}  \pa{   \big[  F_{q}^{-q}\!\pa{-q} -1 \big]^2 -  \big[ F_{q}^{-q} \!\pa{q} \big]^2  }   }
\;,
\label{appendix thermo lim FF facteur forme -q q}
\enq
and finally
\beq
\abs{ \mc{F}_{ \emptyset }^{ \emptyset } }^2=  \mc{A}_+\pac{ F_{\emptyset}^{\emptyset} ,p}\mc{B}\pac{ F_{\emptyset}^{\emptyset} ,p }
\mc{G}_{0;1}^{\pa{0}}\pab{\emptyset}{\emptyset} \;\;
\exp\paa{  i\f{\pi}{2} \pa{ \big[F_{\emptyset}^{\emptyset} \!\pa{-q} \big]^2 -\big[ F_{\emptyset}^{\emptyset}\!\pa{q} + 1\big]^2 }  }  \; .
\label{appendix thermo lim FF facteur forme 0 0}
\enq
The functionals $\mc{B}$, $\mc{A}_{\pm}$ and $\mc{A}_0$ appearing above have been defined in \eqref{definition fonctionnelle B},
\eqref{definition fonctionnelle A+ et kappa} and \eqref{definition fonctionnelle A- et A0}.




\section{The generalized free-fermion summation formulae}
\label{Appendix section Free fermionic correlators}

In this appendix, we establish summation identities allowing one to recast the form factor expansion of an analogue of the field/conjugated-field
two-point function that would appear in a generalized free fermion model in terms of a finite-size determinant. The representation we obtain constitutes
the very cornerstone for deriving various representations for the correlation functions in the interacting case. In particular,
it allows one for an analysis of their asymptotic behavior in the large-distance/long-time regime. We first establish re-summation formulae allowing
one to estimate discreet analogs of singular integrals. This will open the way for obtaining Fredholm determinant like representations out of the
form factor based expansions.

\subsection{Computation of singular sums}
\label{Appendix computation of singular sums}

Let $\xi$ stand for the thermodynamic limit of the counting function \eqref{ecriture limite thermo fction comptage} and $E_-$ be
a non-vanishing and holomorphic function in some open neighborhood $U_{\de}$ (\textit{cf} 
\eqref{definiton voisinage ouvert de R pour holomorphie Gn}) of $\R$ such that
$\Re\pa{\ln E_-^{-2}}$ has, at most, polynomial growth, \textit{ie}
\beq
\abs{ \Re\pac{\ln E_-^{-2}\pa{\la} -i C_1 \la^{k} } } \leq C_2 \abs{ \Re\pa{i \la^{k-1}}} + C_3 \; , \quad \e{for}\;\e{some}
\; C_1 \;, \; C_2 \; , \; C_3 \in \R^{+}
\quad \e{and} \; k>1 \quad \e{uniformly} \; \e{in} \; \la \in U_{\de} \;.
\label{appendix Sing sums contrainte fonction E-}
\enq
We remind that the neighborhood $U_{\de}$ is always taken such that $\xi$ is a  biholomorphism on $U_{\de}$

In the following, we study the below singular sums over  the set $\paa{\mu_a}$:
\beq
\mc{S}_r^{\pa{L}}\!\pac{E_-^{-2}}\pa{\la} = \sul{a \in \mc{B}_L }{} \f{ E_-^{-2} \! \pa{\mu_{a}} }{ 2\pi L \xi^{\prime}\!\pa{\mu_a} \pa{\mu_a-\la}^r}
\qquad \e{with} \; \mu_a \; \e{being} \; \e{the} \;\e{unique} \; \e{solution}\; \e{to} \; \xi\pa{\mu_a}=\tf{a}{L}  \;.
\label{appendix Sing sums ecriture Sr somme discrete}
\enq
The summation runs through the set  $\mc{B}_L=\paa{a \in \mathbb{Z} \; : \; -w_L \leq a \leq w_L}$ where $w_L$ is some $L$-dependent sequence
in $\mathbb{N}$ such that $L=\e{o}\pa{w_L}$ and $ \pa{w_L \cdot L^{-1}}^{k-1} = \e{o}\pa{L}$.

\begin{prop}
\label{Proposition calcul des sommes discrete singulieres}
Let $\mc{N}_q$ be a compact neighborhood of $\intff{-q}{q}$ lying in $U_{\de}$, then
under the above assumptions and provided that $L$ is large enough, one has, uniformly in $\la \in \mc{N}_q$
\beqa
\mc{S}_0^{\pa{L}}\!\pac{E_-^{-2}}\pa{\la} &=&  \Int{ \msc{C}_{bk;L} }{ } \!\!  \f{ \dd \mu }{2\pi} \, E_-^{-2}\!\pa{\mu} \; + \; I_0^{\pa{L}}\!\pac{E_-^{-2}}\pa{\la}
\label{ecrituire somme S0}\\
\mc{S}_1^{\pa{L}}\!\pac{E_-^{-2}}\pa{\la} &=&  \Int{ \msc{C}_{bk;L} }{  } \!\!  \f{ \dd \mu }{2\pi} \, \f{ E_-^{-2}\!\pa{\mu}}{\mu-\la}
			\; - \; i \f{E_-^{-2}\!\pa{\la}}{ \ex{2i\pi L \xi\pa{\la}}-1  }  \; + \; I_1^{\pa{L}}\!\pac{E_-^{-2}}\pa{\la}
\label{ecrituire somme S1}  \\
\mc{S}_2^{\pa{L}}\!\pac{E_-^{-2}}\pa{\la} &=&
\f{ \Dp{} }{ \Dp{}\la } \Int{ \msc{C}_{bk;L} }{ }\!\!   \f{ \dd \mu }{2\pi} \, \f{ E_-^{-2}\!\pa{\mu}}{\mu-\la}
 \; - \; i \f{ \Dp{\la} \pac{E_-^{-2}\!\pa{\la}} }{ \ex{2i\pi L \xi\pa{\la}}-1   }  \; + \;
\pi \f{ E_-^{-2}\!\pa{\la} L \xi^{\prime}\pa{\la} }{ 2 \sin^2\pac{\pi L \xi\pa{\la}} }
 \; + \;  I_2^{\pa{L}}\!\pac{E_-^{-2}}\pa{\la}
\label{ecrituire somme S2} \;.
\eeqa

The integration goes along the curve $\msc{C}_{bk;L}$ depicted on Fig.~\ref{contour pour integral centrales et bords qui contribuent peu}. 
Also, given $r\in \mathbb{N}$,
\beq
I_r^{\pa{L}}\!\pac{E_-^{-2}}\pa{\la} = \Int{ \msc{C}_{\ua;L} }{} \f{ \dd z }{ 2\pi }  \f{ E_-^{-2}\pa{z}  }{ \pa{z -\la}^{r} }
 \f{  1  }{ 1 - \ex{-2i\pi L \xi\pa{z}  } }  \; + \;
\Int{ \msc{C}_{\da;L} }{} \f{ \dd z }{ 2\pi }  \f{ E_-^{-2}\pa{z}  }{ \pa{z -\la}^{r} }
 \f{  1  }{  \ex{ 2i\pi L \xi\pa{z} } -1  }
\; + \; \Int{ \msc{C}_{bd;L} }{ } \f{ \dd z }{ 2\pi } \f{ E_-^{-2}\!\pa{z} }{ \pa{z-\la}^r } \;.
\label{definition facteur IL}
\enq
The contours $\msc{C}_{\ua/\da;L}$ are depicted in Fig.~\ref{contour pour integral IL} whereas $\msc{C}_{bd;L}$  is depicted on
Fig.~\ref{contour pour integral centrales et bords qui contribuent peu}.

The functionals $I_r^{\pa{L}}\!\pac{ E_-^{-2} }\pa{\la}$ are such that
$I_r^{\pa{L}}\!\pac{ E_-^{-2} }\pa{\la}=\e{O}\pa{ \pa{\tf{L}{w_L}}^{k+r-1} }$, uniformly in $\la \in \mc{N}_q$.

\begin{figure}[h]
\begin{center}

\begin{pspicture}(12.5,4)

\psline[linestyle=dashed, dash=3pt 2pt](1.5,2)(4.5,2)


\psdots(1.5,2)(4.5,2)

\rput(1,1.8){$-A_L$}
\rput(4.8,1.8){$B_L$}
\rput(4.9,2.6){$\msc{C}_{\ua;L}$}
\rput(1.1 ,1.1){$\msc{C}_{\da;L}$}


\pscurve(1.5,2)(1.5,1.5)(1.5,1)(3,1)(4.5,1)(4.5,1.5)(4.5,2)(4.5,2.5)(4.5,3)(3,3)(1.5,3)(1.5,2.5)(1.5,2)

\psline[linewidth=2pt]{->}(3,1)(3.1,1)
\psline[linewidth=2pt]{->}(3.1,3)(3,3)


\psline(8,1)(8,3)
\psline(8,3)(11,3)
\psline(11,3)(11,1)
\psline(11,1)(8,1)
\psline[linewidth=2pt]{->}(9.5,3)(9.4,3)
\psline[linewidth=2pt]{->}(9.4,1)(9.5,1)


\psdots(8,2)(11,2)

\rput(8.9,1.8){$-\f{w_L}{L}-\f{1}{2L}$}
\rput(11.8,1.8){$ \f{w_L}{L}+\f{1}{2L}$}

\pscurve[linewidth=1pt]{->}(5.5,3.2)(6.25,3.5)(7.4,3.2)
\rput(6.7,3.7){$\xi$}

\pscurve[linewidth=1pt]{<-}(5.5,1.2)(6.25,0.9)(7.4,1.2)
\rput(6,0.6){$\xi^{-1}$}

\psline[linestyle=dashed, dash=3pt 2pt](11,3)(12,3)
\psline[linestyle=dashed, dash=3pt 2pt](11,1)(12,1)

\psline[linestyle=dashed, dash=3pt 2pt](8,3)(7.5,3)
\psline[linestyle=dashed, dash=3pt 2pt](8,1)(7.5,1)

\rput(11.7,3.2){$ \R + i \a $}
\rput(11.7,.8){$ \R - i \a $}

\end{pspicture}

\caption{Contour $\msc{C}_{\ua;L} \cup \msc{C}_{\da;L}$ lying in $U_{\de}$. \label{contour pour integral IL}}
\end{center}
\end{figure}
\begin{figure}[h]
\begin{center}

\begin{pspicture}(13,3)



\psdots(0.5,1.5)(1.5,1.5)(4.5,1.5)(5.5,1.5)

\rput(0.5,1.7){$-A_L$}
\rput(5.5,1.7){$B_L$}

\rput(1.6,1.2){$-A$}
\rput(4.5,1.2){$A$}

\rput(4.2,2.3){$\msc{C}_{bk;L}$}
\rput(3,1.5){$\mc{N}_{q}$}
\rput(5.9,1){$\msc{C}_{bd;L}$}
\rput(0.1,1){$\msc{C}_{bd;L}$}


\pscurve(0.5,0.5)(0.8,0.5)(1,0.7)(1.5,1.5)(1.7,1.5)(2,1.5)(2.5,2.5)(3,2.5)(3.5,2.5)(4,1.5)(4.2,1.5)(4.5,1.5)(5,0.7)(5.2,0.5)(5.5,0.5)

\pscurve[linestyle=dashed, dash=3pt 2pt](0.5,0.5)(0.6,1)(0.5,1.5)
\pscurve[linestyle=dashed, dash=3pt 2pt](5.5,0.5)(5.4,1)(5.5,1.5)


\psline[linestyle=dotted](2.4,1.2)(3.6,1.2)
\psline[linestyle=dotted](2.4,1.8)(3.6,1.8)
\psline[linestyle=dotted](2.4,1.2)(2.4,1.8)
\psline[linestyle=dotted](3.6,1.2)(3.6,1.8)


\psline[linewidth=2pt]{->}(0.6,1)(0.6,0.95)
\psline[linewidth=2pt]{<-}(5.4,1.1)(5.4,1.05)

\psline[linewidth=2pt]{->}(3,2.5)(3.1,2.5)


\psdots(8.5,1.5)(9.5,1.5)(12.5,1.5)(13.5,1.5)

\rput(8.5,1.3){$-A_L$}
\rput(13.5,1.7){$B_L$}

\rput(9.6,1.2){$-A$}
\rput(12.5,1.2){$A$}

\rput(12.2,2.3){$\msc{C}_{bk;L}$}
\rput(11,1.5){$\mc{N}_{q}$}
\rput(13.9,1){$\msc{C}_{bd;L}$}
\rput(8.1,2){$\msc{C}_{bd;L}$}


\pscurve(8.5,2.5)(8.8,2.5)(9,2.3)(9.5,1.5)(9.7,1.5)(10,1.5)(10.5,2.5)(11,2.5)(11.5,2.5)(12,1.5)(12.2,1.5)(12.5,1.5)(13,0.7)(13.2,0.5)(13.5,0.5)

\pscurve[linestyle=dashed, dash=3pt 2pt](8.5,2.5)(8.6,2)(8.5,1.5)
\pscurve[linestyle=dashed, dash=3pt 2pt](13.5,0.5)(13.4,1)(13.5,1.5)


\psline[linestyle=dotted](10.4,1.2)(11.6,1.2)
\psline[linestyle=dotted](10.4,1.8)(11.6,1.8)
\psline[linestyle=dotted](10.4,1.2)(10.4,1.8)
\psline[linestyle=dotted](11.6,1.2)(11.6,1.8)


\psline[linewidth=2pt]{->}(8.6,2)(8.6,2.05)
\psline[linewidth=2pt]{<-}(13.4,1.1)(13.4,1.05)

\psline[linewidth=2pt]{->}(11,2.5)(11.1,2.5)

\end{pspicture}

\caption{Contours $\msc{C}_{bk;L}$ (solid lines) and $ \msc{C}_{bd;L}$ (dashed lines) in the case of $k$ odd (\textit{lhs})
 and $k$ even (\textit{rhs}) both in the case $C_1<0$.  The dashed lines $\msc{C}_{bd;L}$ are pre-images of the segments
 $\intff{ \eps_{ \upsilon } w_L +  \tf{ \eps_{\upsilon} }{2}}{ \eps_{\upsilon} w_L  + \tf{\eps_{\upsilon}}{2} +i\eps^{\prime} \a }$,
with $\upsilon \in \paa{ \e{l}, \e{r} }$ and  $\eps_{\e{l}}=-1$ and $\eps_{r}=1$. The sign of $\eps^{\prime}$ depends on the left or right boundary, 
the parity of $k$ and the sign of $C_1$.
 \label{contour pour integral centrales et bords qui contribuent peu}}
\end{center}
\end{figure}

\end{prop}

\Proof

Let $\mc{N}_q$ be a compact neighborhood of $\intff{-q}{q}$ in $U_{\de}$. Then, for $L$ large enough it is contained inside
of the contour $\msc{C}_{\ua;L} \cup \msc{C}_{\da;L}$ as depicted in Fig.~\ref{contour pour integral IL}, and thus

\beqa
\mc{S}_1^{\pa{L}}\!\pac{E_-^{-2}}\pa{\la} &=&  - \f{ i E_-^{-2}\!\pa{\la} }{\ex{2i\pi L \xi \pa{\la} } -1 }  +
 \Int{ \msc{C}_{\ua;L} \cup \msc{C}_{\da;L}  }{} \hspace{-2mm}  \f{ E_-^{-2}\pa{z} }{2\pi\pa{z-\la}} \cdot \f{1}{ \ex{2i\pi L \xi\pa{z}}-1}    \dd z
\nonumber \\
&=&
\Int{\msc{C}_{\da;L} }{}  \f{ E_-^{-2}\!\pa{z} }{ 2\pi \pa{z-\la} }\f{1}{ \ex{2i\pi L \xi\pa{z}} -1 } \dd z
 \; + \;   \Int{ \msc{C}_{\ua;L}  }{}  \f{ E_-^{-2}\!\pa{z} }{ 2\pi \pa{z-\la} }
 \paa{ \f{ \ex{2i\pi L \xi\pa{z}} }{ \ex{2i\pi L \xi\pa{z}} -1 }  -1  } \dd z
 \; -  \;  \f{ i  E_-^{-2}\!\pa{\la} }{ \ex{2i\pi L \xi \pa{\la}} -1 }   \nonumber \\
&=& \Int{ \msc{C}_{bk;L} }{  } \!\!  \f{ \dd \mu }{2\pi} \, \f{ E_-^{-2}\!\pa{\mu}}{\mu-\la}
			\; - \; i \f{E_-^{-2}\!\pa{\la}}{ \ex{2i\pi L \xi\pa{\la}}-1  }  \; + \; I_1^{\pa{L}}\!\pac{E_-^{-2}}\pa{\la}   \;.
\label{equation pour la somme simple}
\eeqa
In order to obtain the last line, we have deformed the contour $\msc{C}_{\ua;L}$ into the contour $\msc{C}_{bk;L}\cup\msc{C}_{bd;L}$
as depicted in Fig.~\ref{contour pour integral centrales et bords qui contribuent peu}.
The intermediate points $\pm A$ entering in the definition of $\msc{C}_{bk;L}$ are chosen large (in order to include $\mc{N}_q$) but 
fixed, in the sense that $L$ independent.

The representation
for $\mc{S}_2^{\pa{L}}\!\pac{E^{-2}_-}\pa{\la}$ follows by differentiation. 
The computations for $\mc{S}_0^{\pa{L}}\!\pac{E^{-2}_-}\pa{\la}$
are carried out similarly with the sole difference that there is no pole at $z=\la$.

In now remains to prove the statement relative to the asymptotic behavior in $L$ of the functionals $I_r^{\pa{L}}\!\pac{E_-^{-2}}$.
The main difficulty is that the function $E_-\pa{\la}$ might have an exponential increase when $\la$ belongs to the upper or lower half-plane.
We establish the claimed estimates for the $\msc{C}_{\ua;L}$-part of the contour.
This can be done similarly for $\msc{C}_{\da;L}$, and we leave these details to the reader.

We first perform the change of variables\symbolfootnote[2]{we remind that $\xi$ is a biholomorphism on $U_{\de}$ and that $p^{\prime}>0$ of $\R$.}
$z = \xi^{-1}\!\pa{s} $ and set $u_L= \tf{w_L}{L} + \tf{1}{(2L)}$. The contour of integration is then mapped to the
contour depicted on the $rhs$ of Fig.~\ref{contour pour integral IL}. We stress that the parameter $\a>0$ is chosen in such a way that
$\msc{C}_{\ua;L}\cup\msc{C}_{\da;L}$ lies in $U_{\de}$. The aforementioned change of variables leads to
\bem
\Int{\msc{C}_{\ua;L}}{} \f{\dd z}{2\pi}  \f{E_-^{-2}\pa{z}}{\pa{z-\la}^{r}} \cdot \f{1}{ 1-\ex{-2i\pi L \xi\pa{z}} } =
\Int{ u_L }{- u_L }  \f{\dd s}{2\pi} \;   \f{E_-^{-2}}{\xi^{\prime}}\circ \xi^{-1}\pa{s +i\a}
\f{1}{ \pac{  \xi^{-1}\pa{s+i\a} - \la}^{r}  }  \paa{1-\ex{2\pi L \a } \ex{-2i\pi s L} }^{-1}  \\
+ \; \sul{\eps=\pm}{} \eps \Int{0}{\a} \f{i \dd s}{2\pi}  \f{E_-^{-2}}{\xi^{\prime}}\circ \xi^{-1}\pa{is+ \eps u_L }
\f{1}{ \pac{  \xi^{-1}\pa{ is+\eps u_L } - \la}^{r}  }  \paa{1+ \ex{2\pi s L} }^{-1}  \;.
\end{multline}
We first establish a bound for the integral over the line $ \intff{ - u_L }{ u_L }$.
It follows from the integral equation \eqref{definition eqn integral p} satisfied by $p$
that,
\beq
 p\pa{\la}=\la \pm \pi D - 2 \tf{cD}{\la} + \e{O}\pa{\la^{-2}} \qquad  \e{when} \quad  \Re\pa{\la}\tend \pm \infty \;.
\enq
Hence, uniformly in $0\leq \tau \leq \a$ and for $s \in \R$ large,
\beq
\xi^{-1}\!\pa{s+i\tau}= \psi_s \,+\, 2i\pi \tau \,+\, \e{O}( s^{-2} ) \;  \quad \e{where} \quad \psi_s =  2\pi s-\pi D\pa{1\pm 1} + \f{cD}{ \pi s}  \in \R \;.
\enq
The condition \eqref{appendix Sing sums contrainte fonction E-} implies that there exists constants $C>0$, $C^{\prime}>0$ such that
\beq
\abs{\Re \pac{ \ln E_-^{-2}\!\pa{\la}}} \leq C \abs{\Re\pa{i\la^k}} + C^{\prime} \; , \qquad \e{uniformly} \; \e{in} \; \la \in U_{\de}\;.
\enq
As a consequence, uniformly in $0 \leq \tau \leq \a$,
\beqa
\abs{ \Re\pac{ \ln E_-^{-2} \circ \xi^{-1}\pa{s+i \tau} }  } &\leq & C
\abs{ \Im\pac{ \pa{ \psi_s }^{k}   + 2i\pi k \tau \pa{ \psi_s }^{k-1}  +\e{O}\big(s^{k-2}\big)   }} +C^{\prime}
\nonumber \\
& \leq & C \tau k \pa{2\pi}^k \abs{s}^{k-1} \abs{\Im\pac{1+\e{O}\big(s^{-1}\big)} }  + C^{\prime} \;.
\label{appendix Sing sums estimation brute partie reelle log}
\eeqa

There exists an $s_0$ such that $\big| \e{O} (s^{-1}) \big|<1$ for $\abs{s}\geq s_0$, this uniformly in $0\leq \tau \leq \a$.
Moreover, for such an $s_0$, we define
\beq
C^{\prime \prime} = C^{\prime} + \max \abs{ \Re\pac{ \ln E_-^{-2} \! \circ \xi^{-1} \! \pa{s+i \tau}  }  } \;,
\enq
with the maximum being taken over $\abs{s} \leq s_0$ and $0\leq \tau \leq \a$. Hence, for any $s \in \R$ and $0\leq \tau \leq \a$
\beq
\abs{ \Re\pac{ \ln E_-^{-2} \circ \xi^{-1}\pa{s+i \tau} }  }  \leq  2 k C \a  \pa{2\pi}^k \abs{s}^{k-1}  + C^{\prime \prime } \;.
\enq
Therefore, we obtain the estimate
\bem
\abs{ \Int{u_L }{- u_L }  \f{\dd s}{2\pi}  \f{E_-^{-2}}{\xi^{\prime}}\circ \xi^{-1}\pa{s +i\a}
\f{1}{ \pac{  \xi^{-1}\pa{s+i\a} - \la}^{r}  }  \paa{1-\ex{2\pi L \a } \ex{-2i\pi s L} }^{-1}  }  \\
\leq  \sup_{z \in \xi^{-1}\pa{\R+i\a} } \paa{ \f{ \ex{C^{\prime \prime}} }{ \abs{z-\la}^r \xi^{\prime}\pa{z}  } } \cdot \f{2w_L+1}{2\pi L}  \cdot
 \f{ \ex{2k C \a  \pa{2\pi}^k u_L^{k-1}} }{\ex{2\pi L \a}-1} = \e{O}\pa{L^{-\infty}} \; ,
\end{multline}
where we have used that $\big(w_L\cdot L^{-1} \big)^{k-1} =\e{o}\pa{L}$. It remains to estimate the integral over the lines $\intff{0}{\a}$:
\bem
\abs{ \sul{\eps=\pm}{} \eps \Int{0}{\a} \f{i \dd s}{2\pi}  \f{E_-^{-2}}{\xi^{\prime}}\circ \xi^{-1}\pa{is+ \eps u_L }
\f{1}{ \pa{  \xi^{-1}\pa{ is+\eps u_L } - \la}^{r}  }  \paa{1+ \ex{2\pi s L} }^{-1}   } \leq \\
 \sup_{ \substack{ s \in \intff{0}{\a} \\ \eps \in \paa{\pm 1} } }
 \paa{   \f{ 1 }{  \abs{\xi^{\prime}\circ\xi^{-1}\pa{is + \eps u_L}} \abs{\pac{\xi^{-1}\pa{is+\eps u_L }-\la} }^{r} }     }
\times \f{ \ex{C^{\prime \prime}} }{ \pi }
\Int{0}{\a} \f{ \ex{2C k \pa{2\pi}^k \pa{u_L}^{k-1} \tau} }{ 1+\ex{2\pi L \tau} }  \dd \tau \;.
\label{appendix Sing sums domination integrale bords}
\end{multline}
By making the change of variables $y=L \tau$ and then applying Lebesgue's dominated convergence theorem,
one can convince oneself that the integral in the second line of \eqref{appendix Sing sums domination integrale bords}
is a $\e{O}\big( L^{-1} \big)$.

The last class of integrals to consider stems from integrations along $\msc{C}_{bd;L}$.
In order to carry the estimates, we need to use the finer condition \eqref{appendix Sing sums contrainte fonction E-}.
Here, we only treat the case of $k$ even and $C_1<0$. All other cases are treated very similarly.
An analogous reasoning to \eqref{appendix Sing sums estimation brute partie reelle log} leads, uniformly in $0\leq  \tau \leq \a$ to
\beq
\Re\pac{\ln E_-^{-2}\circ  \xi^{-1}\!\pa{s\pm i \tau }} = \tau \pa{ \mp k \pa{2\pi}^k  C_1 s^{k-1} + \e{O}\big(s^{k-2}\big) } \qquad 
\e{for} \quad \mp s >0 \;.
\enq
There exists $s_0^{\prime}$ such that for $\abs{s}\geq s_0^{\prime}$  one has
$\abs{\e{O}\pa{s^{k-2}}} \leq k \pa{2\pi}^k  \tf{ \abs{C_1 s^{k-1}} }{2}$.
As a consequence, for $\abs{s}\geq s_0^{\prime}$ and $\mp s >0$ 
\beq
\Re\pac{\ln E_-^{-2}\circ  \xi^{-1}\!\pa{s\pm i \tau }} \leq -k \f{ \pa{2\pi}^k }{2} \tau \abs{ C_1 s^{k-1}} \quad
\e{uniformly} \quad 0\leq \tau \leq \a \;.
\enq
Therefore,
\bem
\abs{ \; \Int{ \msc{C}_{bd;L} }{} \f{ \dd \mu }{ 2\pi } \f{ E_-^{-2} \pa{\mu} }{ \pa{\la-\mu}^r } }
= \abs{ \sul{\eps=\pm 1}{} \eps \Int{0}{\a} \f{ \dd \tau}{2i\pi}
\f{\pa{\tf{ E_-^{-2} }{ \xi^{\prime} }} \circ \xi^{-1}\!\pa{\eps u_L -i\eps \tau} }{ \pac{ \xi^{-1}\!\pa{\eps u_L -i\eps \tau}-\la }^r } } \\
\leq 2 \sup_{\substack{\tau \in \intff{0}{\a}  \\ \eps \in \paa{\pm 1}}}
\abs{   \f{ \pac{ \xi^{-1}\!\pa{\eps u_L -i\eps \tau}-\la }^{-r} }{ \xi^{\prime}  \circ \xi^{-1}\!\pa{\eps u_L -i\eps \tau} } }
\Int{ 0}{\a} \f{\dd \tau}{2\pi} \ex{- k \pa{2\pi}^k u_L^{k-1} \abs{C_1} \f{\tau}{2}  }  = \e{O}\pa{u_L^{1-k-r}}  \;.
\end{multline}

\qed


\subsection{The generating function: form factor-like representation}
\label{Appendix Subsection generating function FF sums}

From now on, we assume that the function $E_-$ takes the form $E_-^{-2}\pa{\la}=\ex{ixu\pa{\la}+g\pa{\la}}$ where
$u\pa{\la}$ is given by \eqref{ecritutre representation integrale u} and $g$ is a bounded holomorphic function on
the strip $U_{\de}$ around $\R$ \eqref{definiton voisinage ouvert de R pour holomorphie Gn}.
We also assume that $\nu \in \msc{O}\pa{U_{\de}}$.
We remind that the parameters $\paa{\mu_a}_{a\in\mathbb{Z}}$ (resp. $\paa{\la_a}_{a\in\mathbb{Z}}$)
are defined as the unique solutions to $L \xi\pa{\mu_a}= a$,
(resp. $L\xi_{\nu}\pa{\la_a}=a$), where $\xi$ is given by \eqref{ecriture limite thermo fction comptage} and
$\xi_{\nu}\pa{\la}=\xi\pa{\la}+\tf{\nu\pa{\la}}{L}$. We define the functional  $X_N\pac{\nu, E_-^{2}}$ as
\beq
X_N\pac{\nu, E_-^{2}}  =     \sul{n = 0}{ N +1 } \;
\sul{ \substack{p_1<\dots < p_n \\ p_k \in \mc{B}_L^{\e{ext}} }  }{}\;  \sul{ \substack{h_1<\dots < h_n \\ h_k \in \mc{B}_L^{\e{int}} } }{}
\; \f{ \pl{a=1}{N} E_-^2\pa{\la_a} }{ \pl{a=1}{N+1} E_-^{2}\pa{\mu_{\ell_a}} } \;
\wh{D}_{N} \pab{ \!\! \paa{p_a}_1^n \!\! }{ \!\! \paa{h_a}_1^n \!\! } \pac{ \nu, \xi, \xi_{\nu} }
 \; .
\label{definition Fonction generatrice X_N}
\enq
The functional $\wh{D}_{N,n}$ has been introduced in  \eqref{ecriture explicite fonctionnelle D+}.
The sums in \eqref{definition Fonction generatrice X_N} run through ordered $n-uples$ of integers $p_1<\dots<p_n$ belonging to
$\mc{B}_L^{\e{ext}}=\mc{B}_L \setminus \intn{1}{N+1}$ and through ordered  $n-uples$ of integers $h_1<\dots<h_n$ belonging to
$\mc{B}_L^{\e{in}}= \intn{1}{N+1}$.
Finally, $\mc{B}_L= \paa{ j \in \mathbb{Z} \; : \; -w_L \leq j \leq w_L} $ and the sequence $w_L \sim L^{\f{5}{4}}$.
In particular, when $L\tend +\infty$, $w_L$ grows much faster then $ N $.
The integers $\paa{p_a}$ and $\paa{h_a}$ define the sequence $\ell_1<\dots <\ell_{N+1}$ as explained in
\eqref{definition correpondance entiers ella et particules-trous}.

The functional $X_N\big[ \nu,E_-^{2} \big]$ admits two different representations. On the one hand,
as written in \eqref{definition Fonction generatrice X_N},
$X_N \big[ \nu,E_-^{2} \big]$ is closely related to a form factor expansion of certain two-point functions in generalized free-fermion models.
On the other hand, after some standard manipulations \cite{KorepinSlavnovTimeDepCorrImpBoseGas}, one can also recast $X_N\big[ \nu,E_-^{2} \big]$
in terms of a finite-size determinant which goes to a  Fredholm minor  in
the $N,L \tend +\infty$ limit.

We derive this finite-size determinant representation for $X_N\big[ \nu,E_-^{2} \big]$ below.

\begin{prop}
\label{Proposition representation XN determinant fini et infini}
Under the aforestated assumptions concerning the functions $E_-$ and $\nu$,
the functional $X_N\big[ \nu,E_-^{2} \big]$ admits a finite-size determinant representation
\beq
X_N\pac{\nu,E_-^{2}}  =  \paa{\mc{S}_0^{\pa{L}}\pac{E_-^{-2}} + \f{\Dp{} }{\Dp{}\a}}_{\mid \a=0} \hspace{-3mm} \cdot\;
 \ddet{N}{\de_{k\ell} +   \f{ V^{\pa{L}}\!\pa{\la_k,\la_{\ell}} }{L \xi^{\prime}_{\nu}\!\pa{\la_{\ell}}  }  +
 \a \f{ P^{\pa{L}}\!\pa{\la_k,\la_{\ell}} }{L \xi^{\prime}_{\nu}\!\pa{\la_{\ell}}  }  } \;,
\label{ecriture XN determinant finite size}
\enq
where
\beqa
V^{\pa{L}}\! \pa{\la,\mu} &=& 4  \f{ \sin \pac{\pi\nu\pa{\la}}  \sin \pac{\pi\nu\pa{\mu}} }{ 2i\pi \pa{\la-\mu}} E_-\pa{\mu} E_-\pa{\la}  \cdot
\paa{O^{\pa{L}}\!\pac{\nu,E_-^{-2}}\pa{\la}- O^{\pa{L}}\!\pac{\nu,E_-^{-2}}\pa{\mu}  }  \; ,
\label{definition noyau GSK integrable en taille L finie}  \\
P^{\pa{L}}\! \pa{\la,\mu} &=& 4  \f{ \sin \pac{\pi\nu\pa{\la}}  \sin \pac{\pi\nu\pa{\mu}} }{ 2\pi} E_-\pa{\la} E_-\pa{\mu} \cdot
O^{\pa{L}}\!\pac{\nu,E_-^{-2}}\!\pa{\la}\cdot O^{\pa{L}}\!\pac{\nu,E_-^{-2}}\pa{\mu} \;.
\label{definition projecteur P volume fini}
\eeqa
Also, we have set
\beq
O^{\pa{L}}\!\pac{\nu,E_-^{-2}}\pa{\la} =
i \hspace{-2mm} \Int{ \msc{C}_{bk;L}  }{ } \hspace{-2mm} \f{ \dd \mu }{2\pi } \f{ E^{-2}_{-}\pa{\mu} }{ \mu-\la }  \;
+ \; \f{ E_-^{-2}\pa{\la} }{ \ex{-2i\pi \nu\pa{\la}}-1 }
   \; + \; i I_1^{\pa{L}}\!\pac{E_-^{-2}}\pa{ \la }  \;.
\label{definition E+ en taille L}
%
%
%
\enq
The contour of integration has been depicted on Fig. \ref{contour definition E+L} and $S_0^{\pa{L}}$ (resp. $I_r^{\pa{L}}$)
is given by \eqref{ecrituire somme S0} (resp. \eqref{definition facteur IL}).

\begin{figure}[h]
\begin{center}

\begin{pspicture}(6,2.5)

\pscurve(0,1.2)(0.5,1)(0.7,0.8)(1,0.5)

\psline(1,0.5)(1,2.5)
\psline(1,2.5)(5,2.5)
\psline(5,2.5)(5,0.5)

\psline(5,0.5)(5.2,0.4)(5.5,0.1)(6,0)

\psline[linewidth=2pt]{->}(3,2.5)(3.1,2.5)


\psdots(0,0.5)(1,0.5)(2,0.5)(4,0.5)(5,0.5)(6,0.5)

\rput(0.9,0.2){$-3q$}
\rput(1.9,0.3){$-q$}
\rput(3.8,0.4){$q$}
\rput(5.3,0.7){$3q$}
\rput(-0.3,0.4){$-A_L$}
\rput(6.3,0.4){$B_L$}

\rput(7,2.5){$L \xi\pa{-A_L}=-w_L-\tf{1}{2}$}
\rput(7,1.8){$L \xi\pa{B_L}=w_L+\tf{1}{2}$}

\rput(4.1,1.7){$\wt{\msc{C}}_q$}
\rput(3,0.8){$\msc{C}_q$}
\rput(0.9,2.7){$\msc{C}_{bk;L}$}

\psline[linewidth=2pt]{->}(3.1,1.1)(3,1.1)

\pscurve(1.6,0.5)(1.7,0.9)(2.5,1.1)(3,1.1)(4,1)(4.5,0.5)(4.5,0.3)(4,0.2)(2.5,0.2)(1.7,0.2)(1.6,0.5)

\psline[linewidth=2pt]{->}(3.1,1.5)(3,1.5)

\pscurve[linestyle=dashed](1.3,0.5)(1.4,1.1)(2.3,1.4)(3,1.5)(4,1.4)(4.5,1) (4.8,0.5)(4.8,0.2)(4,0)(2.5,0)(1.7,0)(1.3,0.5)

\end{pspicture}

\caption{Contour $\msc{C}_{bk;L}$ appearing in the definition of $O^{\pa{L}}\big[\nu,E_-^{-2} \big]\!\pa{\la}$,
contour $\msc{C}_q$ (solid line) and contour $\wt{\msc{C}}_q$ (dashed line).
The contour $\msc{C}_{bk;L}$ is such that, for $\abs{\Re{\la}}\geq 4q$ , it stays uniformly away from the real axis.
\label{contour definition E+L}}
\end{center}
\end{figure}
\end{prop}

\Proof
We first recast the sum over the integers $\paa{p_a}$ and $\paa{h_a}$ corresponding to particle-hole like excitations into
the equivalent sum over all possible choices of integers $\ell_a$: $\ell_1<\dots<\ell_{N+1}$
with $\ell_a \in  \mc{B}_L=\mc{B}_L^{\e{int}}\cup\mc{B}_L^{\e{ext}}$, \textit{cf} \eqref{definition correpondance entiers ella et particules-trous}.
As all the sums are finite, there is no problem in permuting the orders of summation. Therefore,
\beq
X_N \big[ \nu, E_-^{2} \big]  =
\sul{ \substack{\ell_1<\dots < \ell_{N+1} \\ \ell_a\in \mc{B}_L }  }{}   \f{ \pl{a=1}{N} E_-^2\pa{\la_a} }{ \pl{a=1}{N+1} E_-^{2}\pa{\mu_{\ell_a}}}
\cdot \wh{D}_{N}\pab{ \!\! \paa{p_a}_1^n \!\! }{ \!\! \paa{h_a}_1^n \!\! } \pac{ \nu, \xi, \xi_{\nu} }
 \; .
\label{ecriture X_N series sur les ells}
\enq
The determinant entering in the definition of $\wh{D}_N$ can be represented as
\beq
\hspace{-3mm} \pl{a=1}{N} \f{ \mu_{\ell_a}-\mu_{\ell_{N+1}}}{\la_a-\mu_{\ell_{N+1}}}  \cdot \ddet{N}{\f{1}{\mu_{\ell_a}-\la_b}}=
\ddet{N+1}{\f{\pa{1-\de_{b,N+1}} }{\mu_{\ell_a}-\la_b} +\de_{b,N+1} }
=  \pa{1+ \f{\Dp{}}{\Dp{}\a}}_{\mid \a = 0} \hspace{-3mm}
\ddet{N}{\f{1}{\mu_{\ell_a}-\la_b}-\f{\a}{\mu_{\ell_{N+1}}-\la_b}}  \;.
\label{equation diverses decompositions determinant}
\enq
There we have used that for any polynomial $Q$ of degree $1$, one has $Q\!\pa{1}=Q\!\pa{0}+Q^{\prime}\!\pa{0}$.

It follows from the above representation that the summand in \eqref{ecriture X_N series sur les ells} is a symmetric function of the $N+1$ summation
variables $\mu_{\ell_a}$ that is moreover vanishing whenever $\ell_k=\ell_{a} \, , \, k \not= a$ .
Therefore, we can replace the summation over the fundamental  simplex
$\ell_1<\dots<\ell_{N+1}$ in the $\pa{N+1}^{\e{th}}$ power Cartesian product $\mc{B}_L^{N+1}$ by
a summation over the whole space $\mc{B}_L^{N+1}$, provided that we divide the result by $\pa{N+1}!$.
Once that the summation domain is symmetric, we can invoke  the antisymmetry of the determinant so as to
replace one of the Cauchy determinants by $\pa{N+1}!$ times the product of its diagonal entries. This last operation produces a separation
of variables \cite{KorepinSlavnovTimeDepCorrImpBoseGas}. Eventually, the result can be recast in the form of a
single $N\times N$ determinant:
\beq
X_N\pac{ \nu , E_-^{2}} = \pl{a=1}{N} \f{ 4 \sin^2 \pac{\pi \nu \pa{\la_a}}  }{ \wh{\xi}^{\prime}_{\nu}\!\pa{\la_a}}
\;\cdot \; \sul{ n \in \mc{B}_L }{} \f{ E_-^{-2}\! \pa{\mu_n} }{2\pi L \xi^{\prime}\!\pa{\mu_n} } \pa{1+ \f{\Dp{}}{\Dp{}\a}}_{\mid \a = 0}
  \det_{N} \pac{  M_{jk}+ \a \wt{P}_{jk}\pa{\mu_n} } \; ,
\enq
with
\beq
M_{k \ell}=  \de_{k , \ell } \f{ E_-^{2}\pa{\la_{\ell}}}{2\pi L} \mc{S}^{\pa{L}}_2\big[E_-^{-2}\big]\pa{\la_{\ell}}
\; +\; \pa{1-\de_{k, \ell}}  \f{ E_-^{2}\pa{\la_{\ell}}}{2\pi L \pa{\la_{k}-\la_{\ell}} }\;
\bigg\{ \mc{S}^{\pa{L}}_1\big[E_-^{-2} \big]\pa{\la_{k}} - \mc{S}^{\pa{L}}_1\big[E_-^{-2}\big]\pa{\la_{\ell}}   \bigg\} \; ,
\enq
$\mc{S}_r^{\pa{L}}$ being given by \eqref{appendix Sing sums ecriture Sr somme discrete},
\eqref{ecrituire somme S0}-\eqref{ecrituire somme S2} and $\wt{P}_{jk}\pa{\mu_n}$ being a $\mu_n$-dependent rank 1 matrix:
\beq
\wt{P}_{jk}\pa{\mu_n} = -\f{E_-^{2}\pa{\la_k}}{\pa{\mu_n-\la_j} } \cdot   \f{ \mc{S}_1^{\pa{L}}\big[E_-^{-2}\big]\pa{\la_k} }{2\pi L }\;.
\enq

Using the fact that $\wt{P}_{jk}\pa{\mu_n}$ is a rank one matrix that contains all the dependence of the determinant on the summation variable $\mu_n$, it
is readily seen that
\beq
\sul{ n \in \mc{B}_L }{} \f{ E_-^{-2}\!\pa{\mu_n} }{2\pi L \xi^{\prime}\!\pa{\mu_n} } \pa{1+ \f{\Dp{}}{\Dp{}\a}}_{\mid \a = 0}
\hspace{-3mm}\cdot \hspace{2mm} \det_{N} \pac{  M_{jk}+ \a \wt{P}_{jk}\pa{\mu_n} }  =
\bigg[ \mc{S}^{\pa{L}}_0\big[E_-^{-2}\big] + \f{\Dp{}}{\Dp{}\a} \bigg]_{\mid \a = 0} 
 \hspace{-3mm} \cdot \; \det_{N} \pac{  M_{jk}+ \a P_{jk}  }
\enq
where
\beq
P_{jk}  = -\f{E_-^{2}\!\pa{\la_k}}{ 2\pi L } \mc{S}^{\pa{L}}_1\big[E_-^{-2}\big]\pa{\la_k}  \cdot 
\mc{S}^{\pa{L}}_1\big[E_-^{-2}\big](\la_j )\;.
\enq
Applying \eqref{ecrituire somme S1}, \eqref{ecrituire somme S2} and then using  that
$L \xi\pa{\la_k}=L\xi_{\nu}\pa{\la_k} - \nu\pa{\la_k}=k- \nu\pa{\la_k}$, we obtain that
\beq
M_{k \ell} \f{E_-\pa{\la_k}}{E_-\pa{\la_{\ell}}} =
            \de_{k\ell} \f{ \xi_{\nu}^{\prime}\pa{\la_{\ell}}}{4 \sin^2\! \pac{ \pi \nu\pa{\la_{\ell}}}  }  \; + \;
          E_-\pa{\la_{\ell}}  E_-\pa{\la_{k}}
			\f{ O^{\pa{L}}\!\pac{\nu,E_-^{-2}}\pa{\la_k} -  O^{\pa{L}}\!\pac{\nu,E_-^{-2}}\pa{\la_{\ell}} }{ 2i\pi L \pa{\la_k -\la_{\ell}} }  \; ,
\label{Calcul Element diagonal matrice M}
\enq
where $O^{\pa{L}}\!\pac{\nu,E_-^{-2}}$ is given by \eqref{definition E+ en taille L}.
Note that we have slightly deformed the form of the contours $\msc{C}_{bk;L}$ in respect to 
Fig.~\ref{contour pour integral centrales et bords qui contribuent peu}. 
Very similarly, we find
\beq
P_{j k } \f{E_-(\la_j)}{E_-\pa{\la_{k}}} =  \f{ E_-\pa{\la_{j}}  E_-\pa{\la_k} }{ 2\pi L}
 O^{\pa{L}}\!\pac{\nu,E_-^{-2}}\pa{\la_k} O^{\pa{L}}\!\pac{\nu,E_-^{-2}}\pa{\la_{j}}  =
\f{  P^{\pa{L}}\!\pa{\la_{j},\la_k}    }{4 L \sin\pi\nu\pa{\la_{j}} \sin\pi\nu\pa{\la_k} } \;.
\enq
where $P^{\pa{L}}\pa{\la_j,\la_k}$ is given by \eqref{definition projecteur P volume fini}.
It then remains to factor out the pre-factors from the determinant. \qed

\subsection{Thermodynamic limit of $X_N\pac{\nu,E_-^{-2}}$}
\label{Appendix Subsection thermo limit of the Fredholm Minor}

\begin{prop}
\label{Proposition limit thermo Fred minor pur}
The thermodynamic limit of $X_N\pac{\nu,E_-^{2}}$ is well defined and can be expressed in terms of a Fredholm determinant minor.
Namely, $X_N\pac{\nu,E_-^{2}}  \limit{\tf{N}{L}}{D} X_{\msc{C}_E^{\pa{\infty}}}\pac{\nu,E_-^{2}}$ with
\beq
 X_{ \msc{C}_E^{\pa{\infty}} }\! \pac{\nu,E_-^{2}}  =
\pa{  \mc{S}_{ \msc{C}_E^{\pa{\infty}} } \!\pac{E_-^{-2}}  + 2 \Int{-q}{q} \f{\dd \la}{\pi} \sin^2\!\pac{\pi \nu\pa{\la}} F_+\pa{\la}E_-\pa{\la}
O_{\msc{C}_E^{\pa{\infty}}}\big[\nu,E_-^{-2} \big]\pa{\la}   }
\cdot \ddet{}{ I+V } \pac{\nu,E_-^{-2}} \;.
\label{appendix ecriture limite thermo XN}
\enq
Here $I+V$ is an integral operator on $\intff{-q}{q}$ acting on $L^{2}\pa{\intff{-q}{q}}$ with a kernel
\beq
V\pa{\la,\mu} = 4  \f{ \sin \pac{\pi\nu\pa{\la}}  \sin \pac{\pi\nu\pa{\mu}} }{ 2i\pi \pa{\la-\mu}}  E_-\pa{\la} E_-\pa{\mu}  \cdot
\paa{O_{ \msc{C}_E^{\pa{\infty}} }\!\pac{\nu,E_-^{-2}}\pa{\la} -O_{\msc{C}_E^{\pa{\infty}} }\!\pac{\nu,E_-^{-2}}\pa{\mu}  }
\label{definition noyau integrable volume infini}
\enq
and the contour $\msc{C}_E^{\pa{w}}$dependent functionals  $O_{\msc{C}_E^{\pa{w}}}\!\pac{\nu,E_-^{-2}}\pa{\la}$ and 
$\mc{S}_{\msc{C}_E^{\pa{w}}}\!\pac{E_-^{-2}}$ are given by
\beq
O_{ \msc{C}_E^{\pa{w}} }\!\pac{\nu,E_-^{-2}}\pa{\la}  =  i \Int{ \msc{C}_E^{\pa{w}} }{} \f{ \dd \mu }{2\pi } \f{ E^{-2}_{-}\!\pa{\mu} }{ \mu-\la }  \; + \;
\f{ E_-^{-2}\!\pa{\la} }{ \ex{-2i\pi \nu\pa{\la}}-1 }    \qquad \e{and} \qquad
\mc{S}_{\msc{C}_E^{\pa{w}}}\!\pac{E_-^{-2}} = \Int{ \msc{C}_E^{\pa{w}} }{} \! \f{\dd \la}{2\pi} \, E_-^{-2}\!\pa{\la} \;.
\label{definition fonction E+ et G volume infini}
\enq
$F_+\pa{\la}$ is the unique solution to the integral equation\symbolfootnote[2]{By no means $F_+$ ought to be confused with the shift function}
\beq
\sin \pac{\pi \nu\pa{\la}} F_+\pa{\la} + \Int{-q}{q} V\pa{\la,\mu} \sin \pac{\pi \nu\pa{\mu}} F_+\pa{\mu} \dd \mu
= \sin \pac{\pi \nu\pa{\la}}  E_-\pa{\la} O_{ \msc{C}_E^{\pa{\infty}} }\!\pac{\nu,E_-^{-2}}\pa{\la}  \;.
\label{definition fonction F+}
\enq
Also, $\msc{C}^{\pa{w}}_E=\msc{C}_E^{\pa{\infty}}\cap\paa{ z \in \Cx \; :\; \abs{\Re\pa{z}}\leq w}$ and $\msc{C}_E^{\pa{\infty}}$
 have been depicted on Fig.~\ref{contour CE et sa restriction CEw}. 
\end{prop}

This representation can be seen as a generalization of the results obtained in \cite{KorepinSlavnovTimeDepCorrImpBoseGas}.
Also, the contour $\msc{C}_E^{\pa{\infty}}$ can be thought of as the $L\tend +\infty$ limit of the contour $\msc{C}_{bk;L}$.

\Proof

It is a direct consequence of the estimates
obtained in appendix \ref{Appendix computation of singular sums}
for $I_r^{\pa{L}} \big[ E_-^{-2} \big]$ together with the fact that
$\ddet{N}{ \de_{k\ell}+ \e{o}\pa{L^{-1}}} \tend 1$ in the case of remainders $\e{o}\big( L^{-1} \big)$  that are uniform in the entries, that
\beq
X_N\pac{\nu,E_-^{2}}  \limit{\tf{N}{L}}{D}  \pa{ \mc{S}_{\msc{C}_E^{\pa{\infty}}}\!\pac{E_-^{-2}} + \f{ \Dp{}}{\Dp{}\a}  }_{\mid \a=0} \hspace{-3mm} \cdot \;\;
\ddet{}{I+V + \a P}
\enq
with $I+V+\a P$ acting on $\intff{-q}{q}$ and
\beq
P\pa{\la,\mu}= \f{2}{\pi} \sin\pac{\pi\nu\pa{\la} }  \sin\pac{\pi\nu\pa{\mu} } E_-\pa{\la}E_-\pa{\mu}
O_{\msc{C}^{\pa{\infty}}_E}\!\pac{\nu,E_-^{-2}}\pa{\la}O_{\msc{C}^{\pa{\infty}}_E}\!\pac{\nu,E_-^{-2}}\pa{\mu}\;.
\enq
Note that there is no problem with the integration over an infinite contour $\msc{C}_E^{\pa{\infty}}$ in
$O_{\msc{C}^{\pa{\infty}}_E}\!\pac{\nu,E_-^{-2}}\pa{\mu}$ and $\mc{S}_{\msc{C}_E^{\pa{\infty}}}\!\pac{E_-^{-2}}$  in as much as $\msc{C}_E^{\pa{\infty}}$ is built precisely
in such a way to ensure the exponential decay of the integrand at infinity.

Using that $P$ is a one dimensional projector, we get that
\beq
\ddet{}{I+V + \a P} = \ddet{}{I+V } \pa{1+ \a \Int{-q}{q} \pa{I+V}^{-1}\!\pa{\la,\mu} P\pa{\mu,\la} \dd \la \dd \mu } \;.
\enq
It then remains to take the $\a$-derivative and use the definition of $F_+\pa{\la}$. \qed

\vspace{2mm}


\subsection{An algebraic representation for the Fredholm minor}

\begin{prop}
\label{Proposition algebraic representation Fredholm minor}
For $L$ large enough, the finite $N$ Fredholm minor $X_N\pac{\nu, E_-^{2}}$ defined in \eqref{ecriture XN determinant finite size} can be represented, through purely algebraic
manipulations,  as the below finite sum:
\bem
X_N\pac{\nu, E_-^{2}} = \sul{n=0}{N} i\f{\pa{-1}^n}{n!} \hspace{-2mm}\sul{ \substack{i_1,\dots, i_n \\ i_a \in  \intn{1}{N} }  }{}
  \Oint{\msc{C}_q}{} \f{\dd^n z }{ \pa{2i\pi}^{2n} }  \Int{ \msc{C}^{\pa{L}} }{} \f{\dd^{n+1} y}{ \pa{2i\pi}^{n+1} }
\pl{k=1}{n+1} \paa{ f^{\pa{L}}\!\pa{y_{k},\nu\pa{y_k}} \cdot E_-^{-2}\!\pa{y_k}}
  \pl{a=1}{n} E^2_-\pa{\la_{i_a}}   \\
\times \pl{k=1}{n} \f{ y_{n+1}-z_k}{\pa{y_{n+1}-\la_{i_k}} \pa{y_k-z_k} } \cdot \ddet{n}{\f{1}{z_a-\la_{i_b}}}
\pl{k=1}{n}
\paa{ \f{ 4 \sin^2\pac{\pi \nu\pa{\la_{i_k}} }   }{  \pa{z_k-\la_{i_k}} L \xi^{\prime}_{\nu}\! \pa{\la_{i_k}}  }}  \;.
\label{ecriture rep XN serie Fredholm discete}
\end{multline}
Above, appear two contours, $\msc{C}_q$ which stands for a small counterclockwise loop around $\intff{-q}{q}$
as depicted on Fig.~\ref{contour definition E+L} and $\msc{C}^{\pa{L}}= \msc{C}_{bk;L}\cup
\msc{C}_{\ua;L}\cup\msc{C}_{\da;L}\cup \msc{C}_{bd;L}\cup\wt{\msc{C}_q}$. Note that $\msc{C}_{\e{bk};L}$ 
is as it has been depicted on Figs.~\ref{contour pour integral IL}-\ref{contour pour integral centrales et bords qui contribuent peu}.
As shown on  Fig.~\ref{contour definition E+L},
$\wt{\msc{C}}_q$ stands for a small counterclockwise loop encircling $\msc{C}_q$. Finally, the function $f^{\pa{L}}\!\pa{y,\nu}$
is supported on $\msc{C}^{\pa{L}}$ and reads
\beq
f^{\pa{L}}\!\pa{y,\nu} = \bs{1}_{\msc{C}_{bk;L}}\!\pa{y}  \; + \;  \f{ 1  }{ 1-\ex{-2i\pi L \xi\pa{y}} } \bs{1}_{\msc{C}_{\ua;L}}\!\pa{y}
\; + \;  \f{ 1  }{ \ex{2i\pi L \xi\pa{y}} - 1 } \bs{1}_{\msc{C}_{\da;L}}\!\pa{y} \; - \;
\f{ 1 }{ \ex{-2i\pi \nu } - 1 }  \bs{1}_{\wt{\msc{C}}_q}\!\pa{y}
 \; +  \;   \bs{1}_{\msc{C}_{bd;L}}\pa{y} \;.
\label{appendix mult dim Fred definition fction fL}
\enq
where $\bs{1}_A$ stands for the indicator function of $A$.
\end{prop}

\Proof

The functional $O^{\pa{L}}\big[\nu, E_-^{-2} \big]\pa{z}$ as defined in \eqref{definition E+ en taille L} is holomorphic in some
sufficiently small open neighborhood of $\intff{-q}{q}$.
Hence, there exists a small counterclockwise loop $\msc{C}_q$ around $\intff{-q}{q}$ (\textit{cf} Fig.~\ref{contour definition E+L})
such that the kernel $V^{\pa{L}}\!\pa{\la,\mu}$ admits the integral representation
\beq
\hspace{-1cm} V^{\pa{L}}\!\pa{\la,\mu}   = 4 \sin\pac{\pi \nu\pa{\la}}  \sin\pac{\pi \nu\pa{\mu}}  E_-\pa{\la}  E_-\pa{\mu}
\Oint{\msc{C}_q}{}   \f{ O^{\pa{L}} \big[ \nu, E_-^{-2}\big] \pa{z}  }{ \pa{z-\la}\pa{z-\mu}  }   \f{ \dd z }{ \pa{2i\pi}^{2} }  \; ,
\quad \e{for} \quad \la, \mu \in \paa{\la_1,\dots,\la_N} \;.
\label{appendix sing sums representation V integral ctr}
\enq
In \eqref{appendix sing sums representation V integral ctr} we have used that $\la_1,\dots, \la_N$ are all inside of $\msc{C}_q$ for $L$
large enough.
We first expand the $N\times N$ determinant appearing in the final expression for $X_N \big[ \nu,E_-^{2} \big]$ into its discreet Fredholm series:
\beq
\ddet{N}{\de_{k\ell} + \f{ V^{\pa{L}}\! \pa{\la_k,\la_{\ell} } }{L\xi_{\nu}^{\prime}\!\pa{\la_{\ell}}} +
\a  \f{ P^{\pa{L}}\! \pa{\la_k,\la_{\ell} } }{L\xi_{\nu}^{\prime}\!\pa{\la_{\ell}}}  }  =
\sul{n=0}{N}  \sul{ \substack{i_1,\dots, i_n \\ i_a \in \intn{1}{N} } }{}
\f{ \ddet{n}{  V^{\pa{L}}\! \pa{ \la_{i_a},\la_{i_b} }  +  \a   P^{\pa{L}}\! \pa{ \la_{i_a},\la_{i_b} } } }
{n! \prod_{k=1}^{n}  \pac{  L \xi^{\prime}_{\nu}\!\pa{\la_{i_k}} }   } \;.
\enq
Next, observe that
\bem
\ddet{n}{  V^{\pa{L}}\! \pa{ \la_{i_a},\la_{i_b} }  +  \a   P^{\pa{L}}\! \pa{ \la_{i_a},\la_{i_b} } } =
\Oint{\msc{C}_q}{} \f{ \dd^n z }{ \pa{2i\pi}^{2n} } \pl{a=1}{n} \Bigg\{
\f{ O^{\pa{L}}\big[ \nu, E_-^{-2} \big]\pa{z_a} }{z_a-\la_{i_a}}  \Bigg\} \\
\times \pl{a=1}{n} \paa{4 \sin^{2}\pac{\pi \nu\pa{\la_{i_a}}} E_-^{2}\pa{\la_{i_a}}   }
\times   \ddet{n+1}{ \ba{cc} \pa{z_a-\la_{i_b}}^{-1}  &  \a \vspace{2mm}  \\
						-iO^{\pa{L}}\big[\nu,E_-^{-2} \big]\pa{\la_{i_b}} & 1 \ea }
\end{multline}
It can be readily seen that for any $z$ belonging to the interior of $\wt{\msc{C}}_{q}$
\beq
O^{\pa{L}}\pac{\nu,E_-^{-2}}\pa{ z } = \Int{\msc{C}^{\pa{L}}}{}  \f{i \dd y }{2\pi} \f{ f^{\pa{L}}\!\pa{y,\nu\pa{y}} }{ y-z } E_-^{-2}\pa{y} \qquad
\e{with} \quad \msc{C}^{\pa{L}} = \msc{C}_{bk;L}\cup\msc{C}_{\ua;L}\cup\msc{C}_{\da;L}\cup \wt{\msc{C}}_q \cup \msc{C}_{bd;L}\;,
\enq
and $f^{\pa{L}}$ is as given by \eqref{appendix mult dim Fred definition fction fL}. Then, using the multilinear structure of a determinant, one gets that
\bem
\paa{   \mc{S}_0^{\pa{L}}\pac{E_-^{-2}} + \f{\Dp{} }{ \Dp{} \a }  }_{\a=0} \hspace{-3mm} \cdot
\ddet{n+1}{ \ba{cc} \pa{z_a-\la_{i_b}}^{-1}  &  \a  \vspace{2mm}\\
						-iO^{\pa{L}}\!\pac{\nu,E_-^{-2}}\pa{\la_{i_a}} & 1 \ea }
 =
\ddet{n+1}{    \ba{cc} \pa{z_a-\la_{i_b}}^{-1} 	& 1  \vspace{2mm}\\	
						  -i O^{\pa{L}}\!\pac{\nu, E_-^{-2} } \pa{\la_{i_b}}  & \mc{S}_0^{\pa{L}}\!\pac{E_-^{-2}} \ea} \\
= \Int{ \msc{C}^{\pa{L}} }{} \!\! \f{ \dd y }{ 2\pi } E_-^{-2}\!\pa{y} f^{\pa{L}}\!\pa{y,\nu\pa{y} }
 \ddet{n+1}{    \ba{cc} \pa{z_a-\la_{i_b}}^{-1} 	& 1  \vspace{2mm}\\	
						  \pa{y -\la_{i_b}}^{-1}   & 1  \ea}
= \Int{ \msc{C}^{\pa{L}} }{} \!\! \f{ \dd y }{ 2\pi } E_-^{-2}\!\pa{y} f^{\pa{L}}\!\pa{y,\nu\pa{y} }
\pl{k=1}{n} \f{ z_k-y }{ \la_{i_k}-y }   \ddet{n}{ \f{1}{z_a-\la_{i_b}} }  \; .
\label{ecriture representation alg det discret}
\end{multline}
This leads to the claim, once upon inserting this representation into the discreet Fredholm series. \qed



\subsection{The Natte series for a Fredholm minor}
\label{subsection Natte series Fred minor}

In this subsection, we recall the form of the Natte series representation for the Fredholm minor \eqref{appendix ecriture limite thermo XN}
involved in the representation of form factor sums in generalized free fermionic models.
We refer the reader to theorem 2.2 and proposition 7.2 of reference \cite{KozTimeDepGSKandNatteSeries} for
further details relative to this Natte series expansion.

\vspace{2mm}

Let $E_-^2=\ex{-ix u\pa{\la} - g\pa{\la}}$ be such that
\begin{itemize}
\item $u$ and $g$ are holomorphic in the  open neighborhood $U_{\tf{\de}{2}}$ of $\R$;
\item $u$ has a unique saddle-point $\la_{0}$ on the real axis which is of order $1$, \textit{ie} $u^{\prime\prime}\!\pa{\la_{0}}<0$;
\item the function $\nu$ is holomorphic in an open neighborhood $\mc{N}_q \subset U_{\tf{\de}{2}}$ of $\intff{-q}{q}$.
\end{itemize}
Also, let $\msc{C}_E^{\pa{w}} =\msc{C}_E^{\pa{\infty}} \cap \paa{ z  \in \Cx \; : \; \abs{\Re z}< w}$.
The contours $\msc{C}_E^{\pa{\infty}}$ and $\msc{C}_{E}^{\pa{w}}$ have been depicted in  Fig.~\ref{contour CE et sa restriction CEw}.

For $w> \abs{\la_0} +q >0$ and $x$ large enough,
the Fredholm minor $X_{\msc{C}_E^{\pa{w}}}\!\big[\nu, E_-^{2} \big]$ defined in \eqref{appendix ecriture limite thermo XN}
admits the below Natte series representation
\bem
X_{\msc{C}_E^{\pa{w}}}\big[\nu, E_-^{2}\big]
= \f{\mc{B}\pac{ \nu,u+i0^+}}{x^{ \nu^2\pa{q} + \nu^2\pa{-q}}}
\ex{\,  \Int{ -q }{ q } \pac{i x u^{\prime}\! \pa{\la} + g^{\prime}\! \pa{\la}  }  \nu\pa{\la}  \dd \la }
\Bigg\{  \f{\mc{A}_0\!\pac{\nu} \bs{1}_{\intoo{q}{+\infty}}\pa{\la_0} }{\sqrt{-2\pi x u^{\prime \prime}\!\pa{\la_{0}}} }
\ex{i x u\pa{\la_{0}}+g\pa{\la_{0}} }
+ \f{\mc{A}_+\!\pac{ \nu,u+i0^{+}} }{ x^{1+2\nu\pa{q}} } \ex{i x u\pa{q}+g\pa{q} }     \\
 + \; \f{\mc{A}_-\!\pac{\nu,u} }{ x^{1-2\nu\pa{-q}} } \ex{i x u\pa{-q}+g\pa{-q} } \; + \;
\sul{n \geq 1}{} \sul{ \mc{K}_n  }{}
\sul{  \mc{E}_n(\vec{k})  }{}  \int_{ \msc{C}_{\eps_{\bs{t}} }^{\pa{w}} }{}
  H_{n;x}^{\pa{\paa{\eps_{\bs{t}}}}}\pa{  \paa{u\pa{z_{\bs{t}}} } ; \paa{z_{\bs{t}}}  } \pac{ \nu }
 \pl{ \bs{t} \in \J{ k } }{} \ex{ \eps_{\bs{t}} g\pa{ z_{\bs{t}} } }      \f{ \dd^n z_{\bs{t}} }{ \pa{2i\pi}^n }
 \Bigg\}   \; .
\label{equation developpement det serie de Natte}
\end{multline}
The $+i0^+$ regularization of $u$ only matters in the time-like regime (where $\abs{\la_{0}}<q$).
The functionals $\mc{B}$, $\mc{A}_{\pm}$ and $\mc{A}_0$ are given respectively by \eqref{definition fonctionnelle B}
\eqref{definition fonctionnelle A+ et kappa} and
\eqref{definition fonctionnelle A- et A0}.
The notations and the structure of the sums appearing in the second line of \eqref{equation developpement det serie de Natte}
 are exactly as explained in theorem \ref{Theorem comportement asympt coeff Taylor limite Thermo}.

The Natte series is convergent for $x$ large enough in as much as, for $n$ large enough,
\beq
\sul{\mc{K}_n}{}\sul{\mc{E}_n(\vec{k}) }{} \bigg| \bigg| H_{n;x}^{\pa{\paa{\eps_{\bs{t}}}}} \pac{\nu}
\pl{ \bs{t} \in \J{ k } }{} \ex{ \eps_{\bs{t}} g }  \bigg|\bigg|
_{L^{1}\big(  \msc{C}^{\pa{\infty}}_{\eps_{\bs{t}}} \big) }   \;    \leq \;\;
 c_2\paf{c_1}{x}^{ n c_3 } \;.
\label{ecriture estimation de norme L1 Hn}
\enq
There $c_1$ and $c_2$ are some $n$-independent constants. They only depend on the values taken by  $u$, and $g$ in some small neighborhood of the base curve
$\msc{C}_E^{\pa{\infty}}$ and by $\nu $ on a small neighborhood of $\intff{-q}{q}$, whereas
\beq
c_3= \f{3}{4} \min\pa{\tf{1}{2},1-2 \max_{\tau=\pm }\abs{\Re\pac{\nu\pa{\tau q}}} -\Ups_{\eps}}  \quad
 \e{where}  \quad \Ups_{\eps} =2 \sup\bigg\{ \abs{ \Re\pac{ \nu\pa{z} - \nu\pa{\tau q}} }^{}_{} \; : \; \abs{z-\tau q} \leq \eps \;, \;  \tau=\pm   \bigg\}  \;.
\nonumber
\enq
Here $\eps>0$ is sufficiently small but arbitrary otherwise. 
We stress that, should these norms change, then so would change the constants $c_1$, $c_2$ and $c_3$ but the overall structure of
the estimates in $x$ would remain.

The Natte series expansion \eqref{equation developpement det serie de Natte}  has a well defined $w\tend +\infty$ limit:
all the concerned integrals are convergent as the functions $H_{n;x}^{\pa{ \{ \eps_{\bs{t}} \} }}$
approach zero exponentially fast in respect to any variable that runs to $\infty$ along $\msc{C}^{\pa{\infty}}_{\eps_{\bs{t}}}$.
Moreover, this limit does not alter in any way the estimates \eqref{ecriture estimation de norme L1 Hn} ensuring the convergence of the Natte series
(the constants $c_1$-$c_3$ are $w$-independent).

We now list several properties of the functions $H_{n;x}^{\pa{\paa{\eps_{\bs{t}}}}}$:
\begin{enumerate}
\item[i)] $H_{n;x}^{\pa{\paa{\eps_{\bs{t}}}}}\pa{ \paa{u\pa{z_{\bs{t}}} } ; \paa{z_{\bs{t}}}  }\pac{\nu}$
is a function of $\paa{u\pa{z_{\bs{t}}}}$ and $\paa{z_{\bs{t}}}$.
It is also a regular functional of $\nu$.
\item[ii)] $ H_{n;x}^{\pa{\paa{\eps_{\bs{t}}}}}\pa{ \paa{u\pa{z_{\bs{t}}} } ; \paa{z_{\bs{t}}}  }\pac{\ga \nu} = \e{O}\pa{\ga^n}$
and the O  holds in the $\big( L^{1}\cap L^{\infty} \big) \big( \msc{C}_{\eps_{\bs{t}}}^{\pa{\infty}} \big)$ sense.
\item[iii)]  $H_{n;x}^{\pa{\paa{\eps_{\bs{t}}}}}$ can be represented as:
\beq
H_{n;x}^{\pa{\paa{\eps_{\bs{t}}}}}\pa{ \paa{u\pa{z_{\bs{t}}} } ; \paa{z_{\bs{t}}}  } \pac{ \nu } =
\wt{H}_{n;x}^{\pa{\paa{\eps_{\bs{t}}}}} \pa{  \paa{\nu\pa{z_{\bs{t}}} }  ; \paa{u\pa{z_{\bs{t}}} } ; \paa{z_{\bs{t}}}  }
\pl{ \bs{t} \in \J{k} }{} \pa{  \varkappa\pac{\nu}\pa{z_{\bs{t}}} }^{-2\eps_{\bs{t}}}
\times
\pl{ \substack{ \bs{t} \in \J{k} \\ \eps_{\bs{t}}=1 } }{} \pa{\ex{-2i\pi \nu\pa{z_{\bs{t}}}} -1}^2 \;.
\label{ecriture dependence serie Natte en nu}
\enq
with $\wt{H}_{n;x}^{\pa{\paa{\eps_{\bs{t}}}}}$ is a holomorphic function for $ \abs{\Re\pa{\nu}} \leq \tf{1}{2}$ and $\varkappa\pac{\nu}\pa{\la}$
is given by \eqref{definition fonctionnelle A+ et kappa}.

\item[iv)] One has $H_{1;x}^{\pa{\paa{\eps_{\bs{t}}}}}=\e{O}\pa{x^{-\infty}}$ and for $n\geq 2$
\beq
\hspace{-5mm}  H_{n;x}^{\pa{\paa{\eps_{\bs{t}}}}} = \e{O}\pa{x^{-\infty}}  +
\sul{b=0}{ \pac{\tf{n}{2}} } \sul{p=0}{b} \sul{ m=b-\pac{\f{n}{2}} }{ \pac{\tf{n}{2}}-b}
%
%
 \paf{ \ex{ix \pac{u\pa{\la_0}-u\pa{-q}} } }{ x^{2 \nu\pa{-q}   }  }^{\bs{\eta} b} \hspace{-1mm}\cdot  \,
\paf{ \ex{ix \pac{u\pa{q}-u\pa{-q}} } }{ x^{2\pac{ \nu\pa{q}+\nu\pa{-q}}}  }^{m-\bs{\eta} p}\hspace{-2mm}\cdot  \;
\sul{\tau \in \paa{\pm 1; 0} }{}  \f{ \mf{e}_{\tau} }{ x^{n-\f{b}{2}}} \, \cdot  \,
\pac{ H_{n;x}^{\pa{\paa{\eps_{\bs{t}}}}} }_{m,p,b,\tau} \;.
\label{ecriture forme detaille fonction Hn}
\enq
\end{enumerate}
The $\e{O}\pa{x^{-\infty}}$ appearing above holds in the $ \big( L^{1}\cap L^{\infty} \big)\big(\msc{C}^{\pa{\infty}}_E\big)$ sense.
In order to lighten the formula, we have dropped the argument-dependent part. However, we do stress that the $\e{O}\pa{x^{-\infty}}$
as well as $\big[ H_{n;x}^{\pa{\paa{\eps_{\bs{t}}}}} \big]_{m,p,b,\tau} $ depend on the same set of variables as $H_{n;x}^{\pa{\eps_{\bs{t}}}}$.
Also, we agree upon $\bs{\eta}=1$ for $\la_0 >q$, $\bs{\eta}=-1$ for $\abs{\la_0}<q$ and
we made use of the shorthand notation
\beq
\mf{e}_{+}= \ex{ix u\pa{q}} x^{-2\nu\pa{q}}\; ,  \mf{e}_{-}= \ex{ix u\pa{-q}} x^{2\nu\pa{-q}}
\qquad \e{and} \quad \mf{e}_0=\pa{1+\bs{\eta}}\ex{ix u\pa{\la_0}} \; .
\enq
Finally, the functions  $\big[ H_{n;x}^{\pa{\paa{\eps_{\bs{t}}}}} \big]_{m,p,b,\tau} $  are only supported on a small
vicinity of the points $\pm q$ and $\la_0$. In such a case, the contour of integration reduces to an integration for 
each variable $z_{\bs{t}}$ to a small circle $\Dp{}\mc{D}_{0;v_{\tau}}$ around $v_{\tau}$ ($v_{\pm} =\pm q$, $v_0=\la_0$). 
Their dependence on $x$ is as follows. If a variable $z_{\bs{t}}$ is integrated in a vicinity of $v_{\tau}$, 
The function  $\big[ H_{n;x}^{\pa{\paa{\eps_{\bs{t}}}}} \big]_{m,p,b,\tau} $ contains a fractional power of 
$x^{ \pm \pac{ 2\nu\pa{z_{\bs{t}}} -\nu\pa{v_{\tau}} } }$, multiplied by a function of 
$z_{\bs{t}}$ which has an asymptotic expansion into inverse powers of $x$. 
This asymptotic expansion holds on $\Dp{}\mc{D}_{0;v_{\tau}}$. The coefficients in this asymptotic expansions contain poles at $z_{\bs{t}}=v_{\tau}$.
By computing the integrals associated to the terms in this asymptotic expansion through the poles at $z_{\bs{t}}=v_{\tau}$
one obtains that function coefficients associated to $x^{-r}$ terms produce, \textit{in fine}, a contribution that is a $\pa{\tf{\ln x}{x}}^r$.
Finally, the structure of these poles is such that, upon computing all the partial derivatives and for any holomorphic 
function $h$ in the vicinity of the points $\pm q$, $\la_0$, one should make the replacement:
\beq
\sul{\bs{t} \in \J{ \, \vec{k} } }{} \eps_{\bs{t}} h\pa{z_{\bs{t}}} \hookrightarrow  \bs{\eta} b \pa{ h\pa{\la_0}-h\pa{-q}} + \pa{m-\bs{\eta}p}\pa{h\pa{q}-h\pa{-q}}
+ \pa{\de_{\tau;1}+\de_{\tau;-1} + \pa{1+\bs{\eta}} \de_{\tau;0}/2 } h\pa{v_{\tau}}\;.
\enq

There is one last property which we conjecture to be true for the detailed representation \eqref{ecriture forme detaille fonction Hn} of 
$H_{n;x}^{\pa{\{\eps_{\bs{t}}\}}}$ but that has not been proven so far.  Namely, 

\begin{conj}
\label{Conjecture structure fine forme detaillee Hn}
For a given $n$ the sum in \eqref{ecriture forme detaille fonction Hn}
only contains those combinations of the integers $m,p,b$ and $\tau$ that satisfy to the constraint
\beq
\pa{m-\bs{\eta} p +\de_{\tau,1}}^2 + \f{b}{2}+ \pa{m+\bs{\eta} \pa{b-p}-\de_{\tau,-1}}^2 \leq n \;.
\label{ecriture conjecture structure forme detaillee Hn}
\enq
\end{conj}




\section{Multidimensional Fredholm series for $\lim_{N\tend +\infty} \rho^{\pa{m}}_{N;\e{eff}}\!\pa{x,t}$}
\label{Appendix multidimensional Fredholm series}

We begin this appendix by deriving the so-called discreet multidimensional Fredholm series representation for $\rho_{N;\e{eff}}^{\pa{m}}\!\pa{x,t}$.
We will prove in theorem \ref{Theorem representation series Fredholm multidim} that this representation has a well defined thermodynamic limit that we denote $\rho_{\e{eff}}^{\pa{m}}\!\pa{x,t}$.
This analysis will allow us to provide (proposition \ref{Proposition permutation limit thermo rho eff})
yet another representation for the thermodynamic limit $\rho_{\e{eff}}^{\pa{m}}\!\pa{x,t}$.
This alternative representation for $\rho_{\e{eff}}^{\pa{m}}\!\pa{x,t}$ is used in subsection \ref{subsection multidim Natte}
so as to construct the multidimensional  Natte series for
$\rho_{\e{eff}}^{\pa{m}}\!\pa{x,t}$.

\begin{theorem}
\label{Theorem representation series Fredholm multidim}
$\rho_{N;\e{eff}}^{\pa{m}}\!\pa{x,t}$ admits a well defined thermodynamic limit $\rho_{\e{eff}}^{\pa{m}}\!\pa{x,t}$ that is given by a multidimensional
Fredholm series
\bem
\hspace{-1cm} \rho_{\e{eff}}^{\pa{m}}\!\pa{x,t} = \sul{n=0}{m}  c \f{ \pa{-1}^n }{n!} \f{ \Dp{}^m }{ \Dp{}\ga^m } \Int{-q}{q} \f{ \dd^n \la }{ \pa{2i\pi}^n}
\Oint{\msc{C}_q}{} \f{ \dd^n z }{ \pa{2i\pi}^n }
\Int{ \msc{C} }{} \! \f{ \dd^{n+1} y }{ \pa{2i\pi }^{n+1}}
\f{ \ex{ix \mc{U}\pa{\paa{\la_a}_1^n; \paa{y_a}_1^{n+1}\mid \ga}} \prod_{k=1}^{n+1} f\!\pa{y_k,\ga \nu\pa{y_k}}  }
{ \prod_{k=1}^{n}\pa{z_k-\la_k}\pa{y_k-z_k} \pa{y_{n+1}-\la_k}  }  \ddet{n}{ \f{ \pa{y_{n+1}-z_k} }{z_a-\la_b } } \\
\times
 \f{ \pl{a=1}{n} \pl{b=1}{n+1} \pa{y_b-\la_a-ic}\pa{\la_a-y_b-ic} }{ \pl{a,b=1}{n+1} \pa{y_a-y_b-ic} \pl{a,b=1}{n}\pa{\la_a-\la_b-ic} }
 \pl{k=1}{n} \paa{ 4\sin^{2}\!\pac{\pi \ga \nu\pa{\la_k}} }
 \left.  \f{ \ddet{n}{ \de_{k\ell} + \ga \wh{V}_{k\ell}\pac{\ga \nu} }   \det_{n} \Big[ \de_{k\ell} + \ga \wh{\ov{V}}_{k\ell}\pac{\ga \nu} \Big]  }
{  \ddet{}{I+\tf{\ga R}{2\pi}}   \det^{2}\pac{I-\tf{K}{2\pi}}}  \right|_{\ga=0} \hspace{-2mm} .
\label{appendix multi dim Fred representation rho eff series Multdim Fred}
\end{multline}
The function $f$ appearing above is supported on the contour $\msc{C}=\msc{C}_E^{\pa{\infty}}\cup\wt{\msc{C}}_q$.
The contour $\msc{C}_q$ is a small loop around $\intff{-q}{q}$ whereas $\wt{\msc{C}}_q$ is a small loop around $\msc{C}_q$.
Both $\msc{C}_q$ and $\wt{\msc{C}}_q$ lie below the curve $\msc{C}_E^{\pa{\infty}}$ as depicted on
Fig.~\ref{contour CE sa restriction CEw et les boucles Cq et Ctile q}.
All of the aforementioned contours lie inside of the strip $U_{\de}$
\eqref{definiton voisinage ouvert de R pour holomorphie Gn}. The function $f$ is supported on $\msc{C}$ and reads
\beq
f\!\pa{y,\nu\pa{y}} = \bs{1}_{\msc{C}_E^{\pa{\infty}} }\!\pa{y}  - \f{1 }{ \ex{-2i\pi \nu\pa{y}} -1 }   \bs{1}_{\wt{\msc{C}}_q}\!\pa{y}   \qquad \e{with} \qquad
\msc{C}= \msc{C}_E^{\pa{\infty}} \cup \wt{\msc{C}}_q \;.
\enq
There $\bs{1}_A$ stands for the indicator function of the set $A$. The function $\nu$ appearing in the $n^{\e{th}}$-summand of
\eqref{appendix multi dim Fred representation rho eff series Multdim Fred} corresponds to the unique solution of the linear integral equation
driven by the resolvent $R$ of the Lieb kernel  $\pa{ ie \; \pac{ I-\tf{K}{2\pi}}\pac{I+\tf{R}{2\pi}} =I } $:
\beq
\nu\pa{\la} \; + \;  \ga \Int{-q}{q} \! \f{\dd \mu}{2\pi} R\pa{\la,\mu} \nu\pa{\mu}
=   -\tf{Z\pa{\la}}{2}  \; + \; \sul{a=1}{n} \phi\pa{\la,\la_a}  \; - \;  \sul{a=1}{n+1} \phi\pa{\la,y_a} \;.
\label{appendix Mul Dim Fred ecriture Int eqn nu}
\enq
Hence, $\nu$ depends on the integration variables
$\la_a$ (with $a=1,\dots,n$) and $y_a$  (with $a=1,\dots,n+1$), \textit{ie} $\nu\pa{\la} \equiv \nu\big(\la \mid \paa{\la_{a}}_1^n ; \paa{y_a}_1^{n+1} \big)$.
We kept this dependence implicit in \eqref{appendix multi dim Fred representation rho eff series Multdim Fred} so as to shorten the formulae.  
The entries of the finite-size determinants are as defined in \eqref{appendix thermo lim FF definition entree V et Vbar chapeau}.
They depend on the same set of auxiliary variables as $\nu$.
Finally, we agree upon
\beq
\mc{U}\!\pa{ \paa{\la_a}_1^n;\paa{y_a}_1^{n+1} \mid \ga } =
\sul{a=1}{n+1} u_0\pa{y_a} - \sul{a=1}{n} u_0\pa{\la_{a}} +\pa{1-\ga} \Int{-q}{q}\!  u_0^{\prime}\pa{\la} \nu\pa{\la}  \dd \la \;.
\enq
\end{theorem}
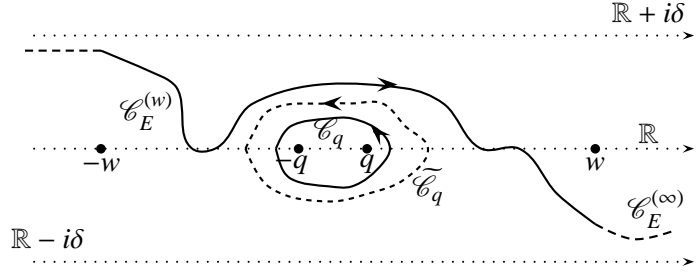
\begin{figure}[h]
\begin{center}

\begin{pspicture}(6,4)

\psline[linestyle=dotted, dash=3pt 2pt]{->}(-1,2)(7.8,2)
\psline[linestyle=dotted, dash=3pt 2pt]{->}(-1,0.5)(7.8,0.5)
\psline[linestyle=dotted, dash=3pt 2pt]{->}(-1,3.5)(7.8,3.5)

\rput(7.2,2.2){$\R$}
\rput(7.2,3.8){$\R + i\de$}
\rput(-0.7,0.8){$\R - i\de$}


\psdots(2.6,2)(3.5,2)(0,2)(6.5,2)

\rput(2.5,1.8){$-q$}
\rput(3.5,1.8){$q$}
\rput(0,1.8){$-w$}
\rput(6.5,1.8){$w$}

\pscurve(0,3.3)(0.7,3)(0.9,2.9)(1,2.8)(1.2,2)(1.5,2)(2,2.6)(4,2.8)(4.5,2.6)(5,2)(5.5,2)(6,1.4)(6.5,1)

\pscurve(2.3,2)(2.4,2.2)(2.5,2.3)(3,2.4)(3.3,2.4)(3.7,2.2)(3.8,2)(3.7,1.8)(3.3,1.5)(3,1.5)(2.5,1.6)(2.4,1.7)(2.3,2)

\pscurve[linestyle=dashed, dash=2pt 2pt](1.9,2)(2,2.2)(2.2,2.5)(3,2.6)(3.3,2.6)(3.7,2.6)(4,2.3)(4.3,2)(4,1.6)(3.5,1.4)(3,1.3)(2.2,1.5)(1.9,2)


\pscurve[linestyle=dashed, dash=3pt 2pt](-1,3.3)(-0.5,3.3)(0,3.3)

\pscurve[linestyle=dashed, dash=3pt 2pt](6.5,1)(7,0.8)(7.5,0.9)

\rput(3,2.2){$\msc{C}_q$}
\rput(4.3,1.5){$\wt{\msc{C}}_q$}

\rput(0.6,2.5){$\msc{C}_E^{\pa{w}}$}

\rput(7.3,1.2){$\msc{C}_E^{\pa{\infty}}$}

\psline[linewidth=2pt]{->}(3.8,2.85)(3.9,2.85)

\psline[linewidth=2pt]{->}(3,2.6)(2.9,2.6)

\psline[linewidth=2pt]{->}(3.6,2.25)(3.55,2.3)

\end{pspicture}

\caption{The contour $\msc{C}_{E}^{\pa{w}}$ consists of the solid line. The contour $\msc{C}_{E}^{\pa{\infty}}$
corresponds to the union of the solid and dotted lines. The loop $\msc{C}_q$ is depicted in solid lines whereas the loop
$\wt{\msc{C}}_q$ is depicted in dotted lines.
\label{contour CE sa restriction CEw et les boucles Cq et Ctile q}}
\end{center}
\end{figure}

\Proof

In order to implement the substitution of the operators $\Dp{\vsg_p}$  and $\Dp{\eta_{j,p}}$
(\textit{cf} section \ref{Subsection Operator ordering for functional translation})
in the representation \eqref{equation exprimant Q cal en terme moyenne XN} we introduce, exactly as it was done in the proof of theorem \ref{Theorem
comportement asympt coeff Taylor limite Thermo},
the functions $\wt{E}_-\pa{\la}$ \eqref{definition fction E-tilde}  (whose definition involves the functions  $\wt{g}=\wt{g}_{1,s}+\wt{g}_{2,r}$
\textit{cf} \eqref{definition fonctions tilde g1 et g2}) as well as $\nu_s$ \eqref{definition fonction nus}
and $\varpi_r$ \eqref{definition fonction varpi r}.

We then consider the discreet Fredholm series representation for $\Dp{\ga}^m X_N\pac{\ga \nu_s, \wt{E}_-^{2}}_{\mid \ga=0}$ obtained in
proposition \ref{Proposition algebraic representation Fredholm minor}. This will allow us to compute
the relevant Taylor coefficients (\textit{cf} subsection \ref{Subsection Operator ordering for functional translation}
equation \eqref{ecriture developpement F nu et g en puissance des aj} and \eqref{definition substitution operatorielle})
arising in the representation \eqref{equation exprimant Q cal en terme moyenne XN} for $\rho_{N;\e{eff}}^{\pa{m}}\! \pa{x,t}$.
One has that
\bem
\f{ \Dp{}^m }{ \Dp{}\ga^m } \paa{ \pl{a=1}{N+1} \wt{E}^{\,2}_-\pa{\mu_a} \cdot \pl{a=1}{N} \wt{E}_-^{-2}\pa{\la_a}  \cdot
 X_N\pac{\ga \nu_s, \wt{E}_-^{\,2} }  \wh{\msc{G}}_{\ga; 2A_L}^{\,\pa{\be}} \pac{\varpi_r}}_{\mid \ga=0}  = \\
\sul{n=0}{m}  i \f{\pa{-1}^n}{n!} \sul{ \substack{i_1,\dots, i_n \\ i_a \in  \intn{1}{N} }  }{}
  \Oint{\msc{C}_q}{} \f{\dd^n z }{ \pa{2i\pi}^{2n} }  \Int{ \msc{C}^{\pa{L}} }{} \f{ \dd^{n+1}y }{ \pa{2i\pi}^{n+1} }
\wt{\mc{L}}^{\pa{m}}_{\Ga^{\pa{L}}}\pac{\mc{F}_{i_1,\dots i_n}  \wh{\msc{G}}_{\ga;2A_L}^{\,\pa{\be}} } \;.
\label{appendix multi Fred ecriture serie discete rho m}
\end{multline}
The contours $\msc{C}^{\pa{L}}$, $\msc{C}_q$  have been defined in proposition \ref{Proposition algebraic representation Fredholm minor}.
We stress that the summation over $n$ in \eqref{appendix multi Fred ecriture serie discete rho m} could have been
stopped at $n=m$ since, prior to taking the $\ga$-derivative at $\ga=0$, the $n^{\e{th}}$ term of the series
\eqref{ecriture rep XN serie Fredholm discete} is a smooth function of $\ga$ that behaves as $\e{O}\pa{\ga^{n}}$. We have set
\beq
\mc{F}_{i_1,\dots, i_n}\pac{\ga \nu_s} =
  \pl{k=1}{n} \paa{ \f{ 4 \sin^2\!\pac{\pi \ga \nu_s \! \pa{\la_{i_k}} }   }{ L \xi^{\prime}_{\ga \nu_s}\!\pa{\la_{i_k}} \pa{z_k-\la_{i_k}}}  }
 \f{  \prod_{k=1}^{n+1} f^{\pa{L}}\!\pa{y_k,\ga \nu_{s}\pa{y_k}}  }{ \pl{k=1}{n} \pa{y_{n+1}-\la_{i_k}}\pa{y_k-z_k}  }
\times  \ddet{n}{ \f{  y_{n+1}-z_j }{ z_j-\la_{i_k} }}   \ex{ix \mc{U}^{\pa{L}}\pa{ \paa{\la_k}; \paa{\mu_k} ; \paa{y_k}  \mid \ga} } \;,
\label{appendix multi Fred definition fnelle F i1 in}
\enq
the function $f^{\pa{L}}\!\pa{y,\nu\pa{y}}$ is given in \eqref{appendix mult dim Fred definition fction fL} and we have set

\beq
\mc{U}^{\pa{L}}\!\pa{\paa{\la_{k}};\paa{\mu_k};\paa{y_k} \mid \ga}= \sul{k=1}{n+1} u\pa{y_k} \; - \; \sul{k=1}{n} u\pa{\mu_{i_k}} \; -\; u\pa{\mu_{N+1}}
\; + \;  \sul{ \substack{ k=1 \\ \not= i_1,\dots, i_n }  }{ N } u\pa{\la_k}-u\pa{\mu_k} \; .
\label{appendix mult Fred definition fction de la phase oscillante}
\enq
Last but not least,
\bem
\wt{\mc{L}}^{\pa{m}}_{\Ga^{\pa{L}}}\pac{\mc{F}_{i_1,\dots i_n} \msc{G}_{\ga;2A_L}^{\pa{\be}} } =
\sul{ \substack{ n_1,\dots, n_s \\ = 0 }  }{ m } \pl{p=1}{s}  \paa{ \f{ a_p^{n_p} }{ n_p! } }
\f{\Dp{}^m}{\Dp{}\ga^m} \Bigg\{  \f{  \prod_{k=1}^{n+1} \ex{\wt{g}_{1,s}\pa{y_k}}  }
{ \ex{\wt{g}_{1,s}\pa{\mu_{N+1} }  } \prod_{k=1}^{n} \ex{\wt{g}_{1,s} (\mu_{i_k} )  } }
\pl{p=1}{s} \paa{ \Ga^{\pa{L}}\pac{\ga \nu_s}\big( t_p \big)  }^{n_p} \\
\times \pl{j=1}{N+1} \ex{-\wt{g}_{2,r} (\mu_j) } \pl{\substack{j=1 \\ \not= i_1,\dots, i_n } }{N} \ex{\wt{g}_{2,r} (\la_j) }
\pl{j=1}{n+1} \ex{\wt{g}_{2,r} (y_j) }
\cdot  \mc{F}_{i_1,\dots ,i_n}\pac{\ga \nu_s}_{ \left| \substack{ \ga=0 \\ \vsg_a=0} \right. } \wh{\msc{G}}_{\ga;2A_L}^{\,\pa{\be}}\pac{\varpi_r}
\Bigg\}_{\mid \ga=0}\;.
\label{appendix mult Fred tilde Lagrange series for Ga L}
\end{multline}

The functional $\Ga^{\pa{L}}$ is evaluated at the discretization points (\textit{cf} definition
\ref{Definition point discretisation}, subsection \ref{Subsection resum effective FF series}
and subsection \ref{Subsection effective FF series})
$t_p$, $p=1,\dots,s$ for the contour $\msc{C}_{out}$
encircling the compact $K_{2q}$. $\msc{C}_{out}$ has been depicted in the $lhs$ of Fig.~\ref{contour exemple de courbes encerclantes}.
The functional $\Ga^{\pa{L}}$ reads
\beq
\Ga^{\pa{L}}{\pac{\nu}}\pa{\mu} = \sul{ \substack{ j=1 \\ j\not= i_1,\dots, i_n} }{N } \phi (\mu,\mu_j) - \phi (\mu,\la_j)
\qquad \e{with} \quad \mu_j \; \e{and}\; \la_j \; \e{defined} \; \e{by}\quad
 \xi (\mu_j)=\tf{j}{L} \;\; \e{and} \;\; \xi_{\nu} ( \la_j ) = \tf{j}{L} \;.
\enq

 We do stress that the variables $y_k$, with $k=1,\dots,n+1$, and $\mu_p$ or $\la_p$ with  $p=1,\dots,N+1$
appearing in \eqref{appendix multi Fred ecriture serie discete rho m}-\eqref{appendix mult Fred tilde Lagrange series for Ga L}
are  all located inside of the compact $K_{2A_L }$, where $A_L$ is such that $L\xi\pa{-A_L}=-w_L-\tf{1}{2}$.
As a consequence, the singularities at  $\la=t_{i,p}$ of the functions $\wt{g}_{2,r}\pa{\la}$ \eqref{appendix mult Fred tilde Lagrange series for Ga L}
are always disjoint from the variables $y_k, \mu_p$ or $\la_{i_b}$.
Indeed,  $t_{1,p}$ and $t_{2,p}$ with $p=1,\dots,r$  stand for  discretization points of the contour $\msc{C}_{out/in}$
appearing in the $rhs$ of Fig.~\ref{contour exemple de courbes encerclantes}, \textit{cf} subsection \ref{Subsection resum effective FF series}.
These two contours are such that $\e{d}\!\pa{ \msc{C}_{out/in},K_{2A_L}}>0$ uniformly in $L$.

According to the prescription that has been adopted in section \ref{Subsection Operator ordering for functional translation},
one has to compute the $m^{\e{th}}$ $\ga$-derivative of representation \eqref{appendix mult Fred tilde Lagrange series for Ga L}
prior to implementing the operator substitution.
For this, consider any smooth function $w\pa{\ga}$ such that $w\pa{\ga}=\e{O}\pa{\ga^n}$
at $\ga=0$. By applying the Faa-d\'{i}-Bruno formula, we get that
\bem
\hspace{-5mm} \f{1}{m!} \f{ \Dp{}^m }{ \Dp{}\ga^m } \Bigg\{ w\pa{\ga} \pl{\substack{a=1 \\ \not= i_1,\dots, i_n } }{N} \!\! \! \ex{\wt{g}_{2,r}\pa{\la_a}}
\; \; \wh{\msc{G}}_{\ga;2A_L}^{\,\pa{\be}}\pac{\varpi_r} \Bigg\}_{\mid \ga=0} =
\sul{ \paa{\ell_a}  }{} \hspace{-1mm}^{\prime } \;
\f{w^{\pa{\ell_{N+1}}}\!\pa{0}}{ \ell_0 ! \; \ell_{N+1} !  } \,
		\hspace{-2mm}  \pl{\substack{p=1 \\ \not =i_1,\dots, i_n}}{N}
\f{1}{\ell_p !} \f{ \Dp{}^{\ell_p} }{ \Dp{}\ga^{\ell_p}}    \paa{ \ex{ \wt{g}_{2,r} ( \la_p ) } }_{\mid \ga=0}
\times \f{ \Dp{}^{\ell_0} }{ \Dp{}\ga^{\ell_0} } \paa{ \wh{\msc{G}}_{\ga;2A_L}^{\,\pa{\be}}\pac{\varpi_r} }_{\mid \ga=0} \\
= \sul{ \paa{\ell_a}  }{} \hspace{-1mm}^{\prime } \; \f{w^{\pa{\ell_{N+1}}}\!\pa{0}}{ \ell_0 ! \ell_{N+1} ! } \,
\sul{ \paa{k_{a,j} } }{} \hspace{-1mm}^{\prime }   \hspace{1mm}\pl{\substack{p=1 \\ \not =i_1,\dots, i_n}}{N}
\Bigg\{
\hspace{-1mm} \left.  \f{ \Dp{}^{ \abs{\bs{k}_p} }  \ex{ \wt{g}_{2,r} (\tau_p ) }  }{ \Dp{}\tau_p^{ \abs{\bs{k}_p} } }     \right|_{\tau_p=\mu_p}
\hspace{-4mm} \times \; \; \pl{j=1}{\ell_p} \bigg( \f{ \la_p^{\pa{j}} }{ j! } \bigg)^{k_{p,j}}      \Bigg\}
\times \f{ \Dp{}^{\ell_0} }{ \Dp{}\ga^{\ell_0} } \paa{ \wh{\msc{G}}_{\ga;2A_L}^{\,\pa{\be}}\pac{\varpi_r}} _{ \mid \ga=0} \;.
\label{appendix mult Fred somme reduire gamma der}
\end{multline}

There the $\prime$ in front of the sums indicates that these are constrained. The first sums runs
through all choices of $N+2$ integers $\ell_p\geq 0$ such that
\beq
\ell_{i_p} = 0\; , \quad \e{for} \quad  p=1,\dots,n \; , \quad  \ell_{N+1} \geq n \quad \e{and} \quad
 \sum_{p=0}^{N+1} \ell_p = m \;.
\enq
 The second sum runs through all the possible choices of
sequences of integers $k_{p,j}$ with
\beq
p=1, \dots, N \quad \e{and} \quad  j=1,\dots , \ell_p \quad  \e{such}  \;  \e{that} \quad  \sum_{j=1}^{\ell_p} j k_{p,j}= \ell_p \;.
\enq
Finally, we agree upon
\beq
\abs{\bs{k}_p}=\sum_{j=1}^{\ell_p}k_{p,j} \quad \e{and} \; \e{have} \; \e{set} \quad
\la_p^{\pa{j}}= \Dp{\ga_p}^j \pac{ \la_p\pa{\ga_p}} \! _{\mid \ga_p=0 } \quad
\e{with} \;\la_p \; \e{defined}\; \e{by}  \quad \xi_{\ga_p\nu_s}\pa{\la_p}=\tf{p}{L} \; .
\label{appendix mult Fred contrainte entiers k}
\enq
In \eqref{appendix mult Fred contrainte entiers k}, we have explicitly insisted on the fact that $\la_p$ is a function of
the parameter $\ga_p$.
By substituting the representation \eqref{appendix mult Fred somme reduire gamma der} on the level of
\eqref{appendix multi Fred ecriture serie discete rho m}-\eqref{appendix mult Fred tilde Lagrange series for Ga L},
one can implement the operator substitution
$a_k \hookrightarrow \Dp{\vsg_k}$ and $b_{j,k} \hookrightarrow \Dp{\eta_{j,k} }$ on the level of
\eqref{appendix mult Fred tilde Lagrange series for Ga L}.

The functionals $\mc{F}_{i_1,\dots,i_n}$ and $\wh{\msc{G}}_{\ga;2A_L}^{\,\pa{\be}}$ are regular in the sense of definition
\ref{Definition Fonctionelle reguliere}. Moreover, as $L$ and hence $2A_L$ are large enough, and $\be_0$  defining $\bs{\wt{U}}_{\be_0}$
is  chosen in such a specific way\symbolfootnote[2]{in particular it depends on L, 
\textit{cf} lemma \ref{Lemme fonctionnelle hat GA pour partie lisse hat FF}. However, $\abs{\ga \be }\cdot L^{-1}$ is still very small 
\textit{cf} lemma \ref{Lemme fonctionnelle hat GA pour partie lisse hat FF}.}  that the constant of regularity
$C_{\msc{G}_{2A_L}}$ of $\wh{\msc{G}}_{\ga;2A_L}^{\,\pa{\be}}$ satisfies \eqref{ecriture condition grandeur constante de regularite GAkappa},
one gets that
\beq
\paa{\vsg_a}_1^{s} \mapsto  \mc{F}_{i_1,\dots,i_n}\pac{\ga \nu_s\pa{ * \mid \paa{\vsg_a}}} \qquad  \e{and} \qquad
\{ \eta_{1,p} \}_1^{r} \cup \{ \eta_{2,p} \}_1^{r}  \mapsto
\wh{\msc{G}}_{\ga;2A_L}^{\,\pa{\be}} \Big[ \varpi_r\big( * \mid \{ \eta_{a,p} \} \big) \Big]
\enq
are holomorphic in respect to $\paa{\vsg_a}_1^{s} \in \mc{N}_0^s$ , $\{ \eta_{1,p} \}_1^{r} \cup \{ \eta_{2,p} \}_1^{r} \in \mc{N}_0^{2r}$,
where $\mc{N}_0$ is an $r$ and $s$ independent neighborhood of $0\in \Cx$.
As the constant of regularity $C_{\msc{G}_{2A_L}}$ is large enough and $\abs{\ga}$
can be taken small enough, the size of the neighborhood $\mc{N}_0$ is large enough in order to ensure the convergence of the series of differential
operators issuing from the exponentials
$\ex{\wt{g}_{1,s}}$ and $\ex{\wt{g}_{2,r}}$, once upon the operator substitution is carried out. In virtue of corollary
\ref{Corollaire translation dans le cas integral et derivee},
and similarly to the summations \eqref{ecriture series des derivees substitues en dehors integral}-\eqref{definition fonctions Omega et Omega Prime},
the action of the translation operators can be computed directly under the integral sign in  \eqref{appendix multi Fred ecriture serie discete rho m} 
(the integration contours being Cartesian products of one dimensional compact curves)
and prior to taking the partial $\tau_p$ or $\ga$-derivatives in \eqref{appendix mult Fred somme reduire gamma der}.
There are also the differential operators arising from the substitutions $a_p \hookrightarrow \Dp{\vsg_p}$
in \eqref{appendix mult Fred tilde Lagrange series for Ga L} for those parameters $a_p$ that are written down explicitly.
The resulting $\Dp{\vsg_p}$-derivatives should appear outside of the integrals that are written down
in \eqref{appendix multi Fred ecriture serie discete rho m}.
However, the integrand of these compactly supported integrals is a continuous function of the integration variables
that is holomorphic in respect to $\{\vsg_p\}_1^s$, this  uniformly in respect to the integration variables. As a consequence, one can exchange the
derivation and integration symbols in this case as well.

Note that the constraints \eqref{appendix mult Fred contrainte entiers k} on the $k_{p,j}$'s
ensure that in \eqref{appendix mult Fred somme reduire gamma der} there is at most $m-n$ integers $k_{p,j}$ that differ from zero.
As a consequence, there will be at most $m$ translation operators in respect to the $\eta_{i,k}$ variables to take into account
once that the operator substitution is made.
More precisely, the substitution $b_{j,k} \hookrightarrow \Dp{\eta_{j,k}}$ shifts the parameters $\eta_{j,k}$ in 
$\varpi_r\,\big( \la, \{ \eta_{j,k}\} \big)$
\eqref{definition fonction varpi r} to the below value
\beq
\eta_{j,k} = \sul{p=1}{n+1} \f{1}{t_{j,k} - y_p} -  \sul{p=1}{n} \f{1}{t_{j,k} -\mu_{i_p}}
-\f{1}{t_{j,k} -\mu_{N+1}}
+ \sul{ \substack{ p=1 \\ \ell_p \not=0} }{N} \f{1}{t_{j,k} - \tau_p}-\f{1}{t_{j,k} - \mu_p} \; ,
\label{appendix mult Fred definition parametres eta}
\enq
where the ulimate sum in \eqref{appendix mult Fred definition parametres eta} only involves $m$ terms at most. 
Under the substitution $a_p \hookrightarrow \Dp{\vsg_a}$, the exponentials in \eqref{appendix mult Fred tilde Lagrange series for Ga L} produce a
translation of the function $\nu_s \hookrightarrow \wt{\nu}_s$, where
\beq
\wt{\nu}_s\pa{\la; \paa{\vsg_a} } = \nu_s\pa{\la; \paa{\vsg_a} } \; +
\; \sul{j=1}{s} \f{ t_{j+1}-t_j }{ 2i\pi\big( t_j-\la\big) } 
\bigg\{ \phi\big( t_j,\mu_{N+1} \big)-\phi\big(t_j,y_{n+1}\big)
\; +\;  \sul{a=1}{n} \phi\big(t_j,\mu_{i_a}\big)- \phi\big(t_j,y_a\big)  \bigg\} \;.
\enq

After carrying out all these manipulations, we are led to the representation
\beq
\rho_{N;\e{eff}}^{\pa{m}}\!\pa{x,t} =\lim_{\be \tend 0 } \lim_{s \tend +\infty} \lim_{r \tend +\infty}
 \sul{n=0}{m}  i \f{\pa{-1}^n}{n!} \sul{ \substack{i_1,\dots, i_n \\ i_a \in  \intn{1}{N} }  }{}
  \Oint{\msc{C}_q}{} \f{\dd^n z }{ \pa{2i\pi}^{2n} }  \Int{ \msc{C}^{\pa{L}} }{} \f{ \dd^{n+1}y }{ \pa{2i\pi}^{n+1} }
\mc{L}^{\pa{m}}_{\Ga^{\pa{L}}}\pac{\mc{F}_{i_1,\dots i_n}  \wh{\msc{G}}_{\ga;2A_L}^{\,\pa{\be}} } \;.
\label{appendix mult Fred rep rho apres subst}
\enq
where $\mc{L}^{\pa{m}}_{\Ga^{\pa{L}}}$ is  a truncated Lagrange series:
\bem
\mc{L}^{\pa{m}}_{\Ga^{\pa{L}}}\pac{\mc{F}_{i_1,\dots i_n} \msc{G}_{\ga;2A_L}^{\pa{\be}} } =
\sul{ \substack{ n_1,\dots, n_s \\ = 0 }  }{ m } \pl{p=1}{s}  \paa{ \f{ 1}{ n_p! } \f{\Dp{}^{n_p}}{ \Dp{} \vsg_p^{n_p} } }
 \sul{ \paa{\ell_a}  }{} \hspace{-1mm}^{\prime } \; \f{m ! }{ \ell_0 ! \ell_{N+1} ! } \,
\f{\Dp{}^{\ell_{N+1}} }{ \Dp{}\ga^{\ell_{N+1}}} \paa{ \pl{p=1}{s} \paa{ \Ga^{\pa{L}}\pac{\ga \wt{\nu}_s}\big( t_p \big)  }^{n_p}
\mc{F}_{i_1,\dots ,i_n}\pac{\ga \wt{\nu}_s} }_{ \mid \ga=0 } \\
\times \sul{ \paa{k_{a,j} } }{} \hspace{-1mm}^{\prime }   \hspace{1mm}\pl{\substack{a=1 \\ \not =i_1,\dots, i_n}}{N}
\pl{j=1}{\ell_a} \paf{ \la_a^{\pa{j}} }{ j! }^{k_{a,j}}
\cdot \f{ \Dp{}^{\ell_0} }{ \Dp{}\ga^{\ell_0} } \cdot 
\pl{\substack{a=1 \\ \not =i_1,\dots, i_n}}{N}  \f{ \Dp{}^{ \abs{\bs{k}_a} }    }{ \Dp{}\tau_a^{ \abs{\bs{k}_a} } }
\times  \paa{ \wh{\msc{G}}_{\ga;2A_L}^{\,\pa{\be}}\pac{ \varpi_r\pa{ * \mid \paa{\eta_{i,k}}} } }_{ \left| \substack{ \ga=0 \\ \tau_a=\mu_a } \right. } \;.
\label{appendix mult Fred forme serie Lagrance apres substitution}
\end{multline}

We now take the $r \tend +\infty$ limit of
\eqref{appendix mult Fred rep rho apres subst}-\eqref{appendix mult Fred forme serie Lagrance apres substitution}.
The very construction of $\varpi_r (\la\mid \{ \eta_{j,k} \} )$ along with the choice of parameters $\eta_{i,k}$
given by \eqref{appendix mult Fred definition parametres eta} associated with the fact that $\msc{G}_{ \ga;2A_L }^{ \pa{\be} }$ is a regular functional
with a sufficiently large regularity constant,
leads to (\textit{cf} proof of proposition \ref{Proposition translation pure avec dependence parametre auxiliaire})
\beq
\lim_{r \tend +\infty} \wh{\msc{G}}_{\ga;2A_L}^{\,\pa{\be}}\pac{\varpi_r} =
\wh{\msc{G}}_{\ga;2A_L}^{\,\pa{\be}}
\pac{H\pabb{ * }{\paa{y_a}_1^{n+1}\cup \paa{\tau_a}_{a : \ell_a\not=0} }{  \paa{\mu_{i_a}}_{1}^{n}\cup\paa{\mu_{N+1}}\cup
\paa{\mu_a}_{a : \ell_a\not=0} }} \; ,
\enq
this uniformly in $y_a$, $\mu_a$ $\la_a$ and $\tau_a$ belonging to $K_{2A_L}$. The function $H$ has been defined 
in \eqref{definition nouvelle fonctionnelle pure G smooth}

 This uniform convergence also holds
in respect to any finite order partial derivative in these parameters.
The uniformness of this limit in respect to the integration parameters occurring in \eqref{appendix mult Fred rep rho apres subst} allows one
to take it directly under the integral sign over a compact domain.

As a consequence, we get that
\bem
\lim_{r\tend +\infty} \sul{ \paa{k_{a,j} } }{} \hspace{-1mm}^{\prime }   \hspace{1mm}\pl{\substack{a=1 \\ \not =i_1,\dots, i_n}}{N}
 \pl{j=1}{\ell_a} \paf{ \la_a^{\pa{j}} }{ j! }^{k_{a,j}}
\f{ \Dp{}^{\ell_0} }{ \Dp{}\ga^{\ell_0} }
\pl{\substack{a=1 \\ \not =i_1,\dots, i_n}}{N}  \f{ \Dp{}^{ \abs{\bs{k}_a} }    }{ \Dp{}\tau_a^{ \abs{\bs{k}_a} } }
\times  \paa{ \wh{\msc{G}}_{\ga;2A_L}^{\,\pa{\be}}\pac{ \varpi_r\pa{ * \mid \{ \eta_{j,k}  \}} } }_{ \left| \substack{ \ga=0 \\ \tau_a=\mu_a} \right.  }  \\
= \pl{\substack{a=0 \\ \not= i_1,\dots,i_n  }}{N} \f{1}{ \ell_a ! } \f{\Dp{}^{\ell_a}}{ \Dp{}\ga_a^{\ell_a} }
\paa{ \wh{\msc{G}}_{\ga_0;2A_L}^{\,\pa{\be}}\pac{H\pabb{ * }{\paa{y_a}_1^{n+1}\cup \paa{\la_a\!\pa{\ga_a}}_{1}^{N} }
{ \paa{\mu_{a}}_1^{N+1} \cup \paa{\la_{i_a}\!\pa{\ga_{i_a}}}_1^{n}  } } }_{ \mid \ga_a=0}\;.
\label{appendix mult Fred limite r infini dans fnelle G hat}
\end{multline}
To get the $rhs$ of this equality we have, in addition to exchanging the limits and derivatives, applied the Faa-d\'{i}-Bruno formula backwards.
The constant of regularity of $\wh{\msc{G}}^{\, \pa{\be}}_{\ga_0;2A_L}$ being large enough, the action of  $\wh{\msc{G}}^{\, \pa{\be}}_{\ga_0;2A_L}$
on $H$ as written in the second line of \eqref{appendix mult Fred limite r infini dans fnelle G hat} is indeed well defined.

After collecting the various $\ga_a$  derivatives into a single one, we arrive to the representation
\beq
\lim_{r\tend+\infty} \mc{L}^{\pa{m}}_{\Ga^{\pa{L}}}\pac{\mc{F}_{i_1,\dots i_n} \wh{\msc{G}}_{\ga;2A_L}^{\,\pa{\be}} } =
\sul{ \substack{ n_1,\dots, n_s \\ = 0 }  }{ m } \pl{a=1}{s}  \paa{ \f{ 1 }{ n_a! } \f{ \Dp{}^{n_a} }{ \Dp{}\vsg_a^{n_a} } }
\f{\Dp{}^m}{\Dp{}\ga^m} \Bigg\{ \pl{a=1}{s} \paa{ \Ga^{\pa{L}}\pac{\ga \wt{\nu}_s}\pa{t_a} }^{n_a}
\cdot  \mc{J}_{i_1,\dots ,i_n}\pac{\ga \wt{\nu}_s} \Bigg\}_{ \left| \substack{ \ga=0 \\ \vsg_a=0} \right. }  \;,
\label{appendix mult Fred Lagrange series for Ga L}
\enq
where we have set
\beq
\mc{J}_{i_1,\dots ,i_n}\pac{\ga \wt{\nu}_s}=\mc{F}_{i_1,\dots ,i_n}\pac{\ga \wt{\nu}_s}
\wh{\msc{G}}_{\ga;2A_L}^{\,\pa{\be}}\pac{H\pabb{ * }{\paa{y_a}_1^{n+1}\cup \paa{\la_a}_{1}^{N} }
{ \paa{\mu_{a}}_1^{N+1} \cup \paa{\la_{i_a} }_1^{n}  } } \;.
\label{appendix mult Fred definition fnelle J i1 in}
\enq
Since, no confusion is possible  on the level of \eqref{appendix mult Fred Lagrange series for Ga L}-\eqref{appendix mult Fred definition fnelle J i1 in},
the $\ga$-dependence of the parameters $\la_p$, $p=1,\dots,N$ is kept implicit again. We also remind that these are functions of 
$\wt{\nu}_s$. 

The truncated $s$-dimensional Lagrange series \eqref{appendix mult Fred Lagrange series for Ga L} together with its $s\tend + \infty$ limit
has been studied in appendix \ref{appendix funct der nonlinear funct Ga}.
It follows from the latter analysis that the $s \tend +\infty$ limit is uniform in respect to the parameters ($\paa{y_k}_1^{n+1}$, $\paa{z_k}_1^n$)
on which $\mc{J}_{i_1,\dots,i_n}$ depends. Therefore, this limit can be taken under the integrals signs.
Similarly, one can exchange the limit with the $m^{\e{th}}$ $\ga$-derivative symbol.
It follows from the results gathered in appendix \ref{appendix funct der nonlinear funct Ga} that
\beq
\lim_{s \tend + \infty} \lim_{r\tend+\infty} \mc{L}^{\pa{m}}_{\Ga^{\pa{L}}}\pac{\mc{F}_{i_1,\dots i_n} \msc{G}_{\ga;2A_L}^{\pa{\be}} } =
\f{ \Dp{}^m }{  \Dp{}\ga^m} \Bigg\{  \mc{J}_{i_1,\dots ,i_n}\pac{\ga \nu^{\pa{L}} }
\cdot \det^{-1}_{\msc{C}_q }\pac{I- \ga \f{\de \Ga^{\pa{L}}\!\pac{\rho}}{\de \rho\pa{\zeta}}\pa{\mu} }_{\rho=\ga\nu^{\pa{L}}}   \Bigg\}_{\mid \ga=0}\;.
\label{appendix mult Fred resultat sommation serie lagrange}
\enq
The answer is expressed with the help of $\nu^{\pa{L}}$, the unique solution
(for $\ga$-small enough) to the non-linear integral equation driven by the functional $\Ga^{\pa{L}}$:
\beq
\nu^{\pa{L}}\pa{\la} = \pa{i\be -\tf{1}{2}}Z\pa{\la} \; -\;  \phi\pa{\la, q} \; + \; \sul{a=1}{N+1} \phi\pa{\la,\mu_a} \;
- \; \sul{a=1}{n+1} \phi\pa{\la,y_a} \;  - \; \sul{ \substack{a=1 \\ \not= i_1,\dots, i_n}}{N} \phi\pa{\la,\la_a} \quad
\e{with} \quad  \left\{ \ba{c}   \la_a =  \xi_{ \ga \nu^{\pa{L}} }^{-1} \! \pa{\tf{a}{L}}  \vspace{2mm}\\
						           \mu_a=\xi^{-1}\!\pa{\tf{a}{L}} \ea  \right. \hspace{-3mm}.
\label{appendix mult Fred NLIE for nuL}
\enq
Also, in \eqref{appendix mult Fred resultat sommation serie lagrange}, appears the Fredholm determinant of the
linear integral operator acting  on a small loop $\msc{C}_q$ around $\intff{-q}{q}$ whose kernel
is given in terms of the functional derivative $\tf{\de \Ga^{\pa{L}}\pac{\rho}\pa{\mu}}{\de \rho\pa{\zeta}}$.
The definition of the functional derivative is given in \eqref{appendix funct der definition funct der}.

Lemma \ref{Proposition evaluation fnelle G hat Reduction vars determinant petit} allows one to reexpress
the functional $\wh{\msc{G}}_{\ga;2A_L}^{\,\pa{\be}}$ appearing \eqref{appendix mult Fred definition fnelle J i1 in}
in the case where the parameters $\la_a$ and $\mu_a$ are defined exactly as in \eqref{appendix mult Fred NLIE for nuL}
in terms of the unique solution $\nu^{\pa{L}}$.

This leads to the below representation:
\bem
\hspace{-1cm}\rho_{N;\e{eff}}^{\pa{m}}\! \pa{x,t}  =  \lim_{\be \tend 0}
\sul{n=0}{m} c \f{\pa{-1}^n}{n!} \f{ \Dp{}^m }{ \Dp{}\ga^m } \hspace{-3mm}\sul{ \substack{i_1,\dots, i_n \\ i_a \in  \intn{1}{N} }  }{}
  \Oint{\msc{C}_q}{} \f{\dd^n z }{ \pa{2i\pi}^{2n} }   \Int{ \msc{C}^{\pa{L}} }{} \hspace{-2mm} \f{\dd^{n+1} y }{ \pa{2i\pi}^{n+1} }
\pl{k=1}{n} \paa{ \f{ 4 \sin^2\!\pac{\pi \ga \nu^{\pa{L}}\! \pa{\la_{i_k}} }   }{ L \xi^{\prime}_{\ga \nu^{\pa{L}}}\!\pa{\la_{i_k}} \pa{z_k-\la_{i_k}}}  }
\ddet{n}{ \f{1}{ z_a-\la_{i_b}} } \\
\times    \pl{k=1}{n} \paf{ y_{n+1}-z_k }{ \pa{y_k-z_k}\pa{y_{n+1}  - \la_{i_k}} }
\f{ \ex{i x \mc{U}^{\pa{L}}\!\pa{\paa{\la_{a}}_1^N;\paa{\mu_a}_1^{N+1};\paa{y_a}_1^{n+1} \mid \ga} }  }
{\det_{\msc{C}_q }\pac{I- \ga \f{\de \Ga^{\pa{L}}\pac{\rho}}{\de \rho\pa{\zeta}}\pa{\mu} }_{\rho=\ga\nu^{\pa{L}}} }
\f{ \pl{a=1}{n} \pl{b=1}{n+1} \pa{y_b-\la_{i_a}-ic}\pa{\la_{i_a}-y_b-ic} }{ \pl{a,b=1}{n+1} \pa{y_a-y_b-ic} \pl{a,b=1}{n}\pa{\la_{i_a}-\la_{i_b}-ic} } \\
\hspace{-1cm}\times \f{ \prod_{k=1}^{n+1}  f^{\pa{L}}\!\pa{y_k, \ga \nu^{\pa{L}}\!\pa{y_k}}  }
{  \det_{N+1}\big[ \Xi^{\pa{\mu}}\pac{\xi} \big]  \det_{N} \big[ \Xi^{\pa{\la}} [\xi_{\ga\nu^{\pa{L}}}] \big] }
\cdot  \pa{ \det_{n}\Big[ \de_{k\ell} + \ga \wh{V}_{\! k\ell} \big[\ga\nu^{\pa{L}} \big]  \Big]
\det_{n}\Big[ \de_{k\ell} + \ga \wh{\ov{V}}_{\! k\ell} \big[\ga\nu^{\pa{L}} \big]  \Big]  } \pa{ \{\la_{i_a}\}_1^n, \{y_a\}_1^{n+1}}
\label{appendix mult Fred Somme discrete rho m avant limite thermo}
\end{multline}
Above we have written down the dependence of both determinants on $\{\la_{i_a}\}$ and $\{y_a\}$ as a common argument.

There is no problem to carry out the analytic continuation in \eqref{appendix mult Fred Somme discrete rho m avant limite thermo}
from $\be \in \bs{\wt{U}}_{\be_0}$ up to $\be=0$: the potential singularities that could appear in the determinants
are canceled by the prefactor $\prod_{k=1}^{n}\sin^2\pac{\pi \ga \nu^{\pa{L}}\! \pa{\la_{i_k}}}$.
From now on, we can thus set $\be=0$

In order to prove the theorem, it remains to take the thermodynamic limit of \eqref{appendix mult Fred Somme discrete rho m avant limite thermo}
at $\be=0$.

\subsubsection*{$L\tend +\infty$ behavior of $\nu^{\pa{L}}$}

It was shown in appendix \ref{appendix funct der nonlinear funct Ga}, equation \eqref{appendix funct der integral req Asympt rho},
that $\nu^{\pa{L}}$ admits a large $L$ asymptotic expansion
$\nu^{\pa{L}}\!\pa{\la} =\nu\big( \la\mid \paa{\la_{i_a}}_1^n;\paa{y_a}_1^{n+1} \big) + \e{O} \big( L^{-1} \big)$.
There the $\e{O}$ is holomorphic and uniform in some open neighborhood of the real axis and the function
$\nu\pa{\la}=\nu\big( \la\mid \paa{\la_{i_a}}_1^n,\paa{y_a}_1^{n+1} \big)$ stands for the unique solution to
the linear integral equation \eqref{appendix Mul Dim Fred ecriture Int eqn nu} (here we have already set $\be=0$).

As all of the functions we deal with are smooth functions of $\nu^{\pa{L}}$, we are thus able to replace everywhere
$\nu^{\pa{L}}$ by $\nu$, up to $\e{O}\pa{L^{-1}}$ corrections.

Building on the large $L$ asymptotics of $\Ga^{\pa{L}}$ and $\nu^{\pa{L}}$ it is shown in
subsection \ref{appendix funct der nonlinear funct Ga}, \eqref{appendix funct der convergence derivee fonctionnelle}-\eqref{appendix funct der convergence determinant derivee fonctionnelle}, that
\beq
\det_{\msc{C}_q }\pac{I- \ga \f{\de \Ga^{\pa{L}}\pac{\rho}}{\de \rho\pa{\zeta}}\pa{\mu} }_{\rho=\ga\nu^{\pa{L}}}
= \ddet{ \intff{-q}{q} }{ I+\ga \f{R}{2\pi}} \pa{1+ \e{O}\pa{L^{-1}}} \;,
\enq
with a $\e{O}$ that has the same uniformness properties as stated before.
Above, we did not insist that the Fredholm determinant $\ddet{}{I+\ga\tf{R}{2\pi}}$ corresponds to an action on $\intff{-q}{q}$.

\subsubsection*{$L\tend +\infty$ limit of $\mc{U}^{\pa{L}}$}

The thermodynamic limit of $\mc{U}^{\pa{L}}\!\pa{\paa{\la_{a}};\paa{\mu_a};\paa{y_a} \mid \ga}$ is readily computed by using that
\beq
\xi\pa{\mu_a}-\xi_{\ga \nu^{\pa{L}} }\!\pa{\la_a} = 0 = -\f{\ga \nu^{\pa{L}}\!\pa{\mu_a}}{L}  \; + \;  \f{p^{\prime}\!\pa{\mu_a}}{2\pi}\pa{\mu_a-\la_a}
\; + \;  \e{O}\pa{L^{-2}}  \;.
\label{appendix mult Fred espacement entre mu et la's}
\enq
The remainder $\e{O}\pa{L^{-2}}$ is uniform in $a\in \intn{1}{N}$ and holomorphic in respect to the variables $y_a$ and $z_a$
belonging to $U_{\tf{\de}{2}}$.
By using the Euler-MacLaurin formula, the linear
integral equation \eqref{appendix Mul Dim Fred ecriture Int eqn nu} satisfied by $\nu$ and the integral representation
\eqref{ecriture limite thermo exposant osc} for $u$ one gets that
\beq
\mc{U}^{\pa{L}}\!\pa{\paa{\la_{a}}_1^{N};\paa{\mu_a}_{1}^{N+1};\paa{y_a}_1^{n+1} \mid \ga} =
\mc{U}\!\pa{\paa{\mu_{i_a}}_1^n;\paa{y_a}_1^{n+1} \mid \ga}  +\e{O}\pa{L^{-1}}
 \;.
\enq
with a $\e{O}$ that, again, is uniform and holomorphic in respect to $\mu_{i_a}$ or  $y_a$
belonging to $U_{\tf{\de}{2}}$. It is also holomorphic in $\Re\pa{\be} \geq 0$.

By using the densification of the parameters $\la_a$ and $\mu_a$ on $\intff{-q}{q}$, it is likewise easy to check that
\beq
\det_{N+1}\big[ \Xi^{\pa{\mu}}\pac{\xi} \big]  \det_{N} \big[ \Xi^{\pa{\la}} [\xi_{\ga\nu^{\pa{L}}}] \big] =
 \det^{2}\pac{ I-\tf{K}{2\pi} }  \cdot \pa{1+\e{O}\pa{L^{-1}}} \;.
\enq

\subsubsection*{$L\tend +\infty$ limit of the remaining terms}

It is also readily seen due to the densification of the parameters $\la_a$ on $\intff{-q}{q}$ that the sums over 
the discreet sets $\la_{i_a}$ can be replaced by integrals over $\intff{-q}{q}$
up to $\e{O}\big( L^{-1} \big)$ corrections.
Finally, it remains to estimate the contributions of the functions $f^{\pa{L}}$. If one focuses on the contributions of the
integrals over $y_a$, $a=1,\dots,n+1$ along the curves $\msc{C}_{\ua / \da;L}$ and $\msc{C}_{bd;L}$, then one readily convinces oneself that
one deals with the type of integrals studied in the proof of proposition \ref{Proposition calcul des sommes discrete singulieres}. Namely, these are
precisely the integrals appearing when deriving the estimates for the functionals $I_{k}^{\pa{L}}$ given in \eqref{definition facteur IL}.
Clearly, each of these integrals can be estimated successively. By repeating word for word the proof
given in proposition \ref{Proposition calcul des sommes discrete singulieres}, one has that each of these integrals produces a
$\e{O}\pa{ \tf{L}{w_L}}=\e{o}\pa{1}$ contribution. Hence, this part of the contour $\msc{C}^{\pa{L}}$ does not contribute to the thermodynamic limit.

As a consequence one obtains the following representation for $\rho^{\pa{m}}_{N;\e{eff}}\!\pa{x,t}$:
\bem
\hspace{-1cm}\rho_{N;\e{eff}}^{\pa{m}}\! \pa{x,t}  =
\sul{n=0}{m} \f{c\pa{-1}^n}{n!} \f{ \Dp{}^m }{ \Dp{}\ga^m } \hspace{-2mm}\Int{-q}{q} \! \dd^{n}\la
  \Oint{\msc{C}_q}{} \f{\dd^n z }{ \pa{2i\pi}^{2n} }   \Int{ \msc{C} }{} \hspace{-2mm} \f{\dd^{n+1} y }{ \pa{2i\pi}^{n+1} }
 \f{ \ex{i x \mc{U}\pa{\paa{\la_{a}}_1^{n};\paa{y_a}_1^{n+1} \mid \ga} }  \prod_{k=1}^{n+1}  f\!\pa{y_k, \ga \nu\!\pa{y_k}}   }
 { \pl{k=1}{n} \pa{z_k-\la_{k}} \pa{y_k-z_k}\pa{y_{n+1}  - \la_{k}}}
 \ddet{n}{ \f{  y_{n+1}-z_b }{z_a-\la_b } }    \\
\times  \f{ \pl{a=1}{n} \pl{b=1}{n+1} \pa{y_b-\la_a-ic}\pa{\la_a-y_b-ic} }{ \pl{a,b=1}{n+1} \pa{y_a-y_b-ic} \pl{a,b=1}{n}\pa{\la_a-\la_b-ic} }
\pl{k=1}{n}\paa{ 4 \sin^2 \big[\pi \ga \nu\! \pa{\la_{k}} \big] }
\f{ \ddet{n}{ \de_{k\ell} + \ga \wh{V}_{k\ell}\big[\ga \nu\big] }   \ddet{n}{ \de_{k\ell} + \ga \wh{\ov{V}}_{k\ell}\big[\ga \nu \big] }  }
{  \ddet{}{I+\tf{\ga R}{2\pi}}   \det^{2}\pac{I-\tf{K}{2\pi}}} \\
\times    \pa{ 1+\e{o}\pa{1} }_{\mid \ga=0} \;.
\label{appendix mult Fred Somme discrete rho m avant limite thermo}
\end{multline}

In order to obtain the representation \eqref{appendix multi dim Fred representation rho eff series Multdim Fred}
for the thermodynamic limit it remains to drop the $\e{o}\pa{1}$ corrections. \qed

We now provide an alternative representation for the thermodynamic limit $\rho_{\e{ext}}^{\pa{m}}\!\pa{x,t}$.

\begin{prop}
\label{Proposition permutation limit thermo rho eff}

The function $\rho_{\e{ext}}^{\pa{m}}\!\pa{x,t}$ admits the representation
\beq
\hspace{-8mm} \rho_{\e{eff}}^{\pa{m}}\!\pa{x,t}= \lim_{w \tend +\infty}\; \lim_{\be \tend 0} \; \lim_{s \tend +\infty} \; \lim_{ r \tend +\infty } \bs{:}
\f{ \Dp{}^m }{ \Dp{}\ga^m }  \Bigg\{  \wh{E}_-^{\,2}\pa{q}
\ex{- \Int{-q}{q} \pac{ix u^{\prime}\!\pa{\la} + \wh{g}^{\, \prime}\!\pa{\la} }\ga \nu_s\pa{\la} \dd \la }
X_{ \msc{C}_E^{\pa{w}} } \pac{\ga \nu_s, \wh{E}_-^{-2}}   \msc{G}_{\ga;2w}^{\pa{\be}}\pac{\varpi_r}    \Bigg\}_{\mid \ga=0} 
\hspace{-2mm} \bs{:} \!\!.
\label{appendix mult Dim Fred ecriture rep. Alt thermo limit rho}
\enq

The contour  $\msc{C}_E^{\pa{w}} = \msc{C}_E^{\pa{\infty}} \cap \paa{z \in \Cx \; : \;  \abs{\Re\pa{z}} \leq w  }$ corresponds to a compact approximation
of $\msc{C}_E^{\pa{\infty}}$ as depicted in Fig.~\ref{contour CE et sa restriction CEw}. It is such that $\underset{w \tend +\infty}{\e{lim}}
\msc{C}_E^{\pa{w}} = \msc{C}_E^{\pa{\infty}}$. The functional $X_{ \msc{C}_E^{\pa{w}} }$ has been defined in \eqref{appendix ecriture limite thermo XN}.

The functional $\msc{G}_{\ga;2w}^{\pa{\be}}\pac{\varpi_r}$ appearing in \eqref{appendix mult Dim Fred ecriture rep. Alt thermo limit rho}
acts on the loop $\msc{C}\pa{K_{2w}}$ and has been defined in lemma \ref{Lemme fonctionnelle GA pour partie lisse FF}.

\end{prop}

The compact approximation $\msc{C}_E^{\pa{w}}$ of the contour $\msc{C}^{\pa{\infty}}_E$
appearing in \eqref{appendix mult Dim Fred ecriture rep. Alt thermo limit rho}
is there to ensure the well-definiteness of the translation operators. Indeed, in the setting discussed in subsection
\ref{Appendice transl fonct Conte fnelle} and appendix \ref{Appendice section functional translation},
the translation operators are, \textit{a priori}, only defined for functionals
that involve the values of their argument on some compact subset of $\Cx$.
As a consequence, \textit{a priori}, the $w\tend +\infty$ limit and $r \tend +\infty$ limit do not commute.

Also, the $\be \tend 0$ limit and the $w \tend +\infty$ limits do not commute. These limit should be understood as follows.
Given $w$ fixed and large enough, one considers the regular functional $\msc{G}_{\ga;2w}^{\pa{\be}}$ as introduced in lemma
\ref{Lemme fonctionnelle GA pour partie lisse FF}. The value of $w$ defines an associated $\be_0 \in \Cx$ and $\wt{\ga}_0>0$
such that $\msc{G}_{\ga;2w}^{\pa{\be}}$ is a regular functional for $\be \in \bs{\wt{U}}_{\be_0}$ and $\abs{\ga}\leq \wt{\ga}_0$
with a regularity constant large enough (in particular satisfying \eqref{ecriture condition grandeur constante de regularite GAkappa}).
These $\Re\pa{\be_0}$ and $\wt{\ga}_0$ are such that $\Re\pa{\be_0} \tend +\infty$  and $\wt{\ga}_0 \tend 0$ when $w\tend +\infty$.

\Proof

Let $\wt{E}_-^{-2}$ be as given in \eqref{definition fction E-tilde}-\eqref{definition fonctions tilde g1 et g2},
$\nu_s$ as in \eqref{definition fonction nus} and $\varpi_r$ \eqref{definition fonction varpi r}.
In order to implement the operator substitution, we first expand the functional $X_{\msc{C}^{\pa{w}}_{E}}\!\big[\nu,\wt{E}_-^{\, 2} \big]$
appearing in the \textit{rhs} of \eqref{appendix mult Dim Fred ecriture rep. Alt thermo limit rho} into a series very similar to the one occurring in
the proof of proposition \ref{Proposition algebraic representation Fredholm minor}.
The sole exception is that, this time, the sums over $\la_{i_a}$'s are directly replaced by integrals over $\intff{-q}{q}$ of the
corresponding variables. Also, the function $f^{\pa{L}}$ (resp. its associated contour $\msc{C}^{\pa{L}}$)
should be replaced by $f$ (resp. $\msc{C}^{\pa{w}}=\msc{C}_E^{\pa{w}} \cup \wt{\msc{C}}_q$).
At the end of the day, one deals with the multi-dimensional Lagrange series below
\beq
\lim_{w \tend +\infty} \lim_{s \tend +\infty} \lim_{r\tend +\infty}  \sul{n=0}{m} i\f{\pa{-1}^n}{n!}
\Int{-q}{q} \f{\dd^n \la}{\pa{2i\pi}^n} \;  \Oint{\msc{C}_q}{} \f{ \dd^n z}{\pa{2i\pi}^n}  \Int{\msc{C}^{\pa{w}}}{} \hspace{-1mm}
\f{ \dd^{n+1} y }{ \pa{2i\pi}^{n+1} }  \wt{\mc{L}}^{\pa{m}}_{\Ga}\!\pac{ \mc{F} \msc{G}_{\ga;2w}^{\pa{\be} } }  \;.
\label{appendix multi dim Fred serie Lagrange qui donne rhe eff}
\enq
The functional $\mc{F}$ appearing above reads
\beq
\hspace{-5mm} \mc{F}\pac{\ga \nu_s}=    \f{  \prod_{k=1}^{n+1} f\!\pa{y_k, \ga \nu_s\!\pa{y_k} }  } { \pl{k=1}{n} \pa{z_k-\la_k} \pa{y_k-z_k} }
\pl{k=1}{n}  \paf{ y_{n+1}-z_k }{ y_{n+1}-\la_k } \cdot \ddet{n}{ \f{ 4 \sin^2\!\pac{ \pi \ga \nu_s\!\pa{\la_k} }  }{z_a-\la_k} }
\ex{- ix \int_{-q}^{\,q}  u^{\prime}\!\pa{\la} \ga \nu_s\pa{\la} \dd \la  }
\ex{-ix u\pa{q}}
\f{  \pl{a=1}{n} \ex{-ix u\pa{\la_a}}  }{ \pl{a=1}{n+1} \ex{-ix u\pa{y_a}}  } \;.
\nonumber
\enq
And we have set
\bem
\wt{\mc{L}}^{\pa{m}}_{\Ga}\pac{\mc{F}\msc{G}_{\ga;2w}^{\pa{\be}}} =
\sul{ \substack{ n_1,\dots, n_s \\ = 0 }  }{ m } \pl{p=1}{s}  \paa{ \f{ \big[a_p \big]^{n_p} }{ n_p! } }
\f{\Dp{}^m}{\Dp{}\ga^m} \Bigg\{  \f{  \prod_{k=1}^{n+1} \ex{\wt{g}_{1,s}\pa{y_k}}  }
{ \ex{\wt{g}_{1,s}\pa{ q }  } \prod_{k=1}^{n} \ex{\wt{g}_{1,s}\pa{\la_{k} }  } }
\pl{p=1}{s} \paa{ \Ga\pac{\ga \nu_s}\big( t_p \big)  }^{n_p} \\
\times
\f{  \prod_{k=1}^{n+1} \ex{\wt{g}_{2,r}\pa{y_k}}  }
{ \ex{\wt{g}_{2,r}\pa{ q }  } \prod_{k=1}^{n} \ex{\wt{g}_{2,r}\pa{\la_{k} }  } }
\ex{-\Int{-q}{q} \wt{g}^{\prime}_{2,r}\pa{\la} \ga \nu_s\pa{\la} \dd \la }
\cdot  \mc{F}\pac{\ga \nu_s}_{ \left| \substack{ \ga=0 \\ \vsg_a=0} \right. } \msc{G}_{\ga;2w}^{\pa{\be}}\pac{\varpi_r}
\Bigg\}_{\mid \ga=0}\;.
\label{appendix mult Fred tilde Lagrange series for Ga  fnell lineaire}
\end{multline}

The functional $\Ga$ is evaluated at the discretization points
$t_p$, $p=1,\dots,s$ for the contour $\msc{C}_{out}$
appearing in the $lhs$ of Fig.~\ref{contour exemple de courbes encerclantes}.

One can implement the operator substitution on the level of \eqref{appendix mult Fred tilde Lagrange series for Ga  fnell lineaire} as
it was done in the proof of theorem \ref{Theorem representation series Fredholm multidim}. The well-foundedness of these
manipulations (in particular the justification of the exchange of various limits, partial derivatives and integrals over compact
contours) is justified along very similar lines. Once upon taking the $r\tend +\infty$ limit
we end-up with the below multidimensional Lagrange series
\beq
\mc{L}^{\pa{m}}_{\Ga}\pac{\mc{F}\msc{G}_{\ga;2w}^{\pa{\be}}} =
\sul{ \substack{ n_1,\dots, n_s \\ = 0 }  }{ m } \pl{p=1}{s}  \paa{ \f{ 1 }{ n_p! }  \f{ \Dp{}^{n_p} }{ \Dp{}\vsg_p^{n_p} } }
\f{\Dp{}^m}{\Dp{}\ga^m} \Bigg\{  \pl{p=1}{s} \paa{ \Ga\pac{\ga \wt{\nu}_s}\big( t_p \big)  }^{n_p}
\cdot  \mc{F}\pac{\ga \wt{\nu}_s}_{ \left| \substack{ \ga=0 \\ \vsg_a=0} \right. } \msc{G}_{\ga;2w}^{\pa{\be}}\pac{\varpi}
\Bigg\}_{\mid \ga=0}\;.
\label{appendix mult Fred Lagrange series for Ga  fnell lineaire}
\enq
There, we agree upon
\beq
\wt{\nu}_s\pa{\la; \paa{\vsg_a} } = \nu_s\pa{\la; \paa{\vsg_a} } \; +
\; \sul{b=1}{s} \f{ t_{b+1}-t_b }{ 2i\pi\pa{t_b-\la} }
\Bigg\{ \phi\pa{t_b,q} + \sul{a=1}{n} \phi\pa{t_b,\la_a} \; -\;  \sul{a=1}{n+1} \phi\pa{t_b,y_a}  \Bigg\} \;.
\enq
Also, the function $\varpi$ is to be considered as a functional of $\wt{\nu}_s$
\beq
\varpi\pac{\wt{\nu}_s}\pa{\la} = \sul{k=1}{n+1} \f{1}{\la-y_k} \; - \;  \f{1}{\la-q}  \; - \; \sul{k=1}{n} \f{1}{\la-\la_k}
-\Int{-q}{q} \f{\ga \wt{\nu}_s\pa{\tau} }  {\pa{\tau-\la}^2 }   \dd \tau \;.
\enq

The multidimensional Lagrange series \eqref{appendix mult Fred Lagrange series for Ga  fnell lineaire}
has been studied in appendix \ref{appendix func der affine functional}.
Its $s \tend +\infty$ limit is uniform in respect to the auxiliary parameters $\paa{\la_a}_1^n$, $\paa{z_a}_1^n$
and $\paa{y_a}_1^{n+1}$. Hence, just as in the proof of theorem \ref{Theorem Multidimensional Lagrange series},
one is allowed to exchange the $s \tend + \infty$ limit with the integration over the compact contours.
One can then apply the results of appendix \ref{appendix func der affine functional}
leading to
\beq
 \underset{s \tend +\infty }{ \e{lim} }  \mc{L}^{\pa{m}}_{\Ga}\!\pac{\mc{F} \msc{G}_{\ga;2w}^{\pa{\be}} }   =
\f{ \Dp{}^m }{  \Dp{} \ga^m }  \Bigg\{ \f{ \mc{F}\pac{\ga \nu} \msc{G}_{\ga;2w}^{\pa{\be}}\pac{\varpi\pac{\nu}} }{  \ddet{\intff{-q}{q}}{ I+\tf{\ga R}{2\pi}  }  } \Bigg\}_{\mid \ga=0}  \hspace{-2mm}.
\enq
The function $\nu$ appearing above is the unique solution  to the linear integral equation 
\beq
\nu\pa{\la} \; + \;  \ga \Int{-q}{q} \! \f{\dd \mu}{2\pi} R\pa{\la,\mu} \nu\pa{\mu}
=  \pa{i\be  -\tf{1}{2} } Z\pa{\la} \; + \; \sul{a=1}{n} \phi\pa{\la,\la_a}  \; - \;  \sul{a=1}{n+1} \phi\pa{\la,y_a} \;.
\enq
One can build on this result so as to simplify the obtained expression.
The expression for the functional function $\msc{G}_{\ga;2w}^{\pa{\be}}\big[\varpi[\nu ]\big]$ is
simplified with the help of lemma \ref{Proposition evaluation fonctionnelle G super caligraphique}.

By using the linear integral equation satisfied by $\nu$
together with the representation of $u$ in terms of $\phi$ and $u_0$
\eqref{ecritutre representation integrale u}, we get that the oscillating factor present in $\mc{F}\pac{\ga \nu}$
 coincides with the one appearing  in theorem \ref{Theorem representation series Fredholm multidim}:
\beq
\sul{a=1}{n+1} u\pa{y_{a}} \; - \; \sul{a=1}{n} u\pa{\la_a} \; - \;  u\pa{q} \; - \; \ga \Int{-q}{q} u^{\prime}\!\pa{\la} \nu\pa{\la} \dd \la
= \mc{U}\pa{\paa{\la_a}_1^n, \paa{y_a}_1^{n+1} \mid \ga} -2i\be p_F\;.
\enq
We are thus led to the below representation for the \textit{rhs} of \eqref{appendix mult Dim Fred ecriture rep. Alt thermo limit rho}
\bem
\hspace{-7mm}  \lim_{w \tend +\infty}\; \lim_{\be \tend 0} \ex{2x \be p_F} \sul{n=0}{m} \f{c\pa{-1}^n}{n!} \f{ \Dp{}^m }{ \Dp{}\ga^m } \hspace{-2mm}\Int{-q}{q} \! \dd^{n}\la
  \Oint{\msc{C}_q}{} \f{\dd^n z }{ \pa{2i\pi}^{2n} }   \Int{ \msc{C}^{\pa{w}} }{} \hspace{-2mm} \f{\dd^{n+1} y }{ \pa{2i\pi}^{n+1} }
 \f{ \ex{i x \mc{U}\pa{\paa{\la_{a}}_1^{n};\paa{y_a}_1^{n+1} \mid \ga} }  \prod_{k=1}^{n+1}  f\!\pa{y_k, \ga \nu\!\pa{y_k}}   }
 { \pl{k=1}{n} \pa{z_k-\la_{k}} \pa{y_k-z_k}\pa{y_{n+1}  - \la_{k}}}
 \ddet{n}{ \f{  y_{n+1}-z_b }{z_a-\la_b } }    \\
\hspace{-5mm}\times  \f{ \pl{a=1}{n} \pl{b=1}{n+1} \pa{y_b-\la_a-ic}\pa{\la_a-y_b-ic} }{ \pl{a,b=1}{n+1} \pa{y_a-y_b-ic} \pl{a,b=1}{n}\pa{\la_a-\la_b-ic} }
\pl{k=1}{n}\pac{ 4 \sin^2\!\pac{\pi \ga \nu\! \pa{\la_{k}} } }
\f{ \ddet{n}{ \de_{k\ell} + \ga \wh{V}_{k\ell}\pac{ \ga \nu} }   \ddet{n}{ \de_{k\ell} + \ga \wh{\ov{V}}_{k\ell}\pac{\ga \nu} }  }
{  \ddet{}{I+\tf{\ga R}{2\pi}}   \det^{2}\pac{I-\tf{K}{2\pi}}} \;.
\end{multline}
Tha auxiliary arguments of the entries $\wh{V}_{k\ell}\pac{\nu} $ and $\wh{\ov{V}}_{k\ell}\pac{\nu}$ are undercurrent by those of $\nu$. 

One can carry out the analytic continuation from $\be \in \bs{\wt{U}}_{\be_0}$ up to $\be=0$
as the potential singularities of the two determinants are canceled by the pre-factors
$\prod_{k=1}^{n} \sin^2\!\pac{\pi \ga \nu\! \pa{\la_{k}} }$.

There is no problem to take the $w\tend +\infty$ limit of the above integrals. Indeed $\msc{C}_E^{\pa{\infty}}$ is chosen in such a way that
$\ex{ixu\pa{y_a}}$, $a=1,\dots,n+1$ is decaying exponentially fast in $y_a$ when $y_a \tend \infty$ along $\msc{C}_E^{\pa{\infty}}$. As the rest of the integrand is
a $\e{O}\pa{y_a^n}$, $a=1,\dots,n+1$ at infinity, the integrals along $\msc{C}_E^{\pa{\infty}}$ are convergent.
Once upon taking the $\be \tend 0$ and the $w\tend +\infty$ limits, we recover
the representation given in \eqref{appendix multi dim Fred representation rho eff series Multdim Fred}.
\qed





\section{Functional Translation operator}
\label{Appendice section functional translation}

In this appendix, we build a convenient for our purposes representation of a functional translation. Our representation applies
to sufficiently regular classes of  functionals  acting on holomorphic functions. Our construction utilizes multidimensional Lagrange
series (see eg.~\cite{AizenbergYuzhakovIntRepAndMultidimensionalResidues}).

\subsection{Lagrange series}

\begin{theorem} \cite{AizenbergYuzhakovIntRepAndMultidimensionalResidues}
\label{Theorem Multidimensional Lagrange series}

Let $\mc{D}_{0,r}=\paa{z \in \Cx \; : \; \abs{z} < r}$. Assume that

\begin{itemize}
\item $\vp_j\!\big( \{  \vsg_a \}_1^s\big)$, $j=1,\dots, s$ and $f\!\big( \{  \vsg_a \}_1^s\big)$ are holomorphic functions of $\{\vsg_a\}_1^s$ belonging to the Cartesian
product $\mc{D}_{0,r}^{s}$;
\item there exists a series of radii $r_j<r $ such that for $\abs{\vsg_j}= r_j$, $j=1,\dots, s$, one has
$\abs{ \vp_j\pa{\paa{\vsg_a}} } < r_j$.
\end{itemize}
Then, the multidimensional Lagrange series
\beq
\mc{L}_s =   \sul{ \substack{n_1,\dots , n_s\\ \in \mathbb{N}} }{}
 \pl{r=1}{s} \paa{ \f{1}{n_r!} \f{\Dp{}^{n_r}}{ \Dp{}\vsg_r^{n_r} } } \cdot
\; \pl{r=1}{s}\left.  \vp_r^{n_r}\!\pa{ \paa{\vsg_a} }  \cdot
 f\pa{\paa{\vsg_a} }    \;    \right|_{\vsg_{p}=0}   \;
\nonumber
\enq
is convergent and its sum is given by
\beq
\mc{L}_s=   \f{  f\pa{\paa{z_a} }   }
{ \ddet{s}{  \de_{jk} -  \f{ \Dp{} }{ \Dp{}\vsg_k } \vp_j\pa{ \paa{\vsg_a} }   } _{ \mid \paa{\vsg_a}=\paa{z_a} }   } \;.
\label{appendix funct der resultat series lagrange discrete}
\enq
Above, $\pa{z_1,\dots,z_s}$ stands for the unique solution to the system $z_j =\vp_j\pa{\paa{z_a}}$  such that $\abs{z_j}<r$ for all $j$.
The uniqueness and existence of this solution is part of the conclusion of this theorem.

\end{theorem}

\subsection{Some preliminary definitions}

Throughout this appendix, $M$ and $K$ will always stand for two compacts of $\Cx$ such that $K \subset \e{Int}\pa{M}$,
$M$ has $n$ holes (\textit{ie} $\Cx\subset M$ has $n$ bounded connected components) and
$\Dp{}M$ can be realized as  disjoint union of $n+1$ smooth Jordan curves\symbolfootnote[2]{we remind that $\ga_a$
satisfies  $\ga_a\pa{0}=\ga_a\pa{1}$ and ${\ga_a}_{\mid \intfo{0}{1}}$ is injective. }
$\ga_a : \intff{0}{1} \tend \Dp{}M =\amalg_{a=1}^{n+1} \ga_a\pa{\intff{0}{1}}$.

Let $h$ be a holomorphic function on $M$ and set
\beq
f_s\big( \la \mid \{ \vsg_{a,p} \} \big) = 
\sul{p=1}{s} \sul{a=1}{n+1} \f{  (t_{a,p+1}-t_{a,p} ) }{2i\pi  ( t_{a,p}-\la ) } \, \vsg_{a,p}
 \; + \; \sul{p=1}{s} \sul{a=1}{n+1} \f{  (t_{a,p+1}-t_{a,p}) }{2i\pi (t_{a,p}-\la ) } \, h (t_{a,p})  \;,
\label{appendix funct der fonction f_s definition}
\enq
The points $t_{a,p}$ correspond to the discretization points for $\Dp{}M$ associated with the Jordan curves $\ga_a$, as given in definition
\ref{Definition point discretisation}. It follows readily that the function $ \la \mapsto f_s\!\pa{\la \mid \{ \vsg_a\} }$ is holomorphic in $\la\in K$. Moreover, given any holomorphic function $\nu\pa{ \la,\bs{y} } \in \msc{O}\pa{ M \times W_y}$
where $W_y$ is a compact in $\Cx^{\ell_y}$, $\ell_y \in \mathbb{N}$, one has that
\beq
f_s\big( \la \mid \{ \nu( t_{a,p}, \bs{y} ) \} \big) \underset{ s \tend +\infty }{ \longrightarrow  }
\Int{\Dp{}M}{} \f{\nu\pa{\zeta,\bs{y}}+h\pa{\zeta}}{2i\pi\pa{\zeta-\la}} \dd \zeta = \nu\pa{ \la,\bs{y} } +h\pa{\la}
\qquad \e{uniformly} \; \e{in} \; \la \in K  \; \e{and} \;  \bs{y} \in W_y \;.
\label{appendix funct der convergence uniform approximant fs}
\enq
This convergence holds since $\pa{\zeta, \la,\bs{y} }  \mapsto \tf{\nu\pa{\zeta,\bs{y}}+h\pa{\zeta}}{\pa{\zeta-\la}}$ is uniformly continuous on
$\Dp{}M\times K\times W_y$.

We recall that, given a holomorphic function $h$ on $M$ (and hence also on some open neighborhood of $M$), and $S$  a subset of $M$,
we denote $\norm{h}_S=\sup_{s\in S}\abs{h\pa{s}}$.

\subsection{Pure translations}
\label{subsection Pure translations}

We are now in position to establish a representation for translation operators for functionals acting on holomorphic functions.

\begin{prop}
\label{Proposition translation pure avec dependence parametre auxiliaire}
Let $\mc{F}\pac{\cdot}\pa{\bs{z}}$, $\bs{z} \in W_z \subset \Cx^{\ell_z}$ be a regular functional in respect to
the pair $\pa{M, K}$ and let the functions $f_s$, $\nu$ and $h$  as well as the compacts $M$ and $K$ be defined as above.
Then, for any $\pa{m,k_1,\dots,k_{\ell_y}} \in \mathbb{N}^{\ell_y+1}$
\beq
\lim_{s\tend + \infty} \pl{j=1}{\ell_y} \f{ \Dp{}^{k_j} }{ \Dp{} y_{j}^{k_j}} \cdot
\pl{p=1}{s} \pl{a=1}{n+1} \ex{\nu\pa{t_{a,p},\bs{y}} \Dp{\vsg_{a,p} } } \cdot \left.  \f{ \Dp{}^m }{\Dp{}\ga^m}
\mc{F}\pac{ \ga f_s\big( * \mid \{ \vsg_{a,p} \} \big) }\pa{\bs{z}} \right|_{\vsg_{a,p}=0}=
\pl{j=1}{\ell_y} \f{ \Dp{}^{k_j} }{ \Dp{} y_{j}^{k_j}} \f{ \Dp{}^m }{\Dp{}\ga^m}
\mc{F}\pac{ \ga \nu\pa{*,\bs{y}} + \ga h\pa{*} } \pa{\bs{z}}\;.
\nonumber
\enq
Above, the $\cdot$ inside of the argument of indicates the running variable on which the functional $\mc{F}\pac{\cdot }\pa{\bs{y}}$ acts.
This convergence holds uniformly in $\pa{\ga, \bs{y},\bs{z}}$ belonging to compact subsets of 
$\mc{D}_{0,\ga_0}\times \e{Int}\, \big( W_y \big)\times \e{Int}\!\pa{W_z}$, where
\beq
3 \ga_0 =  \f{C_{\mc{F}}}{ 2\norm{\nu}_{M\times W_y} + \norm{h}_M } \;   \f{ \pi  \e{d}\!\pa{\Dp{}M,K} }{ \abs{\Dp{} M}   +  2\pi \, \e{d}\!\pa{\Dp{}M,K}  } \;,
\label{appendix funct der definition ga0 pour translation pures}
\enq
$\abs{\Dp{}M}$  stands for the length of $\Dp{}M$, $\e{d}\!\pa{\Dp{}M,K}$ for the distance of $K$ to $\Dp{}M$ and
$C_{\mc{F}}>0$ is the constant of regularity of $\mc{F}$.
Finally, $\mc{D}_{0,\ga_0} = \paa{ z \in \Cx \; : \; \abs{z}\leq \ga_0}$.

\end{prop}

\Proof
We first consider the case $m=0$ and $k_1=\dots = k_{\ell}=0$. We assume that $s$
is taken large enough so that
\beq
\sul{p=1}{s} \sul{a=1}{n+1} \abs{t_{a,p} - t_{a,p+1}} \leq 2  \abs{\Dp{}M}  \;.
\enq
Then, for $\abs{\ga} < 2 \ga_0$ and $\abs{\vsg_{a,p} } \leq 2\norm{\nu}_{M\times W_y} $, one has
\beq
\abs{ \ga f_s\pa{\la\mid \{ \vsg_{a,p} \} }} \leq  \abs{\ga} \bigg( \sup_{a,p}\abs{ \vsg_{a, p} }  + \norm{h}_M  \bigg)
\times \sul{p=1}{s}\sul{a=1}{n+1}  \f{\abs{t_{a,p+1}-t_{a,p} } }{2\pi \abs{\la-t_{a,p} } }
\leq    \abs{\ga}  \pa{ \sup_{a,p}\abs{ \vsg_{a, p} }  + \norm{h}_M }  \f{ 2 \abs{\Dp{}M}   }{2\pi\,  \e{d}\!\pa{\Dp{}M,K}}  < 
\f{2}{3} C_{\mc{F}} \;,
\label{appendix funct der bornage fN}
\enq
%
%
%
Hence,
\beq
\pa{\ga, \{ \vsg_{a,p} \},\bs{z}}  \mapsto \mc{F}\pac{ \ga f_s\big( * \mid \{ \vsg_{a,p} \} \big) }\pa{\bs{z}} \quad \e{is}\; \e{holomorphic} \;  \e{in}
\qquad \pa{\ga, \{ \vsg_{a,p} \},\bs{z}} \in \mc{D}_{0,2\ga_0} \times \mc{D}_{0,2\norm{\nu}_M}^{s\pa{n+1}} \times W_z \; ,
\enq
this for any $s$ large enough. As a consequence, the below multi-dimensional Taylor series is convergent
uniformly in $\pa{\ga,\bs{y},\bs{z} } \in \mc{D}_{0,2\ga_0} \times W_y\times W_z$ and
\bem
\hspace{-5mm}
\pl{p=1}{s}\pl{a=1}{n+1} \ex{\nu\pa{t_{a,p},\bs{y} } \Dp{\vsg_{a,p}} } \;
\left.  \mc{F}\pac{ \ga f_s\Big( * \mid \{ \vsg_{a,p} \} \Big)  }\pa{\bs{z}} \right|_{\vsg_{a,p}=0}  \\
\equiv 
\sul{n_{a,p} \geq 0 }{+\infty} \pl{p=1}{s}\pl{a=1}{n+1}  \Bigg\{ \f{ \big[ \nu\big( t_{a,p},\bs{y} \big)  \big]^{n_{a,p}}  }{ (n_{a,p}) !}
 \f{\Dp{}^{n_{a,p}}}{\Dp{}\vsg_{a,p}^{n_{a,p}}} \Bigg\} \cdot
\left.  \mc{F}\pac{ \ga f_s\Big( * \mid \{ \vsg_{a,p} \}  \Big) } \pa{\bs{z}} \right|_{\vsg_{a,p}=0} 
= \mc{F}\pac{ \ga f_s\big( * \mid \{ \nu\big( t_{a,p},\bs{y} \big) \} \big)  }\pa{\bs{z}}  \;.
\label{appendix funct der serie Taylor simple a resommer}
\end{multline}
Moreover, for any $\bs{y}\in W_y$ and $\ga \in \ov{\mc{D}}_{0,2\ga_0}$,
one has the bound $\norm{\ga f_s\big( * \mid \{ \nu\big( t_{a,p},\bs{y} \big) \} \big)  }_K 
+ \abs{\ga} \pa{\norm{\nu\pa{\cdot,\bs{y}} }_K + \norm{h}_K}<C_{\mc{F}}$.
As a consequence, by \eqref{Appendice transl fonct Conte fnelle}
\bem
\norm{ \mc{F}\big[ \ga f_s\big( * \mid \{ \nu\big( t_{a,p},\bs{y} \big) \} \big) \big]\pa{\bs{z}} -
\mc{F}\pac{\ga \nu\pa{*,\bs{y} } + \ga h\pa{*} }\pa{\bs{z}} }_{ \ov{\mc{D}}_{0,2\ga_0}\times W_y\times W_z} \\
\leq \ga_0 C^{\prime} \norm{ f_s\big(\la \mid \{ \nu(t_{a,p},\bs{y} ) \} \big) - \nu\pa{\la,\bs{y} }-h\pa{\la} }_{K\times W_y} \limit{s}{+\infty} 0 \;,
\nonumber
\end{multline}
due to \eqref{appendix funct der convergence uniform approximant fs}. The norm in the first line is computed in respect to
$\pa{\ga,\bs{y},\bs{z}}\in \ov{\mc{D}}_{0,2\ga_0}\times W_y\times W_z$. The one in the second line in respect to
$\pa{\la,\bs{y}}\in K\times W_y$. We insisted explicitly on the variable-dependence of the functions so as to make this fact clear.

It remains to show that the convergence also holds uniformly on all compacts of 
$\mc{D}_{0,2\ga_0}\times \e{Int}\,(W_y)\times \e{Int}\pa{W_z} $
when considering partial derivatives in respect to $\ga, y_1,\dots,y_{\ell_y}$
of finite total order.

One can exchange any such partial derivatives with the Taylor series in \eqref{appendix funct der serie Taylor simple a resommer}
in as much as its partial sums define a sequence of holomorphic functions that is
uniformly convergent on $\ov{\mc{D}}_{0,2\ga_0}\times W_y\times W_z$. The same arguments can be applied to the sequence of
holomorphic functions $\mc{F}\big[ \ga f_s\big( * \mid \{ \nu\big( t_{a,p},\bs{y} \big) \} \big) \big]\pa{\bs{z}}$ . \qed

\begin{cor}
\label{Corollaire translation dans le cas integral et derivee}
Assume that the conditions and notations of proposition \ref{Proposition translation pure avec dependence parametre auxiliaire} hold.
Let $\msc{C}^{\pa{\ell_y}}=\msc{C}_1\times \dots \times \msc{C}_{\ell_y}$ and $\wt{\msc{C}}^{\pa{\ell_z}}=\wt{\msc{C}}_1\times \dots \times \wt{\msc{C}}_{\ell_z}$ be Cartesian products of compact curves in $\Cx$ such that
 $\msc{C}^{\pa{\ell_y}} \subset \e{Int}\,(W_y)$  and  $\wt{\msc{C}}^{\pa{\ell_z}} \subset \e{Int}\pa{W_z}$.
Then one has
\bem
\hspace{-7mm} \lim_{s\tend + \infty}  \sul{ n_{a,p}=0 }{+\infty}
\pl{p=1}{s} \pl{a=1}{n+1}  \paa{ \f{1}{ (n_{a,p}) !} \f{ \Dp{}^{n_{a,p}} }{ \Dp{} \vsg_{a,p}^{n_{a,p}}  }  }  \cdot
\Int{\msc{C}^{ \pa{\ell_y}} }{ } \hspace{-2mm}  \dd^{\ell_y} \bs{y} \hspace{-2mm}
\Int{ \wt{\msc{C}}^{ \pa{\ell_z}} }{ } \hspace{-2mm}  \dd^{\ell_z} \bs{z}
\pl{j=1}{\ell_y} \f{ \Dp{}^{k_j} }{ \Dp{}y_{j}^{k_j}} \cdot
\pl{p=1}{s} \pl{a=1}{n+1} \pac{\nu\pa{ t_{a,p}, \bs{y} }}^{n_{a,p}}
\left.  \f{ \Dp{}^m }{\Dp{}\ga^m}
\mc{F}\big[ \ga f_s\big(* \mid \{ \vsg_{a,p} \} \big) \big]\pa{\bs{y},\bs{z}} \right|_{\vsg_{a,p}=0} \hspace{-3mm}   \\
= \Int{\msc{C}^{ \pa{\ell_y}} }{ } \hspace{-2mm}  \dd^{\ell_y} \bs{y} \hspace{-2mm}
\Int{ \wt{\msc{C}}^{ \pa{\ell_z}} }{ } \hspace{-2mm}  \dd^{\ell_z} \bs{z}
\pl{j=1}{\ell_y} \f{ \Dp{}^{k_j} }{ \Dp{}y_{j}^{k_j}} \f{ \Dp{}^m }{\Dp{}\ga^m}
\mc{F}\pac{ \ga \nu\pa{*,\bs{y}} + \ga h\pa{*} } \pa{\bs{y},\bs{z}}  \;.
\end{multline}
this uniformly in $\ga$ belonging to compact subsets of $\mc{D}_{0,\ga_0}$.

\end{cor}

Note that if $\mc{F}$ depends on a third set of variables belonging to a compact, the results hold as well in respect to this third set
uniformly on the compact.

\Proof

Proposition \ref{Proposition translation pure avec dependence parametre auxiliaire} allows
one to conclude, in virtue of the uniform convergence of the sequences, that for $\ga$ belonging to compact subsets
of $\mc{D}_{0,\ga_0}$ one has the equality
\bem
\hspace{-7mm}\lim_{s\tend + \infty}  \sul{ n_{a,p}=0 }{+\infty}
\Int{\msc{C}^{ \pa{\ell_y}} }{ } \hspace{-2mm}  \dd^{\ell_y} \bs{y} \hspace{-2mm}
\Int{ \wt{\msc{C}}^{ \pa{\ell_z}} }{ } \hspace{-2mm}  \dd^{\ell_z} \bs{z}
\pl{p=1}{s} \pl{a=1}{n+1}  \paa{ \f{1}{n_{a,p}!} \f{ \Dp{}^{n_{a,p}} }{ \Dp{} \vsg_{a,p}^{n_{a,p}}  }  } \cdot
\pl{j=1}{\ell_y} \f{ \Dp{}^{k_j} }{ \Dp{}y_{j}^{k_j}} \cdot
\pl{p=1}{s} \pl{a=1}{n+1} \pac{\nu\pa{ t_{a,p}, \bs{y} }}^{n_{a,p}}
\left.  \f{ \Dp{}^m }{\Dp{}\ga^m}
\mc{F}\big[ \ga f_s\big(* \mid \{ \vsg_{a,p} \} \big) \big] \pa{\bs{y},\bs{z}} \right|_{\vsg_{a,p}=0} \hspace{-3mm} \\
=
\Int{\msc{C}^{ \pa{\ell_y}} }{ } \hspace{-2mm}  \dd^{\ell_y} \bs{y} \hspace{-2mm}
\Int{ \wt{\msc{C}}^{ \pa{\ell_z}} }{ } \hspace{-2mm}  \dd^{\ell_z} \bs{z}
\pl{j=1}{\ell_y} \f{ \Dp{}^{k_j} }{ \Dp{}y_{j}^{k_j}} \f{ \Dp{}^m }{\Dp{}\ga^m}
\mc{F}\pac{ \ga \nu\pa{*,\bs{y}} + \ga h\pa{*} } \pa{\bs{y},\bs{z}} \;.
\label{appendix func der permutation derivee somme integrale transl pure}
\end{multline}
The integrals occurring in the first line of \eqref{appendix func der permutation derivee somme integrale transl pure} are over compact curves
and the integrand is smooth in respect to the integration variables $\pa{\bs{y},\bs{z}}$ and the auxiliary parameters $\vsg_{a,p}$.
As a consequence, the partial $\vsg_{a,p}$-derivatives can be pulled outside of the integration symbols. \qed




\subsection{Weighted translation}
\label{subsection weighted translations}

One can generalize the notion of functional translation with the help of multi-dimensional Lagrange series and consider more
complex objects. For this purpose, we need to introduce some more definitions. Also, from now on we only focus on the
case of a compact $M$ without holes.

\vspace{3mm}
\noindent Let $\Ga\pac{\cdot}\pa{\mu}$ be a one parameter family of functionals such that:

\begin{itemize}
\item There exists a constant $C_{\Ga}>0$ such that if $\nu\pa{\la,\bs{y}}$ is holomorphic in $\pa{\la, \bs{y} } \in M \times W_y$,
with $W_y \subset \Cx^{\ell_y}$  and $\norm{\nu}_{K\times W_y} <C_{\Ga}$  then $\pa{\la,\bs{y}} \mapsto \Ga\pac{\nu\pa{*,\bs{y} }}\pa{\la}$
is holomorphic in $M\times W_y$.
\item There exists a contour $\msc{C}$ in $\e{Int}\pa{K}$ such that for $\norm{\rho}_K+\norm{\tau}_K \leq C_{\Ga}$ one has
\beq
\Ga\pac{\rho}\pa{\mu} - \Ga\pac{\tau}\pa{\mu}  = \Int{ \msc{C} }{}   \pa{\rho-\tau}\pa{\zeta}
\f{\de \Ga\pac{\nu}}{\de \nu\pa{\zeta}}\pa{\mu} _{\mid {\nu=\tau} }  \dd \zeta
 \;  + \;  \e{o}\pa{\norm{\rho-\tau}_K} \;.
\label{appendix funct der definition funct der}
\enq
$ \tf{ \de \Ga \pac{\nu} \pa{\mu}}{ \de \nu \pa{\zeta}}$ will be called the functional derivative of $\Ga$.
This functional derivative is such that, for any $\tau$ holomorphic on $M$ with $\norm{\tau}_K < C_{\Ga}$, there exists an open neighborhood
$\mc{V}\pa{\msc{C}}$ of the contour $\msc{C}$ appearing in \eqref{appendix funct der definition funct der} such that
\beq
\pa{\mu,\zeta} \mapsto \f{\de \Ga\pac{\nu}}{\de \nu\pa{\zeta}}\pa{\mu} \mid_{\nu=\tau} \quad \e{is} \; \e{holomorphic} \; \e{in}
\pa{\mu,\zeta} \in M\times \mc{V}\pa{\msc{C}} \; .
\enq

\item There exists a constant $ C^{\prime}_{\Ga} >0$ such that for $\norm{\tau}_K+\norm{\nu}_{K}\leq C_{\Ga}$ one has
\beq
\norm{ \Ga\pac{\nu}\pa{\mu} }_{M} \leq  C^{\prime}_{\Ga} \norm{\nu}_K
\qquad \e{and} \qquad
\norm{ \Ga\pac{\tau}\pa{\mu}-\Ga\pac{\nu}\pa{\mu} }_M \leq
C^{\prime}_{\Ga}   \norm{\nu-\tau}_K \; .
\label{appendix funct der proprietes fnelle Ga}
\enq

\end{itemize}

The properties of the functional $\Ga\pac{\cdot}\pa{\la}$ ensure the solvability of an associated integral equation

\begin{lemme}
\label{Lemme preuve unicite existance sols NLIE avec Gamma}
Let the compacts $M,K$  and the one parameter family of functional $\Ga\pac{\cdot} \pa{\la}$ be as defined above. 
Let $h \in \msc{O}\pa{M}$ and $r, \;\ga_0$  be such that  
\beq
2\ga_0 \pa{r + \norm{h}_M} \pa{1+  \f{2 \abs{\Dp{}M}  }{ 2\pi \e{d}\!\pa{\Dp{}M,K} } }  \leq  \f{r \, \e{min}\pa{1,C_{\Ga}} }{2\pa{r+C_{\Ga}^{\prime}}} \;,  \quad \e{and} \quad
2 C^{\prime}_{\Ga} \ga_0 \pa{1+  \f{ 2 \abs{\Dp{}M}  }{ 2 \pi \e{d}\!\pa{\Dp{}M,K} } } < \f{\e{min}\pa{1,C_{\Ga}}}{2} \; ,
\label{appendix func der definition gamma0 et r}
\enq
Then for $\abs{\ga} \leq \ga_0$, there exists a unique solution $\rho$ to the equation 
$\rho\pa{\la}=\Ga\pac{\ga \rho\!\pa{ * } + \ga h\!\pa{ * } }\!\pa{\la}$.
This solution is holomorphic in $\pa{\la,\ga} \in  M \times \ov{\mc{D}}_{0,\ga_0}$ and such that $\norm{\rho}_M <r$. 

\end{lemme}

\Proof

Suppose that $\rho$ and $\rho^{\prime}$ are two solutions.  Then for $\abs{\ga}<2 \ga_0$ one has,
by construction of $\ga_0$, that $\abs{\ga}\norm{\rho+h}_K + \abs{\ga}\norm{\rho^{\prime}+h}_K< C_{\Ga}$. As a consequence,
\beq
\norm{\rho-\rho^{\prime}}_M = \norm{ \Ga\pac{\ga\pa{\rho+h}} -\Ga\pac{\ga\pa{\rho^{\prime}+h}} }_M \leq 2 C^{\prime}_{\Ga} 
\ga_0 \norm{\rho-\rho^{\prime}}_M < \norm{\rho-\rho^{\prime}}_M \;.
\enq
Therefore,  $\rho=\rho^{\prime}$ on $M$, this uniformly in $\abs{\ga} \leq 2\ga_0$.

In order to prove the existence, one considers the sequence of holomorphic functions on $M$: $h_0=h$ and, for $n\geq 1$
$h_n\!\pa{\la}= h\!\pa{\la} +\Ga\pac{\ga h_{n-1}\pa{\cdot}}\!\pa{\la}$.
It is readily seen  by straightforward induction that, for all $n \in \mathbb{N}$ and $\abs{\ga} \leq \ga_0$,
\beq
\ga_0\norm{h_n}_K \leq \tf{C_{\Ga}}{2}  \qquad \e{and} \qquad \norm{h_{n+1}-h_{n}}_M \leq \tf{\norm{h_n-h_{n-1}}_M}{2} \; .
\enq
Hence $h_n$ is a Cauchy sequence in the space of holomorphic functions on
$\e{Int}\pa{M} \times \ov{\mc{D}}_{0;\ga_0}$. It is thus convergent to some holomorphic function $\wt{h}$ 
on $\e{Int}\pa{M} \times \ov{\mc{D}}_{0;\ga_0} $.
Since  $\wt{h}\pa{\la}=h\pa{\la} + \Ga\big[ \ga \wt{h}  \big]\pa{\la}$, it can  be analytically 
continued to a holomorphic function on $M \times \ov{\mc{D}}_{0;\ga_0} $.
Then, the function $\rho = \wt{h}-h$ solves $\rho\pa{\la}=\Ga\pac{\ga\pa{\rho+h}}\pa{\la}$.
It also follows that then $\norm{\rho}_M < r$
\qed

\begin{prop}
\label{Proposition construction derivee fnelle complexe}
Let $f_s$ be as in \eqref{appendix funct der fonction f_s definition}
and assume that the functional $\Ga\pac{\rho}\pa{\mu}$  satisfies to the assumptions given above.
Let $\mc{F}\!\pac{\cdot}\!\pa{\bs{z}}$, with $\bs{z}\in W_z \subset \Cx^{\ell_z}$,
be a regular functional in the sense of definition \ref{Definition Fonctionelle reguliere}. Set
\beq
\mc{L}_{\Ga}\pa{\ga,\bs{z} } =  \lim_{s\tend +\infty} \; \bs{:} \left.
 \pl{r=1}{s} \ex{\Ga\pac{ \ga f_s\pa{* \mid \{ \vsg_p \}}}\pa{t_r} \Dp{\vsg_r}}\;
\mc{F} \pac{\ga f_s\pa{ * \mid \{ \vsg_p \} }}\pa{\bs{z}} \right|_{\vsg_p=0} \hspace{-2mm}\bs{:} \; ,
\label{appendix funct der formule L cal Gamma cas generique}
\enq
where  $\bs{:} \cdot \bs{:}$ indicates that the expression is ordered in such a way that all the partial derivatives appear to the left
(cf. subsection \ref{Subsection Operator ordering for functional translation})  and $t_r$ are the discretization points of $\Dp{}M$.

\vspace{2mm}
Then, there exists $\ga_0>0$ such that $\mc{L}_{\Ga}\pa{\ga,\bs{z}}$ defines a holomorphic function of 
$\pa{\ga,\bs{z}} \in \ov{\mc{D}}_{0,\ga_0}\times \e{Int} \pa{W_z} $.

\noindent The convergence of the \textit{rhs} of \eqref{appendix funct der formule L cal Gamma cas generique} to $\mc{L}_{\Ga}\pa{\ga,\bs{z}}$
is uniform on compact subsets of $\ov{\mc{D}}_{0,\ga_0}\times \e{Int} \pa{W_z} $, and this in respect to
any partial $\ga$ or $\bs{z}$-derivative of finite order. $\mc{L}_{\Ga}\pa{\ga,\bs{z}}$ is given by
\beq
\mc{L}_{\Ga}\pa{\ga,\bs{z}}  =  \f{ \mc{F}\pac{ \ga \rho}\pa{\bs{z}} }
{  \det_{\msc{C}} \bigg[ I-\ga \f{ \de \Ga\pac{\nu}}{\de \nu\pa{\mu}}\pa{\la}  \bigg]_{\mid \nu=\ga \rho}  }
\quad \e{with} \; \rho  \; \e{being} \; \e{the} \; \e{unique} \; \e{solution} \; \e{to} \quad \rho\pa{\la}=h\pa{\la} + \Ga\pac{ \ga \rho}\pa{\la} \;.
\label{appendix funct der resultat somme Lagrange}
\enq
In the denominator appears the  Fredholm determinant of the \underline{linear}  integral operator acting  on the contour $\msc{C}$
with an integral kernel $\tf{ \de \Ga\pac{\nu}\pa{\la} }{\de \nu\pa{\mu}}_{\mid \nu=\ga \rho}$. The contour $\msc{C}$
is defined in  \eqref{appendix funct der definition funct der}.

\end{prop}

\Proof

We have, by definition, $\mc{L}_{\Ga}\pa{\ga,\bs{z}} =  \lim_{s\tend +\infty} \mc{L}_s\pa{\ga,\bs{z}}$ with
\beq
\mc{L}_s\pa{\ga,\bs{z}} =   \sul{ \substack{n_1,\dots , n_s\\ \in \mathbb{N}} }{}
 \pl{r=1}{s} \paa{ \f{1}{n_r!} \f{\Dp{}^{n_r}}{ \Dp{}\vsg_r^{n_r} } } \cdot
\left. \;  \pl{r=1}{s} \Ga^{n_r}\pac{ \ga f_s\pa{* \mid \{ \vsg_p \} }}\pa{t_r}  \cdot
\mc{F}^{} \pac{\ga f_s\pa{* \mid \{\vsg_p\}}}\pa{\bs{z}}    \;    \right|_{\vsg_{a}=0}   \hspace{-3mm}.
\label{appendic func der serie laplace mult dim ordre s}
\enq
The above series representation for $\mc{L}_s\pa{\ga,\bs{z}}$ corresponds to a particular case of a multidimensional Lagrange series.

We start by checking the convergence conditions. Let $C$ denote a common constant of regularity for the functionals $\mc{F}$ and $\Ga$, \textit{ie}
 for any $\nu\pa{\la,\bs{y}} \in \msc{O}\pa{M\times W_y}$ , with  $ W_y \subset \Cx^{\ell_y}$ such that $  \norm{\nu}_{M\times W_y} \leq C$
 one has
\beq
 \mc{F}\pac{\nu\pa{*,\bs{y}}}\pa{\bs{z}} \in \msc{O}\pa{W_y\times W_z} \quad \e{and} \quad
\Ga\pac{\nu\pa{*,\bs{y}}}\pa{\la} \in \msc{O}\pa{M\times W_y} \;.
\enq
Then let $r>0$ and $\ga_0>0$ be as given by \eqref{appendix func der definition gamma0 et r} but with $C_{\Ga}$ being replaced
with $C$. Let $s$ be large enough so that
$\sum_{a=1}^{s} \abs{t_a-t_{a+1}} \leq 2 \abs{\Dp{} M}$ and $\abs{\ga}\leq 2\ga_0$.
It then follows from \eqref{appendix funct der bornage fN} that $\norm{\ga f_s\pa{\cdot \mid \paa{\vsg_a} }}_K < C$
for $\abs{\vsg_a}<r$.
It is also easy to see that for $\abs{\vsg_p} \leq r $ and for any $t_p \in \Dp{}M$, one has
$\abs{\Ga\big[ \ga f_s \big(* \mid \{\vsg_p\} \big) \big] \pa{t_k}} \leq \tf{r}{2}$.
Therefore,
\begin{itemize}
\item $\Ga\,\big[ \ga f_s \big(* \mid \{ \vsg_p \} \big) \big]\pa{t_k}$, $k=1,\dots, s$ and $\mc{F}\,\big[ \ga f_s \big(* \mid \{\vsg_p\} \big) \big] \pa{ \bs{z} }$
are holomorphic functions of $\{\vsg_a\}$ in $\mc{D}_{0,r}^s$;
\item for $\abs{\vsg_k}= 3r/4$ with  $k \in \intn{1}{s}$ one has
$\abs{ \Ga\big[ \ga f_s \big(* \mid \{\vsg_p\} \big) \big] \pa{t_k} } \leq \tf{r}{2} < 3r/4$.
\end{itemize}
Hence, according to theorem  \ref{Theorem Multidimensional Lagrange series}, the multidimensional Lagrange series
is convergent and its sum is given by
\beq
\mc{L}_s\pa{\ga,\bs{z}}=   \f{  \mc{F} \pac{\ga f_s\pa{* \mid \{ \tau_p \}}}\pa{ \bs{z} }  }
{ \ddet{s}{ \de_{jk}-  \f{ \Dp{} }{\Dp{}\vsg_k}\Ga\,\big[ \ga f_s \big(* \mid \{\vsg_p\} \big) \big]\big( t_j \big)   }_{\mid \vsg_p=\tau_p}  }
\label{appendix funct der resultat series lagrange discrete}
\enq
where $\pa{\tau_1,\dots,\tau_s}$ is the unique solution to the system $\tau_j =\Ga\big[\ga f_s\big(* \mid \{ \tau_p \} \big) \big](t_j)$
with  $\abs{\tau_j}<r$ for all $j$.

It is easy to see that, in fact, $\mc{L}_s\pa{\ga,\bs{z}}$  is a uniform limit of holomorphic functions of
$\pa{\ga, \bs{z}} \in \mc{D}_{0,2\ga_0}\times W_z$. Therefore, $\mc{L}_s\pa{\ga,\bs{z}}$ is holomorphic on all compact
subsets of $\mc{D}_{0,2\ga_0}\times W_z$. Moreover, there one can permute any partial $\ga$ or $\bs{z}$-derivatives with the
summations in \eqref{appendic func der serie laplace mult dim ordre s}.
It is also clear for the previously obtained bounds that, $\mc{L}_s\pa{\ga,\bs{z}}$ is well defined for any $s$ large enough and this independently of the
choice of the points $t_k$.

\vspace{2mm}
\noindent We now show that its $s \tend +\infty$ limit exists and then we will compute it.
It is readily inferred from the integral representation
\beq
\tau_j = \Int{ \abs{\zeta_a} = r }{}  \;
\f{ \zeta_j }{  \pl{p=1}{s} \paa{ \zeta_p- \Ga\pac{\ga f_s\pa{* \mid \paa{\zeta_a}}} (t_p)  } } \dd^s \zeta
\enq
that $\tau_j\equiv \tau_j\pa{\ga}$, $j=1,\dots, s$, solving the system $\tau_j =\Ga\big[ \ga f_s\big(* \mid \{ \tau_p \} \big) \big](t_j)$,
 is a holomorphic function of $\ga$ for $\abs{\ga}\leq \ga_0$.
Hence, the function $\rho_s\pa{\la;\ga}= \Ga\big[\ga f_s\big(* \mid \{ \tau_p \}\big) \big]\pa{\la}$ is holomorphic in
$\pa{\la,\ga}  \in M \times \ov{\mc{D}}_{0,\ga_0}$. Also, by construction,
\beq
\rho_{s}\, ( t_j; \ga )=\tau_j\pa{\ga} \quad \e{and} \quad \norm{\rho}_{M\times \mc{D}_{0,\ga_0}} <r \; .
\enq
Now let $\rho$ be the unique solution to $\rho\pa{\la}=\Ga\pac{\ga \rho + \ga h }\pa{\la}$ with $\norm{\rho}_M\leq r$, 
as follows from lemma \ref{Lemme preuve unicite existance sols NLIE avec Gamma}.

Then, keeping the $\ga$ dependence implicit, we consider
\beq
\rho\pa{\la}-\rho_s\pa{\la} = \Ga\pac{\ga \pa{\rho+h} }\pa{\la} - \Ga\big[\ga f_s\big( * \mid \{\rho(t_p) \} \big) \big]\pa{\la} +
\underbrace{ \Ga\big[ \ga f_s\big(* \mid \{ \rho(t_p) \} \big) \big]\pa{\la}
		- \Ga\big[ \ga f_s\big( * \mid \{ \rho_s(t_p) \} \big) \big]\pa{\la} }_{\psi_s\pa{\la}} \;.
\label{appendix funct der difference rho rhoN}
\enq
As $\norm{ \rho_s }_M \leq r $, it  follows that
\beq
\norm{\psi_s}_M \leq   C_{\Ga}^{\prime} \ga_0 \norm{ f_s\big( * \mid \{ \rho(t_p) \} \big) - f_s\big(* \mid \{ \rho_s(t_p) \} \big) }_K
\leq \f{ 2 C_{\Ga}^{\prime} \ga_0  \abs{\Dp{} M} }{2\pi \e{d}\pa{\Dp{}M,K}}  \norm{\rho-\rho_s}_M < \f{1}{2} \norm{\rho-\rho_s}_M  \;.
\enq
Hence, $\norm{\rho-\rho_s-\psi_s}_M \geq \tf{\norm{\rho-\rho_s}_M}{2}$. On the other hand, it follows from \eqref{appendix funct der difference rho rhoN}
that
\beq
\norm{ \rho-\rho_s-\psi_s }_M =  \norm{ \Ga\pac{\ga \rho + \ga h }- \Ga\pac{ \ga f_s\big( * \mid\{ \rho(t_p) \} \big] }  }_M
\leq  \ga_0  C_{\Ga}^{\prime}  \norm{ \rho + h - f_s\big(* \mid\{\rho(t_p)\} \big) }_K
\underset{s \tend +\infty}{\longrightarrow} 0
\enq
Therefore $\rho_s$ converges uniformly to $\rho$ on $M$. Hence, in virtue of the
regularity of $\mc{F}$,
\beq
\mc{F}\big[ \ga f_s\big( * \mid \{\rho_s(t_p) \}\big) \big] \pa{\bs{z}}  \underset{s \tend +\infty}{\longrightarrow} \mc{F}\pac{\ga \rho}\pa{\bs{z}}
\quad
\e{uniformly}\;\e{in} \;  \pa{\ga,\bs{z}} \in \ov{\mc{D}}_{0,\ga_0} \times W_z \;.
\enq

It remains to compute the limit of the determinant. It follows from the functional derivative property
\eqref{appendix funct der definition funct der} that
\beq
 \f{ \Dp{} }{\Dp{}\vsg_k}\Ga\big[ \ga f_s\big(* \mid \{ \vsg_p \}\big) \big] (t_j) _{ \mid \{ \vsg_p \} = \{\tau_p\} } =
 \ga  \pa{t_{k+1}- t_k} \Int{ \msc{C} }{} \f{\dd \mu}{ 2i\pi } \f{1}{ t_k-\mu }
\left. \f{ \de \Ga }{ \de \nu\pa{\mu} }  \pac{ \nu+ \ga  f_s\big(* \mid \{ \tau_p\} \big) } (t_j) \right|_{\nu=0} \;.
\enq
By expanding the determinant appearing in \eqref{appendix funct der resultat series lagrange discrete}
into its discreet Fredholm series we get
\beq
\det_{s} \bigg[ \de_{jk}- \f{ \Dp{} }{\Dp{}\vsg_k}\Ga\big[ \ga f_s\big(* \mid \{ \vsg_p \}\big) \big] (t_j)   \bigg]_{ \mid \paa{\vsg_p}=\paa{\tau_p}} =
\sul{p=0}{s} \f{\pa{-\ga}^p}{p!} \Int{\msc{C}}{} \hspace{-1mm} \dd^p \! \mu  \; 
\ddet{p}{A_s\big( \mu_q,\mu_{\ell} \big) }
\enq
with
\bem
A_s\pa{\mu_q,\mu_{\ell}}=\sul{k=1}{s} \f{t_{k+1}-t_k}{2i\pi\pa{t_k-\mu_{\ell}}}
\left. \f{ \de  } { \de \nu(\mu_q) } \Ga\big[ \nu+ f_s\big( * \mid \{ \rho_s (t_p) \} \big) \big]\, (t_j) \right|_{\nu=0}
 \\
\underset{s \tend +\infty}{ \longrightarrow} \Int{\Dp{}M}{} \f{\dd \zeta}{2i\pi} \f{1}{\zeta-\mu_{\ell}}
\left. \f{ \de \Ga\pac{ \nu+ \ga \rho } }
{\de \nu(\mu_q) } \pa{\zeta}  \right|_{\nu=0}
= \left. \f{ \de \Ga\pac{ \nu+ \ga \rho}  } {\de \nu(\mu_q) } \pa{\mu_{\ell}} \right|_{\nu=0}
\end{multline}
The above convergence is uniform in $ ( \mu_{q}, \mu_{\ell} ) \in \msc{C} \times \msc{C}$.
Therefore, by elementary estimates, we obtain that the determinant of
interest does indeed converge to the Fredholm determinant given in \eqref{appendix funct der resultat somme Lagrange}, this
uniformly in $\abs{\ga}\leq \ga_0$.

Therefore, we obtain that $\mc{L}_s\pa{\ga,\bs{z}}$ is a sequence of holomorphic functions on 
$\ov{\mc{D}}_{0,\ga_0}\times \e{Int}\pa{W_z}$.
that converges uniformly. As a consequence,  $\mc{L}_{\Ga}\pa{\ga,\bs{z}}$ is holomorphic on every compact subset of 
$\ov{\mc{D}}_{0,\ga_0}\times \e{Int}\pa{W_z}$
and one can permute any partial-$\ga$ or $\bs{z}$ derivative of finite order with the $s \tend +\infty$ limit on these compacts.

\qed

\subsection{Examples}

We now treat two examples that are of direct interest for the resummation of the form factor series.
In the below examples, $\phi\pa{\la,\mu}$ refers to the dressed phase \eqref{definition eqn int Z et phi}.
We remind that it is holomorphic on $U_{\de}\times U_{\de}$.
In the following, the compacts $K$ and $M \subset U_{\de}$ are such that $\intff{-q}{q}$ is contained in their interior.
 We will also consider functions $h$ that are holomorphic
on $M$.

\subsubsection{$\Ga\pac{\rho}\pa{\mu}$ as a linear functional of $\rho$}
\label{appendix func der affine functional}

Let $\Ga\pac{\rho}\pa{\la}=   \int_{-q}^{q}   \Dp{\la}\phi\pa{\mu,\la} \rho\pa{\mu}\;  \dd \mu $. Then, given a regular functional
$\mc{F}\!\pac{\cdot}\!\pa{\bs{z}}$, $\bs{z}\in W_z \subset \Cx^{\ell_z}$, there exists $\ga_0>0$ such that
for $\pa{\ga , \bs{z}} \in  \ov{\mc{D}}_{0,\ga_0} \times W_z$
\beq
\mc{L}_{\Ga}\pa{\ga,\bs{z}} =  \f{ \mc{F}\pac{\rho}\pa{\bs{z}} }{ \ddet{\intff{-q}{q}}{ I- \ga \Dp{\la} \phi }  } \qquad \e{with}
\quad \rho\pa{\la}-\ga \Int{-q}{q} \Dp{\la}\phi\pa{\mu,\la} \rho\pa{\mu} \dd \mu  = h\pa{\la} \;.
\enq
The limit defining $\mc{L}_{\Ga}\pa{\ga,\bs{z}}$ as in \eqref{appendix funct der resultat somme Lagrange} is uniform
in respect to such parameters $\bs{z} \in W_z$ and  $\abs{\ga}\leq \ga_0$.

\Proof

In order to apply proposition \ref{Proposition construction derivee fnelle complexe}, one should check the assumptions on the functional $\Ga$.
It is readily seen that, independently of the norm of $\nu$ and $\rho$
\beq
\norm{\Ga\pac{\nu}}_M \leq  2q \norm{\Dp{\la}\phi}_{M\times M} \norm{\nu}_K  \quad \e{and} \quad
\norm{ \Ga\pac{\nu} - \Ga\pac{\rho} }_M \leq 2q \norm{\Dp{\la}\phi}_{M\times M} \norm{\nu-\rho}_K
\quad \e{and} \quad
\f{\de \Ga\pac{\tau}}{\de \tau\pa{\zeta}}\pa{\mu} = \pa{\Dp{\mu}\phi}\pa{\zeta,\mu} \;.
\nonumber
\enq
The validity of the holomorphicity conditions is readily checked by standard derivation under the integral theorems.
One is thus in position to apply proposition \ref{Proposition construction derivee fnelle complexe} and the claim follows. \qed

\subsubsection{Non-linear functional $\Ga^{\pa{L}}\pac{\rho}\pa{\mu}$}
\label{appendix funct der nonlinear funct Ga}

We now treat the case of the non-linear functional below
\beq
\Ga^{\pa{L}}\!\pac{\rho}\pa{\mu} =  \sul{j \in J}{} \phi\big( \mu,\mu_j \big)-\phi\big( \mu,\la_j \big)
=
\sul{j \in J}{}  \Oint{ \msc{C}_q }{}  \phi\pa{\mu,\om}
\paa{ \f{ \xi^{\prime}\pa{\om} }{ \xi\pa{\om}- \tf{j}{L}  }
					- \f{ \xi^{\prime}_{\rho}\pa{\om}  }{ \xi_{\rho}\pa{\om}- \tf{j}{L}  }      }
				\f{\dd \om }{ 2i\pi } 	\;.
\label{appendix funct der definition fnelle Gamma L}
\enq
There $J=\intn{1}{N}\setminus \paa{i_1,\dots, i_n}$, $\xi$ is given by \eqref{ecriture limite thermo fction comptage},
$\xi_F=\xi+\tf{F}{L}$ and $0\leq \tf{j}{L}\leq D$
with $\tf{N}{L} \tend D$. Finally, $\msc{C}_q$ is a small counterclockwise Jordan curve around $\intff{-q}{q}$ such that
$\e{Int}\pa{K}\supset\msc{C}_q$.
Note that $\mu_a$, resp. $\la_a$, appearing in \eqref{appendix funct der definition fnelle Gamma L} stand for the unique solutions
to  $\xi\pa{\mu_a}=\tf{a}{L}$, resp. to $\xi_{\rho}\pa{\la_a}=\tf{a}{L}$.

\begin{prop}
\label{Proposition serie Lagrange avec fonctionnelle Gamma L}
Let $\mc{F}\!\pac{\cdot}\!\pa{\bs{z}}$, $\bs{z}\in W_z \subset \Cx^{\ell_z}$ be a regular functional
and assume that $N,L$ are large (and such that $\tf{N}{L} \tend D$).
Then, there exists $\ga_0>0$ such that for $\pa{\ga , \bs{z}} \in  \ov{\mc{D}}_{0,\ga_0} \times W_z$
and $L$ large enough
\beq
\mc{L}_{\Ga^{\pa{L}} }\pa{\ga,\bs{z}} =  \f{ \mc{F}\big[ \ga \rho^{\pa{L}} \big] \!\pa{ \bs{z} } }
{ \det_{\msc{C}_q} \bigg[ I-\ga \f{\de \Ga^{\pa{L}}\!\pac{\nu}}{\de \nu\pa{\zeta}}\pa{\mu} \bigg]_{\nu=\ga \rho^{\pa{L}}} } \;\;,\qquad
%
%
%
\enq
where $\rho^{\pa{L}}$ is the unique solution to 
$\rho^{\pa{L}}\!\pa{\la} = h\!\pa{\la} + \Ga^{\pa{L}}\big[ \ga \rho^{\pa{L}} \big]\pa{\la}$.
This solution is such that 
\beq
 \rho^{\pa{L}}\pa{\la}=\rho\pa{\la} + \e{O}\big(L^{-1}\big) \;, 
\enq
 where $\rho$ solves the linear integral equation
$\pa{I+\f{\ga R}{2\pi}} \cdot \rho = h $ and the $\e{O}\big(L^{-1}\big)$ is a holomorphic function of $\ga$ and $\la$. 
Moreover, this estimate and holds uniformly 
in $\abs{\ga} \leq \ga_0$ and $\la \in U_{\de}$.
Finally,
\beq
\left. \f{\de \Ga^{\pa{L}} }{\de \nu\pa{\zeta}}\pac{\nu}\pa{\mu}  \right|_{\nu=\ga \tau} = \f{ \big( \Dp{\zeta} \phi \big) \!\pa{\mu,\zeta} }{2i\pi L}
\sul{j \in J}{} \f{1}{\xi_{\ga\tau}\pa{\zeta}-\tf{j}{L}} \; .
\enq
Above $I+\tf{R}{2\pi}$ stands for the resolvent of the Lieb kernel acting on $\intff{-q}{q}$. And one has
\beq
\ddet{\msc{C}_q}{I-\ga \f{\de \Ga^{\pa{L}}\pac{\nu}}{\de \nu\pa{\zeta}}\pa{\mu} }_{\nu=\ga \rho^{\pa{L}}} =
\ddet{\intff{-q}{q}}{I+\tf{\ga R}{2\pi}} \cdot \pa{1+\e{O} \paf{1}{L} } \;.
\enq
\end{prop}

\Proof

In order to apply proposition \ref{Proposition construction derivee fnelle complexe}, we ought to check that $\Ga^{\pa{L}}$
satisfies to all the necessary conditions. For this we observe that
\beq
\eps = \inf_{  }  \abs{\xi\pa{\om}-\la} >0
\quad \e{where}\; \e{the} \; \e{inf} \; \e{is} \; \e{taken}\; \e{for} \quad  \om  \in \msc{C}_q  \, , \;\;  \la \in \intff{0}{D} \;.
\label{appendix funct der definition cste epsilon}
\enq
Then, we choose a constant $C>0$ and consider $L$ large enough so that $C < \tf{\eps L}{2}$. It then follows that the functions 
 $\xi_{\nu}\pa{\om} - \tf{j}{L}$, for $j=1,\dots,N$,  have no zeroes on $\msc{C}_q$ and some immediate neighborhood thereof
provided that $\nu\pa{z,\bs{y}} \in \msc{O}\pa{M\times W_y}$ with $W_y \subset \Cx^{\ell_y}$ and $\norm{\nu}_{M\times W_y} <C$. 
It then follows by the derivation under the integral sign theorems
that $\Ga^{\pa{L}}\pac{\nu\pa{ * ,\bs{y}}}\pa{\mu}$ is holomorphic in $\pa{\mu,\bs{y}} \in M\times W_y$.

In order to establish bounds on $\Ga^{\pa{L}}\pac{ \nu\pa{*, \bs{y}} }\pa{\mu}$  for $\nu$ holomorphic and
such that $\norm{\nu}_{M\times W_y} < C$, it is convenient to represent
\beq
\f{ \xi^{\prime}\pa{\om} }{ \xi\pa{\om}- \tf{j}{L}  }
					- \f{ \xi^{\prime}_{ \nu }\pa{\om}  }{ \xi_{ \nu }\pa{\om}- \tf{j}{L}  }
= - \f{\nu^{\prime}\pa{\om}}{L \pa{ \xi\pa{\om}- \tf{j}{L} } }  \; + \;
 \Int{0}{ \nu\pa{\om} }  \f{\dd t }{ L } \; \f{ \xi^{\prime}_{\nu}\pa{\om} }{ \pa{ \xi_{ t}\pa{\om}- \tf{j}{L} }^2 }\;.
\enq
As $\msc{C}_q \subset \e{Int}\pa{K}$, there exists a constant $c_1>0$ such that  for any function $\nu$  holomorphic on $K$, one has
$\norm{\nu^{\prime}}_{ \msc{C}_q} \leq c_1 \norm{\nu}_K$. Also,
\beq
 \inf_{  }  \abs{\xi_{t}\pa{\om}-s} >\tf{\eps}{2}
\quad \e{where}\; \e{the} \; \e{inf} \; \e{is} \; \e{taken}\; \e{for} \quad  \om  \in \msc{C}_q  \, , \;\;  s \in \intff{0}{D}  \; \e{and} \;
\abs{t}< \tf{\eps L}{2} \;.
\enq
Hence, for any $\nu \in \msc{O}\pa{ M \times W_y }$ such that $\norm{\nu}_{K \times W_y}<C  < \tf{\eps L }{2}$
\bem
\abs{ \Ga^{\pa{L}}\pac{\nu\pa{ * , \bs{y} }}\pa{\mu} } \leq  \norm{\phi}_{M\times M}  \f{\abs{J}}{2\pi}  \abs{ \msc{C}_q }
\paa{ \f{c_1}{L\eps} \norm{\nu}_{K\times W_y}   + \norm{\xi^{\prime}_{\nu}}_{ \msc{C}_q \times W_y } \cdot   \f{  \norm{\nu}_{K\times W_y} }{ L \pa{\tf{\eps}{2}}^2  } } \\
\leq   \norm{\phi}_{M\times M}  \f{N \abs{ \msc{C}_q } }{2\pi L \eps} c_1
\paa{1+ \f{4}{\eps} \pac{ \norm{\xi}_K +\tf{\eps}{2} } }    \;\; \norm{\nu}_{K\times W_y} \;.
\end{multline}
This provides an estimate  for the constant $C_{\Ga^{\pa{L}}}^{\prime}$ entering in the bounds for $\norm{ \Ga^{\pa{L}}\pac{\nu} }_{M \times W_y}$.
Next one has
\beq
\Ga^{\pa{L}}\!\pac{\rho}\pa{\mu} - \Ga^{\pa{L}}\!\pac{\tau}\pa{\mu} =  \sul{j \in J}{} \Oint{ \msc{C}_q  }{}
\phi\pa{\mu,\om} \paa{    \f{  \pa{\tau^{\prime}-\rho^{\prime}}\pa{\om}  }{ \xi_{\tau}\pa{\om}-\tf{j}{L}  }
+ \f{ \pa{\rho-\tau}\pa{\om} \xi_{ \tau }^{\prime} \pa{\om}  }{ \pa{\xi_{\tau} \pa{\om}-\tf{j}{L}}^2  }  } \f{\dd \om }{ 2i\pi L }
\; + \; \mc{R}^{\pa{L}}\pa{\mu} \;.
%
%
%
%
%
\enq
Where,
\beq
\mc{R}^{\pa{L}}\pa{\mu} = \sul{j \in J}{} \Oint{ \msc{C}_q}{} \phi\pa{\mu,\om} \paa{
\f{ \pa{\rho-\tau}\pa{\om} \pa{\rho-\tau}^{\prime} \pa{\om}  }{ L^2 \pa{\xi_{\tau} \pa{\om}-\tf{j}{L}}^2  }
+ 2\Int{\tau\pa{\om}}{\rho\pa{\om}} \f{ \dd t }{L^2} \f{ \pa{t-\rho\pa{\om}} \xi_{ \rho}^{\prime} \pa{\om}  }{ \pa{\xi_{t} \pa{\om}-\tf{j}{L}}^3  }
		}   \f{\dd \om}{2i\pi}  = \e{O}\pa{\norm{\rho-\tau}_K^{2}} \; ,
\enq
this uniformly in $\mu \in M$ and $L$ large enough. Therefore,
\beq
\left. \f{\de \Ga^{\pa{L}} }{\de \nu\pa{\zeta}} \pac{\ga  \nu}\pa{\mu}  \right|_{\nu= \ga\rho} =
\f{\ga }{2i\pi L} \pa{\Dp{\zeta}\phi}\pa{\mu,\zeta} \sul{j \in J }{} \f{1 }{  \xi_{\ga \rho}\pa{\zeta}-\tf{j}{L} }\;.
\label{appendix funct der form explicite derivee fonct}
\enq
It follows that there exists a sufficiently small open neighborhood $\mc{V}\pa{\msc{C}_q}$ of $\msc{C}_q$ such that
the functional derivative is holomorphic in $\pa{\mu,\zeta} \in M\times \mc{V}\pa{\msc{C}_q}$.
Moreover, we get that there exists an $L$-independent constant $\wt{C}_2$ such that
$\norm{\Ga^{\pa{L}}\pac{\rho}-\Ga^{\pa{L}}\pac{\tau}}_{M} \leq \wt{C}_2 \norm{\rho-\tau}_{K}$

 \vspace{2mm}
We are now in position to apply proposition \ref{Proposition construction derivee fnelle complexe}. It follows that $\mc{L}_{\Ga^{\pa{L}}}$
can be expressed in terms of the unique solution $\rho^{\pa{L}}$ to 
$\rho^{\pa{L}}\!\pa{\mu} = h\pa{\mu} + \Ga^{\pa{L}}\big[\ga \rho^{\pa{L}}\big]\pa{\mu}$
with $\norm{\rho^{\pa{L}}}_M <r $ uniformly in $\abs{\ga} \leq \ga_0$.

This means that,
\beq
\left. \f{\de \Ga^{\pa{L}}\! \pac{\ga \nu} }{\de \nu\pa{\zeta}} \pa{\mu}   \right|_{\nu=\rho^{\pa{L}}}  \underset{ L\tend +\infty }{\longrightarrow }
\ga  \pa{\Dp{\zeta}\phi}\pa{\mu,\zeta}  \Int{-q}{q} \f{  \dd u }{2i\pi}  \f{ \xi^{\prime}\pa{u} }{ \xi\pa{\zeta}-\xi\pa{u} }
\qquad \e{uniformly} \; \e{in}  \quad \pa{\mu,\zeta} \in \msc{C}_q^2 \;.
\label{appendix funct der convergence derivee fonctionnelle}
\enq
In this limit,  the contour integral $\msc{C}_q$ in the Fredholm determinant can be computed and
since the Fredholm determinant of a trace class operator is continuous in respect to the trace class norm (which is bounded by the
sup norm in the case of integral operators acting on compact contours)
\beq
\ddet{\msc{C}_q}{ I- \tf{\de \Ga^{\pa{L}}\!\pac{\ga \nu}\pa{\mu} }{\de \nu\pa{\zeta}} \mid_{\nu=\rho^{\pa{L}}}  }
\underset{ L\tend +\infty }{\longrightarrow }
\ddet{\intff{-q}{q}}{ I-\ga \Dp{\la}\phi\pa{\mu,\la}} = \ddet{\intff{-q}{q}}{I+\tf{\ga R}{2\pi}} \;.
\label{appendix funct der convergence determinant derivee fonctionnelle}
\enq
Where $R$ is the resolvent of the Lieb kernel.

We now characterize the leading behavior of the solution $\rho^{\pa{L}}$ when $N,L \tend +\infty$.
By repeating the type of manipulations carried our previously, and using that $\rho^{\pa{L}}$ is bounded
on $K$ uniformly in $L$, we get that the non-linear integral equation for $\rho^{\pa{L}}$ takes the form
\beq
\rho^{\pa{L}}\!\pa{\mu} = h \pa{\mu} \; +\;
\f{\ga}{L} \sul{j\in J}{} \Oint{\msc{C}_q}{} \f{\dd \om}{2i\pi} \rho^{\pa{L}}\!\pa{\om}
	\f{ \pa{\Dp{\om}\phi}\pa{\mu,\om}}{ \xi\pa{\om}-\tf{j}{L}  }
\; \; +	 \; \e{O}\pa{ \f{1}{L}}  \,.
\nonumber
\enq
There the $\e{O}$ is uniform in $\mu \in U_{\de}$. The Riemann sum can be estimated by using the Euler-McLaurin formula
and the uniform boundedness of $\rho^{\pa{L}}$ on $K$. After carrying out the resulting contour integral over $\msc{C}_q$ we obtain
\beq
\rho^{\pa{L}}\!\pa{\mu} = h\pa{\mu} \; -\;
\ga \Int{-q}{q} \f{\dd s}{2\pi} R\pa{\mu,s} \rho^{\pa{L}}\!\pa{s} + \e{O}\pa{L^{-1}} \;.
\label{appendix funct der integral req Asympt rho}
\enq
The $\e{O}$ appearing in \eqref{appendix funct der integral req Asympt rho} is holomorphic in $\mu \in U_{\de}$.
Indeed, $\rho^{\pa{L}}$ just as all the other terms in \eqref{appendix funct der integral req Asympt rho}
are holomorphic on $U_{\de}$.
This proves that $\rho^{\pa{L}}$ admits an asymptotic expansion in $L$ such that $\rho^{\pa{L}}\!\pa{\om}=\rho\pa{\om} + \e{O}\pa{L^{-1}}$, where
$\rho$ is the solution to the integral equation  $\pa{I+\ga \tf{R}{2\pi} } \cdot \rho = h$ .
As $\rho^{\pa{L}}$ and $\rho$ are both holomorphic in $U_{\de}$, so is the remainder. Moreover, one can convince oneself that
this $\e{O}$ is uniform in $\mu \in U_{\de}$. Therefore, the regularity  of the functional $\mc{F}\pac{\cdot}\pa{\bs{z}}$
 \eqref{Appendice transl fonct Conte fnelle} leads to $\mc{F}\big[\ga\rho^{\pa{L}}\big]\pa{\bs{z}} = \mc{F}\pac{\ga \rho}\pa{\bs{z}} + \e{O}\pa{L^{-1}}$. \qed


\end{document}